\documentclass{book}

\usepackage{afterpage}

\usepackage[disable,textsize=tiny,prependcap
tion,backgroundcolor=red!20]{todonotes}
\reversemarginpar

\usepackage[b5paper,margin=2cm,inner=2cm,top=3cm,bottom=3cm,footskip=1.25cm]{geometry}
\usepackage[acronym,toc,nonumberlist,ucmark=true,symbols,nogroupskip,automake]{glossaries-extra}
\setglossarystyle{altlist}

\usepackage[nottoc,numbib]{tocbibind}
\usepackage{amsfonts}
\usepackage{bm}
\usepackage{enumitem}
\usepackage[labelfont=bf,textfont=small]{caption}
\usepackage[labelfont=bf,textfont=small]{subcaption}
\usepackage[percent]{overpic}
\usepackage{booktabs}

\usepackage{algpseudocode,algorithm,algorithmicx}
\usepackage{mathtools}
\usepackage{siunitx}
\usepackage{amssymb}
\usepackage{amsthm}
\usepackage{centernot}
\usepackage{ragged2e}

\usepackage{import}

\usepackage{tikz}
\usetikzlibrary{positioning,shapes.misc}

\usepackage{fancyhdr}
\newcolumntype{L}[1]{>{\raggedright\arraybackslash\hspace{0pt}}m{#1}}
\newcolumntype{C}[1]{>{\centering\arraybackslash\hspace{0pt}}m{#1}}
\newcolumntype{R}[1]{>{\raggedleft\arraybackslash\hspace{0pt}}m{#1}}

\usepackage{amsmath}
\usepackage{mathtools} 
\usepackage{amsfonts} 
\usepackage{amsthm}
\usepackage{amssymb}
\usepackage{amsfonts}
\usepackage{mathrsfs}
\usepackage{bbm}
\usepackage{graphicx}
\usepackage{caption}
\usepackage{bookmark,hyperref}
\usepackage{12many}
\usepackage{setspace}
\usepackage{multirow}
\usepackage{arydshln}
\usepackage{braket}
\usepackage{dirtytalk}
\usepackage{MnSymbol}

\theoremstyle{definition}

\newcommand{\epsL}{\varepsilon_{1}}
\newcommand{\epsR}{\varepsilon_{2}}
\newcommand{\epsLR}{\varepsilon_{i}}

\newcommand{\spinfourfour}{\text{Spin$(4,4)$}}
\newcommand{\spintwotwo}{\text{Spin$(2,2)$}}
\newcommand{\sofourfour}{\text{SO$(4,4)$}}
\newcommand{\sotwotwo}{\text{SO$(2,2)$}}

\newcommand{\sofour}{\text{SO$(4)$}}

\newcommand{\spinfour}{\text{Spin$(4)$}}
\newcommand{\sotwo}{\text{SO$(2)$}}
\newcommand{\sutwo}{\text{SU$(2)$}}
\newcommand{\sltwo}{\text{SL$(2)$}}

\newcommand{\psymbol}[2]{\genfrac{}{}{0pt}{1}{#1}{#2}}
\makeglossaries

\tikzset{northround/.style={append after command={
   \pgfextra
        \draw[sharp corners, fill=#1, draw=none, very thick]%
    (\tikzlastnode.west)%
    [rounded corners=5pt] |- (\tikzlastnode.north)%
    [rounded corners=5pt] -| (\tikzlastnode.east)%
    [rounded corners=0pt] |- (\tikzlastnode.south)%
    [rounded corners=0pt] -| (\tikzlastnode.west);
   \endpgfextra}}}

\definecolor{myturq}{RGB}{79,205,207}
\tikzstyle{mybox} = [draw=myturq, fill=blue!0, very thick,
    inner sep=10pt, inner ysep=10pt]
\tikzstyle{fancytitle} =[northround=myturq, text width=.94\textwidth, inner xsep=10.42pt]

\begin{document}


\frontmatter
\thispagestyle{empty}

\onehalfspacing

\begin{center}
    
\null\vspace{45mm}

{\huge\textbf{Strings on freely acting orbifolds}}

\vspace{5mm}

{\LARGE
Spectra, moduli spaces and branes }

\vspace{25mm}

\end{center}

\normalsize

\clearpage\newpage

\thispagestyle{empty}

\null
\vfill

\begin{justify}
PhD thesis, Utrecht University, December 2025

\vspace{10mm}

\noindent ISBN: 978-94-93483-43-9

\vspace{10mm}

\noindent DOI: 10.33540/3264

\vspace{10mm}

\noindent Cover design by: Wouter Fris

\vspace{3mm}

\noindent About the cover:  The illustration depicts the flat two-dimensional \textit{pillowcase} orbifold (right), defined as the global quotient of the two-torus (left) by the reflection involution on its canonical coordinates, together with three different types of strings propagating on it. There are four points on the torus fixed under the orbifold action, and these correspond to the four corners of the pillowcase orbifold. The red string is a closed string and the blue string is a winding mode. Both strings are closed on the torus and on the orbifold, and are in the untwisted orbifold sector. The green string is closed only on the orbifold; it is a twisted string centered around a fixed point. \end{justify}

\clearpage\newpage

\thispagestyle{empty}

\onehalfspacing

\begin{center}


{\huge
\textbf{Strings on freely acting orbifolds}}

\vspace{5mm}
{\LARGE
Spectra, moduli spaces and branes }

\vspace{15mm}

{\huge {Snaren op vrij werkende orbifolds}}

\vspace{4mm}
{\Large
Spectra, moduliruimten en branes }

\vspace{3mm}
{\normalsize (met een samenvatting in het Nederlands)}

\vspace{20mm}

{\Large
Proefschrift}

\vspace{10mm}

{\large
\begin{justify}
ter verkrijging van de graad van doctor aan de 
Universiteit Utrecht
op gezag van de
rector magnificus, prof. dr. ir. W. Hazeleger,
 ingevolge het besluit van het College voor Promoties 
in het openbaar te verdedigen op maandag 26 januari 2026 des middags te 12.15 uur
\end{justify}
}

\vspace{10mm}

{\large door} 
\vspace{8mm}

{\Large Georgios Gkountoumis}

\vspace{8mm}

{\large geboren op  22 Januari 1993}

\vspace{1mm}

{\large te Amarousio, Attica, Griekenland}
\end{center}

\newpage

\thispagestyle{empty}

\noindent\textbf{Promotor:}

\noindent Prof. dr. S.J.G. Vandoren

\vspace{3mm}
\noindent\textbf{Copromotor:}

\noindent dr. E. Plauschinn

\vspace{3mm}
\noindent\textbf{Beoordelingscommissie:}

\noindent Prof. dr. G.R. Cavalcanti

\noindent Prof. dr. T.W. Grimm

\noindent Prof. dr. E. Kiritsis

\noindent Prof. dr. M. Trigiante

\noindent Dr. I. Valenzuela

\vfill


\vspace{5mm}


\vspace{5mm}


\clearpage 

\thispagestyle{empty}
\null 
\vspace{10mm}
\begin{center}


    $ \Sigma \tau o \upsilon \varsigma $  $\gamma o \nu \varepsilon \acute{\iota} \varsigma $  $\mu o \upsilon,$ 
    
  $\gamma \iota \alpha$ $\acute{o}\lambda\alpha$ $\acute{o}\sigma\alpha$ $\acute{\varepsilon}\kappa\alpha\nu\alpha\nu$  $\gamma \iota \alpha$ $\mu\acute{\varepsilon}\nu\alpha$
    
    \vspace{2mm}
    
    $\kappa \alpha \iota$
    
   \vspace{2mm}
    
    $\sigma\tau\eta$ $\mathrm{X}\rho\upsilon\sigma\acute{\alpha}\nu\vartheta\eta,$ 
    
    $\pi o \upsilon$ $\acute{\eta}\tau\alpha\nu$ $\delta\acute{\iota}\pi\lambda\alpha$ $\mu o \upsilon$ $\acute{o}\lambda\alpha$ $\alpha\upsilon\tau\acute{\alpha}$ $\tau \alpha$ $\chi\rho\acute{o}\nu\iota\alpha$
\end{center}
\clearpage
\newpage
\thispagestyle{empty}

\null

\newpage
\thispagestyle{empty}
\newpage

\thispagestyle{plain}
\pagenumbering{roman}
\setcounter{page}{1}

\onehalfspacing

\noindent{\Large {\textbf{List of publications}}}

\vspace{5mm}

\noindent Chapters \ref{chap:fao} and \ref{chap:spectrum} of this thesis are based on:

\vspace{3mm}

\noindent \cite{Gkountoumis:2023fym} G. Gkountoumis, C. Hull, K. Stemerdink and S. Vandoren: \textit{Freely acting orbifolds
of type IIB string theory on $T^5$}
, \textit{JHEP} \textbf{08} (2023) 089, [arXiv: 2302.09112],

\vspace{2mm}

\noindent \cite{Gkountoumis:2024boj} G. Gkountoumis: \textit{Asymmetric $\mathbb{Z}_4$ orbifolds of type IIB string theory revisited}, \textit{JHEP} \textbf{11}
(2024) 136, [arXiv: 2404.12962].

\vspace{5mm}

\noindent  Chapter \ref{chap:mod5} of this thesis is based on:

\vspace{3mm}

\noindent \cite{Gkountoumis:2023fym} G. Gkountoumis, C. Hull, K. Stemerdink and S. Vandoren: \textit{Freely acting orbifolds
of type IIB string theory on $T^5$}
, \textit{JHEP} \textbf{08} (2023) 089, [arXiv: 2302.09112],

\vspace{2mm}

\noindent \cite{Gkountoumis:2024dwc} G. Gkountoumis, C. Hull and S. Vandoren: \textit{Exact moduli spaces for $\mathcal{N}$ = 2, D = 5 freely
acting orbifolds}, \textit{JHEP} \textbf{07} (2024) 126, [arXiv: 2403.05650].

\vspace{5mm}

\noindent Chapter \ref{chap:swampland} of this thesis is based on:

\vspace{2mm}

\noindent \cite{Gkountoumis:2025btc} G. Gkountoumis, C. Hull, G. Nian and S. Vandoren: \textit{Duality and Infinite
Distance Limits in Asymmetric Freely Acting Orbifolds}, \textit{JHEP} \textbf{09} (2025) 198, [arXiv:2506.11699].

\vspace{5mm}

\noindent Chapter \ref{chap:branes} of this thesis is based on: 

\vspace{2mm}

\noindent \cite{GHSV2} G. Gkountoumis, C. Hull, K. Stemerdink and S. Vandoren:  \textit{D-branes on freely acting orbifolds}, In
preparation.

\clearpage 
\thispagestyle{plain}

\noindent Paper \cite{Gkountoumis:2023fym} and parts of chapter \ref{chap:branes} have also appeared in the thesis of K. Stemerdink. Paper \cite{Gkountoumis:2025btc} will be also part of the upcoming thesis of G. Nian. Before moving on, we make the following comments:

\vspace{5mm}

\noindent \cite{Gkountoumis:2023fym}: The author focused more on the calculations of sections 3 and 4, while K. Stemerdink focused more on sections 2 and 5.

\vspace{3mm}

\noindent Chapter \ref{chap:branes}: The results of section \ref{sec:branesupersymmetry} were obtained jointly with K. Stemerdink. The author performed most of the calculations of section \ref{stapproach}, while K. Stemerdink performed most of the computations is section \ref{branes from supergravity}.

\vspace{3mm}

\noindent \cite{Gkountoumis:2025btc}: The results of sections 2 and 4 were obtained together with G. Nian. The author performed most of the computations in section 3.

\clearpage 
\thispagestyle{empty}
\onehalfspacing
\include{frontmatter}
\tableofcontents
\clearpage



\mainmatter

\chapter{Introduction}
\label{chap:introduction}
\lhead[\textit{\chaptername\ \thechapter\quad \leftmark}]{}
\rhead[]{\textit{\thesection\quad \rightmark}}

\section{Quantum gravity}

Currently, there are four known fundamental forces in nature: gravity, electromagnetism, and the weak and strong nuclear forces. Gravity is described by Einstein's theory of general relativity, while the other three forces are described by the standard model of particle physics. Both theories have been extensively examined and repeatedly verified by observations and experiments. Arguably, they are the two most successful theories in modern physics.

General relativity is a \textit{classical} theory that explains gravity at large scales, at which the effect of the other three forces is insignificant. Some of the most notable predictions of general relativity include the existence of black holes and gravitational waves, as well as the bending of light near massive objects. On the other hand, the standard model is a \textit{quantum} theory that describes interactions of matter at small scales, at which gravity can be neglected. One of the greatest successes of the standard model is the prediction of the existence of various particles such as the Higgs boson, the $W$ and $Z$ bosons and the gluon.

Obviously, the challenging question is the following: how can we describe a physical phenomenon for which all four forces should be taken into account? In order to answer this question we need to formulate a quantum theory of gravity. Presently, the most prominent candidate for a consistent theory of \textit{quantum gravity} is string theory. Moreover, string theory is the only theory that provides a unified quantum description of all fundamental forces.

For the purposes of this thesis we will work in the framework of string theory. Before we dive into technical details, we will give a brief introduction to string theory and we will present the basic concepts that are necessary for the subsequent chapters. A thorough discussion on the topics of the following sections can be found in the standard string theory textbooks \cite{becker2006string,blumenhagen2012basic,johnson2002d,kiritsis2019string,1polchinski1998string,polchinski1998string,Green:1987sp,Green:1987mn}.

\section{String theory}
\label{stringsintro}
The fundamental idea of string theory is that matter can be described by one-dimensional objects, which are called \textit{strings}. These strings oscillate in various ways and each oscillation represents a different particle. This is the analogue of a string in a musical instrument, which produces distinct notes as it vibrates in different ways. 

\begin{figure}[H]
    \centering
    \includegraphics[width=0.8\linewidth]{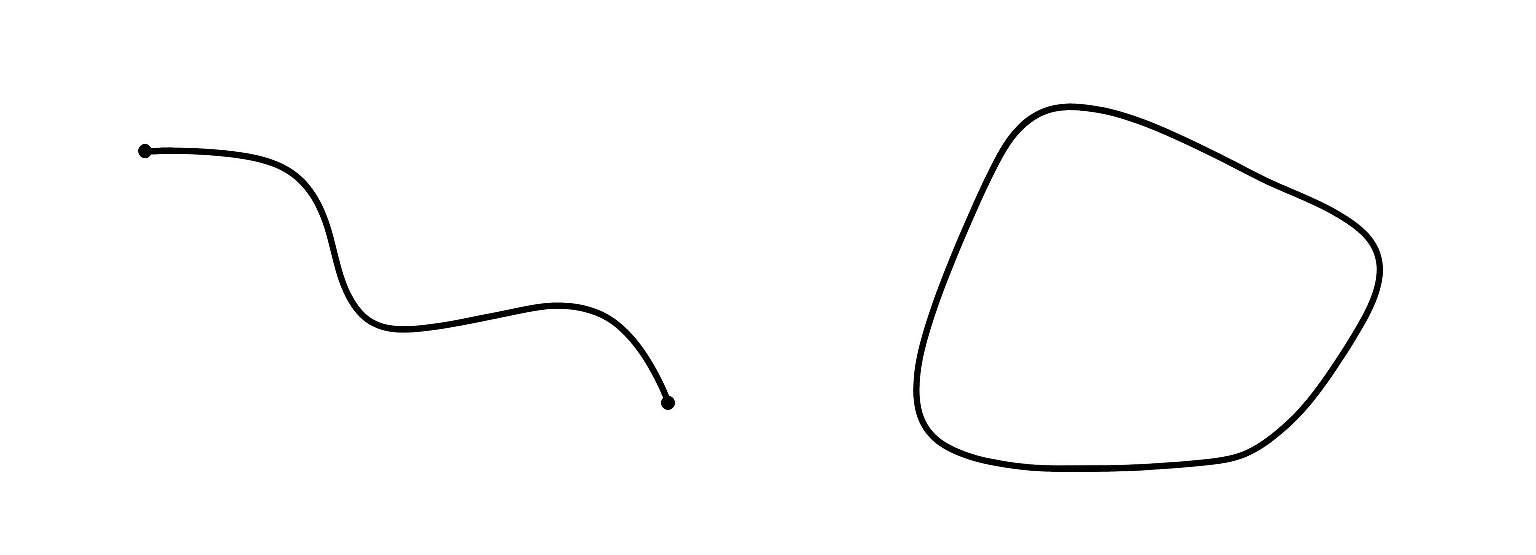}
    \caption{\textit{An open string with two end-points (left),  and a closed string forming a loop (right).}}
   \label{open-closed-fig}
\end{figure}

Strings can be either open or closed. Open strings have two end-points, while closed strings have no end-points and form loops, as depicted in \autoref{open-closed-fig}.
One of the prominent features of a closed string is that it contains a vibrational state that describes the \textit{graviton}, which is the quantum particle mediating the gravitational force. Hence, string theory is a theory of \textit{quantum gravity}. Regarding open strings, their oscillation modes can describe particles of the standard model such as the photon, which is the particle mediating the electromagnetic force. Moreover, open strings can join to form closed strings, and closed string can split to create open strings. Therefore, gravity and particle physics are naturally unified in string theory.

\begin{figure}[H]
   \centering
    \includegraphics[width=0.7\linewidth]{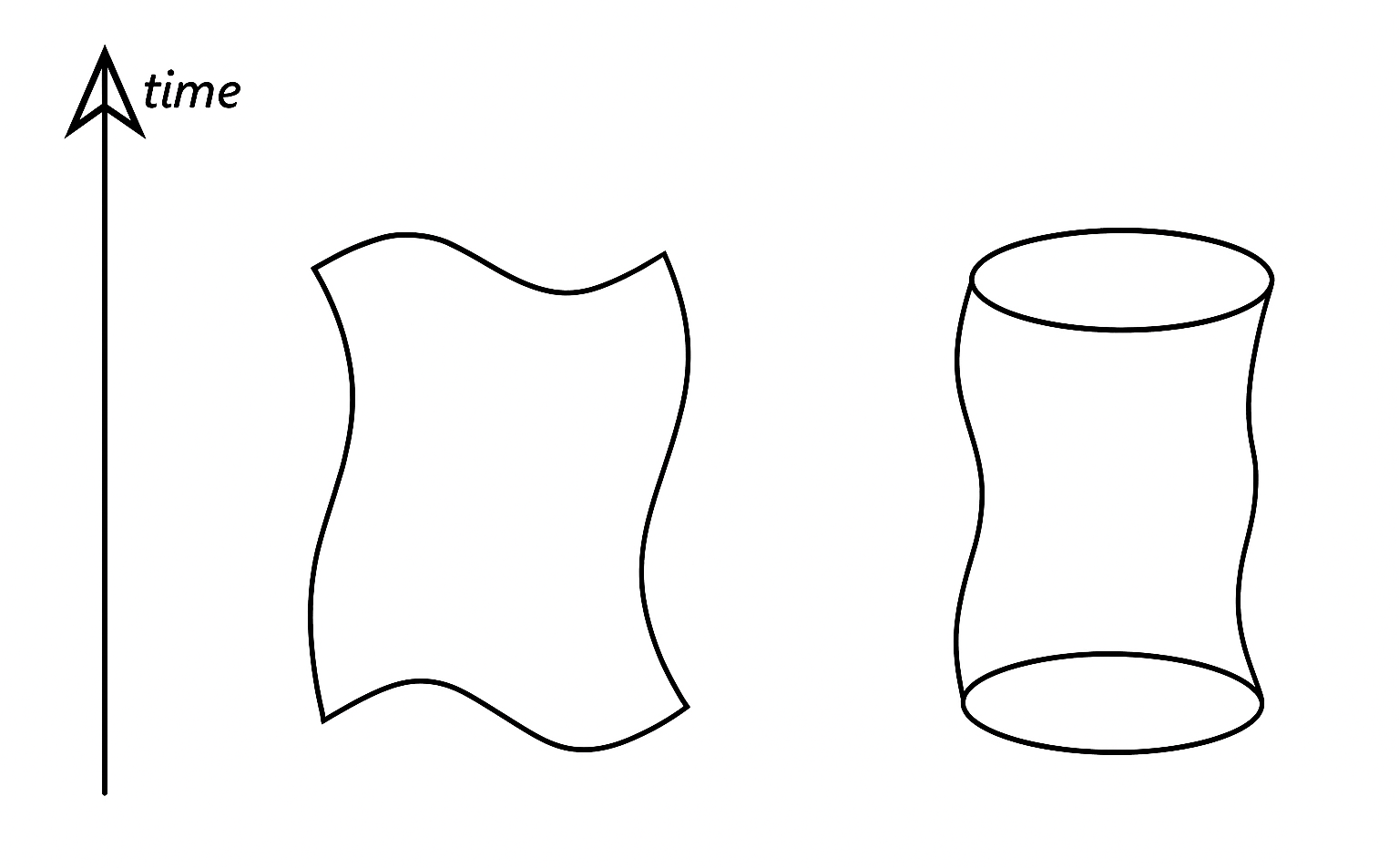}
    \caption{\textit{An open string world-sheet (left), and a closed string world-sheet (right). Time evolves upwards}.}
    \label{worldsheet}
\end{figure}

Strings are dynamical objects, meaning that they evolve in spacetime. For now, let us assume that strings live in a $D$-dimensional Minkowski spacetime, denoted by $\mathbb{R}^{1,D-1}$. In the context of string theory, the spacetime in which strings live  is usually referred to as the \textit{target space}. As a string moves in spacetime it sweeps out a two-dimensional surface, which is called the string \textit{world-sheet}. A closed string traces out a world-sheet with the topology of a cylinder, whereas an open string traces out a world-sheet with the topology of a strip, as shown in \autoref{worldsheet}. 

The string world-sheet can be parameterised by a timelike coordinate $\sigma^1$ and a spacelike coordinate $\sigma^2$. Given this parametrisation, we can define the \textit{embedding} functions $X^{\mu}(\sigma^1,\sigma^2)$, $\mu=0,\ldots,D-1$, which describe the position of each point of the string world-sheet in the target space. In a more mathematical language, the functions $X^{\mu}(\sigma^1,\sigma^2)$ provide a map from the string world-sheet into the target space.

\subsection{Bosonic string theory}

Thus far, we have seen that string theory contains open and closed strings, which are dynamical objects. The dynamics of a string can be described by the \textit{Nambu-Goto} action, which is given by the area $A$ of the string world-sheet $\Sigma$ that is swept out by the string, namely
\begin{equation}
    S_{\text{NG}} = -T \int_{\Sigma}\mathrm{d}A=-T \int_{\Sigma}\mathrm{d}^2\sigma\, \sqrt{-\text{det}_{\alpha\beta}\frac{\partial X^{\mu}}{\partial \sigma^{\alpha}}\frac{\partial X^{\nu}}{\partial \sigma^{\beta}}\eta_{\mu\nu}}\,.
    \label{ngaction}
\end{equation}
Here,  $\sigma^{\alpha}=(\sigma^1,\sigma^2)$ are the coordinates on the world-sheet and $\eta_{\mu\nu}$  is the metric on the target space, which we take to be $\mathbb{R}^{1,D-1}$. Also, $T$ is the tension of the string, which has dimensions of mass per unit length. The string tension is related to the string length scale by the following expression
\begin{equation}
    T=\frac{1}{2\pi\alpha'}\,.
\end{equation}
Note that $\sqrt{\alpha'}$ has dimensions of length, and this is the natural length appearing in string theory.  

The presence of the square root in the Nambu-Goto action makes calculations challenging. Fortunately, there exists another action that describes the string dynamics and eliminates the square root. This the \textit{Polyakov} action, which reads
\begin{equation}
    S_{\text{P}}=-\frac{1}{4\pi\alpha'}\int_{\Sigma}\mathrm{d}^2\sigma\,\sqrt{-h}h^{\alpha\beta}\partial_{\alpha}X^{\mu}\partial_{\beta}X^{\nu}\eta_{\mu\nu}\,.
    \label{polaction}
\end{equation}
In this form, the string action contains one additional field, which is the metric on the world-sheet, denoted by $h_{\alpha\beta}$. However, this additional field can be replaced in the Polyakov action by its equation of motion. This replacement gives back the Nambu-Goto action, meaning that the  actions \eqref{ngaction} and \eqref{polaction} are classically equivalent. 

Using the symmetries of the Polyakov action, the world-sheet metric $h_{\alpha\beta}$ can be replaced by the metric on a two-dimensional flat space, that is $\eta_{\alpha\beta}$. Then, the equation of motion for the string embedding functions $X^{\mu}$ is the free wave equation, that is 
\begin{equation}
    \partial_{\alpha}\partial^{\alpha}X^{\mu}=0\,.
    \label{wave equation}
\end{equation}
Besides the equation of motion we need to specify the boundary conditions that strings should obey. For a closed string we impose the following periodicity condition
\begin{equation}
    X^{\mu}(\sigma^1,\sigma^2+2\pi) = X^{\mu}(\sigma^1,\sigma^2)\,.
    \label{periodic bc}
\end{equation}
In the case of an open string, we can impose either Neumann (N), or Dirichlet (D) boundary conditions at each of the two end-points of the string:
\begin{equation}
    \begin{aligned}
      & \text{N}:\qquad \partial_{\sigma^2}X^{\mu}(\sigma^1,\sigma^2)|_{\sigma^2=0,\pi} = 0\,.\\
      & \text{D}:\qquad \partial_{\sigma^1} X^{\mu}(\sigma^1,\sigma^2)|_{\sigma^2=0,\pi} = 0\,.
    \end{aligned}
    \label{open bc}
\end{equation}
Neumann boundary conditions allow the end-point of a string to move freely, while Dirichlet boundary conditions imply that the end-point is fixed in spacetime.

Now, we have all the necessary ingredients in order to solve the string equation of motion. For a closed string, the most general solution of the wave equation \eqref{wave equation} that is compatible with the periodicity condition \eqref{periodic bc} is
\begin{equation}
    X^{\mu}(\sigma^1,\sigma^2) = X^{\mu}_L(\sigma^+) + X^{\mu}_R(\sigma^-)\,,
\end{equation}
where we have defined $\sigma^{\pm}=\sigma^1\pm\sigma^2$. The functions $X^{\mu}_{L,R}$ are arbitrary functions of $\sigma^{\pm}$, they describe left and right-moving waves and are usually referred to as left and right-movers, respectively. An explicit expression of these functions is given in terms of Fourier modes as follows
\begin{equation}
    \begin{aligned}
          &X^{\mu}_{L}(\sigma^+)= \frac{1}{2}x^{\mu}+\frac{1}{2}\alpha'p^{\mu}\sigma^++i\sqrt{\frac{\alpha'}{2}} \sum_{n\in \mathbb{Z}^*}\frac{1}{n}\tilde{\alpha}^{\mu}_n e^{-in\sigma^+}\,,\\
    &X^{\mu}_{R}(\sigma^-)=\frac{1}{2}x^{\mu}++\frac{1}{2}\alpha'p^{\mu}\sigma^-+i\sqrt{\frac{\alpha'}{2}} \sum_{n\in\mathbb{Z}^*}\frac{1}{n}\alpha^{\mu}_n e^{-in\sigma^-}\,.
    \end{aligned}
    \label{closedoscillators}
\end{equation}
Here, the zero modes $x^{\mu}$, $p^{\mu}$ are the centre of mass position of the string and the total spacetime momentum of the string, respectively. The modes $\tilde{\alpha}^{\mu}_n$, ${\alpha}^{\mu}_n$ create and annihilate the various string vibrations. Note that the left and right-moving modes $\tilde{\alpha}^{\mu}_n$, ${\alpha}^{\mu}_n$ are completely independent.

Regarding open strings, there are four different solutions to the wave equation \eqref{wave equation} depending on which boundary condition \eqref{open bc} each of the string end-points obeys. The Fourier expansions of the open string solutions are similar with those of the closed string \eqref{closedoscillators}, so we omit writing them down. However, we would like to highlight one crucial difference between closed and open string solutions. The boundaries of open strings reflect left to right-movers, and the other way around. Hence, for an open string, the modes $\tilde{\alpha}^{\mu}_n$, ${\alpha}^{\mu}_n$ are related to each other and there is only one set of independent modes.

String states, namely different string oscillations, can be constructed by acting on the string vacuum state with creation operators. The collection of all string states is called the \textit{string spectrum}. As we have already mentioned in the beginning of this section, the string spectrum contains the graviton. However, the string spectrum also contains a \textit{tachyon}, which is a particle with imaginary mass; this indicates that the vacuum of the theory is unstable. Moreover, the string spectrum contains only bosons and no fermions at all. 

Instability of the vacuum and absence of fermions are two of the major drawbacks of string theory described by the Polyakov action \eqref{polaction}, which is known as \textit{bosonic} string theory. The two aforementioned issues render bosonic string theory unrealistic. Nevertheless, vacuum stability and presence of fermions can be achieved by considering \textit{superstring} theory, which is a generalization of bosonic string theory. 

\subsection{Superstring theory}
\label{superstrings}

Superstring theory enjoys a symmetry that bosonic string theory does not. This symmetry is known as \textit{supersymmetry} and it relates bosons to fermions, postulating that each boson has a fermionic superpartner and vice versa. Supersymmetry can be realised on the string world-sheet or on spacetime. Here, we will focus on strings with world-sheet supersymmetry, which do not necessarily possess spacetime supersymmetry. It is important to mention that the absence of tachyons and the presence of spacetime fermions is guaranteed by spacetime supersymmetry.

The superstring action is given by a supersymmetric extension of the Polyakov action \eqref{polaction}, in which the Dirac action for $D$ massless fermions $\psi^{\mu}(\sigma^1,\sigma^2)$ is added. Considering the flat metric on the world-sheet, the superstring action reads
\begin{equation}
    S=-\frac{1}{4\pi} \int_{\Sigma}\mathrm{d}^2\sigma\left(\frac{1}{\alpha'}\partial_{\alpha}X_{\mu}\partial^{\alpha}X^{\mu} + i\overline{\psi}^{\mu}\gamma^{\alpha}\partial_{\alpha}\psi_{\mu}\right)\,,
\end{equation}
where $\gamma^{\alpha}$, $\alpha=1,2$, represent the two-dimensional Dirac matrices. The equation of motion for $X^{\mu}$ is still given by the wave equation \eqref{wave equation}, while for $\psi^{\mu}$ we obtain the two-dimensional Dirac equation:
\begin{equation}
    \gamma^{\alpha}\partial_{\alpha}\psi^{\mu}=0\,.
    \label{psiequation}
\end{equation}
Of course, the equations of motion are accompanied by boundary conditions. In the case of closed superstrings, fermions can satisfy either periodic or anti-periodic boundary conditions, which are usually referred to as Ramond and Neveu-Schwarz boundary conditions, respectively. To be precise
\begin{equation}
    \begin{aligned}
     & \text{Ramond:}\qquad\qquad \quad\psi^{\mu}(\sigma^1,\sigma^2+2\pi)=  \psi^{\mu}(\sigma^1,\sigma^2)\,.\\
      &\text{Neveu-Schwarz:}\qquad \,\psi^{\mu}(\sigma^1,\sigma^2+2\pi)= -\psi^{\mu}(\sigma^1,\sigma^2)\,.
    \end{aligned}
\end{equation}
In the case of open superstrings, analogous boundary conditions can be imposed at each of the two end-points of the string, along with Neumann or Dirichlet boundary conditions. 

Similarly with the solutions of the wave equation \eqref{wave equation}, the solutions of the Dirac equation \eqref{psiequation} can be expanded in terms of Fourier modes, which can be interpreted as fermionic creation and annihilation operators. Then, the superstring spectrum can be constructed by acting on the vacuum state with bosonic\footnote{These are the modes $\tilde{\alpha}^{\mu}_n, \alpha^{\mu}_n$ encountered in \eqref{closedoscillators}.} and/or fermionic creation operators.

Remarkably, there is not a unique superstring spectrum because there are various consistent superstring theories. One reason for this is that fermions can be added to the world-sheet in many ways. For instance, adding both left and right-moving fermions of opposite handedness gives rise to \say{type IIA} superstring theory. If the fermions are of same handedness, the resulting theory is called \say{type IIB}. Adding only right-moving fermions results in \say{heterotic} superstring theory. Moreover, by applying modifications on type II and heterotic superstring theories other consistent superstring theories can be constructed. For the purposes this thesis, we will focus on type II superstring theories.

Type II superstring theories are spacetime supersymmetric, so their spectra contain both spacetime bosons and fermions and are tachyon-free. Hence, these theories resolve the two major issues of the bosonic string. Nevertheless, type II superstrings are not flawless; they are supersymmetric and live consistently in ten-dimensional target spaces. However, our observed world is four-dimensional and supersymmetry has not been detected in the universe. So, let us now explain how we can \say{hide} the extra dimensions and \say{break} supersymmetry, in order to make type II superstrings realistic and suitable for phenomenology.

\section{Compactification}

\begin{figure}[H]
   \centering
    \includegraphics[width=0.7\linewidth]{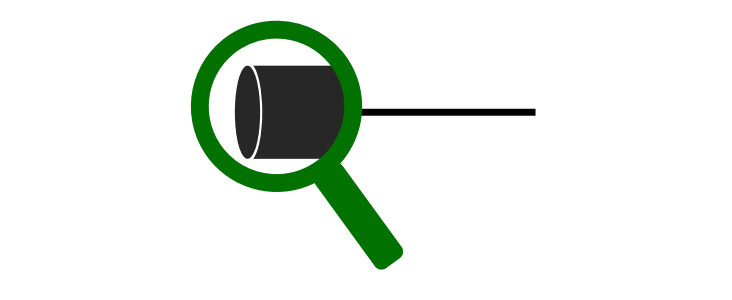}
    \caption{\textit{A two-dimensional cylinder of tiny radius appears to be a one-dimensional line. The picture is taken from} \cite{hetLam:2020goj}.}
    \label{cylinder}
\end{figure}

The fundamental idea of compactification is that the extra dimensions that are required for the consistency of superstring theories have a finite and very small length, so that they can not be observed\footnote{Of course, the concept of compactification can be also applied in other fields of physics and mathematics.}. A very simple and intuitive way to understand the concept of compactification is to consider a two-dimensional cylinder. If the radius of the cylinder is tiny, or equivalently, if the cylinder is observed from far away, it appears to be a one-dimensional line. This is illustrated in \autoref{cylinder}.
The concept of compactification was first introduced back in 1920 when Kaluza and Klein tried to unify gravity and electromagnetism by compactifying five-dimensional gravity on a circle. For our purposes, the general procedure of type II superstring theory compactification suggests that the ten-dimensional target space $\mathcal{M}^{1,9}$ can be split into a non-compact external spacetime $\mathcal{M}^{1,9-d}$ and a compact internal manifold $\mathcal{M}^{d}$ as
\begin{equation}
    \mathcal{M}^{1,9} = \mathcal{M}^{1,9-d}\times \mathcal{M}^d\,,
\end{equation}
where $d$ is the number of compactified dimensions.

Depending on the compactification manifold, spacetime supersymmetry can be preserved or broken. For example, compactification on a $d$-dimensional torus $T^d$ preserves supersymmetry, while compactification on Calabi-Yau manifolds generically breaks supersymmetry. In this thesis we will focus on compactifications on toroidal orbifolds, which can break supersymmetry partially or completely.

\subsection{Orbifolds}

Given a Riemannian manifold $\mathcal{M}$ with a discrete isometry group $G$, an orbifold $\mathcal{O}$ can be obtained as the quotient space 
\begin{equation}
    \mathcal{O} = \mathcal{M}/G\,.
\end{equation}
Then, for an element $g\in G$ and a point $x\in \mathcal{M}$, the points $x$ and $g\cdot x$ are equivalent in the orbifold. Now, if the action of $G$ leaves one or more points in $\mathcal{M}$ invariant, the corresponding quotient space exhibits singularities and is referred to as \textit{non-freely acting} orbifold. On the other hand, if $G$ leaves no points in $\mathcal{M}$ invariant, the resulting quotient space is a smooth manifold, called \textit{freely acting} orbifold.

\begin{figure}[H]
   \centering
    \includegraphics[width=0.5\linewidth]{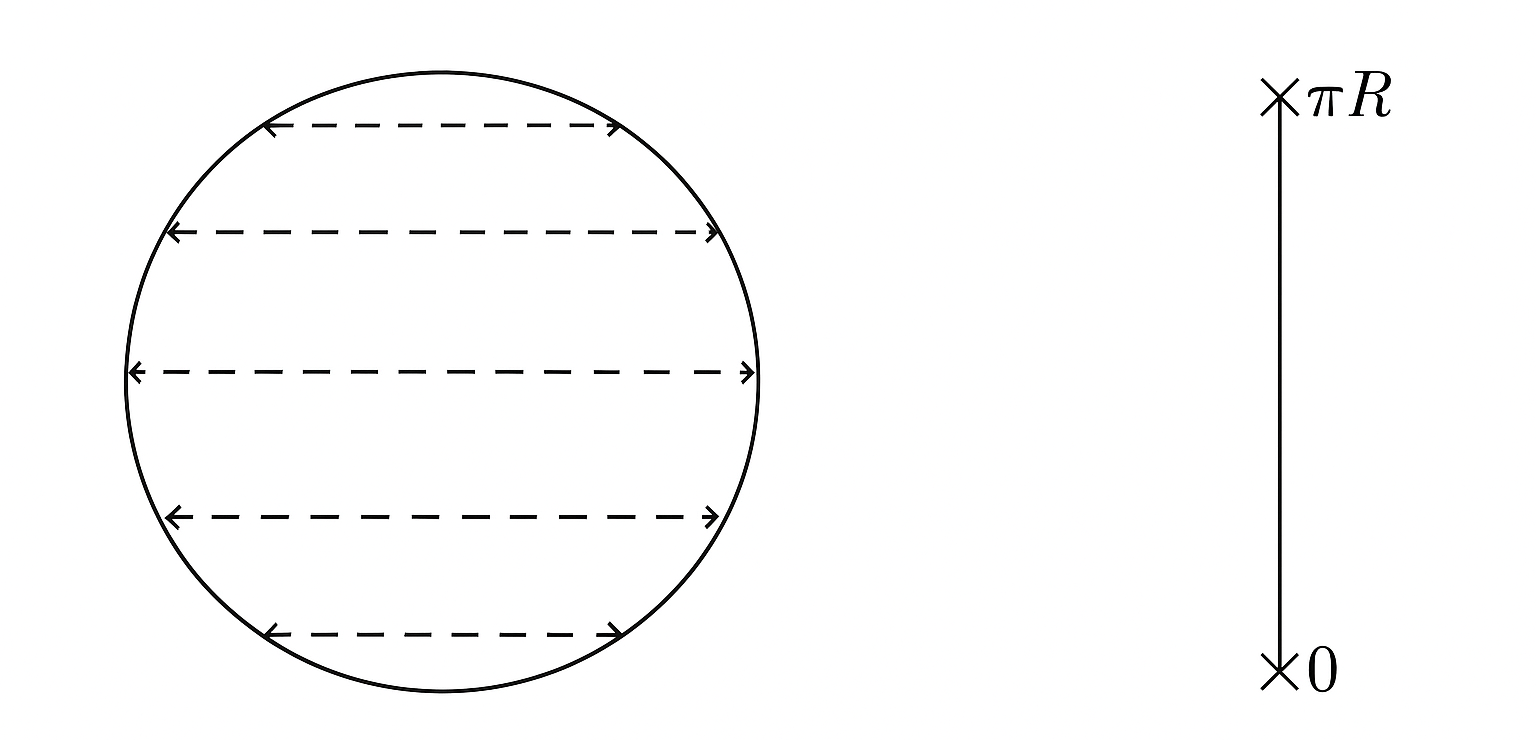}
    \caption{\textit{The orbifold $S^1/\mathbb{Z}_2$ (right figure). Here, the $\mathbb{Z}_2$ action is the reflection $z\to -z$. The picture is taken from} \cite{bassant}.}
    \label{orbicircle}
\end{figure}

\begin{figure}[H]
   \centering
    \includegraphics[width=0.45\linewidth]{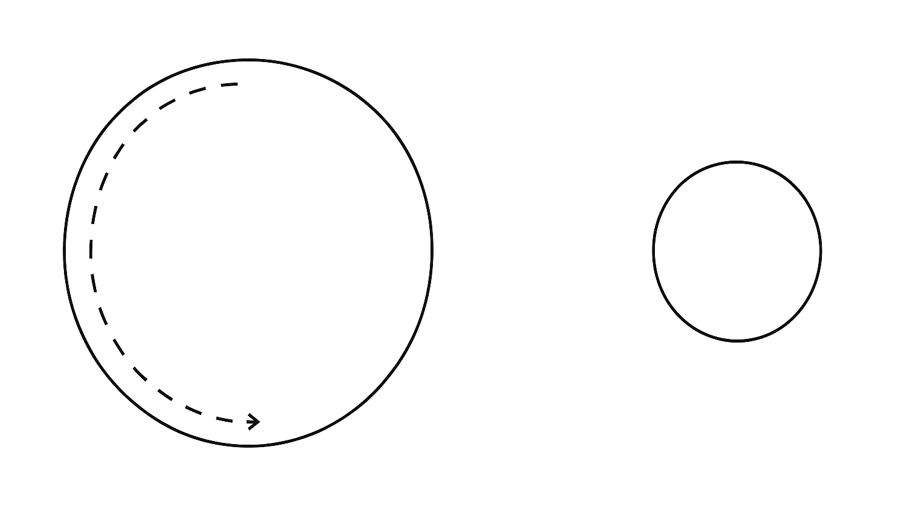}
    \caption{\textit{The orbifold $S^1/\mathbb{Z}_2'$ (right figure). Here, the $\mathbb{Z}_2'$ action is the translation $z\to z+\pi R$. The picture is taken from} \cite{bassant}.}
    \label{orbicircle2}
\end{figure}

As an example of a non-freely acting orbifold, let us consider a circle $S^1$ of radius $R$, parametrised by $z\in [0,2\pi R]$. The circle enjoys a discrete $\mathbb{Z}_2$ symmetry given by the reflection $z\to -z$. This $\mathbb{Z}_2$ action leaves the points $z=0$ and $z=\pi R$ invariant. The resulting quotient space $S^1/\mathbb{Z}_2$ is a line segment with a fixed point at each end, as shown in \autoref{orbicircle}.
Another discrete $\mathbb{Z}_2$ symmetry of the circle is the translation $z\to z+\pi R$, which, for clarity, will be denoted here by $\mathbb{Z}_2'$. This $\mathbb{Z}_2'$ action leaves no points invariant and the quotient space $S^1/\mathbb{Z}_2'$ is a freely acting orbifold. Actually, this orbifold is simply a circle of radius $R/2$, as can be seen in \autoref{orbicircle2}.
Regarding orbifold compactifications of string theory, we mention here that strings propagating on an orbifold $\mathcal{M}/G$ can close up to a transformation $g\in G$, namely 
\begin{equation}
     X^{\mu}(\sigma^1,\sigma^2+2\pi) = gX^{\mu}(\sigma^1,\sigma^2)\,.
\end{equation}
These modified boundary conditions define new sectors in the theory, which are called \textit{twisted} sectors. Note that for $g=1$, the usual periodic boundary conditions \eqref{periodic bc} are recovered. The sector characterized by $g=1$ is referred to as the \textit{untwisted} sector.

Finally, depending on the group element $g\in G$ supersymmetry can be partially or completely broken. Also, non-freely acting orbifolds break supersymmetry explicitly, while freely acting orbifolds break supersymmetry spontaneously. We will discuss these issues in more detail in the forthcoming chapters.

\subsection{Dualities}

As we have already mentioned in section \ref{superstrings}, there are many consistent superstring theories. One of the greatest achievements in string theory was the discovery of maps that connect the various superstring theories to each other. These maps are called \textit{dualities} and imply that seemingly different theories may actually describe the same physics!

One of the most well-understood dualities is the target space duality, or T-duality. This is a duality that relates string theories that live on different target spaces. As an example, T-duality implies that type IIA superstring theory compactified on a circle of radius $R$ is equivalent to type IIB superstring theory compactified on a circle of radius $\alpha'/R$. Also, there exist T-duality transformations that map a theory into itself. These transformations form a symmetry group, which is the T-duality group. For type II theories compactified on a $d$-dimensional torus the T-duality group is $\text{Spin}(d,d;\mathbb{Z})$.

Another type of string duality is the strong-weak duality, called S-duality. This duality connects weakly coupled theories to strongly coupled theories by changing the string coupling $\lambda_s$ to $1/\lambda_s$. In the case of type IIB theory, this transformation is actually a symmetry of the theory contained in the larger symmetry group $\text{SL}(2;\mathbb{Z})$, which is called the S-duality group.

Finally, the last string duality that we would like to mention is the unified duality, or U-duality. This duality contains both T and S-dualities. For instance, regarding type II theories compactified on a $d$-dimensional torus, with $d\geq 5$, the U-duality group is the maximal discrete subgroup of the maximally non-compact exceptional group $E_{d+1(d+1)}$.

\section{Supergravity}

Instead of working with the entire superstring spectrum, we may sometimes focus only on the dynamics of the lightest degrees of freedom. This low-energy limit of superstring theory is called \textit{supergravity}. Supergravity is a field theory that combines supersymmetry and general relativity and can be understood as an effective field-theoretic description of superstring theory.

Similarly with superstring theory, there is not a unique supergravity theory, as all consistent superstring theories give rise to consistent low-energy effective field theories. These effective theories naturally live in higher-dimensional spacetimes and enjoy supersymmetry. 

In order to deal with the higher-dimensional spacetimes, we can once again employ the idea that the extra dimensions should be very small and compact. Regarding supersymmetry, there exist mechanisms that can lead to supersymmetry breaking in supergravity theories. In this thesis we will focus on a mechanism that is called \textit{Scherk-Schwarz} reduction, and was first introduced by Scherk and Schwarz in \cite{Scherk:1978ta,Scherk:1979zr}.

The Scherk-Schwarz reduction is a mechanism that leads to spontaneous supersymmetry breaking in supergravity. We wish to highlight here that Scherk-Schwarz supergravity theories can be embedded in string theory by using freely acting orbifolds. The connection between Scherk-Schwarz reduction and freely acting orbifolds of string theory will be discussed in detail in the following chapter.

\subsection{Swampland and landscape}
As mentioned above, all consistent string theories give rise to consistent supergravity theories. However, the converse is not always true, meaning that not all supergravity theories can be embedded consistently in string theory, or more generally in quantum gravity. Therefore, it is natural to ask which are the criteria that a supergravity theory should meet such that it can be consistently uplifted in quantum gravity.

This question was first posed in \cite{vafa2005string} and since then the \textit{swampland program} aims to answer it. Many consistency criteria have been derived from string theory, or by directly studying the low-energy theories and using quantum gravity arguments. These criteria have not been proven yet, hence they are usually referred to as \textit{swampland conjectures}.

Finally, in the context of the swampland program, all supergravity theories that can be lifted consistently in quantum gravity are considered to be in the \textit{landscape}, while those that cannot are regarded to be in the \textit{swampland}.

\section{D-branes}

As we have seen in the previous sections, strings can be closed or open. Regarding open strings, each of their end-points can satisfy  either Neumann or Dirichlet boundary conditions \eqref{open bc}. Recall that Neumann boundary conditions imply that the end-point of a string can move freely, while Dirichlet boundary conditions dictate that the end-point lies at some constant position in spacetime.

Let us now consider an open string in a ten-dimensional Minkowski spacetime $\mathbb{R}^{1,9}$, with both end-points satisfying Neumann boundary conditions in the time dimension and in $p$ spatial dimensions, and Dirichlet boundary conditions in the rest. These boundary conditions suggest that the end-points of this string can only move in a $(p+1)$-dimensional hypersurface in spacetime. This hypersurface is called a D$p$-brane, where D stands for Dirichlet and $p$ is the number of spatial dimensions of the brane. 

\begin{figure}[H]
    \centering
    \includegraphics[width=0.45\linewidth]{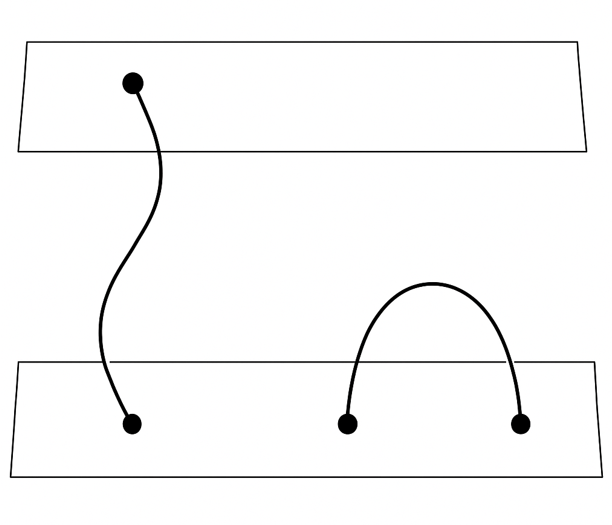}
    \caption{\textit{An open string with each end-point moving on a different D-brane (left), and an open string with both end-points lying on the same D-brane (right).}}
   \label{branesopcl}
\end{figure}

Of course, the two end-points of an open string may not obey the same boundary conditions in all spacetime dimensions. In this case, the end-points are allowed to move on separate D-branes. Different open strings configurations are shown in \autoref{branesopcl}. 

Finally, we wish to mention here that D-branes should not be simply regarded as hypersurfaces on which open strings can end. D-branes are actually dynamical objects of string theory and their dynamics are governed by a generalization of the Nambu-Goto action \eqref{ngaction}.

\section{Motivation and outline}

Presently, the most well-studied and well-understood examples of string theory vacua arise from geometric compactifications such as compactifications on Calabi-Yau manifolds. In order to gain a deeper understanding of all possible stringy vacua, it is necessary to examine more exotic corners of the string landscape. In this dissertation, with this purpose in mind, we will study string theory on freely acting orbifolds, with emphasis on asymmetric constructions, which provide examples of non-geometric compactifications of string theory.

The majority of geometric compactifications of string theory preserve (some amount of) supersymmetry. However, supersymmetry has not been detected in experiments yet, so we should naturally try to realise string theories with broken supersymmetry. A way to accomplish this is by using freely acting orbifolds. In this thesis, we will present explicit examples of string theories compactified on freely acting orbifolds, which preserve no supersymmetry.

We should highlight here that non-supersymmetric compactifications of string theory usually suffer from instabilities manifested by the presence of tachyons in their spectrum. Hence, it is essential to construct stable, tachyon-free non-supersymmetric models. Here, we will make a first step towards this direction, and we will illustrate that tachyon-free non-supersymmetric string-theoretic models can be realised by utilising freely acting orbifolds.

Another common characteristic of geometric compactifications is that they come with a large number of massless scalar fields, i.e. moduli. Since the expectation values of these scalar fields determine the values of physical quantities, such as gauge and Yukawa couplings, it is essential to stabilise the moduli. In this thesis, we will try to tackle the issue of moduli stabilisation by compactifying string theory on freely acting asymmetric orbifolds, in which most of the moduli can be frozen.

Regarding the unfrozen massless scalar fields, they parametrise the moduli space of the compactification. Generally, for theories preserving eight (or less) supersymmetries, the classical metric on the moduli space can receive corrections at the quantum level. Then, thorough understanding of the moduli space would demand very challenging calculations of quantum corrections. In this dissertation, we will use an alternative approach in order to deal with the issue of computing quantum corrections; we will construct models, based on freely acting asymmetric orbifolds, in which the classical result is exact, as no quantum corrections arise.

Now, as a theory of quantum gravity, string theory should be able to provide answers to questions about the cosmological constant such as why is it positive and near vanishing. Unfortunately, a realisation of a de Sitter vacuum in string theory is very challenging, and almost all string-theoretic examples yield anti-de Sitter vacua. Hence, it is natural to ask, if a positive cosmological constant can be generated in string compactifications on freely acting orbifolds. In this thesis, we will engage with this issue both classically and at the quantum level.

Finally, one of the greatest achievements in string theory is the microscopic description of black holes: black holes in string theory can be described by D-branes. Therefore, in order to study black holes in an orbifolded background, it is essential to determine the D-brane spectrum of the orbifolded theory; this will be the last topic of this thesis.

All of the above issues are discussed in detail in the subsequent chapters of this thesis. First, in chapters \ref{chap:fao} and \ref{chap:spectrum} we explain the construction of freely acting (asymmetric) orbifolds and we discuss various examples of type IIB string theory compactified on such spaces, preserving partial or no supersymmetry. Then, in chapters \ref{chap:mod5} and \ref{chap:swampland} we determine the moduli spaces of our orbifold constructions and we study aspects of the swampland program in the context of freely acting orbifolds. Finally, in chapter \ref{chap:branes} we discuss the D-brane spectrum of freely acting orbifolds. We conclude with a summary and directions for future research in chapter \ref{chap:conclusion}.

\clearpage
\thispagestyle{empty}

\chapter{Freely acting orbifolds}
\label{chap:fao}

In this chapter, we explain the general construction of freely acting orbifolds and we analyse the various conditions for the existence of a consistent orbifold. First, in section \ref{orbiandss} we discuss the connection between freely acting orbifolds and Scherk-Schwarz reduction. This is essential for the description of the orbifold in terms of duality twists, which is the subject of section \ref{sec:Orbifold constructions}. Then, in section \ref{The orbifold action and partition function} we discuss another approach to the orbifold construction; we will refer to this as the lattice approach. The duality twists and lattice approach are two complementary methods for the orbifold construction. We will use both of them and we will discuss their relationship in this chapter. Finally, in section \ref{sec:Partition function} we will present the general formalism for the construction of the partition function of freely acting  orbifolds.

\section{Orbifolds and Scherk-Schwarz reduction}
\label{orbiandss}

Spontaneous breaking of supersymmetry in supergravity can be realized via Scherk-Schwarz reductions on a circle with a twist by a supergravity duality symmetry \cite{Scherk:1978ta,Scherk:1979zr,Cremmer:1979uq}. The string theory uplifts of these are typically reductions of string theories with a duality twist \cite{Dabholkar:2002sy} which can in some cases be 
viewed as compactifications on torus bundles over a circle but in general can give
 reductions on non-geometric T-folds or U-folds \cite{Hull:2004in}. For the case in which the duality twist is a T-duality, these become freely acting orbifolds at special points of the moduli space \cite{Kiritsis:1997ca,Dabholkar:2002sy}. Our interest here lies mostly in partial supersymmetry breaking, which can easily be accommodated with the Scherk-Schwarz mechanism and its stringy uplift. This yields a variety of string theories in lower dimensions with various amounts of supersymmetry. Complete breaking of supersymmetry without the presence of tachyons can be achieved as well, and we will present some new examples of such string theories. Typically they suffer from known problems of generating either a cosmological constant that is too large, or extra dimensions that are too large. This renders these models not realistic for phenomenology, though still interesting for present purposes, as we shall discuss. 
 
Scherk-Schwarz reductions over a circle $S^1$ from $D+1$ to $D$ dimensions arise when the fields in $D+1$ dimensions pick up a monodromy around the circle contained in the continuous duality symmetry group $G$ of the $D+1$ dimensional supergravity. The ansatz for the fields is
\begin{equation}
    \psi(x^\mu,y)=g(y)[\psi(x^\mu)]\ ,\qquad g(y)=\exp \Big(\frac{My}{2\pi R}\Big)\ ,
\end{equation}
where $g\in G$, $y$ is the coordinate on $S^1$ with radius $R$ and $M$, which is usually referred to as the mass matrix, is an element of the Lie algebra of $G$. The fields then pick up a monodromy matrix $e^{M}$ going around the $S^1$.

To fully specify the reduction, we need to also choose a spin structure on the circle; this will be discussed in more detail later on. Scherk-Schwarz reductions of supergravity theories yield gauged supergravities in $D$ dimensions, and when the (classical) moduli space in $D+1$ dimensions is a coset $G/K$ with $K$ compact, the scalar potential on the moduli space in $D$ dimensions is positive definite. If furthermore the monodromy matrix is $G$-conjugate to a rotation matrix in $K$, there exist stable or at least marginally stable Minkowski vacua that may preserve some of the supersymmetry \cite{Dabholkar:2002sy}. The quadratic fluctuations around the vacuum then yield the mass spectrum with masses determined by the parameters in the monodromy matrix. 

Scherk-Schwarz reductions can   be generalised to string theory \cite{Dabholkar:2002sy, Hull:1998vy} (for some early references on this topic see \cite{Rohm:1983aq,Kounnas:1988ye,Ferrara:1987es,Ferrara:1988jx,Kiritsis:1997ca}). In string theory  the global symmetry group $G$ of the low-energy effective action is broken to the discrete U-duality group $G(\mathbb{Z})$ which is a symmetry of the non-perturbative string theory \cite{Hull:1994ys} 
and the monodromy is required to be in $G(\mathbb{Z})$ \cite{Dabholkar:2002sy,Hull:1998vy} giving non-linear quantization conditions on the mass matrix. Now, if the quantized monodromy is conjugate to a rotation, the construction is an orbifold \cite{Dabholkar:2002sy}. Each choice of a monodromy matrix corresponds to a choice of orbifold and yields a particular quantization condition on the masses. From the point of view of the string landscape, it is therefore important to study the mass spectra allowed by these quantization conditions. When the scalar potential in supergravity leads to a mass spectrum that does not obey these quantization conditions, it belongs to the swampland. One of the aims of this chapter is to study this question in detail for a large class of freely acting orbifolds.

 For our purposes, we will consider monodromies contained in the T-duality subgroup of $G(\mathbb{Z})$. Such supergravity vacua can be realized in string theory as freely acting orbifolds. 
 We focus on the case of type IIB string theory on $T^5$ and 
 the orbifold actions we discuss here involve a T-duality action on $T^4$ together with a shift along the remaining $S^1$. The six-dimensional supergravity theory obtained by compactification on $T^4$ has symmetries given by
 $G=\text{Spin}(5,5)$ and $K=[\text{Spin}(5)\times \text{Spin}(5)]/ \mathbb{Z}_2$. The U-duality symmetry is then $G(\mathbb{Z})=\text{Spin}(5,5;\mathbb{Z})$ and the T-duality subgroup in which the monodromy lies is $\text{Spin}(4,4;\mathbb{Z})$.
In addition to the $\text{Spin}(5,5;\mathbb{Z})$ U-duality, the string theory compactified on $T^4$ also has a local symmetry given by the double cover $\tilde K$ of $K$, $\tilde K=[\text{Spin}(5)\times \text{Spin}(5)]$ \cite{hull2007generalised}: while the local symmetry acts through $K$ on the bosons, the double cover is 
  needed for the action on the fermions, requiring a generalized spin structure \cite{hull2007generalised}. We will work in a gauge in which the local symmetry is fixed, so that the U-duality transformations act on the fermions through compensating $\tilde K$ transformations.

  If the monodromy acts as a diffeomorphism on $T^4$, this corresponds to a symmetric orbifold of the IIB string, so that the compactification geometry is a $T^4$ bundle over $S^1$ \cite{Hull:2004in}. If the monodromy acts as a T-duality on $T^4$, this constructs a T-fold background which is realized as an asymmetric orbifold at a special point in the moduli space \cite{Dabholkar:2002sy,Hull:2020byc}. Asymmetric orbifolds are interesting for moduli stabilization because more of the moduli  are fixed than in the symmetric case. Furthermore, by combining the T-duality action on $T^4$ with a shift on $S^1$ one ensures that no extra moduli arise from the twisted sector (see e.g. \cite{dine1997new,Angelantonj:2006ut,Anastasopoulos:2009kj,Condeescu:2013yma} for some references). The original references for symmetric and asymmetric orbifolds are \cite{dixon1985strings,dixon1986strings} and \cite{Narain:1986qm,narain1991asymmetric} respectively. 
  These references mainly focus on the lattice approach, which  is particularly useful when constructing partition functions, as we will discuss. However, for making contact with the Scherk-Schwarz reductions and finding the low-energy supergravity theory, the language of monodromies and duality twists is more useful. These reductions correspond to orbifolds by symmetries which are freely acting and these have some important differences with the non-freely acting ones. In this thesis, we will use these two complementary approaches and discuss their relationship.

  The T-duality transformations on $T^4$ can be realised as diffeomorphisms of a doubled torus $T^8$, so that the T-fold construction described above can be viewed as a bundle of the doubled torus $T^8$ over a circle \cite{Hull:2004in}. More general constructions are possible in which the dependence on the momentum on the circle introduced by the monodromy is generalised to include dependence on the string winding number on the circle. String theory on the circle  can be formulated  by introducing a dual coordinate conjugate to the winding number so that the circle is promoted to a doubled circle.  Then the general construction has separate monodromies which are introduced on both the circle and the dual circle; these two monodromies  are given by two commuting T-duality transformations \cite{Dabholkar:2005ve}. This gives a construction that is not even locally geometric and can be understood in terms of a doubled torus bundle over the doubled circle \cite{Hull:2007jy,Hull:2009sg} (such constructions are sometimes referred to as having \say{R-flux}). The corresponding orbifolds involve a shift on both the circle and the dual circle, introducing phases depending on both momentum and winding \cite{Dabholkar:2005ve}. Further possibilities arise with shifts on several circles, see e.g. \cite{dabholkar1999string}. In this thesis we will focus on the cases in which there is a shift on just one circle and 
  no shift on the dual circle.

  We will carefully discuss the orbifold action on the fermions. The theory has further discrete symmetries such as $(-1)^{F_s}$ where $F_s$
is the spacetime fermion number, together with refinements of this into chiral fermion numbers. We will see that we need to consider monodromies such as ${\mathcal M}(-1)^{F_s}$ where
${\mathcal M}$ is in $\text{Spin}(4,4;\mathbb{Z})$. Note that for a monodromy $(-1)^{F_s}$
(with ${\mathcal M}=1$), the orbifold by this together with a shift on the circle amounts to choosing the anti-periodic spin structure for the circle. This corresponds to a non-supersymmetric string compactification which can be made free of tachyons. We will discuss explicit examples of such compactifications in this chapter.

\section{Duality twists}
\label{sec:Orbifold constructions}
The orbifolds we are interested in have target spaces of the form 
\begin{equation}\label{background}
\mathbb{R}^{1,4}\times  S^1\times T^4\ , \end{equation}
identified under the action of a $\mathbb{Z}_p$ symmetry.
This requires being at a point in the moduli space at which the fields on $T^4$ are invariant under the action of 
$\mathbb{Z}_p$. We then orbifold by a $\mathbb{Z}_p$ given by this $T^4$ symmetry combined with a shift by $2\pi \mathcal{R}/p$ on the circle $S^1$ with radius $\mathcal{R}$ which makes the orbifolds freely acting. Freely acting orbifolds have no fixed points and, at generic points in the moduli space, all states coming from the twisted sectors are massive. Furthermore, as compared to non-freely acting orbifolds, supersymmetry is spontaneously broken instead of being explicitly broken, manifested by the fact that gravitini  become massive instead of being projected out. 

The symmetric orbifolds arise when the $\mathbb{Z}_p$ action on $T^4$ is a geometric discrete symmetry of $T^4$, generated by a diffeomorphism on $T^4$, i.e. by an element in GL(4;$\mathbb{Z}$);  this generator is the monodromy of the corresponding duality twist. The requirement that the monodromy is in the discrete group GL(4;$\mathbb{Z}$)
restricts the rank $p$ of the group $\mathbb{Z}_p$. All possible values of $p$ have been classified in \cite{Erler:1992ki} and for  $T^4$ these are $p=2,3,4,5,6,8,10,12$ and $24$.  Not all values of $p$ yield supersymmetric string theories, and it is worth mentioning that symmetric orbifolds of $T^4$ of rank $p=5,8,10,12$ and 24 break all supersymmetry.

For the asymmetric orbifolds, the $\mathbb{Z}_p$ group   acts as a T-duality transformation on the $T^4$ CFT.
The T-duality group for superstrings on $T^4$ is Spin$(4,4;\mathbb{Z})$, a discrete subgroup
of the double cover Spin$(4,4)$ of $\text{SO}(4,4)$, as the D-brane states transform as a spinor representation of $\text{Spin}(4,4)$ \cite{Hull:1994ys}. Then, in this case, the
 monodromy matrix lies in Spin$(4,4;\mathbb{Z})$.\footnote{When the monodromy matrix is not in the T-duality subgroup of the U-duality group, one has generalized orbifolds that quotient by a non-perturbative symmetry \cite{Dabholkar:2002sy}. Such orbifolds don't have a CFT description on the worldsheet and we will not consider them in this thesis.} The background fields, namely the  torus metric $G$ and the two-form $B$-field, can be combined  into a matrix $E=G+B$. T-duality transforms $E$ to a new background $E'$ through a fractional linear transformation\footnote{For details on how T-duality acts on the background fields we refer to the classic review \cite{giveon1994target} and e.g. \cite{tan2015t,satoh2017lie}. }. Consistency of the asymmetric orbifold then requires that the $\mathbb{Z}_p$ transformation is a symmetry under which $E'=E$. This can be achieved only for special values of the  moduli, which are therefore stabilized in these non-geometric constructions. Requiring modular invariance puts severe constraints on which asymmetric orbifolds are allowed and in particular restricts the values of $p$. Nevertheless, asymmetric orbifolds allow for more possibilities compared to the symmetric ones (see also \cite{harvey1988quasicrystalline,nibbelink2017t}). We will return to these issues when we discuss specific examples.

\subsection{Moduli space and quantization conditions}
\label{sec:quantization conditions}

For a coset space $G/K$, the points are cosets: $[g]=gK$, where $g\in G$. The stabilizer of the origin $[1]=K$ is a subgroup $K_0\subset G$, so that the stabilizer of a point $[g]$ is the conjugate subgroup $K_g=gK_0g^{-1}$. We are interested in those points $[\bar g]$ for 
which there is a $\mathbb{Z}_p$ symmetry preserving $[\bar g]$  which is also in the duality group $G(\mathbb{Z})$, so that 
$G(\mathbb{Z})\cap K_{\bar g}$ is non-trivial and contains the $\mathbb{Z}_p$. 
This is a hard condition to solve in general, and $G(\mathbb{Z})\cap K_{\bar{g}}$ is non-trivial only for special points.
Such a fixed point is necessarily at a minimum of the associated Scherk-Schwarz potential and gives a stable Minkowski vacuum \cite{Dabholkar:2002sy}.

If the generator of the $\mathbb{Z}_p$ symmetry is $\mathcal{M}\in G(\mathbb{Z})$, then 
\begin{equation}\label{conjugational}
\mathcal{M} = {\bar g} \tilde{\mathcal{M}} {\bar g}^{-1} \,; \qquad\quad \tilde{\mathcal{M}}\in K_0 \subset G \,, \qquad {\bar g}\in G \,.
\end{equation}
At a point in the moduli space fixed under the action of such a $\mathbb{Z}_p$ symmetry, the corresponding background field configuration on $T^4\times S^1$ is preserved by the $\mathbb{Z}_p$ symmetry. 
If 
 $   {\mathcal M}=e^{M'}\ $,
then
\begin{equation}
    \tilde{\mathcal M}=e^{M}\ ,
\end{equation}
where 
\begin{equation}
    M={\bar g}^{-1} M'{\bar g}
\end{equation}
will be referred to as the mass-matrix: it is the matrix of masses that appear in the effective action.  In this way, a monodromy  $ {\mathcal M}\in G(\mathbb{Z})$ satisfying $ {\mathcal M}^p=1$ defines a matrix $\tilde{\mathcal{M}}\in K_0$ that also satisfies
$\tilde{\mathcal{M}}^p=1$ and so generates a  $\mathbb{Z}_p$  subgroup of $K$.
The action on fermions is through an element of the double cover $\tilde K$ of $K$ and this in general can generate a  $\mathbb{Z}_{2p}$  subgroup of $\tilde K$.

Monodromies ${\mathcal{M}}, {\mathcal{M}}' $ that are $G(\mathbb{Z})$-conjugate, i.e.
\begin{equation}
    {\mathcal{M}}' =g{\mathcal{M}}g^{-1}\,, \qquad { g}\in G (\mathbb{Z})\,,
\end{equation}
define the same theory. Then the distinct orbifolds correspond to conjugacy classes of monodromies \cite{Hull:1998vy}. This conjugation changes the fixed point $\bar g \to \bar g'= g \bar g $ but leaves 
$\tilde{\mathcal{M}}$ unchanged.
Changing $\bar g$ to $\bar g k$ with $k\in K_0$ gives another representative of the coset $[\bar g]$ and transforms 
$\tilde{\mathcal{M}}$ to $k\tilde{\mathcal{M}}k^{-1}$. Then the monodromy only defines the rotation matrix $\tilde{\mathcal{M}}$ and the mass matrix $M$ up to conjugation by an element of $K_0$.

\subsection{Type IIB string theory compactified on $T^4$ }

For  type IIB string theory compactified on $T^4$, 
 $G=\text{Spin}(5,5)$ and $K=[\text{Spin}(5)\times \text{Spin}(5)]/ \mathbb{Z}_2$ with double cover $\tilde K=[\text{Spin}(5)\times \text{Spin}(5)]$.
 We will focus on the T-duality subgroup $\text{Spin}(4,4)\subset \text{Spin}(5,5)$ which is a perturbative symmetry that can be realised in the world-sheet formulation.
In the classical supergravity theory, the 1/2-BPS 0-branes in $6D$ obtained by compactification on $T^4$ are in the 16-dimensional spinor representation of $\text{Spin}(5,5)$.
In the quantum string theory, the charges of these 0-branes are quantized and take values in a 16-dimensional charge lattice. The U-duality group $G(\mathbb{Z})=\text{Spin}(5,5;\mathbb{Z})$ is the discrete subgroup of $\text{Spin}(5,5)$
preserving this lattice. Under the subgroup $\text{Spin}(4,4)\subset \text{Spin}(5,5)$, the 16 0-brane charges decompose into an 8-dimensional vector representation $\textbf{8}_v$ of $\text{Spin}(4,4)$
corresponding to the NS-NS 0-branes from 4  momentum modes and 4 winding modes on $T^4$, together with an 8-dimensional chiral spinor representation $\textbf{8}_s$ of $\text{Spin}(4,4)$
corresponding to the R-R 0-branes arising from D1-branes and D3-branes wrapping the $T^4$.
Then the charges transform as the $\textbf{8}_v+\textbf{8}_s $ representation of $\text{Spin}(4,4)$, and $\text{Spin}(4,4;\mathbb{Z})$ is the discrete subgroup of  $\text{Spin}(4,4)$ preserving the charge lattice.
 
 At a point $[g]$ in the moduli space that is fixed under a $\mathbb{Z}_p$ symmetry, we compactify on a further $S^1$ and orbifold by the $\mathbb{Z}_p$ symmetry of the $T^4$ combined with a shift 
 on the $S^1$. The orbifold group is then generated by $ \mathcal{M}$ combined with a shift of the $S^1 $ coordinate 
 by $2\pi\mathcal{R}/p$.
This corresponds to a compactification on $T^4$ followed by a compactification on $S^1 $ with a duality twist by the T-duality transformation $\mathcal{M}\in G(\mathbb{Z})$, so that there is a monodromy
$\mathcal{M}$ on the $S^1$.
Then $\mathcal{M}\in\text{Spin}(4,4;\mathbb{Z})$ is required to satisfy
\begin{equation}\label{conjugation}
\mathcal{M} = g \tilde{\mathcal{M}} g^{-1} \,; \qquad \tilde{\mathcal{M}}\in   [(\spinfour\times\spinfour )
/ \mathbb{Z}_2]_0
\subset\text{Spin}(4,4) \,, \qquad g\in\text{Spin}(4,4) \,.
\end{equation}
Note that $\text{Spin}(4,4)$ is the double cover of $\text{SO}(4,4)$ and $[\spinfour\times\spinfour]
/ \mathbb{Z}_2$ is a double cover of the $\text{SO}(4)\times \text{SO}(4)$ subgroup of $\text{SO}(4,4)$. 
In the world-sheet theory, one $\text{SO}(4)$ factor acts as a rotation on the left-movers and the other acts as a rotation on the right-movers. The restriction to $\text{Spin}(4,4;\mathbb{Z})$ ensures that the periodicity condition of the bosonic coordinates on $T^4$ is preserved. The group $\tilde K=\spinfour\times\spinfour
 $ is a quadruple cover of the $\text{SO}(4)\times \text{SO}(4)$.

The elements $ \pm 1 $ of $\spinfour$ both project to the 
identity $ 1 $ of $\text{SO}(4)$, while the four elements
$ (1 , 1)$, $ (1 ,- 1)$, $ (-1 , 1)$, $ (-1 ,- 1)$ of $\spinfour\times\spinfour$ all project to the 
identity $ (1 , 1)$ of $\text{SO}(4)\times \text{SO}(4)$, exhibiting the quadruple cover. Then
\begin{equation}
    \text{SO}(4)\times \text{SO}(4)\cong
 \frac {   \spinfour\times\spinfour}
{
\mathbb{Z}_2 \times \mathbb{Z}_2
}\,,
\label{equation 2.7 ref}
\end{equation}
with the subgroup $\mathbb{Z}_2 \times \mathbb{Z}_2$ consisting of
 the elements
$ (1 , 1)$, $ (1 ,- 1)$, $ (-1 , 1)$, $ (-1 ,- 1)$.
The duality group $K$ is the double cover of \eqref{equation 2.7 ref}, given by
\begin{equation}
   K=
 \frac {   \spinfour\times\spinfour}
{
\mathbb{Z}_2 
}\,,
\end{equation}
with the $\mathbb{Z}_2 $ generated by $ (-1 ,- 1)$.

All fermions in the theory transform as representations of $\spinfour\times\spinfour$
while the bosons transform as representations of $\text{SO}(4)\times \text{SO}(4)$.
The element $ (-1 ,- 1)$ of $\spinfour\times\spinfour$ leaves
all bosons invariant but multiplies each spacetime fermion by $-1$. This means that it acts as 
$(-1)^{F_s}$, where ${F_s}$ is the spacetime fermion number.
Next, $ (-1 , 1)$ acts as $-1$ on fermions transforming under the $\spinfour_L$ subgroup of 
$\spinfour_L\times\spinfour_R$ and so acts as $(-1)^{F_L}$, where $F_L$ is the corresponding fermion number. Similarly, $ (1 ,- 1)$ 
acts as $-1$ on fermions transforming under the $\spinfour_R$ subgroup of 
$\spinfour_L\times\spinfour_R$ and so acts as $(-1)^{F_R}$. Note that $F_s$ should not be confused with the world-sheet fermion number and $F_L,F_R$ should not be confused with the left and right-moving world-sheet fermion numbers, although they are of course related via the GSO projection.
Thus, there are four possible lifts to $\spinfour\times\spinfour$
of the identity in $\text{SO}(4)\times \text{SO}(4)$, given by $1,(-1)^{F_s}$,$(-1)^{F_L}$,$(-1)^{F_R}$.
Of these, $(-1)^{F_s}$ is represented trivially in $\text{Spin}(4,4;\mathbb{Z})$
by $\tilde {\mathcal{M}}= 1$, while $(-1)^{F_L}$,$(-1)^{F_R}$ are represented in 
$\text{Spin}(4,4;\mathbb{Z})$ by $\tilde {\mathcal{M}}= -1$.

We can now specify the orbifolds we will be considering here.
We choose a $ \mathcal{M} \in \text{Spin}(4,4;\mathbb{Z})$ satisfying $ \mathcal{M}^p=1$, for some $p$, and take the monodromy to be
$ \mathcal{M}$.
This then determines an element $\tilde{\mathcal{M}}\in   [(\spinfour\times\spinfour )
/ \mathbb{Z}_2]_0$ via \eqref{conjugation}. To define the transformation of the fermions, we then choose a 
  lift of  $\tilde{\mathcal{M}}$ to the double cover 
  $\hat{\mathcal{M}}\in   \spinfour\times\spinfour $.
  It is also possible to include a fermionic twist, in which case 
  the monodromy becomes $ \mathcal{M}(-1)^{F_s}$, and similarly for $F_{L/R}$.
Note that if $p$ is odd, $ (\mathcal{M}(-1)^{F_s})^p=(-1)^{F_s}$, so that this generates a 
$\mathbb{Z}_{2p}$ symmetry, while for even $p$ it generates a $\mathbb{Z}_{p}$ symmetry. 

Practically, it is often useful to define the orbifold by a choice of matrix $\hat{\mathcal{M}}\in   \spinfour\times\spinfour $ which gives the orbifold action on all the fields. This matrix is only determined up to $\spinfour\times\spinfour$ conjugation, and by conjugation we can bring it to a standard form in a convenient maximal torus $\text{SO}(2)^4$
\begin{equation}
    \hat{\mathcal{M}}= ({\mathcal{M}}_L,{\mathcal{M}}_R)\in \spinfour _L\times\spinfour _R \,,
    \label{mhat}
\end{equation}
with 
\begin{equation}
    {\mathcal{M}}_L=\begin{pmatrix}
    R(m_1) & 0  \\
    0 & R(m_3) 
    \end{pmatrix} \,,\qquad {\mathcal{M}}_R=\begin{pmatrix}
    R(m_2) & 0  \\
    0 & R(m_4) 
    \end{pmatrix} 
     \label{mhat2}\,,
\end{equation}
where
we use the notation $R(x)=\begin{psmallmatrix}\cos x & \,\,\,\,-\sin x \\ \sin x & \,\,\,\,\cos x \end{psmallmatrix}$ for a two by two rotation matrix. Here each matrix acts in the $(\textbf{2},\textbf{0})+(\textbf{0},\textbf{2})$ representation of $\spinfour\cong\text{SU}(2)\times \text{SU}(2)$. The matrices $\mathcal{M}_{L/R}$ act on spinors on $T^4$ such as the internal part of the R-vacua,  as we discuss in the next section, see \eqref{transformation of ramond vacua}.
Thus, the monodromy is specified by four angles $m_i$, and then the key step is determining what angles are allowed, i.e. for which choices of the angles $m_i$ there is a monodromy matrix $ \mathcal{M} \in \text{Spin}(4,4;\mathbb{Z})$.

 These $m_i$ are the same parameters that were used in the supergravity analysis of \cite{Hull:2020byc}, so they can be used to make contact with the results that were obtained there.  Each of the mass parameters $m_i$ gives the mass of exactly two of the eight gravitini, so the amount of preserved supersymmetry in the orbifold can be tuned by choosing the $m_i$ to be zero or non-zero.
Our orbifolds preserve $\mathcal{N}=2r$ supersymmetry, where $r$ is the number of $m_i$ that is zero (or a multiple of $2\pi$).
For $\mathcal{N}=4$ supersymmetry with two of the $m_i$ non-zero, there are two possibilities depending on whether the twist is chiral with both of the non-zero $m_i$ in either ${\mathcal{M}}_L$ or in ${\mathcal{M}}_R$, or non-chiral with one of the non-zero $m_i$ in ${\mathcal{M}}_L$ and one in ${\mathcal{M}}_R$. A chiral twist (e.g. with $m_2=m_4=0$) leads to a (1,1) supergravity theory, as the massive multiplets are (1,1) massive supermultiplets in the terminology of \cite{Hull:2000cf}. Regarding the non-chiral twists, we refer to the twist with $m_3=m_4=0$ as (0,2) theory as the massive multiplets are (0,2) massive supermultiplets, and we refer to the twist with $m_1=m_2=0$ as a (2,0) theory as the massive multiplets are (2,0) massive supermultiplets. More details on the supergravity aspects will be discussed in section \ref{Supergravity} (see also \cite{Hull:2020byc}).

Using the isomorphism $\spinfour\cong\text{SU}(2)\times \text{SU}(2)$ (see appendix \ref{app: group theory} and in particular \eqref{embedding SO4}) the matrix $\hat{\mathcal{M}}\in   \spinfour\times\spinfour $ projects onto a matrix 
\begin{equation}
    {\mathcal{M}}_\theta= ({\mathcal{N}}_L,{\mathcal{N}}_R)\in  \text{SO}(4)_L\times \text{SO}(4)_R \ ,
    \label{rottheta}
\end{equation}
 which (by conjugation) can be brought to the standard form 
\begin{equation}
    {\mathcal{N}}_L=\begin{pmatrix}
    R(\theta_L) & 0  \\
    0 & R(\theta'_L) 
    \end{pmatrix} \,,\qquad
    {\mathcal{N}}_R=\begin{pmatrix}
    R(\theta_R) & 0  \\
    0 & R(\theta'_R) 
    \end{pmatrix} \ ,
    \label{standarformM}
\end{equation}
for four angles $\theta_L,\theta '_L,\theta_R,\theta'_R$ which are related to the $m_i$ by 
\begin{equation}
\begin{alignedat}{4}
\theta_L &=  {m_1+m_3}  \,, &\qquad\quad \theta_R &= {m_2+m_4}  \,, \\[3pt]
\theta '_L &= {m_1-m_3}  \,, &\qquad\quad \theta'_R &= {m_2-m_4}  \,.
\end{alignedat}
\label{theta's}
\end{equation}
Then e.g. $m_1= \frac 1 2 (\theta_L+\theta'_L)$ so that taking $\theta'_L\to \theta'_L+2\pi$ takes $m_1 $ to $m_1+\pi$ and $R(m_1)$ to $-R(m_1)$.
Note that $(m_1,m_2,m_3,m_4)$, $(m_1+\pi,m_2,m_3+\pi,m_4)$, $(m_1,m_2+\pi,m_3,m_4+\pi)$, $(m_1+\pi,m_2+\pi,m_3+\pi,m_4+\pi)$ all give the same angles 
 $\theta_L,\theta '_L,\theta_R,\theta'_R$ (mod $2\pi$), exhibiting the quadruple cover.

\subsection{An ansatz}

Solving in general for all possible integer-valued T-duality elements that are conjugate to a rotation is a  difficult problem. Here we consider a special case for which results are known in the literature. 
For the case $G= \text{SL}(2,\mathbb{R})$, $K=\text{SO}(2)$ all
monodromies satisfying
\begin{equation}\label{conjugationsl2}
\mathcal{M} = g \tilde{\mathcal{M}} g^{-1} \,; \qquad\quad
\mathcal{M}\in\text{SL}(2,\mathbb{Z})\,,\qquad
\tilde{\mathcal{M}}\in\sotwo \,, \qquad g\in\sltwo \,
\end{equation}
are given in \cite{Dabholkar:2002sy,DeWolfe:1998eu}.
The $\tilde{\mathcal{M}}\in\sotwo$ for which this is possible are rotations by angles 
\begin{equation}\label{quantalpha}
\alpha\in\big\{0,\pm\tfrac{\pi}{3},\pm\tfrac{\pi}{2},\pm\tfrac{2\pi}{3},\pi\big\} \quad {\rm mod}\,\, 2\pi\ .
\end{equation}
Such rotations then generate a $\mathbb{Z}_2$, $\mathbb{Z}_3$, $\mathbb{Z}_4$ or $\mathbb{Z}_6$ subgroup of $\sltwo$ generated by $\tilde{\mathcal{M}}\in\sotwo$ and these in turn yield a $\mathbb{Z}_2$, $\mathbb{Z}_3$, $\mathbb{Z}_4$ or $\mathbb{Z}_6$ subgroup of $\text{SL}(2,\mathbb{Z})$
generated by $\mathcal{M}$.

The group $\text{SO}(4,4)$ has a subgroup $\text{SO}(2,2)\times \text{SO}(2,2)$ as well as a subgroup $\text{SO}(4)\times \text{SO}(4)$, and the group theory regarding these two subgroups is very similar.
Now, $\spinfourfour$ has a subgroup that is a double cover of this $\text{SO}(2,2)\times \text{SO}(2,2)$
given by 
\begin{equation}
  \frac {\spintwotwo\times\spintwotwo}   {\mathbb{Z}_2}\subset \spinfourfour\,.
\end{equation}
As
\begin{equation}\label{sl2subgroups}
\sltwo^2 \cong \spintwotwo\,,
\end{equation}
we see that $\spinfourfour$ has a subgroup
\begin{equation}\label{sl2subgroups monodromy}
\frac {\sltwo^4}   {\mathbb{Z}_2}
 \subset \spinfourfour \,.
\end{equation}
The quotient by ${\mathbb{Z}_2}$ reflects the fact that the element $(-{1},-{1},-{1},-{1})\in {\sltwo^4} $ maps to the identity in $\spinfourfour$.
We will then restrict our monodromy to lie in this subgroup
and use the known results for $\sltwo$. We   stress that these are not   all possible $[\spinfour\times\spinfour]
/ \mathbb{Z}_2$ rotations that can be conjugated to integer-valued elements of the T-duality group $\text{Spin}(4,4;\mathbb{Z})$; but these are the ones that follow from our ansatz.

Restricting   the  subgroup $\sltwo^4  /\mathbb{Z}_2 \subset \spinfourfour$  to a subgroup of 
$\text{Spin}(4,4;\mathbb{Z})$, restricts us to a discrete subgroup of $\sltwo^4$ and we need to check what subgroup arises in this way.
The $\textbf{8}_v $ of $\text{Spin}(4,4)$ restricts to a $(\textbf{4},\textbf{1})+(\textbf{1},\textbf{4})$ of $\spintwotwo\times\spintwotwo $ and as the vector representation $\textbf{4}$ of $\spintwotwo$
is the $(\textbf{2},\textbf{2}) $ representation of $\sltwo^2$, the $\textbf{8}_v$ is in the
$$(\textbf{2},\textbf{2},\textbf{1},\textbf{1})+(\textbf{1},\textbf{1},\textbf{2},\textbf{2})$$
representation of $\sltwo^4$. The $\textbf{8}_s$ representation of Spin(4,4) is in the 
$$(\textbf{1},\textbf{2},\textbf{2},\textbf{1})+(\textbf{2},\textbf{1},\textbf{1},\textbf{2})$$
representation of $\sltwo^4$. Thus the 16 charges are in the 
\begin{equation}(\textbf{2},\textbf{2},\textbf{1},\textbf{1})+(\textbf{1},\textbf{1},\textbf{2},\textbf{2})+(\textbf{1},\textbf{2},\textbf{2},\textbf{1})+(\textbf{2},\textbf{1},\textbf{1},\textbf{2})\,.
\label{chargereps}
\end{equation}
Choosing 16 basis vectors $e_i$ for the 16-dimensional charge lattice, the allowed charges are $n^i e_i$ where $n^i$ is a 16-vector of integers, $n^i\in \mathbb{Z}^{16}$.
The group $\sltwo^4$ acts on the integers $n^i$ in the representation (\ref{chargereps})
and will preserve the lattice if restricted to the discrete subgroup 
$\text{SL}(2,\mathbb{Z})^4$.

Then the monodromy $\mathcal{M}\in\text{SL}(2,\mathbb{Z})^4/\mathbb{Z}_2$ 
is conjugate to a rotation in $\text{SO}(2)^4$ specified (up to conjugation) by four angles $\alpha _i\in \big\{0,\pm\tfrac{\pi}{3},\pm\tfrac{\pi}{2},\pm\tfrac{2\pi}{3},\pi\big\}$ mod $2\pi$, where $i=1,\ldots,4$. 
The element $(-1,-1,-1,-1)\in {\sltwo^4} $ with trivial monodromy corresponds to
$ \alpha_1=\alpha_2=  \alpha_3=\alpha_4 $; this will play a role in the ${\cal N}=6$ supersymmetric cases discussed below.

The angles $\alpha _i$ are related to the angles $\theta_i$ by (for details see appendix \ref{Ap:embedding})
\begin{equation}
\begin{alignedat}{4}
\theta_L &=  {\alpha_1+\alpha_3}  \,, &\qquad\quad \theta_R &= {\alpha_1-\alpha_3}  \,, \\[3pt]
\theta '_L &= {\alpha_2+\alpha_4}  \,, &\qquad\quad \theta'_R &= {\alpha_2-\alpha_4}  \,.
\end{alignedat}
\label{thetaalph's}
\end{equation}
This is similar in form to (\ref{theta's}) and so we see again that the $\text{SO}(2)^4$ parameterised by the $\alpha_i$ provides a quadruple cover of the $\text{SO}(2)^4$ parameterised by the $\theta_i$.
Comparing \eqref{thetaalph's} with (\ref{theta's}), we see that
\begin{equation}
\begin{alignedat}{4}
{m_1+m_3} &=  {\alpha_1+\alpha_3}  \,, &\qquad\quad {m_2+m_4} &= {\alpha_1-\alpha_3}  \,, \\[3pt]
{m_1-m_3}  &= {\alpha_2+\alpha_4}   \,, &\qquad\quad {m_2-m_4} &= {\alpha_2-\alpha_4}  \,.
\end{alignedat}
\label{malph}
\end{equation}
All the equations relating angles hold modulo $2\pi$. The equations (\ref{malph}) then determine the mass parameters $m_i$, giving one solution as
\begin{equation}
\begin{aligned}\label{m's in terms of alpha's}
m_1&=\tfrac{1}{2}(\alpha_1+\alpha_2+\alpha_3+\alpha_4) \,,\qquad\quad
&m_2=\tfrac{1}{2}(\alpha_1+\alpha_2-\alpha_3-\alpha_4) \,,\\[4pt]
m_3&=\tfrac{1}{2}(\alpha_1-\alpha_2+\alpha_3-\alpha_4) \,,\qquad\quad
&m_4=\tfrac{1}{2}(\alpha_1-\alpha_2-\alpha_3+\alpha_4) \,.
\end{aligned}
\end{equation}
For a given set of $\alpha_i$, the complete set of solutions is given by the 
 $(m_1,m_2,m_3,m_4)$ in (\ref{m's in terms of alpha's}), together with $(m_1+\pi,m_2,m_3+\pi,m_4)$, 
  $(m_1,m_2+\pi,m_3,m_4+\pi)$, $(m_1+\pi,m_2+\pi,m_3+\pi,m_4+\pi)$. All these yield the same angles 
 $\theta_L,\theta '_L,\theta_R,\theta'_R$, giving a quadruple cover.
Then for a given set of $\alpha_i$, there are four possible choices of $\hat {\mathcal{M}} $. Denoting the canonical $\hat {\mathcal{M}} $ (given by (\ref{mhat}),(\ref{mhat2}) with the $m$'s given by (\ref{m's in terms of alpha's})) by $\hat {\mathcal{M}}(m) $, the four choices are
$\hat {\mathcal{M}}(m) $, $\hat {\mathcal{M}} (m)(-1)^{F_s}$, $\hat {\mathcal{M}} (m)(-1)^{F_L}$ and $\hat {\mathcal{M}}(m)(-1)^{F_R} $.

The results just stated come from requiring agreement of the two different parameterisations of $\mathcal{M}_\theta$ in terms of the $m$'s or the $\alpha$'s respectively. Half of this ambiguity is lifted by requiring agreement for the two different parameterisations of $\tilde {\mathcal{M}}$. Then  a given set of $\alpha$'s determines a  
$\tilde {\mathcal{M}}$, which we denote $\tilde {\mathcal{M}}(\alpha)$ and is conjugate to a monodromy $ {\mathcal{M}}(\alpha)$.
For this,
there remain two possible choices of $\hat {\mathcal{M}} $, 
which are $\hat {\mathcal{M}}(m) $ and $\hat {\mathcal{M}} (m)(-1)^{F_s}$, corresponding to a choice of generalized spin structure (see the discussion of fermionic monodromies below). We choose the generalized spin structure so that 
the monodromy ${\mathcal{M}}(\alpha)$ gives the twist $\hat {\mathcal{M}}(m) $
and ${\mathcal{M}}(\alpha)(-1)^{F_s}$ gives the twist $\hat {\mathcal{M}}(m) (-1)^{F_s}$.

It will be useful to note that (\ref{m's in terms of alpha's}) can be inverted to give 
\begin{equation}
\begin{aligned}
\alpha_1&=\tfrac{1}{2}(m_1+m_2+m_3+m_4) \,,\qquad\quad
&\alpha_2=\tfrac{1}{2}(m_1+m_2-m_3-m_4) \,,\\[4pt]
\alpha_3&=\tfrac{1}{2}(m_1-m_2+m_3-m_4) \,,\qquad\quad
&\alpha_4=\tfrac{1}{2}(m_1-m_2-m_3+m_4) \,.
\end{aligned}
\label{alpha's in terms of m's}
\end{equation}
As the allowed values for each of the $\alpha_i$ are $\big\{0,\pm\tfrac{\pi}{3},\pm\tfrac{\pi}{2},\pm\tfrac{2\pi}{3},\pi\big\}$ mod $2\pi$, the allowed values of the $m_i$ can be found by taking linear combinations of these. It will often be useful to rewrite these parameters as
\begin{equation}\label{miN}
m_i = \frac{2\pi N_i}{p} \,.
\end{equation}
Here the $N_i$ are integers, and $p$ is the smallest positive integer such that all four $m_i$ can be written like this. This relation defines the quantization condition on the mass parameters and guarantees that
the integer $p$ is the order of the monodromy matrix, $\hat {\cal M}^p=1$.

As follows from \eqref{m's in terms of alpha's}, the quantization of the $\alpha$'s allows for the values $p=2,3,4,6,8$, $12$ and $24$. Notice that the values $p=5$ and $p=10$ found in \cite{Erler:1992ki} and mentioned in the beginning of this section do not appear in our list. This is because the duality twists from our ansatz lie in a particular $\text{SL}(2)^4$ subgroup given by \eqref{sl2subgroups monodromy}, and the values $p=5$ and $10$ do not arise from this subgroup.

 In \autoref{allowed orbifolds} we list which values of $p$ yield orbifolds that preserve a certain amount of supersymmetry. The possible values of $p$ are determined based on the particular ansatz \eqref{sl2subgroups monodromy}. Recall that our orbifolds preserve $\mathcal{N}=2r$ supersymmetry where $r$ is the number of $m_i$ that is zero (mod $2\pi$).  Notice that fixing $p$ does not fix the orbifold. For a given $p$, there can be more than one possibility. An example is $\mathcal{N}=4\,(0,2)$, with $p=4$, for which there can be both a symmetric and an asymmetric orbifold. 
\renewcommand{\arraystretch}{1.4}
\begin{table}[h]
\centering
\begin{tabular}{|c|c|c|}
\hline
\;Preserved supersymmetry\; & \;(A)symmetric\; & \;Possible $\mathbb{Z}_p$ orbifold ranks\; \\ \hline\hline
$\mathcal{N}=6$ & A & $p=2,3$ \\ \hline
 \multirow{2}{*}{$\mathcal{N}=4\;(0,2)$} & S & $p=2,3,4,6$ \\
& A & $p=3,4,6,12$ \\ \hline
$\mathcal{N}=4\;(1,1)$ & A & $p=2,3,4,6,12$ \\ \hline
$\mathcal{N}=2$ & A & $p=2,3,4,6,12$ \\ \hline
 \multirow{2}{*}{$\mathcal{N}=0$} & S & $p=2,3,4,6,8,12,24$ \\
& A & $p=3,4,6,8,12,24$ \\ \hline
\end{tabular}
\captionsetup{width=.83\linewidth}
\caption{\textit{All possible orbifolds following from the ansatz \eqref{sl2subgroups monodromy}. Here we indicate the amount of preserved supersymmetry and the rank of the orbifold. We also indicate which orbifolds are symmetric (S, $m_1=m_2$ and $m_3=m_4$) or asymmetric (A). The $\mathcal{N}=4\,(0,2)$ is defined from having $m_3=m_4=0$ whereas the $(1,1)$ is defined from $m_2=m_4=0$, up to trivial permutations exchanging $m_{1,2}$ with $m_{3,4}$. Notice the absence of $p=2$ for asymmetric $\mathcal{N}=4\,(0,2)$ orbifolds. In this case we have $m_{1,2}=\pm \pi$ but $+\pi$ is the same as $-\pi$, and so this is a symmetric orbifold in disguise.}}
\label{allowed orbifolds}
\end{table}
\renewcommand{\arraystretch}{1}
\subsection{Special cases and examples}

The action of the group $\text{SL}(2)^4$ on $T^4$ we are considering splits it into the product of   $T^2\times T^2$ with the $\text{SL}(2)^2$ parameterised by $\alpha_1,\alpha _3$ acting on one $T^2$ and the   $\text{SL}(2)^2$ parameterised by $\alpha_2,\alpha _4$ acting on the other $T^2$. 

\subsubsection*{Fermionic monodromies $(-1)^{F_s}$,$(-1)^{F_L}$,$(-1)^{F_R}$}
\label{fermionic monodromies}

Here we consider monodromies such that ${\mathcal{M}}_\theta=1 $
(i.e. all angles $\theta_i=(\theta_L,\theta'_L,\theta_R,\theta'_R)$ are $0$ mod $2\pi$),
so that the NS-NS sector is invariant and the twist only acts on fermions and on the R-R sector. For $\theta_i=0$ mod $2\pi$, each of the $m_i$ and each of the $\alpha _i$ must be either $0$ or $\pi $ (mod $2\pi$).
Consider for instance the twist 
\begin{equation}
     m_1=m_3=\pi, \quad m_2=m_4=0 \ ,
\end{equation}
so that $\theta_L=2\pi, \theta'_L=0$ and $\theta_R=\theta'_R=0$. The $\alpha_i$ are given by $\alpha_1=\alpha_3=\pi,\quad \alpha_2=\alpha_4 = 0$.
This lifts to a monodromy on the double cover
\begin{equation}
    \hat{\mathcal{M}}= ({\mathcal{M}}_L,{\mathcal{M}}_R)=(- 1, 1)\ ,
\end{equation}
so that $ \hat{\mathcal{M}}=(-1)^{F_L}$. Choosing instead $m_1=m_3=0$ and $m_2=m_4=\pi$, we get $\hat{\mathcal{M}}= ( 1, -1)$, so that $ \hat{\mathcal{M}}=(-1)^{F_R}$.

We now consider a monodromy with
\begin{equation}
m_1=m_2=m_3=m_4=\pi\,.
\end{equation}
Then
\begin{equation}
    \hat{\mathcal{M}}= ({\mathcal{M}}_L,{\mathcal{M}}_R)=(- 1, -1)\,,
\end{equation}
so that $ \hat{\mathcal{M}}=(-1)^{F_s}$. This has a trivial projection to $\tilde K$: $\tilde {\mathcal{M}}=1$.
This then corresponds to ${\mathcal{M}}=1$ so that orbifold with $m_1=m_2=m_3=m_4=\pi$
has
a monodromy ${\mathcal{M}}(-1)^{F_s}=(-1)^{F_s}$.

The orbifold of the type IIB string by $(-1)^{F_s}$ gives the non-supersymmetric type 0B string \cite{Dixon:1986iz,Seiberg:1986by}. Here we are combining the $(-1)^{F_s}$ with a half-shift on the $S^1$ to give an interesting $\mathbb{Z}_2$ orbifold breaking all supersymmetry, which can be made tachyon-free. We will discuss it further in the next.

\subsubsection*{Symmetric orbifold of $T^2$}

Choosing
\begin{equation}
     \alpha_2= \alpha_3=\alpha_4 = 0 \qquad\Rightarrow\qquad
   \theta_L=\theta_R = \alpha_1; \qquad \theta'_L=\theta'_R=0\,.
\end{equation}
Then (\ref{m's in terms of alpha's}) gives
\begin{equation}
    m_1=m_2= m_3=m_4= \frac{\alpha_1}{2} \,.
\end{equation}
This is a symmetric orbifold with $\alpha_1$ parameterizing an $\text{SO}(2)$ acting as a rotation on one $T^2$. 
For $\alpha_1 =2\pi/n $, ${\mathcal{M}}^n=1$,
as ${\mathcal{M}}^n$ is a rotation through $2\pi$, but $\hat {{\mathcal{M}}}^n=-1$, as the double cover of the $2\pi $ rotation acts as $-1$ on spinors. Thus, this constitutes a $\mathbb{Z}_{p}$ orbifold with $p=2n$. From the allowed values of $\alpha_1$, we can read of the allowed values of $n$, which are $n=1,2,3,4,6$, such that $p=2,4,6,8,12$. As all $m_i\ne 0$, this orbifold breaks all supersymmetry.

\subsubsection*{Symmetric orbifold of $T^4$}

Choosing
\begin{equation}
    \alpha_3=\alpha_4 = 0 \qquad\Rightarrow\qquad \theta_L= \theta_R=  \alpha_1, \qquad \theta'_L= \theta'_R=  \alpha_2
     \,,
\end{equation}
and (\ref{m's in terms of alpha's}) gives
\begin{equation}
m_1=m_2=\tfrac 1 2 (\alpha_1+\alpha_2), \qquad m_3=m_4=\tfrac 1 2 (\alpha_1-\alpha_2)\,.
\end{equation}
Recall that the allowed values for each of the $\alpha_1,\alpha_2$ are $\big\{0,\pm\tfrac{\pi}{3},\pm\tfrac{\pi}{2},\pm\tfrac{2\pi}{3},\pi\big\}$ mod $2\pi$.
For most choices of $\alpha_{1,2}$, all $m_i\ne 0$, so this orbifold breaks all supersymmetry. When $\alpha_1=\alpha_2$, we have $\mathcal{N}=4$ supersymmetry of type $(0,2)$.

\subsubsection*{Example: a $\mathbb{Z}_{24}$ orbifold}
\label{a Z24 orbifold}
We now   give an explicit example of a $\mathbb{Z}_{24}$ orbifold arising as above, i.e. as a symmetric orbifold of $T^4$. Note that symmetric $\mathbb{Z}_{24}$ orbifolds always break all supersymmetry. The example that we choose is
\begin{equation}\label{parameters Z24}
    \alpha_1=\frac{\pi}{2},\quad \alpha_2=\frac{\pi}{3},\quad \alpha_3=\alpha_4 = 0 \qquad\Rightarrow\qquad m_1=m_2=\frac{5\pi}{12}, \quad m_3=m_4=\frac{\pi}{12} \,.
\end{equation}
The monodromies in the SL$(2,\mathbb{Z})$ subgroups \eqref{sl2subgroups monodromy} that the $\alpha$'s rotate in are known. They can be found e.g. in \cite{Dabholkar:2002sy}. The ones corresponding to $\alpha_1$ and $\alpha_2$ respectively read
\begin{equation}\label{sl2monodromies Z24}
    \begin{pmatrix}
    0 & -1 \\
    1 & 0
    \end{pmatrix} \,, \qquad
    \begin{pmatrix}
    0 & -1 \\
    1 & 1
    \end{pmatrix}\,.
\end{equation}
The first one is simply a rotation matrix over an angle $\alpha_1={\pi}/{2}$, and the second one is conjugate to a rotation matrix over an angle $\alpha_2={\pi}/{3}$ via a conjugation \`a la \eqref{conjugationsl2}:
\begin{equation}
    \begin{pmatrix}
    0 & -1 \\
    1 & 1
    \end{pmatrix} = \sqrt{\frac{2}{\sqrt{3}}}\begin{pmatrix}
    1 & 0 \\
    -\tfrac{1}{2} & \tfrac{\sqrt{3}}{2}
    \end{pmatrix} \cdot \begin{pmatrix}
    \cos \tfrac{\pi}{3} & -\sin \tfrac{\pi}{3}\\
    \sin \tfrac{\pi}{3} & \cos \tfrac{\pi}{3}
    \end{pmatrix} \cdot \left[ \sqrt{\frac{2}{\sqrt{3}}}\begin{pmatrix}
    1 & 0 \\
    -\tfrac{1}{2} & \tfrac{\sqrt{3}}{2}
    \end{pmatrix} \right]^{-1} \;.
\end{equation}
We can use appendix \ref{isomorphism so22} to map the $\sltwo$ matrices \eqref{sl2monodromies Z24} properly to $\sotwotwo$ matrices, which can then be combined into an $\sofourfour$ element. This yields the monodromy
\begin{equation}\label{monodromy Z24}
    \mathcal{M} =\begin{pmatrix}
    0 & -1 & 0 & 0 & 0 & 0 & 0 & 0 \\
    1 & 0 & 0 & 0 & 0 & 0 & 0 & 0 \\
    0 & 0 & 0 & -1 & 0 & 0 & 0 & 0 \\
    0 & 0 & 1 & 1 & 0 & 0 & 0 & 0 \\
    0 & 0 & 0 & 0 & 0 & -1 & 0 & 0 \\
    0 & 0 & 0 & 0 & 1 & 0 & 0 & 0 \\
    0 & 0 & 0 & 0 & 0 & 0 & 1 & -1 \\
    0 & 0 & 0 & 0 & 0 & 0 & 1 & 0
    \end{pmatrix} \in\sofourfour  \,.
\end{equation}
This monodromy is written in $\tau$-frame (using the language from appendix \ref{app: eta and tau frame}), meaning that the group $\sofourfour$ consists of matrices that preserve the metric
\begin{equation}
    \tau = \begin{pmatrix}
    0 & 1_{4\times4} \\
    1_{4\times4} & 0 
    \end{pmatrix} \,.
\end{equation}
We see that, in this frame, the monodromy is integer-valued as it should be\footnote{This is the case because the T-duality group works on integer-valued charges (winding and momentum numbers) in $\tau$-frame.}. 
Notice that the monodromy acts as a diffeomorphism on $T^4$, as it should, since we consider a symmetric orbifold. The geometric group GL($4,\mathbb{Z}$) is embedded in the T-duality group SO(4,4,$\mathbb{Z}$) as
\begin{equation}
   \mathcal{M}=  \begin{pmatrix}
    g & 0 \\
    0 & g^{-t}
    \end{pmatrix}\,,\qquad g = \begin{pmatrix}
    0 & -1 & 0 & 0 \\
    1 & 0 & 0 & 0  \\
    0 & 0 & 0 & -1  \\
    0 & 0 & 1 & 1  \\
    \end{pmatrix}\,,\qquad g \in \text{GL}(4,\mathbb{Z})\,.
\end{equation}
Note that the monodromy in \eqref{monodromy Z24} generates an orbit of rank 12: $\mathcal{M}^{12}=1$. 
The action on the fermions is through the matrix $\hat{\mathcal{M}}$ given by (\ref{mhat}),(\ref{mhat2}) with the $m$'s in \eqref{parameters Z24} and this generates an orbit of rank 24 as it satisfies $\hat{\mathcal{M}}^{24}=1$.

\subsubsection*{Asymmetric orbifold of $T^2$}

Choosing
\begin{equation}
    \alpha_2=\alpha_4 = 0 \qquad\Rightarrow\qquad \theta_L= \alpha_1+\alpha_3 ,\qquad
   \theta_R= \alpha_1-\alpha_3 ; \qquad \theta'_L=\theta'_R=0 \,,
\end{equation}
and (\ref{m's in terms of alpha's}) gives
\begin{equation}
  m_1=m_3=\tfrac{1}{2}(\alpha_1+\alpha_3), \qquad m_2=m_4=\tfrac{1}{2}(\alpha_1-\alpha_3)\,.
\end{equation}
As discussed previously,    $\alpha_1,\alpha_3\in\big\{0,\pm\tfrac{\pi}{3},\pm\tfrac{\pi}{2},\pm\tfrac{2\pi}{3},\pi\big\}$ mod $2\pi$.
For example, choosing
$\alpha_1=\pi/2,\alpha_3=\pi/3$ gives $m_1=m_3=5\pi/12$ and $m_2=m_4=\pi/12$ giving a $\mathbb{Z}_{24}$ orbifold. 
Again, as all $m_i\ne 0$, this orbifold breaks all supersymmetry. We can preserve some supersymmetry e.g. by choosing $\alpha_1=\alpha_3$, so that $m_2=m_4=0$ and there is $\mathcal{N}=4$ supersymmetry of type $(1,1)$.

\subsubsection*{Chiral orbifold of $T^4$}
Choosing
\begin{equation}
  \alpha_1=\alpha_3,\quad  \alpha_2=\alpha_4  \qquad\Rightarrow\qquad \theta_L= 2\alpha_1  ,\qquad \theta'_L= 2\alpha_2,\qquad 
   \theta_R=\theta'_R=0 \,,
   \label{chiral orbifold alphas}
\end{equation}
and (\ref{m's in terms of alpha's}) gives
\begin{equation}
  m_1=\alpha_1+\alpha_2,\qquad
  m_3=\alpha_1-\alpha_2, \qquad m_2=m_4=0\,.
  \label{chiral orbifold m's}
\end{equation}
This type of orbifold can preserve either $\mathcal{N}=6$ or $\mathcal{N}=4$ supersymmetry. We now look at these cases separately.

\subsubsection*{$\mathcal{N}=6$ supersymmetric orbifold}
\label{n=6 allows only p=2 or 3}

For $\mathcal{N}=6$ supersymmetry, precisely one of the $m_i$ should be non-zero. Choosing this to be $m_1\ne 0$ with $m_2=m_3=m_4=0$ requires (using \eqref{chiral orbifold alphas},\eqref{chiral orbifold m's})
\begin{equation}
  \alpha_1=\alpha_2=  \alpha_3=\alpha_4  = \tfrac 1 2  m_1
  \qquad\Rightarrow\qquad \theta_L= \theta'_L= 2\alpha_1 =m_1 ,\qquad  
   \theta_R=\theta'_R=0 \,,
\end{equation}
so that this is a chiral orbifold. Recall that
$\alpha_1=m_1/2\in\big\{0,\pm\tfrac{\pi}{3},\pm\tfrac{\pi}{2},\pm\tfrac{2\pi}{3},\pi\big\}$ mod $2\pi$.

The factor of two in the relation between $\alpha_i$ and $m_1$ is important.
The case $\alpha_1 =\pi$ gives $m_1=2\pi$ as well as $\theta_L= \theta'_L=2 \pi$, so that $\hat {\mathcal {M}}=1$ and $ \mathcal {M}=1$ so that 
the monodromy is trivial. This reflects the fact that
the monodromy is not in ${\sltwo^4}$ but in ${\sltwo^4}/   {\mathbb{Z}_2}$. Similarly, $\alpha _1=\pi/2$ gives a $\mathbb{Z}_2$ symmetry instead of the $\mathbb{Z}_4$ that might have been expected. 
Finally, $\alpha_1={\pi}/{3}$ gives a $\mathbb{Z}_3$ symmetry instead of a $\mathbb{Z}_6$ symmetry, while $\alpha_1={2\pi}/{3}$
also gives a $\mathbb{Z}_3$ symmetry. Thus, the only possible values of $p$ from our ansatz \eqref{sl2subgroups monodromy} are 2 and 3. However, as we will discuss later, it is possible to construct a consistent $\mathbb{Z}_4$ model, preserving $\mathcal{N}=6$ supersymmetry in five dimensions. (The construction will be based on the lattice approach, which we will discuss in section \ref{lattice construction}.)

\subsubsection*{$\mathcal{N}=4$ chiral supersymmetric orbifold}
\label{n=4 chiral orbis}

For an orbifold with $\mathcal{N}=4$ chiral supersymmetry we need to turn on two mass parameters. Requiring $m_1,m_3\neq 0$ and $m_2=m_4=0$ gives (using \eqref{chiral orbifold alphas},\eqref{chiral orbifold m's})
\begin{equation}
\begin{aligned}
 & \alpha_1=\alpha_3= \tfrac 1 2 (m_1+m_3), \quad  
  \alpha_2=\alpha_4  = \tfrac 1 2 (m_1-m_3)
  \quad\Rightarrow\\
  &\theta_L= m_1+m_3, \quad   \theta'_L= m_1-m_3 ,\quad  
   \theta_R=\theta'_R=0 \,.
   \end{aligned}
\end{equation}
This choice of mass parameters leads to a (1,1) theory.

\subsubsection*{Non-chiral supersymmetric orbifolds of $T^4$}
We discuss two examples of this type, preserving $\mathcal{N}=4$ and $\mathcal{N}=2$ supersymmetry respectively.
\subsubsection*{$\mathcal{N}=4$ non-chiral supersymmetric orbifold}
\label{non-chiral N=4}
Requiring $m_1,m_2\neq 0$ and $m_3=m_4=0$ gives (using \eqref{alpha's in terms of m's})
\begin{equation}
\begin{aligned}
  &\alpha_1=\alpha_2= \tfrac 1 2 (m_1+m_2), \quad  
  \alpha_3=\alpha_4  = \tfrac 1 2 (m_1-m_2)
  \quad\Rightarrow\\
  &\theta_L=\theta'_L= m_1, \quad   
   \theta_R=\theta'_R=m_2 \,.
   \end{aligned}
\end{equation}
This choice of mass parameters leads to a (0,2) theory. In general, this is an asymmetric orbifold, but for the special choice $m_1=m_2$, the orbifold is symmetric.

\subsubsection*{$\mathcal{N}=2$ supersymmetric orbifold}
\label{non-chiral N=2}
For $\mathcal{N}=2$, we turn on three mass parameters and the remaining one is zero. Requiring e.g. $m_4=0$ gives (using \eqref{alpha's in terms of m's})
\begin{equation}
\begin{aligned}
\alpha_1&=\tfrac{1}{2}(m_1+m_2+m_3) \,,\qquad\quad
\alpha_2&=\tfrac{1}{2}(m_1+m_2-m_3) \,,\\[4pt]
\alpha_3&=\tfrac{1}{2}(m_1-m_2+m_3) \,,\qquad\quad
\alpha_4&=\tfrac{1}{2}(m_1-m_2-m_3) \,.
\end{aligned}
\end{equation}
so that
\begin{equation}
 \theta_L= m_1+m_3, \quad   \theta'_L= m_1-m_3, \qquad   
   \theta_R=\theta'_R=m_2 \,.
\end{equation}
Notice that this is always an asymmetric orbifold. We will discuss some explicit examples of non-chiral orbifolds in section \ref{closed string spectrum}.

\section{Lattice approach}
\label{The orbifold action and partition function}

\subsection{Lattices and tori}

Our starting point was a torus compactification on a square torus $\mathbb{R}^4/\mathbb{Z}^4$ with periodic torus coordinates so that the
BPS 0-brane charge lattice was preserved by Spin$(4,4;\mathbb{Z})$.
The  moduli were packaged into the background metric and antisymmetric tensor gauge fields  on the torus and the fixed point under the action of the monodromy was at a point $[\bar g]$ in the moduli space.
Acting with a duality transformation $\bar g$ in $G$ moves the fixed point to the origin and diagonalizes the action of the monodromy on the fields. However, it also deforms the torus, so that it is no longer a square torus and the boundary conditions of the torus coordinates are changed, so that the left-moving coordinates take values on a torus $\mathbb{R}^4/\Lambda_l$ for some lattice $\Lambda_l$ and the right-moving coordinates take values on a torus $\mathbb{R}^4/\Lambda_r$ for some lattice $\Lambda_r$. 
The left-moving momenta $p_L$ take values in the lattice dual to $\Lambda_l$ and the right-moving momenta $p_R$ take values in the lattice dual to $\Lambda_r$. Then the vectors $(p_L,p_R)$ build an 8-dimensional even, self-dual Lorentzian lattice, which is known as the Narain lattice \cite{narain1989new} and we will denote by $\Gamma^{4,4}$. The sublattices $\Lambda_l,\Lambda_r$ of $\Gamma^{4,4}$ are invariant under the action of the ${\mathbb{Z}} _p$ symmetry that is used in the orbifold. These sublattices are associated with root lattices of Lie algebras\footnote{For a thorough discussion on lattices we refer to \cite{lerchie1989lattices}.}.

For the ansatz of the last section, the monodromy is in 
\begin{equation}
  \frac {\spintwotwo\times\spintwotwo}   {\mathbb{Z}_2}\subset \spinfourfour \ .
\end{equation}
This means that the $T^4$ can be regarded as $T^2\times T^2$ with one $\spintwotwo$ factor in the monodromy acting on the first $T^2$ and the other acting on the second $T^2$.
In this case, the two four-dimensional lattices $\Lambda_l,\Lambda_r$ must each decompose into the sum of two 2-dimensional lattices: $\Lambda_l=\Lambda_1\oplus \Lambda_2$ and similarly for $\Lambda_r$.
Each 2-dimensional lattice must then be $A_2$  or $A_1\oplus A_1\cong D_2$\footnote{$D_n$ and $A_n$ denote the root lattices of $\text{SO}(2n)$ and $\text{SU}(n+1)$ respectively.}.
However, below we will also consider other 4-dimensional lattices $\Lambda_l,\Lambda_r$ that fall outside the scope of the ansatz of section \ref{sec:Orbifold constructions}. This allows further possible values of $p$, e.g. $p=5$, and in some cases it is necessary for modular invariance, as we will show in detail. In order to handle these more general cases, it will be convenient to use the lattice approach for the construction of orbifolds.

\subsection{Lattice construction}
\label{lattice construction}

The construction of orbifolds based on the lattice approach was first introduced in  \cite{Narain:1986qm,narain1991asymmetric} and it is particularly useful when constructing asymmetric orbifolds. Recall that, upon compactification on $T^4$ the left and right-moving momenta take values in a Narain lattice $\Gamma^{4,4}$. At special points in the moduli space we can construct a Narain lattice admitting purely left and right-moving  symmetries as
\begin{equation}
    \Gamma^{4,4}(\mathcal{G}) \equiv \{(p_L,p_R)|\,p_L \in \Lambda_W(\mathcal{G}),\, p_R \in \Lambda_W(\mathcal{G}),\, p_L-p_R \in \Lambda_R(\mathcal{G})\}\,.
\end{equation}
Here $\mathcal{G}$ is a Lie algebra of rank four and $\Lambda_W(\mathcal{G})$, $\Lambda_R(\mathcal{G})$ denote the weight and root lattices of $\mathcal{G}$, respectively. In order to construct a freely acting orbifold we compactify on an additional $S^1$ and the Narain lattice that we consider is
\begin{equation}
    \Gamma^{5,5}=\Gamma^{4,4}(\mathcal{G})\oplus\Gamma^{1,1}\,.
\end{equation}
Now, the orbifold acts as a rotation on $\Gamma^{4,4}(\mathcal{G})$ and as a shift on $\Gamma^{1,1}$. Here we only consider rotations $\mathcal{M}_{\theta}=(\mathcal{N}_L,\mathcal{N}_R) \in \text{SO}(4)_L\times \text{SO}(4)_R \subset \text{SO}(4,4)\,$ that do not mix left and right-movers (see discussion around \eqref{rottheta}). 
 Consistency of the asymmetric orbifold requires that the rotation is in the automorphism group of the lattice\footnote{A discussion on consistent rotations can be found e.g.\ in \cite{lerchie1989lattices}, cf. appendix B.}. Since the orbifold acts as a shift on $\Gamma^{1,1}$, it leaves $\Gamma^{1,1}$ invariant. On the other hand, $\Gamma^{4,4}(\mathcal{G})$ is not in general invariant under rotations $\mathcal{M}_{\theta}$. If there exists a sublattice $I\subset \Gamma^{4,4}(\mathcal{G})$ that is invariant under the orbifold action, it is given by
\begin{equation}
    I \equiv \{ p \in \Gamma^{4,4}(\mathcal{G})\,|\, \mathcal{M}_{\theta}\cdot p = p\}\,.
    \label{invariant lattice I}
\end{equation}
Then the complete  sublattice that is invariant under the 
orbifold action is 
\begin{equation}
    \hat{I}=I\oplus \Gamma^{1,1}\,.
    \label{invariant lattice hatI}
\end{equation}
Finally, we have to ensure that our models satisfy modular invariance. This can be verified if the following conditions hold \cite{vafa1986modular,Baykara:2023plc}
\begin{equation}
    p \sum_{i=3}^4 \tilde{u}_i \,\in\, 2\mathbb{Z} \qquad\text{and}\qquad p \sum_{i=3}^4 u_i \,\in\, 2\mathbb{Z} \,,
    \label{modular conditions on twist vectors}
\end{equation}
where $p$ is the orbifold rank. Also, if the rank of the orbifold is even, we check the additional condition\footnote{In some cases it is possible to construct consistent orbifolds even if this condition is not satisfied \cite{Harvey:2017rko}.} for $(p_L,p_R)\in \Gamma^{4,4}(\mathcal{G})$ \begin{equation}\label{modcond2}
   p_L \mathcal{N}_L^{\,p/2} p_L - p_R \mathcal{N}_R^{\,p/2} p_R \,\in\, 2\mathbb{Z}\,. 
\end{equation}

\subsection{The world-sheet fields}
\label{sec:the world-sheet fields}
We are now ready to discuss the action of the orbifold on the worldsheet variables. We split the bosonic coordinates as $X^M \rightarrow (\hat{X}^{\hat{\mu}},Y^m) \rightarrow (X^\mu, Z, Y^m)$, where $Y^m$ ($m=1,\ldots,4$) are the $T^4$ coordinates, $Z$ is the circle coordinate, $X^\mu$ ($\mu=0,\ldots,4$) are the $\mathbb{R}^{1,4}$ coordinates, and $\hat{X}^{\hat{\mu}}$ ($\hat{\mu} = 0,\ldots,5$) are the coordinates on $\mathbb{R}^{1,4}\times S^1$. We often work in complex coordinates on the torus, which we denote by $W^i = \tfrac{1}{\sqrt{2}}(Y^{2i-1}+iY^{2i})$ with $i=1,2$. On-shell, the worldsheet coordinates split into  left and right-moving parts as
\begin{equation}
W^i(\sigma^1,\sigma^2) = {W}_{L}^i(\sigma^1+\sigma^2) + W_{R}^i(\sigma^1-\sigma^2) \,.
\end{equation}
Here, $\sigma^1$ and $\sigma^2$ are the coordinates on the worldsheet which we always take to be of Lorentzian signature.
We denote the oscillators of all bosonic coordinates by $\tilde{\alpha}^M_n$ and $\alpha^M_n$ where the tilde indicates a left-mover, and we use different indices ($\hat{\mu}$, $\mu$, $z$, $m$ or $i$) according to the above decomposition. The fermionic modes of the superstring are denoted by $\tilde{b}^M_n$ and $b^M_n$ with a similar index structure. In the case of complex modes, we use a bar to denote the complex conjugate.

Now that we have set up our notation, we are ready to present the orbifold action. It is most easily stated in terms of the matrix $\tilde{\mathcal{M}}$ in \eqref{conjugation}, parametrized by the four mass parameters $m_i$. It works on the bosonic torus coordinates with asymmetric rotations
\begin{equation}\label{orbiaction2}
\begin{aligned}
{W}_{L}^1 \;&\rightarrow\; e^{i(m_1+m_3)}\: {W}_{L}^1 \,, \\
{W}_{L}^2 \;&\rightarrow\; e^{i(m_1-m_3)}\: {W}_{L}^2 \,, \\
W_{R}^1 \;&\rightarrow\; e^{i(m_2+m_4)}\: W_{R}^1 \,, \\
W_{R}^2 \;&\rightarrow\; e^{i(m_2-m_4)}\: W_{R}^2 \,,
\end{aligned}
\end{equation}
and with the same action on the fermionic torus coordinates. 
In addition, symmetric orbifolds correspond to $m_1=m_2$ and $m_3=m_4$. Furthermore, the rotations on the torus are accompanied by a shift along the circle coordinate
\begin{equation}\label{shift}
Z \;\rightarrow\; Z + 2\pi \mathcal{R} / p \,,
\end{equation}
which makes the orbifold freely acting. Here $\mathcal{R}$ is the circle radius ($Z\sim Z+2\pi \mathcal{R}$) and $p$ is the orbifold rank. Due to this shift, states that carry momentum in the $Z$-direction obtain a phase $e^{2\pi i n / p}$ under the orbifold action, where $n$ is the momentum number of the state. 

We mention here that the shift along the circle coordinate \eqref{shift}  can also be represented, in the basis of momentum ($n$) and winding ($w$) numbers, by a shift vector 
\begin{equation}
    v=\begin{pmatrix}
        \frac{1}{p}\\
        0
    \end{pmatrix}\,.
    \label{shiftvector1}
\end{equation}
In this basis, a vector $P$ of the lattice $\Gamma^{1,1}$, associated with $S^1$, can be written as
\begin{equation}
    P=\begin{pmatrix}
        w\\
        n
    \end{pmatrix}\,.
    \label{pvector1,1}
\end{equation}
Adding any lattice vector $P$ to the shift vector $v$  will give the same orbifold. Furthermore, $v$ and $-v$ specify physically equivalent orbifolds as the sign of $v$ is changed by the reflection $Z\to -Z$; such a reflection is an element of the T-duality group $\text{Spin}(5,5,\mathbb{Z})$. Note that changing $v$ to $-v$ has the same effect as replacing the monodromy
$\mathcal{M}$ with $\mathcal{M}^{-1}$. If $\mathcal{M}$ is expressed in terms of a mass matrix $M$ by $\mathcal{M}=e^{M}$, then this amounts to changing the sign of the mass matrix $M$.

Let us now discuss the orbifold action on the Neveu-Schwarz (NS) and Ramond (R) vacua, which we denote by $\ket{0}_{{L}/{R}}$ and $|s_1,s_2,s_3,s_4\rangle_{{L}/{R}}$ respectively, where $s_{\kappa} = \pm \tfrac{1}{2}$. The subscript ${L}/{R}$ is used to distinguish the left and the right-moving vacua. We choose the GSO projection in such a way that both R-vacua   satisfy 
\begin{equation}
    \sum_{\kappa=1}^4 s_{\kappa} \,\in\, 2\mathbb{Z}\,.
\end{equation}
The NS-vacua are spacetime scalars and are invariant under the orbifold action. On the other hand the R-vacua are $10D$ spinors and we know how they transform under rotations, so in particular under the orbifold action. In general we have
\begin{equation}\label{spinorrotation}
|s_1,s_2,s_3,s_4\rangle \;\;\rightarrow\;\; \exp \left( 2\pi i \sum_{\kappa=1}^4 v_{\kappa} \, S_{\kappa} \right) |s_1,s_2,s_3,s_4\rangle = e^{2\pi i \,\vec{v} \cdot \vec{s}}\; |s_1,s_2,s_3,s_4\rangle \,,
\end{equation}
where the $S_{\kappa} = J_{2\kappa-1,2\kappa}$ are the Cartan generators of the little group SO$(8)$ with eigenvalues $s_{\kappa}$. The $v_{\kappa}$ denotes a rotation in the $2\kappa-1$ and $2\kappa$ directions over an angle $2\pi v_{\kappa}$. Whenever the rotation works asymmetrically on left and right-movers, the formula above applies to spinors in each sector individually. In this case we use $\tilde{u}_{\kappa}$ and $u_{\kappa}$ for the left and right-moving rotation parameters respectively. Using this notation, we read off from \eqref{orbiaction2} that our orbifold action is a rotation with
\begin{equation}
\begin{alignedat}{4}
\tilde{u}_3 &= \frac{m_1+m_3}{2\pi} \,, &\qquad\quad u_3 &= \frac{m_2+m_4}{2\pi} \,, \\[3pt]
\tilde{u}_4 &= \frac{m_1-m_3}{2\pi} \,, &\qquad\quad u_4 &= \frac{m_2-m_4}{2\pi} \,.
\end{alignedat}
\label{u's}
\end{equation}
and the other rotation parameters ($\tilde{u}_{1,2}$ and ${u}_{1,2}$) equal to zero. These rotation parameters are subject to the constraint
\begin{equation}
   u_{3,4}=\frac{n_{3,4}}{p}\,, \qquad n_{3,4}\in \mathbb{Z}\,,
\end{equation}
and similarly for $\tilde{u}_{3,4}$, as follows from the quantization condition on the mass parameters \eqref{miN}.
These parameters are related to the angles introduced in the last section by
\begin{equation}
\begin{alignedat}{4}
\tilde{u}_3 &= \frac{\theta_L}{2\pi} \,, &\qquad\quad u_3 &= \frac{\theta_R}{2\pi} \,, \\[3pt]
\tilde{u}_4 &= \frac{\theta'_L}{2\pi} \,, &\qquad\quad u_4 &= \frac{\theta'_R}{2\pi} \,.
\end{alignedat}
\end{equation}
As we can see from \eqref{spinorrotation}, the orbifold action on the R-vacua depends only on the values of $s_3,s_4$. We introduce the following notation for the possible values of these spins:
\begin{equation}
\begin{aligned}
|a_1\rangle_{L/R} &= \big|s_1,s_1,\tfrac{1}{2},\tfrac{1}{2}\big\rangle_{L/R} \,, \\
|a_2\rangle_{L/R} &= \big|s_1,s_1,-\tfrac{1}{2},-\tfrac{1}{2}\big\rangle_{L/R} \,, \\
|a_3\rangle_{L/R} &= \big|s_1,-s_1,\tfrac{1}{2},-\tfrac{1}{2}\big\rangle_{L/R} \,, \\
|a_4\rangle_{L/R} &= \big|s_1,-s_1,-\tfrac{1}{2},\tfrac{1}{2}\big\rangle_{L/R} \,.
\end{aligned}
\label{defRvac}
\end{equation}
Here the relative sign between $s_1$ and $s_2$ is fixed by the GSO projection. By using \eqref{spinorrotation} and \eqref{u's} we find that the orbifold action on each of these is
\begin{equation}
\begin{alignedat}{4}
|a_1\rangle_{L} \;&\rightarrow \; e^{im_1} \, |a_1\rangle_{L} \,, \qquad\qquad
&|a_1\rangle_{R} \;& \rightarrow\; e^{im_2} \, |a_1\rangle_{R} \,, \\
|a_2\rangle_{L} \;&\rightarrow \; e^{-im_1} \, |a_2\rangle_{L} \,, \qquad\qquad
&|a_2\rangle_{R} \;& \rightarrow\; e^{-im_2} \, |a_2\rangle_{R} \,, \\
|a_3\rangle_{L} \;&\rightarrow \; e^{im_3} \, |a_3\rangle_{L} \,, \qquad\qquad
&|a_3\rangle_{R} \;& \rightarrow\; e^{im_4} \, |a_3\rangle_{R} \,, \\
|a_4\rangle_{L} \;&\rightarrow \; e^{-im_3} \, |a_4\rangle_{L} \,, \qquad\qquad
&|a_4\rangle_{R} \;& \rightarrow\; e^{-im_4} \, |a_4\rangle_{R} \,.
\end{alignedat}
\label{transformation of ramond vacua}
\end{equation}
Requiring that the orbifold action on the R-vacua is of order $p$ yields the additional conditions
\begin{equation}\label{psum_u_even}
   p \sum_{i} \tilde{u}_i \,\in\, 2\mathbb{Z} \qquad\text{and}\qquad p \sum_{i} {u}_i \,\in\, 2\mathbb{Z} \,.
\end{equation}
This also follows easily from the earlier analysis on the quantization conditions of the mass parameters, e.g. using \eqref{u's} together with \eqref{miN}, one finds $p(u_3+u_4)=pm_2/\pi=2N_2\in 2\mathbb{Z}$ and $p(\tilde{u}_3+\tilde{u}_4)=2N_1\in2\mathbb{Z}$. Since $pu_i\in\mathbb{Z}$, one also finds that $p(u_3-u_4)\in2\mathbb{Z}$. The same also holds for $\tilde u$. An instructive example is the $\mathbb{Z}_{24}$ orbifold (cf \ref{a Z24 orbifold}) with $m_1=m_2=5\pi/12$ and $m_3=m_4=\pi/12$, which corresponds to $u_3=1/4$ and $u_4=1/6$ (see also \autoref{tab breaking all}) and  leads to $p(u_3+u_4)=10$. We mention here that the conditions \eqref{psum_u_even}  are necessary for modular invariance of the partition function (see the discussion around \eqref{modular conditions on twist vectors}). 

Furthermore, if $\pm {u}_3\pm {u}_4=0$ mod 2 for some choice of signs, half of the right-moving supersymmetries are preserved in the orbifold. Essentially, this means that either $m_2$ or $m_4 = 0$ mod $2\pi$ and two of the four gravitini coming from the NS-R sector remain massless. On the other hand, if the above condition is not met, all right-moving supersymmetries are broken. Exactly the same argument holds for $\tilde{u},\,m_{1,3}$ and the left-moving supersymmetries.

Finally, in order for strings to close in our geometry, they need to satisfy the boundary conditions
\begin{equation}\label{boundaryconditions}
\begin{aligned}
&X^\mu(\sigma^1, \sigma^2+2\pi) = X^\mu(\sigma^1, \sigma^2) \,, \\
&Z(\sigma^1, \sigma^2+2\pi) = Z(\sigma^1, \sigma^2) + 2\pi \mathcal{R} \,(w + k/p) \,, \\
&W_{L}^1(\sigma^1, \sigma^2+2\pi) = \big(e^{i(m_1+m_3)}\big)^k \:W_{L}^1(\sigma^1, \sigma^2) \,,\\
 &W_{R}^1(\sigma^1, \sigma^2+2\pi) = \big(e^{i(m_2+m_4)}\big)^k \:W_{R}^1(\sigma^1, \sigma^2) \,, \\
&W_{L}^2(\sigma^1, \sigma^2+2\pi) = \big(e^{i(m_1-m_3)}\big)^k \:W_{L}^2(\sigma^1, \sigma^2) \,,\\
&W_{R}^2(\sigma^1, \sigma^2+2\pi) = \big(e^{i(m_2-m_4)}\big)^k \:W_{R}^2(\sigma^1, \sigma^2) \,.
\end{aligned}
\end{equation}
Here $w \in \mathbb{Z}$ is the winding number along the $S^1$ (we omit winding modes on the torus here for simplicity of the formulae) and $k = 0,\ldots,p-1$ is an integer that distinguishes between the various sectors. We have the untwisted sector for $k=0$, and $p-1$ twisted sectors for the other values of $k$ in which case the string closes only under application of the orbifold action.

\section{The partition function}
\label{sec:Partition function}

Now that we have defined our orbifold action, we would like to construct the one-loop partition function (we follow the conventions of \cite{Blumenhagen:2013fgp}; for some other references and recent examples, see e.g. \cite{aoki2004construction,nibbelink2021worldsheet}). In general, the starting point for the partition function is
\begin{equation}
    Z(\tau,\bar{\tau})=\text{Tr}\Big[q^{(L_0-\frac{c}{24})}\bar q^{(\bar{L}_0-\frac{\bar c}{24})}\Big]\ ,\qquad q=e^{2\pi i\tau}\ ,
\end{equation}
where $\tau=\tau_1+i\tau_2$ is the complex structure modulus of the torus. In an orbifold, we have twisted sectors and projectors in each of the sectors onto invariant states. Therefore, the trace over the Hilbert space decomposes according to
\begin{equation}
    Z(\tau,\bar \tau)=\frac{1}{p}\sum_{k,l=0}^{p-1}{Z}[k,l](\tau,\bar \tau)\ ,
\end{equation}
where, as mentioned before, $p$ is the orbifold rank and $k$ characterizes the various sectors. In addition, $l$ implements the orbifold projection in each sector\footnote{If we denote the orbifold group element by $g$, with $g^p=1$, then the projection operator takes the form $P=\frac{1}{p}(1+g+g^2+\cdots + g^{p-1})$. }. Furthermore, for our models the partition function will factorize into the following pieces (we omit writing the $\tau$ dependence for simplicity of the notation)
\begin{equation}
    {Z}[k,l]= {Z}_{\mathbb{R}^{1,4}}\,  {Z}_{S^1}[k,l]  {Z}_{T^4}[k,l]  {Z}_F[k,l]\,.
    \label{partition function first}
\end{equation}
Here ${Z}_{\mathbb{R}^{1,4}}$ is the contribution to the partition function from the non-compact bosons, ${Z}_{S^1}[k,l]$ and ${Z}_{T^4}[k,l]$ refer to the compact bosons on $S^1$ and $T^4$ respectively and ${Z}_F[k,l]$ is the fermionic contribution to the partition function.

In the remainder of this section, we construct the various parts of the partition function and discuss modular invariance of the full partition function. For symmetric orbifolds, showing modular invariance is rather easy, as the individual pieces in \eqref{partition function first} will have the same properties under the modular group (or be invariant), in such a way that the sum over $k$ and $l$ is modular invariant. For asymmetric orbifolds, one must take care of the possible phases that will arise in left and right-moving sectors under modular transformations, and show case by case that it all combines into a full modular invariant partition function $Z$.

First, we consider the bosonic piece of the partition function. The contribution from the three non-compact bosons (we work in lightcone gauge) to the partition function is
\begin{equation}
    {Z}_{\mathbb{R}^{1,4}}= \left(\sqrt{\tau_2}\,\eta\,\bar{\eta}\right)^{-3}\,.
\end{equation}
This term is invariant under both $\mathcal{T}$ and $\mathcal{S}$ modular transformations. (Modular functions and transformations are discussed in appendix \ref{ap B}.)

To compute the contribution to the partition function from the compact boson on $S^1$, recall that due to the shift along the circle coordinate, momentum states pick up a phase $e^{2\pi i n/p}$. In addition, the boundary condition of the circle coordinate \eqref{boundaryconditions} implies that in the twisted sectors fractional winding modes can appear. Combining these, we can write  
\begin{equation}
{Z}_{S^1}[k,l]=  \frac{1}{\eta\,\bar{\eta}}\sum_{n,w \in \mathbb{Z}}e^{\frac{2\pi i n}{p}l}\, q^{\frac{\alpha'}{4}P_{R}^2(k)}\, \bar{q}^{\frac{\alpha'}{4}P_{L}^2(k)}\,,
\end{equation}
where
\begin{equation}
     P_{{L}/{R}}(k)=\frac{n}{\mathcal{R}}\pm  \frac{\big(w+\frac{k}{p}\big)\mathcal{R}}{\alpha'}\,.
\end{equation}
${Z}_{S^1}[k,l]$ can be written in a manifestly modular invariant form by performing a Poisson resummation (cf. \eqref{Poissonresum}) over the momentum number $n$. We find
\begin{equation}
     {Z}_{S^1}[k,l]=\frac{\mathcal{R}}{\sqrt{\alpha'}\sqrt{\tau_2}\,\eta\,\bar{\eta}}\sum_{n,w \in \mathbb{Z}}e^{-\frac{\pi \mathcal{R}^2}{\alpha'\tau_2}\left|n-\frac{l}{p}+\left(w+\frac{k}{p}\right)\tau\right|^2}\,.
\end{equation}
Note that the circle partition function consists of four building blocks: ${Z}_{S^1}[0,0]$, ${Z}_{S^1}[0,l]$, ${Z}_{S^1}[k,0]$ and ${Z}_{S^1}[k,l]$. Of course, ${Z}_{S^1}[0,0]$ corresponds simply to a circle compactification and is invariant under both $\mathcal{T}$ and $\mathcal{S}$ modular transformations. The remaining blocks obey the following modular transformations
\begin{equation}
\begin{aligned}
   {Z}_{S^1}[k,l]\xrightarrow{\mathcal{T}}  \frac{\mathcal{R}}{\sqrt{\alpha'}\sqrt{\tau_2}\,\eta\,\bar{\eta}}&\sum_{n,w \in \mathbb{Z}}e^{-\frac{\pi \mathcal{R}^2}{\alpha'\tau_2}\left|(n+w)-\frac{l-k}{p}+\left(w+\frac{k}{p}\right)\tau\right|^2}= {Z}_{S^1}[k,l-k]\,,\\
{Z}_{S^1}[k,l]\xrightarrow{\mathcal{S}} \frac{\mathcal{R}}{\sqrt{\alpha'}\sqrt{\tau_2}\,\eta\,\bar{\eta}}&\sum_{n,w \in \mathbb{Z}}e^{-\frac{\pi \mathcal{R}^2}{\alpha'\tau_2}\left|w+\frac{k}{p}+\left(n+\frac{l}{p}\right)\tau\right|^2}={Z}_{S^1}[l,-k]\,.
\end{aligned}
\label{circle modular transf}
\end{equation}
Also, it is straightforward to verify the following properties
\begin{equation}
    {Z}_{S^1}[k,l]={Z}_{S^1}[-k,-l] ={Z}_{S^1}[-k+p,-l]={Z}_{S^1}[-k,-l+p]={Z}_{S^1}[-k+p,-l+p]\,.
    \label{circle ids}
\end{equation}
The above transformation rules \eqref{circle modular transf} and properties \eqref{circle ids} will be combined with similar ones from the $T^4$ and the fermions to ensure modular invariance of the full partition function.

Next, we discuss the contribution coming from the $T^4$. We consider left and right-movers separately, since the orbifold can act asymmetrically on the torus coordinates. For clarity of the partition function, it is convenient to parametrize the orbifold action by two twist vectors
$\tilde{u}=(0,0,\tilde{u}_3,\tilde{u}_4)$ and $u=(0,0,u_3,u_4)$, with $\Tilde{u}_i, u_i$ as given in \eqref{u's}. (For a discussion on twist vectors see e.g. \cite{ibanez2012string,font2005introduction}.) By this, we mean that
\begin{equation}\label{orbiaction}
\begin{aligned}
{W}_{L}^1 \;&\rightarrow\; e^{2\pi i\tilde{u}_3}\: {W}_{L}^1 \,, \\
{W}_{L}^2 \;&\rightarrow\; e^{{2\pi i\tilde{u}_4}}\: {W}_{L}^2 \,, \\
W_{R}^1 \;&\rightarrow\; e^{{2\pi i{u}_3}}\: W_{R}^1 \,, \\
W_{R}^2 \;&\rightarrow\; e^{2\pi i{u}_4}\: W_{R}^2 \,,
\end{aligned}
\end{equation}
and the coordinates of the non-compact dimensions are not rotated. Note that symmetric orbifolds correspond to $\tilde{u}=u$. Now, let us focus on the oscillator modes and postpone the discussion of the lattice sum over momenta and windings. The oscillator part of the $T^4$ partition function factorizes into left and right-moving pieces as
\begin{equation}
    Z_{T^4}[k,l] =  \widetilde{\mathcal{Z}}_{T^4}[\tilde{\theta}^k,\tilde{\theta}^l] \otimes \mathcal{Z}_{T^4}[\theta^k,\theta^l]\,.
\end{equation}
Here $\theta $ is the generator of the orbifold group. $\theta^k$ refers to twisted sectors where the torus coordinates obey boundary conditions of the form $W_{R}^i(\sigma^1,\sigma^2+2\pi)= \theta^k\, W_{R}^i(\sigma^1,\sigma^2)$  and  $\theta^l$ characterizes the orbifold action: $W_{R}^i \to \theta^l\,W_{R}^i$ ($\tilde{\theta}^k,\tilde{\theta}^l$ correspond to the left-movers). Similarly with the circle, ${Z}_{T^4}[0,0]$ corresponds simply to compactification on $T^4$ and is invariant under both $\mathcal{T}$ and $\mathcal{S}$ modular transformations.

First, we present the right-moving torus partition function. In the untwisted sector ($k=0$) it is given by 
\begin{equation}
   \mathcal{Z}_{T^4}[\mathbf{1},\theta^l]=q^{-\frac{2}{12}}\prod^4_{i=3} \prod_{n=1}^{\infty}(1-q^n\,e^{ 2\pi i  l u_i})^{-1} (1-q^n\,e^{-2\pi il u_i})^{-1} \,,
   \label{torus one}
\end{equation}
and can be rewritten in a more convenient form as 
\begin{equation}
   \mathcal{Z}_{T^4}[\mathbf{1},\theta^l]=\prod^4_{i=3}2 \sin(\pi lu_i)\, \frac{\eta} {\vartheta\Big[\psymbol{ \frac{1}{2}}{ -\frac{1}{2}+lu_i}\Big]}\ .
   \label{torus l}
\end{equation}
By performing successively $\mathcal{S}$ and $\mathcal{T}$ modular transformations, we find the partition function in a twisted sector labeled by $k$ ($k\neq 0$), which reads\footnote{Here we omit an irrelevant constant phase coming from the $\mathcal{T}$ transformation because it is always cancelled by left-moving contributions in both symmetric and asymmetric orbifolds.}
\begin{equation}
   \mathcal{Z}_{T^4}[\theta^k,\theta^l]=e^{-\pi i \sum\limits_{i=3}^4 \left(k u_i l u_i\right)}e^{\pi i \sum\limits_{i=3}^4 \left(k u_i-\frac{1}{2}\right)} {\chi} [\theta^k,\theta^l]\prod^4_{i=3}\,\frac{\eta}{\vartheta\Big[\psymbol{ \frac{1}{2}-ku_i}{ -\frac{1}{2}+lu_i}\Big]}\ ,
   \label{general torus partition}
\end{equation}
where
\begin{equation}
   \chi[\theta^k,\theta^l]= \prod^4_{i=3}2 \sin(\pi \text{\footnotesize{gcd}}(k,l) u_i)
   \label{fixed points}
\end{equation}
\footnote{ $\text{\footnotesize{gcd}}(k,l)$ denotes the greatest common divisor of $k,l$ with the convention $\text{\footnotesize{gcd}}(a,0)=\text{\footnotesize{gcd}}(0,a)=a$.}is the number of simultaneous \say{chiral} fixed points\footnote{The orbifolds that we consider have fixed points on the $T^4$. However, due to the shift on the circle, there are no points left invariant under the full orbifold action.} of $\theta^k$ and $\theta^l$. We note here that equation \eqref{fixed points} is valid for $ku_{3,4} \notin \mathbb{Z}$. If there exists $j\in [3,4]$ such that $ku_j \in \mathbb{Z}$\footnote{An example is the orbifold with $\tilde{u}=u=(0,0,0,\frac{1}{2})$. From \eqref{psum_u_even} it follows that $p=4$, such that $k=0,1,2,3$. For $k=2$, one then has $ku_4\in \mathbb{Z}$. The mass parameters in this case are $m_1=m_2=-m_3=-m_4=\pi/2.$}, $ \chi[\theta^k,\theta^l]$ should be divided by $2\sin(\pi l u_j)$ for $l\neq 0$, and replaced by $\prod_{i\neq j,ku_i \notin \mathbb{Z}}2 \sin(\pi k u_i)$ for $l=0$  (see \cite{katsuki1990zn} for a relevant discussion). Furthermore, under modular transformations \eqref{general torus partition} transforms as
\begin{equation}
    \begin{aligned}
     & \mathcal{Z}_{T^4}[\theta^k,\theta^l] \xrightarrow{\mathcal{T}} e^{-\frac{\pi i}{6}\lambda} \mathcal{Z}_{T^4}[\theta^k,\theta^{l-k}]\,,\\
     &\mathcal{Z}_{T^4}[\theta^k,\theta^l] \xrightarrow{\mathcal{S}}  e^{-\frac{\pi i}{2}\lambda}  \mathcal{Z}_{T^4}[\theta^l,\theta^{-k}]\,,
    \end{aligned}
    \label{bosonic modular transformation}
\end{equation}
where $\lambda$ is the number of $u_i\notin\mathbb{Z}$. The partition function for the left-movers is simply obtained by substituting $q\to \bar{q}, \eta\to \bar{\eta}, \vartheta\to \bar{\vartheta}$ and $u\to \tilde{u}$, and obeys the transformations \eqref{bosonic modular transformation} but with phases of opposite sign. Finally, notice that if the orbifold acts trivially on ${W}_{R}^{1,2}$, i.e. $u_3,u_4 \in \mathbb{Z}$, equation \eqref{torus one}, or equivalently \eqref{torus l}, simply becomes $\eta^{-4}$. 

Regarding the zero modes, as we have already discussed in section \ref{lattice construction}, 
there might be an invariant sublattice $I\subset \Gamma^{4,4}$ (cf. \eqref{invariant lattice I}) of left and/or right-moving momenta. The form of the invariant sublattice depends on the particular orbifold action.  Moreover, if there exists an invariant lattice, it will contribute to the partition function in the twisted sectors with a multiplicative factor that is equal to its volume \cite{Narain:1986qm,narain1991asymmetric}. We will return to this issue when we discuss explicit examples.

For the construction of the fermionic partition function we combine the non-compact and compact fermions in one expression and we consider left and right-movers separately. In the NS-sector the right-moving fermionic partition function in a sector labeled by $k$ reads 
\begin{equation}
    \mathcal{Z}_{\text{NS}}[\theta^k,\theta^l]=\frac{1}{2}e^{\pi i \sum\limits_{i=3}^4\left(k u_i l u_i\right)}\bigg[\left(\frac{\vartheta_3}{\eta}\right)^2\prod^4_{i=3}\frac{{\vartheta[\psymbol{ku_i}{ -lu_i}]}}{\eta}-e^{\pi i \sum\limits_{i=3}^4  k u_i}\left(\frac{\vartheta_4}{\eta}\right)^2\prod^4_{i=3}\frac{\vartheta[\psymbol{ku_i}{ -\frac{1}{2}-lu_i}]}{\eta}\bigg]\ .
    \label{partition NS}
\end{equation}
In the R-sector we have
\begin{equation}
    \mathcal{Z}_{\text{R}}[\theta^k,\theta^l]=\frac{1}{2}e^{\pi i \sum\limits_{i=3}^4\left(k u_i l u_i\right)}\bigg[\left(\frac{\vartheta_2}{\eta}\right)^2\prod^4_{i=3}\frac{{\vartheta[\psymbol{\frac{1}{2}+ku_i} {-lu_i}]}}{\eta}+e^{\pi i \sum\limits_{i=3}^4 k u_i}\left(\frac{\vartheta_1}{\eta}\right)^2\prod^4_{i=3}\frac{{\vartheta[\psymbol{\frac{1}{2}+ku_i} {-\frac{1}{2}-lu_i}]}}{\eta}\bigg]\ .
    \label{partition R}
\end{equation}
By combining the above, we find the right-moving fermionic partition function in a sector labeled by $k$, which reads
\begin{equation}
    \mathcal{Z}_{{F}}[\theta^k,\theta^l]=\mathcal{Z}_{\text{NS}}[\theta^k,\theta^l]-\mathcal{Z}_{\text{R}}[\theta^k,\theta^l]\,,
\end{equation}
and transforms under modular transformations as
\begin{equation}
    \begin{aligned}
      & \mathcal{Z}_F[\theta^k,\theta^l] \xrightarrow{\mathcal{T}} e^{\frac{4\pi i}{6}} \mathcal{Z}_F[\theta^k,\theta^{l-k}]\,,\\
       & \mathcal{Z}_F[\theta^k,\theta^l] \xrightarrow{\mathcal{S}}   \mathcal{Z}_F[\theta^l,\theta^{-k}]\,.
    \end{aligned}
    \label{fermionic modular transformation}
\end{equation}
For later convenience, we rewrite expressions \eqref{partition NS} and \eqref{partition R} in terms of infinite sums as\footnote{In the literature, this is usually referred to as \say{bosonization}. For an alternative construction of the partition function see \cite{condeescu2012asymmetric,Condeescu:2013yma}.} 
\begin{equation}
   \mathcal{Z}_{\text{NS,R}}[\theta^k,\theta^l]= \frac{1}{\eta^4}\, e^{\pi i \sum\limits_{i=3}^4\left(k u_i l u_i\right)}\sum_{{r}} q^{\frac{1}{2}({r}+k {u})^2}\,e^{-2\pi i l[({r}+k {u})\cdot{u}]}\,.
   \label{fermionic infinite sums}
\end{equation}
Here $r=(r_1,r_2,r_3,r_4)$ is an SO(8) weight vector with each component ${r}_i\in \mathbb{Z}$ in the NS-sector and ${r}\in \mathbb{Z}+\tfrac{1}{2}$ in the R-sector. The GSO projection is $\sum_{i=1}^4 r_i \in 2\mathbb{Z}+1$ in the NS-sector and $\sum_{i=1}^4 r_i \in 2\mathbb{Z}$ in the R-sector. Finally, the left-moving fermionic partition function is obtained by substituting $q\to \bar{q}, \eta\to \bar{\eta}, \vartheta\to \bar{\vartheta}, u\to \tilde{u}$  and $r \to \tilde{r}$, where $\tilde{r}=(\tilde{r}_1,\tilde{r}_2,\tilde{r}_3,\tilde{r}_4)$, and transforms as in \eqref{fermionic modular transformation} but with a phase of opposite sign. 

As a last comment here, we observe from \eqref{bosonic modular transformation} and \eqref{fermionic modular transformation} that if we consider only the right-movers, we do not obtain a modular invariant partition function. Of course, modular invariance can be achieved by taking also into account the contribution from the left-movers. This can be easily verified in the case of symmetric orbifolds because the expression for the left-moving partition function is essentially the complex conjugate of the right-moving one. Consequently, the constant phases cancel out, as the partition function is the tensor product of left and right-movers. However, this argument does not hold for asymmetric orbifolds. Therefore, one shall carefully examine modular invariance for each asymmetric orbifold construction. In the next, we will address this issue by discussing specific examples.

\clearpage

\thispagestyle{empty}

\chapter{Closed string spectrum}
\label{chap:spectrum}

In this chapter, we focus on the construction of closed string states in the orbifold untwisted and twisted sectors. In section \ref{closed string spectrum}, we present the general formalism for the construction of the orbifold spectrum. Then, in section \ref{explicitexamples} we present explicit examples of orbifold spectra with $\mathcal{N}=6,4,2$ or $0$ supersymmetry in five dimensions. Next, in section \ref{Supergravity}, we make the connection between freely acting orbifolds and Scherk-Schwarz supergravity manifest by constructing the lowest lying orbifold states. Finally, in section \ref{sec:bad examples} we discuss the classical and one-loop scalar potential of the effective theory, which can be expressed in terms of supertraces.

\section{General formalism}
\label{closed string spectrum}
In this section, we present the general formalism that we use in order to obtain the closed string spectrum that arises from our orbifold constructions. In general, we treat the untwisted and twisted sectors separately. In other words, we fix $k$ and then we sum over $l$ and divide by the orbifold rank $p$ in order to implement the orbifold projection. Furthermore, we  focus on the lowest excited states. Consequently, we expand the $\vartheta$-functions coming from the bosonic contributions as well as all the $\eta$-functions and we keep only the lowest order terms. For the expansion of the partition function, we consider general twist vectors of the form $\tilde{u}=(0,0,\tilde{u}_3,\tilde{u}_4)$ and $u=(0,0,u_3,u_4)$ that act non-trivially along the $T^4$ directions. Also, we omit writing down irrelevant factors of $\tau_2$.

First, we consider  the untwisted sector, i.e. the sector with $k=0$ boundary conditions, in which the partition function can be expanded as (omitting for now factors of ${\tau_2}$)
\begin{equation}
    {Z}[0,l]= (q\bar{q})^{-\frac{1}{2}}\sum_{n,w \in \mathbb{Z}}e^{\frac{2\pi i n}{p}l}\, q^{\frac{\alpha'}{4}P_{R}^2(0)}\, (\bar{q})^{\frac{\alpha'}{4}P_{L}^2(0)}\sum_{{r},\tilde{{r}}} q^{\frac{1}{2}{r}^2}\,(\bar{q})^{\frac{1}{2}\tilde{r}^2}e^{2\pi il (\tilde{r}\cdot \tilde{u}-r\cdot u)}\, \left(1+\cdots\right)\,,
    \label{generic partition untwisted}
\end{equation}
where the dots denote contributions from higher excited oscillator states. We present in \autoref{tablemasslessstates} the NS and R-sector weight vectors for the states of the lowest level that survive the GSO projection (all of these are massless in the absence of momentum and/or winding modes). Furthermore, we table their representations under both the massless little group SO$(3)\approx\text{SU}(2)$ and the massive little group SO$(4)\approx\text{SU}(2)\times\text{SU}(2)$ in five dimensions. The latter is important when adding momenta or windings such that the states become massive.

\renewcommand{\arraystretch}{2}
\begin{table}[h!]
\centering
 \begin{tabular}{|c|c|c|c|}
    \hline
    Sector &  $\tilde{r}, {r}$  & SO(3) rep & SO(4) rep\\
    \hline
    \hline
   \multirow{3}{*}{NS}  & $(\underline{\pm 1,0},0,0)$ & $\textbf{3}\oplus \textbf{1}$ & $(\textbf{2},\textbf{2})$\\
    \cline{2-4}
    &$(0,0,\pm 1,0)$& 2$\,\times\,\textbf{1}$ & 2$\,\times\,(\textbf{1},\textbf{1})$\\
    \cline{2-4}
   & $(0,0,0,\pm 1)$ & 2$\,\times\,\textbf{1}$ & 2$\,\times\,(\textbf{1},\textbf{1})$\\
    \hline
    \hline
    \multirow{4}{*}{R}  & $(\pm\frac{1}{2},\pm\frac{1}{2},\frac{1}{2},\frac{1}{2})$ & $\textbf{2}$ & $(\textbf{2},\textbf{1})$\\
    \cline{2-4}
    & $(\pm\frac{1}{2},\pm\frac{1}{2},-\frac{1}{2},-\frac{1}{2})$ & $\textbf{2}$ & $(\textbf{2},\textbf{1})$\\
    \cline{2-4}
    & $(\underline{\frac{1}{2},-\frac{1}{2}},\frac{1}{2},-\frac{1}{2})$ & $\textbf{2}$ & $(\textbf{1},\textbf{2})$\\
    \cline{2-4}
    & $(\underline{\frac{1}{2},-\frac{1}{2}},-\frac{1}{2},\frac{1}{2})$ & $\textbf{2}$ & $(\textbf{1},\textbf{2})$\\
   \hline
    \end{tabular}
\captionsetup{width=.9\linewidth}
\caption{\textit{Here we write down all the weight vectors for states that are massless in the absence of momentum and/or winding modes, including their representations under the massless and massive little groups in 5D. We write down both left-moving and right-moving weight vectors, where underlining denotes permutations.}}
\label{tablemasslessstates}
\end{table}
\renewcommand{\arraystretch}{1}

\noindent We construct string states by tensoring the left and right-moving weight vectors from \autoref{tablemasslessstates}. For the construction of states, we use the rules
\begin{equation}
\begin{aligned}
\textbf{3}\otimes \textbf{3} = \textbf{5} \oplus \textbf{3} \oplus \textbf{1} \,, \qquad\quad \textbf{2}\otimes \textbf{2} = \textbf{3} \oplus \textbf{1} \,, \qquad\quad \textbf{3}\otimes \textbf{2} = \textbf{4} \oplus \textbf{2} \,,
\end{aligned}
\end{equation}
for tensoring SU$(2)$ representations.  In addition, we table the (massless and massive) representations that correspond to various supergravity fields in five dimensions in \autoref{table5Dfieldreps}.

\renewcommand{\arraystretch}{1.2}
\begin{table}[h!]
\centering
\begin{tabular}{|c|c|}
\hline
\;Massless field\; & \;SO$(3)$ rep\; \\ \hline\hline
$g_{\mu\nu}$ & \textbf{5} \\
$\psi_\mu$ & \textbf{4} \\
$A_\mu$ & \textbf{3} \\
$\chi$ & \textbf{2} \\
$\phi$ & \textbf{1} \\ \hline
\end{tabular}
\hspace{1.5cm}
\begin{tabular}{|c|c|}
\hline
\;\,Massive field\,\; & SO$(4)$ rep \\ \hline\hline
$B_{\mu\nu}^+$ / $B_{\mu\nu}^-$ & $(\textbf{3},\textbf{1})$ / $(\textbf{1},\textbf{3})$ \\
$\psi_\mu^+$ / $\psi_\mu^-$ & \;\:$(\textbf{2},\textbf{3})$ / $(\textbf{3},\textbf{2})$\:\; \\
$A_\mu$ & $(\textbf{2},\textbf{2})$ \\
$\chi^+$ / $\chi^-$ & $(\textbf{1},\textbf{2})$ / $(\textbf{2},\textbf{1})$ \\
$\phi$ & $(\textbf{1},\textbf{1})$ \\ \hline
\end{tabular}
\captionsetup{width=.83\linewidth}
\caption{\textit{Here we show the various massless and massive 5D fields and their representations under the appropriate little group.}}
\label{table5Dfieldreps}
\end{table}
\renewcommand{\arraystretch}{1}
\noindent In general, a state carries a non-trivial orbifold charge, given by the phase $e^{2\pi i l (\tilde{r}\cdot \tilde{u}-r\cdot u)}$, and its degeneracy is 
\begin{equation}
    D(k=0)=\frac{1}{p}\sum_{l=0}^{p-1}e^{2\pi il [(\tilde{r}\cdot \tilde{u}-r\cdot u)+\frac{n}{p}]}\,.
    \label{degeneracy untwisted}
\end{equation}
A charged state will be projected out of the orbifold spectrum when we perform the summation over $l$. However, we can fix this issue by adding appropriate momentum modes on the circle  to the state, such that the orbifold charge is cancelled. This also means that the state will become massive, since momentum modes contribute to the mass of a state; we will discuss this in detail later. Finally, if $(\tilde{r}\cdot \tilde{u}-r\cdot u)\in \mathbb{Z}$, the orbifold charge is trivial. States with trivial charge survive the orbifold projection and remain massless. As follows from \eqref{degeneracy untwisted}, the degeneracy of orbifold invariant states in the untwisted sector is 1.

The masses of left and right-moving states can be read off from the exponents of $\bar{q}$ and $q$, respectively. In particular, the mass formulae read
\begin{equation}
\begin{aligned}
    &\frac{\alpha'm^2_L(0)}{2}=\frac{1}{2}\tilde{r}^2+\frac{\alpha'}{4}P^2_L(0)-\frac{1}{2}+\tilde{N}\,,\\
    &\frac{\alpha'm^2_R(0)}{2}=\frac{1}{2}{r}^2+\frac{\alpha'}{4}P^2_R(0)-\frac{1}{2}+N\,.
\end{aligned}
 \label{untwisted masses}
 \end{equation}
 Here, $\tilde{N}$ and $N$ are integers, which refer to the bosonic occupation number of higher excited left and right-moving oscillator states, respectively. In the untwisted sector, the lightest states satisfy $\tilde{N}=N=0$. 
 
Regarding the construction of states in twisted sectors, the procedure is similar with the untwisted sector. The expansion of the partition function in a twisted sector labelled by $k$ yields
\begin{equation}
    \begin{aligned}
    {Z}[k,l]=&{\chi} [\theta^k,\theta^l]\,\tilde{\chi}[\tilde{\theta}^k,\tilde{\theta}^l] (q\bar{q})^{-\frac{1}{2}}\,e^{i(\tilde{\varphi}-\varphi)}q^{E_k}\,(\bar{q})^{\tilde{E}_k}\sum_{n,w \in \mathbb{Z}}e^{ \frac{2\pi i n}{p}l}\, q^{\frac{\alpha'}{4}P_{R}^2(k)}\, (\bar{q})^{\frac{\alpha'}{4}P_{L}^2(k)}\,\times\\
    & \sum_{r,\tilde{r}} q^{\frac{1}{2}({r}+k u)^2}\, (\bar{q})^{\frac{1}{2}(\tilde{r}+k \tilde{u})^2}\,e^{2\pi il (\tilde{r}\cdot \tilde{u}-r\cdot u)}\, e^{2\pi il k (\tilde{u}^2-u^2)}\, \left(1+\cdots\right)\,,
    \end{aligned}
    \label{generic partition twisted}
\end{equation}
where
\begin{equation}
    \varphi= 2\pi  \sum_{u_i\notin \mathbb{Z}}\left(\frac{1}{2}-k u_i\right)l u_i
    \label{shift phase factor}
\end{equation}
is a phase arising from bosonic contributions and
\begin{equation}
    E_k = \sum_{u_i\notin \mathbb{Z}}\frac{1}{2}ku_i (1-ku_i)\,
    \label{shifted energy}
\end{equation}
is a shift to the zero point energy. Note that if $ku_i>1$, we should substitute $ku_i\to ku_i-1$ in \eqref{shifted energy}, so that we actually compute the energy of the lowest order terms. This leads to the same modification of the phase in \eqref{shift phase factor}, together with an overall shift $\varphi\to \varphi +\pi$. We mention here that in the formulae \eqref{fixed points}, \eqref{shift phase factor} and \eqref{shifted energy} we consider $u_i>0$, as sending $u_i\to-u_i$, for $i=3$ and/or $4$, leaves the bosonic piece of the partition function invariant\footnote{Sending only  $u_3\to-u_3$, or  $u_4\to-u_4$ affects the fermionic partition function, as it changes the GSO projection in the R-sector. Sending both $u_3\to-u_3$ and $u_4\to-u_4$ leaves the fermionic partition function invariant.}. The expressions for $\tilde{\chi}[\tilde{\theta}^k,\tilde{\theta}^l],\tilde{\varphi}, \tilde{E}_k$ are simply obtained by substituting $u\to\tilde{u}$. Finally, the degeneracy of a state in a $k$ twisted sector is given by (see also \cite{ibanez1988heterotic,font1990construction} for a relevant discussion)
\begin{equation}
   D(k)= \frac{1}{p}\sum_{l=0}^{p-1}{\chi} [\theta^k,\theta^l]\,\tilde{\chi}[\tilde{\theta}^k,\tilde{\theta}^l] e^{2\pi il [(\tilde{r}\cdot \tilde{u}-r\cdot u)+k(\tilde{u}^2-u^2)+\frac{n}{p}]+i(\tilde{\varphi}-\varphi)}\,.
   \label{degeneracy twisted}
\end{equation}
However, if the orbifold acts trivially on ${W}_{{L/R}}^{1}$ and/or ${W}_{{L/R}}^{2}$, this degeneracy should be modified because an invariant lattice of left and/or right-moving momenta, $I\subset \Gamma^{4,4}$, can appear. In particular, one should also divide \eqref{degeneracy twisted} by the volume of the invariant momentum sublattice  \cite{Narain:1986qm,narain1991asymmetric}. Consistency of the orbifold requires that the degeneracy of every state should be an integer number.

In the twisted sectors the mass formulae read
\begin{equation}
\begin{aligned}
    &\frac{\alpha'm^2_L(k)}{2}=\frac{1}{2}(\tilde{r}+k\tilde{u})^2+\frac{\alpha'}{4}P^2_L(k)+\tilde{E}_k-\frac{1}{2}+\tilde{N}\,,\\
    &\frac{\alpha'm^2_R(k)}{2}=\frac{1}{2}({r}+k{u})^2+\frac{\alpha'}{4}P^2_R(k)+{E}_k-\frac{1}{2}+N\,,
\end{aligned}
 \label{twisted masses}
 \end{equation}
As in the untwisted sector, $\tilde{N}$ and $N$ refer to the bosonic occupation number of higher excited left and right-moving oscillator states, respectively. However, in the twisted sectors $\tilde{N}$ and $N$ are not integers because the twisted boundary conditions of the $T^4$ coordinates \eqref{boundaryconditions} result in a shift of the moding of the corresponding oscillators. In particular, the moding of the bosonic oscillators along the torus directions read (see e.g. \cite{font2005introduction}) 
\begin{equation}
\begin{aligned}
    &\tilde{\alpha}^1_{n-k\tilde{u}_3}\,, \quad \bar{\tilde{\alpha}}^1_{n+k\tilde{u}_3}\,, \quad{\alpha}^1_{n+k{u}_3}\,,\quad \bar{\alpha}^1_{n-k{u}_3}\,,\qquad n\in\mathbb{Z}\,,\\
    &\tilde{\alpha}^2_{n-k\tilde{u}_4}\,, \quad \bar{\tilde{\alpha}}^2_{n+k\tilde{u}_4}\,, \quad{\alpha}^2_{n+k{u}_4}\,,\quad \bar{\alpha}^2_{n-k{u}_4}\,,\qquad n\in\mathbb{Z}\,.
    \end{aligned}
    \label{shifted oscillators}
\end{equation}
As an example, consider a generic state in a $k$ twisted sector denoted by $\ket{\tilde{r},r}_k$. We can act on this state with a left-moving creation operator, i.e. $\tilde{\alpha}^i_{-k\tilde{u}_i}\ket{\tilde{r},r}_k$. Then, $\tilde{N}= k\tilde{u}_i$, and the degeneracy of the state \eqref{degeneracy twisted} is modified by the addition of a phase $e^{2\pi i l\tilde{u}_i}$ (see e.g. \cite{font2005introduction}). 

Finally, we mention here that the spectrum in an orbifold $k$-twisted sector is identical with the spectrum of the $(p-k)$-twisted sector. This is due to the fact that the partition function of the $k$-twisted sector is equal to the partition function of the $(p-k)$-twisted sector.

\section{Explicit examples}
\label{explicitexamples}

In this section, we present explicit orbifold models preserving $\mathcal{N}=6,4,2$ or $0$ supersymmetry in $D=5$.  We treat both symmetric and asymmetric models. We discuss the untwisted orbifold spectra and verify that they match exactly with the Scherk-Schwarz supergravity spectra obtained in \cite{Hull:2020byc}. In addition, we  construct purely stringy states arising from the orbifold twisted sectors. Regarding the non-supersymmetric orbifolds, we  focus on the twisted sectors where tachyons can appear and we find a critical value for the orbifold circle radius above which the spectrum is tachyon-free.

\subsection{$\mathcal{N}=6$}
\label{n=6orbis}

In this section, we discuss models with $\mathcal{N}=6$ supersymmetry in five dimensions. In the class of orbifolds studied here, there are in fact only four candidate models preserving $\mathcal{N}=6$ supersymmetry in $D=5$, and these are orbifolds of rank $2,3,4$ and $6$. Orbifolds of rank $2$ and $3$ have been discussed in \cite{dabholkar1999string,bianchi2022perturbative}. Here we will review the construction of the rank $2$ model and we will also construct a novel rank $4$ model. It is worth mentioning here that no consistent construction of a rank $6$ model is known. 

In the next, without loss of generality, we will construct orbifolds with twist vectors of the form $\tilde{u}=(0,0,\tilde{u}_3,\tilde{u}_4)$ and $u=(0,0,0,0)$, where $u_3, u_4\neq 0$ and $\pm \tilde{u}_3\pm \tilde{u}_4=0$ mod 2 for some choice of signs. This ensures that we break half of the left-moving supersymmetries. This choice of twist vectors also implies that the left-moving momenta are projected out, while there exists an invariant sublattice of right-moving momenta $\Lambda_r \subset \Gamma^{4,4}$, which will be determined in what follows.

\subsection*{An asymmetric $\mathbb{Z}_2$, $\mathcal{N}=6$ orbifold}
\label{sec:N=6}

Here, we present an example of a $\mathbb{Z}_2$ orbifold, breaking 8 left-moving supersymmetries, with twist vectors $\tilde{u}=\left(0,0,\tfrac{1}{2},\tfrac{1}{2}\right)$ and $u=(0,0,0,0)$, or in terms of the mass parameters $\vec{m}=(\pi,0,0,0)$.\footnote{Here we employ the notation $\vec{m}=(m_1,m_2,m_3,m_4)$.} The appropriate lattice to consider, which is invariant under the $\mathbb{Z}_2$ asymmetric rotation, is
\begin{equation}
    \Gamma^{5,5}=\Gamma^{4,4}(D_4)\oplus\Gamma^{1,1}\,.
    \label{latticeN6}
\end{equation}
Now, let us check the conditions for modular invariance. Fist, we compute
\begin{equation}
    p_L\mathcal{N}_Lp_L - p_R\mathcal{N}_Rp_R=-\left(p_L^2+p_R^2\right)\,,
\end{equation}
which is even for $p_L-p_R \in \Lambda_R(D_4)$. Thus, condition \eqref{modcond2} is met. Also, it can be easily verified that the twist vectors satisfy \eqref{modular conditions on twist vectors} (for this orbifold $p=2$). So, modular invariance is ensured. In the following, we will also construct the partition function and check explicitly that it is modular invariant.

In order to determine the invariant sublattice, we notice that our orbifold acts trivially on the right-movers, while left-movers obtain a non-zero phase under the orbifold action. Hence, the invariant sublattice $I$ is spanned by the vectors $(0,p_R)$, which combined with the condition ${p}_L-{p}_R\in \Lambda_R(D_4)$ yields ${p}_R\in \Lambda_R(D_4)\subset \Lambda_W (D_4)$.  The associated lattice sum is\footnote{A thorough discussion on lattices and theta functions can be found in \cite{conway2013sphere}.} 
\begin{equation}
    \Theta_{D_4}(\tau)= \sum_{P\in D_4}q^{\frac{1}{2}P^2}= \frac{1}{2}\left(\vartheta_3(\tau)^4+\vartheta_4(\tau)^4\right)\,.
\end{equation}
Recall that the orbifold acts as a shift on $\Gamma^{1,1}$ and leaves this lattice invariant. So, the complete orbifold invariant lattice is
\begin{equation}
    \hat{I}=\Lambda_R(D_4) \oplus\Gamma^{1,1}\ .
\end{equation}
Now, we have all the necessary ingredients to construct the partition function. Using the techniques described in section \ref{sec:Partition function} we find
\begin{equation}
    \begin{aligned}
       Z[0,1]=Z_{\mathbb{R}^{1,4}}{Z}_{S^1}[0,1]& \left(\frac{\bar{\eta}}{\bar{\vartheta_2}}\right)^2\frac{1}{\bar{\eta}^4} [(\bar{\vartheta_3}\bar{\vartheta_4})^2-(\bar{\vartheta_4}\bar{\vartheta_3})^2-(\bar{\vartheta_2}\bar{\vartheta_1})^2-(\bar{\vartheta_1}\bar{\vartheta_2)^2}] \,\times\\
      &  \frac{1}{{\eta}^4}\Theta_{D_4}(\tau)\frac{1}{{\eta}^4}
      [(\vartheta_3)^4-(\vartheta_4)^4-(\vartheta_2)^4-(\vartheta_1)^4]\,.\quad \xrightarrow{\mathcal{S}}
    \end{aligned}
\end{equation}
\begin{equation}
    \begin{aligned}
     Z[1,0]=Z_{\mathbb{R}^{1,4}}{Z}_{S^1}[1,0]& \left(\frac{\bar{\eta}}{\bar{\vartheta_4}}\right)^2\frac{1}{\bar{\eta}^4}  [(\bar{\vartheta_3}\bar{\vartheta_2})^2+(\bar{\vartheta_4}\bar{\vartheta_1})^2-(\bar{\vartheta_2}\bar{\vartheta_3})^2+(\bar{\vartheta_1}\bar{\vartheta_4)^2}] \,\times\\
    & \frac{1}{2{\eta}^4}\Theta_{D_4^*}(\tau)\frac{1}{{\eta}^4}
      [(\vartheta_3)^4-(\vartheta_4)^4-(\vartheta_2)^4-(\vartheta_1)^4]\,.\quad \xrightarrow{\mathcal{T}}
    \end{aligned}
\end{equation}
\begin{equation}
    \begin{aligned}
     Z[1,1]=Z_{\mathbb{R}^{1,4}}{Z}_{S^1}[1,1]& \left(\frac{\bar{\eta}}{\bar{\vartheta_3}}\right)^2\frac{1}{\bar{\eta}^4}  [(\bar{\vartheta_3}\bar{\vartheta_1})^2+(\bar{\vartheta_4}\bar{\vartheta_2})^2-(\bar{\vartheta_2}\bar{\vartheta_4})^2+(\bar{\vartheta_1}\bar{\vartheta_3)^2}] \,\times\\
    & \frac{1}{2{\eta}^4}\Theta_{D_4^*}(\tau+1)\frac{1}{{\eta}^4}
      [(\vartheta_3)^4-(\vartheta_4)^4-(\vartheta_2)^4-(\vartheta_1)^4]\,.
    \end{aligned}
\end{equation}
The above pieces of the partition function satisfy the following modular transformations
\begin{equation}
     {{\mathcal{T}}} \lcirclearrowright Z[0,1]      \xleftrightarrow{\mathcal{S}} Z[1,0]  \xleftrightarrow{\mathcal{T}} Z[1,1] \rcirclearrowleft  {\mathcal{S}}\,,
     \label{explicit modular orbits}
\end{equation}
which ensure modular invariance. It is worth mentioning here that the above $\mathbb{Z}_2$ chiral twist is a symmetry of the $(A_1)^4$ lattice as well\footnote{$(A_1)^4$ is a shorthand notation for $A_1 \oplus A_1 \oplus A_1 \oplus A_1$. }. However, if the $(A_1)^4$ lattice is chosen, the condition \eqref{modcond2} is not satisfied and modular invariance is lost. In particular, for the model based on the $(A_1)^4$ lattice, performing a modular $\mathcal{S}$ transformation on $Z[1,1]$ does not give back $Z[1,1]$, due to a sign difference in the lattice sum. Furthermore, under two consecutive $\mathcal{T}$ transformations neither $Z[1,0]$ nor $Z[1,1]$ get back to themselves, which in turn means that level-matching is not satisfied \cite{vafa1986modular}. The fact that $(A_1)^4$ fails level-matching was also discussed in \cite{blumenhagen1999orientifolds}. In addition, the model constructed in \cite{sen1995dual} based on the $(A_1)^4$ lattice suffers from this problem, as  was pointed out in \cite{dabholkar1999string}.

We now move on to the construction of the closed string spectrum. First, we work out the massless spectrum in the untwisted sector. Recall that for doing so we do not add momentum or winding modes on the circle. We use \eqref{generic partition untwisted} with $n=w=0$, $p=2$, $\tilde{u}=\left(0,0,\tfrac{1}{2},\tfrac{1}{2}\right)$ and $u=(0,0,0,0)$. Thus, we obtain\footnote{Here and below we will omit writing down the the $T^4$ lattice sum because, for this model, states carrying $T^4$ momenta are not the lowest excited states.}
\begin{equation}
    {Z}[0,l]= (q\bar{q})^{-\frac{1}{2}}\sum_{{r},\tilde{{r}}} q^{\frac{1}{2}{r}^2}\,(\bar{q})^{\frac{1}{2}\tilde{r}^2}e^{\pi il (\tilde{r}_3+\tilde{r}_4)}\, \left(1+\cdots\right)\,.
    \label{N=6 massless untwisted}
\end{equation}
We observe that orbifold invariant states have to satisfy $\tilde{r}_3+\tilde{r}_4=0$ mod 2. By examining \autoref{tablemasslessstates} we find the following invariant states

{\noindent NS-NS sector:}
\begin{equation}
    \begin{aligned}
     (\underline{\pm 1,0},0,0) &\otimes (\underline{\pm 1,0},0,0)=\textbf{5}\oplus3\times\textbf{3}\oplus2\times\textbf{1}\\
     (\underline{\pm 1,0},0,0) &\otimes (0,0,\underline{\pm 1,0})=4\times\textbf{3}\oplus4\times\textbf{1}
    \end{aligned}
\end{equation}
NS-R sector:
\begin{equation}
    \begin{aligned}
        (\underline{\pm 1,0},0,0) &\otimes \pm (\pm\tfrac{1}{2},\pm\tfrac{1}{2},\tfrac{1}{2},\tfrac{1}{2})= 2\times \textbf{4}\oplus 4\times \textbf{2}  \\
        (\underline{\pm 1,0},0,0) &\otimes \pm(\underline{\tfrac{1}{2},-\tfrac{1}{2}},\tfrac{1}{2},-\tfrac{1}{2})=2\times \textbf{4}\oplus 4\times \textbf{2}
       \end{aligned}
\end{equation}
R-NS sector:
\begin{equation}
    \begin{aligned}
      \pm (\underline{\tfrac{1}{2},-\tfrac{1}{2}},\tfrac{1}{2},-\tfrac{1}{2})&\otimes (\underline{\pm 1,0},0,0) = 2\times \textbf{4}\oplus 4\times \textbf{2}\\
      \pm (\underline{\tfrac{1}{2},-\tfrac{1}{2}},\tfrac{1}{2},-\tfrac{1}{2}) &\otimes (0,0,\underline{\pm 1,0})=8\times \textbf{2}
    \end{aligned}
\end{equation}
R-R sector
\begin{equation}
    \begin{aligned}
      \pm (\underline{\tfrac{1}{2},-\tfrac{1}{2}},\tfrac{1}{2},-\tfrac{1}{2})&\otimes \pm (\pm\tfrac{1}{2},\pm\tfrac{1}{2},\tfrac{1}{2},\tfrac{1}{2})=4\times \textbf{3}\oplus 4 \times \textbf{1}\\
     \pm  (\underline{\tfrac{1}{2},-\tfrac{1}{2}},\tfrac{1}{2},-\tfrac{1}{2})&\otimes \pm (\underline{\tfrac{1}{2},-\tfrac{1}{2}},\tfrac{1}{2},-\tfrac{1}{2})=4 \times \textbf{3}\oplus 4\times \textbf{1}
     \end{aligned}
\end{equation}
All together, we find the graviton, 15 vectors, 14 scalars, 6 gravitini and 20 dilatini. These massless fields fit into the $\mathcal{N}=6$ gravity multiplet. Regarding the notation, underlining denotes permutations, e.g. $(0,0,\underline{\pm 1,0})$ corresponds to $(0,0,1,0), (0,0,0,1), (0,0,-1,0)$ and $(0,0,0,-1)$. Now, we move on to the massive spectrum. For the construction of massive states, we take combinations from table \autoref{tablemasslessstates} that obtain a non-zero phase under the orbifold action and we cancel this phase by adding momentum modes on the circle. Whenever we add momentum and/or winding to a state, we denote it only on the left-movers by $(\tilde{r}_1,\tilde{r}_2,\tilde{r}_3,\tilde{r}_4;n,w)$. Once again, we use \eqref{generic partition untwisted} with the same values as in \eqref{N=6 massless untwisted}, with the exception that we allow $n \neq 0$. We obtain
\begin{equation}
     {Z}[0,l]=(q\,\bar{q})^{-\frac{1}{2}}\sum_{n \in \mathbb{Z}}e^{ {\pi i n}l}\, (q\,\bar{q})^{\frac{\alpha'n^2}{4\mathcal{R}^2}}\,\sum_{{r},\tilde{r}}  q^{\frac{1}{2}{r}^2} (\bar{q})^{\frac{1}{2}\tilde{r}^2} e^{\pi il (\tilde{r}_3+\tilde{r}_4)}\, \left(1+\cdots\right)\,.
\end{equation}
We find the following massive states

{\noindent NS-NS sector:}
\begin{equation}
    \begin{aligned}
     \pm(0,0,\underline{1,0};-1)&\otimes  (\underline{\pm 1,0},0,0)= 4\times(\textbf{2},\textbf{2})\\
    \pm (0,0,\underline{1,0};-1)&\otimes (0,0,\underline{\pm1,0})= 16\times(\textbf{1},\textbf{1})
    \end{aligned}
\end{equation}
NS-R sector:
\begin{equation}
    \begin{aligned}
     \pm (0,0,\underline{1,0};-1)&\otimes\pm(\pm\tfrac{1}{2},\pm\tfrac{1}{2},\tfrac{1}{2},\tfrac{1}{2})=8\times (\textbf{2},\textbf{1})\\
      \pm  (0,0,\underline{1,0};-1)&\otimes\pm(\underline{\tfrac{1}{2},-\tfrac{1}{2}},\tfrac{1}{2},-\tfrac{1}{2})=8\times (\textbf{1},\textbf{2})
    \end{aligned}
\end{equation}
R-NS sector:
\begin{equation}
    \begin{aligned}
    \pm   (\pm\tfrac{1}{2},\pm\tfrac{1}{2},\tfrac{1}{2},\tfrac{1}{2};-1)&\otimes(\underline{\pm 1,0},0,0)=2 \times (\textbf{3},\textbf{2})\oplus 2 \times (\textbf{1},\textbf{2})\\
      \pm (\pm\tfrac{1}{2},\pm\tfrac{1}{2},\tfrac{1}{2},\tfrac{1}{2};-1)&\otimes(0,0,\underline{\pm1,0})=8\times (\textbf{2},\textbf{1})
    \end{aligned}
\end{equation}
R-R sector:
\begin{equation}
    \begin{aligned}
     \pm (\pm\tfrac{1}{2},\pm\tfrac{1}{2},\tfrac{1}{2},\tfrac{1}{2};-1) &\otimes\pm(\pm\tfrac{1}{2},\pm\tfrac{1}{2},\tfrac{1}{2},\tfrac{1}{2})=4\times(\textbf{3},\textbf{1})\oplus 4 \times (\textbf{1},\textbf{1})\\
      \pm  (\pm\tfrac{1}{2},\pm\tfrac{1}{2},\tfrac{1}{2},\tfrac{1}{2};-1)&\otimes\pm (\underline{\tfrac{1}{2},-\tfrac{1}{2}},\tfrac{1}{2},-\tfrac{1}{2})=4\times(\textbf{2},\textbf{2})
    \end{aligned}
\end{equation}
In total, we find 2 gravitini (\textbf{3},\textbf{2}), 4 tensors (\textbf{3},\textbf{1}), 8 vectors (\textbf{2},\textbf{2}), 26 dilatini, $16\times(\textbf{2},\textbf{1})$ and $10\times(\textbf{1},\textbf{2})$, and 20 scalars (\textbf{1},\textbf{1}). All these fields have mass (cf. \eqref{untwisted masses})
\begin{equation}
    \frac{\alpha' m^2_{L}}{2}=\frac{\alpha' m^2_{R}}{2}=\frac{\alpha'}{4\mathcal{R}^2} \quad\Rightarrow \quad m= \left|\frac{1}{\mathcal{R}}\right|\,,
\end{equation}
due to the contribution of the $n=\pm1$ momentum mode on the circle, and fit into a complex (1,2) BPS supermultiplet with the representations
\begin{equation}
(\textbf{3},\textbf{2})\oplus2\times(\textbf{3},\textbf{1})\oplus4\times(\textbf{2},\textbf{2})\oplus5\times(\textbf{1},\textbf{2})\oplus8\times(\textbf{2},\textbf{1})\oplus10\times(\textbf{1},\textbf{1})\,.
\label{bps reps n=6}
\end{equation}
All massive and massless states are constructed such that the combination between left and right-movers ensures zero phase. However, we can add to all these states
a trivial phase $e^{(\pi i l)2\mathbb{Z}}$ simply by adding $2\mathbb{Z}$ momentum modes along the circle. In this way we can construct Kaluza-Klein towers on the circle, where the contribution from each (even) momentum mode to the mass of the state is $\left|{2\mathbb{Z}}/{\mathcal{R}}\right|$. Furthermore, we can generally identify the orbifold radius $\mathcal{R}$, with the Scherk-Schwarz radius $R$, by $\mathcal{R}=p{R}$, where $p$ is the orbifold rank. In our $\mathbb{Z}_2$ example, this means that we can write the mass of the BPS supermultiplet and the contribution from the Kaluza-Klein towers as $|{1}/{2R}|$ and $|{\mathbb{Z}}/{R}|$ respectively. This entire spectrum arising from our orbifold construction in the untwisted sector matches exactly with the one found in \cite{Hull:2020byc} from the Scherk-Schwarz reduction on the level of supergravity.

Besides the untwisted spectrum, we are also interested in finding the lightest states in the twisted sectors (in a $\mathbb{Z}_2$ orbifold there is only one such sector for $k=1$). In order to construct the twisted spectrum we use \eqref{generic partition twisted}-\eqref{shifted energy} (with $p=2$, $k=1$, $\tilde{u}=\left(0,0,\tfrac{1}{2},\tfrac{1}{2}\right)$ and $u=(0,0,0,0)$). We obtain
\begin{equation}
     Z[1,l]=2q^{-\frac{1}{2}}(\bar{q})^{-\frac{1}{4}}\,\sum_{{n,w\in \mathbb{Z}}}e^{{\pi i n}l} q^{\frac{\alpha'}{4}P_{R}^2(1)} (\bar{q})^{\frac{\alpha'}{4}P_{L}^2(1)}\,\sum_{{r},\tilde{r}}  q^{\frac{1}{2}{r}^2}(\bar{q})^{\frac{1}{2}(\tilde{r}+\tilde{u})^2} e^{{\pi}il(\tilde{r}_3+\tilde{r}_4+1)}(1+\cdots)\,.
\end{equation}
Note that all states come with a multiplicative factor of 2, which is precisely the number of chiral fixed points\footnote{Note that in the $k=1$ sector the expression $\prod_{i}2\sin(\pi \text{\footnotesize{gcd}}(1,l)\tilde{u}_i)$ simply becomes $\prod_{i}2\sin(\pi \tilde{u}_i)$.} $4 \sin^2({\pi}/{2})=4$, divided by the volume of the invariant sublattice ${\text{Vol}(D_4)}=2$. The weight vectors for the lightest right-moving states are, again, given in \autoref{tablemasslessstates} because the orbifold acts trivially on the right-movers. On the contrary, the weight vectors for the lightest left-moving states are listed in \autoref{k=1 N=6 states}. Since the twisted states are in general massive, we only write down the representation of these states under the massive little group in five dimensions.
\renewcommand{\arraystretch}{2}
\begin{table}[h!]
\centering
 \begin{tabular}{|c|c|c|}
    \hline
    Sector &  $\tilde{r}$ & SO(4) rep\\
    \hline
    \hline
  NS & $(0,0,\underline{-1,0})$ & 2$\,\times\,(\textbf{1},\textbf{1})$\\
  \hline
   R  & $(\pm\frac{1}{2},\pm\frac{1}{2},-\frac{1}{2},-\frac{1}{2})$ & $(\textbf{2},\textbf{1})$\\
   \hline
    \end{tabular}
\captionsetup{width=.9\linewidth}
\caption{\textit{Here we list the weight vectors of the lightest left-moving states in the $k=1$ twisted sector of the $\mathbb{Z}_2$, $\mathcal{N}=6$ orbifold and their representations under the massive little group in 5D.}}
\label{k=1 N=6 states}
\end{table}
\renewcommand{\arraystretch}{1}

\noindent In the twisted sector, orbifold invariant states have to satisfy $\tilde{r}_3+\tilde{r}_4+1=0$ mod 2. We list below the states that we find in each sector

{\noindent NS-NS sector:}
\begin{equation}
\begin{aligned}
 (0,0,\underline{-1,0}) &\otimes (\underline{\pm 1,0},0,0)= 2 \times (\textbf{2},\textbf{2})\\
     (0,0,\underline{-1,0}) &\otimes  (0,0,\underline{\pm1,0})= 8\times (\textbf{1},\textbf{1})
\end{aligned}
\end{equation}
NS-R sector:
\begin{equation}
    \begin{aligned}
       (0,0,\underline{-1,0})&\otimes\pm (\pm\tfrac{1}{2},\pm\tfrac{1}{2},\tfrac{1}{2},\tfrac{1}{2})=4\times (\textbf{2},\textbf{1})\\
         (0,0,\underline{-1,0})&\otimes\pm(\underline{\tfrac{1}{2},-\tfrac{1}{2}},\tfrac{1}{2},-\tfrac{1}{2})=4\times (\textbf{1},\textbf{2})
    \end{aligned}
\end{equation}
R-NS sector:
\begin{equation}
    \begin{aligned}
        (\pm\tfrac{1}{2},\pm\tfrac{1}{2},-\tfrac{1}{2},-\tfrac{1}{2})&\otimes(\underline{\pm 1,0},0,0)=(\textbf{3},\textbf{2})\oplus (\textbf{1},\textbf{2})\\
       (\pm\tfrac{1}{2},\pm\tfrac{1}{2},-\tfrac{1}{2},-\tfrac{1}{2})&\otimes(0,0,\underline{\pm1,0})=4\times (\textbf{2},\textbf{1})
    \end{aligned}
\end{equation}
R-R sector:
\begin{equation}
    \begin{aligned}
            (\pm\tfrac{1}{2},\pm\tfrac{1}{2},-\tfrac{1}{2},-\tfrac{1}{2}) &\otimes\pm (\pm\tfrac{1}{2},\pm\tfrac{1}{2},\tfrac{1}{2},\tfrac{1}{2})=2\times (\textbf{3},\textbf{1})\oplus 2 \times (\textbf{1},\textbf{1})\\
        (\pm\tfrac{1}{2},\pm\tfrac{1}{2},-\tfrac{1}{2},-\tfrac{1}{2})&\otimes \pm (\underline{\tfrac{1}{2},-\tfrac{1}{2}},\tfrac{1}{2},-\tfrac{1}{2})=2\times (\textbf{2},\textbf{2})
    \end{aligned}
\end{equation}
Remember that all these states have degeneracy 2. In total we find 2 gravitini (\textbf{3},\textbf{2}), 4 tensors (\textbf{3},\textbf{1}), 8 vectors (\textbf{2},\textbf{2}), 26 dilatini, $16\times(\textbf{2},\textbf{1})$ and $10\times(\textbf{1},\textbf{2})$, and 20 scalars (\textbf{1},\textbf{1}). All these fields have mass $\left|{\mathcal{R}}/{2\alpha'}\right|$ due to the $\tfrac{1}{2}$-winding on the circle (cf. \eqref{twisted masses}), and fit into a complex (1,2) BPS supermultiplet, as in \eqref{bps reps n=6}.

\subsection*{An asymmetric $\mathbb{Z}_4$, $\mathcal{N}=6$ orbifold}

Here, we consider a $\mathbb{Z}_4$ asymmetric orbifold preserving 24 supersymmetries in five dimensions, that is $\mathcal{N}=6$ in $D=5$. The twist vectors for this orbifold are given by $\tilde{u}=\left(0,0,\frac{1}{4},\frac{1}{4}\right)$ and $u=\left(0,0,0,0\right)$, and the corresponding mass parameters are $\vec{m}=\left(\frac{\pi}{2},0,0,0\right)$. The appropriate lattice to consider is
\begin{equation}
    \Gamma^{5,5}=\Gamma^{4,4}(D_4)\oplus\Gamma^{1,1}\,,
\end{equation}
and the conditions for modular invariance can be can be verified similarly with the $\mathbb{Z}_2$ example of the previous section. The orbifold invariant lattice is\footnote{The invariant sublattice can be determined using the same arguments as in the $\mathbb{Z}_2$ example of section \ref{sec:N=6}.}
\begin{equation}
    \hat{I}=\Lambda_R(D_4) \oplus\Gamma^{1,1}\ .
\end{equation}
The partition function can be constructed as described in section \ref{sec:Partition function}. Here we will only write down the the contribution to the partition function from the compact bosons on $T^4$ and from the fermions, which we collectively denote by $\left({Z}_{T^4}{Z}_F\right)[k,l]$. We find
\begin{equation}
\begin{aligned}
    \left({Z}_{T^4}{Z}_F\right)[0,1]=  \frac{\bar{\eta}^2}{ \bar{\vartheta}[\psymbol{{1}/{2}}{-{1}/{4}}]^2}\frac{1}{\bar{\eta}^4} &\big[(\bar{\vartheta_3})^2\bar{\vartheta}[\psymbol{0}{-{1}/{4}}]^2 - (\bar{\vartheta_4})^2\bar{\vartheta}[\psymbol{0}{-{3}/{4}}]^2 - (\bar{\vartheta_2})^2\bar{\vartheta}[\psymbol{{1}/{2}}{-{1}/{4}}]^2- \\
    &(\bar{\vartheta_1})^2\bar{\vartheta}[\psymbol{{1}/{2}}{-{3}/{4}}]^2\big]
  \times\,\frac{1}{{\eta}^4}\Theta_{D_4}(\tau){\mathcal{Z}}_F[0,0]\,. \quad \xrightarrow{\mathcal{S}}
  \end{aligned}
\end{equation}
\begin{equation}
\begin{aligned}
    \left({Z}_{T^4}{Z}_F\right)[1,0]=  \frac{i\bar{\eta}^2}{ \bar{\vartheta}[\psymbol{{1}/{4}}{-{1}/{2}}]^2}\frac{1}{\bar{\eta}^4} &\big[(\bar{\vartheta_3})^2\bar{\vartheta}[\psymbol{1/4}{0}]^2 +i(\bar{\vartheta_4})^2\bar{\vartheta}[\psymbol{1/4}{-{1}/{2}}]^2 - (\bar{\vartheta_2})^2\bar{\vartheta}[\psymbol{{3}/{4}}{0}]^2 + \\
    &i(\bar{\vartheta_1})^2\bar{\vartheta}[\psymbol{{3}/{4}}{-{1}/{2}}]^2\big] \times\,\frac{1}{2{\eta}^4}\Theta_{D_4^*}(\tau){\mathcal{Z}}_F[0,0]\,. \quad \xrightarrow{\mathcal{T}}
  \end{aligned}
\end{equation}
\begin{equation}
\begin{aligned}
    \left({Z}_{T^4}{Z}_F\right)[1,3]=  \frac{i\bar{\eta}^2}{ \bar{\vartheta}[\psymbol{{1}/{4}}{{1}/{4}}]^2}\frac{1}{\bar{\eta}^4} &\big[(\bar{\vartheta_3})^2\bar{\vartheta}[\psymbol{1/4}{-3/4}]^2 -i(\bar{\vartheta_4})^2\bar{\vartheta}[\psymbol{1/4}{-{1}/{4}}]^2 - (\bar{\vartheta_2})^2\bar{\vartheta}[\psymbol{{3}/{4}}{-3/4}]^2 -\\
    &i(\bar{\vartheta_1})^2\bar{\vartheta}[\psymbol{{3}/{4}}{-{1}/{4}}]^2\big] \times\,\frac{1}{2{\eta}^4}\Theta_{D_4^*}(\tau+1){\mathcal{Z}}_F[0,0]\,. \quad \xrightarrow{\mathcal{T}}
  \end{aligned}
\end{equation}   
\begin{equation}
\begin{aligned}
    \left({Z}_{T^4}{Z}_F\right)[1,2]=  \frac{i\bar{\eta}^2}{ \bar{\vartheta}[\psymbol{{1}/{4}}{0}]^2}\frac{1}{\bar{\eta}^4} &\big[(\bar{\vartheta_3})^2\bar{\vartheta}[\psymbol{1/4}{-1/2}]^2 -i(\bar{\vartheta_4})^2\bar{\vartheta}[\psymbol{1/4}{0}]^2 - (\bar{\vartheta_2})^2\bar{\vartheta}[\psymbol{{3}/{4}}{-1/2}]^2 - \\
    &i(\bar{\vartheta_1})^2\bar{\vartheta}[\psymbol{{3}/{4}}{0}]^2\big] \times\,\frac{1}{2{\eta}^4}\Theta_{D_4^*}(\tau){\mathcal{Z}}_F[0,0]\,. \quad \xrightarrow{\mathcal{T}}
  \end{aligned}
\end{equation}   
\begin{equation}
\begin{aligned}
     \left({Z}_{T^4}{Z}_F\right)[1,1]=  \frac{i\bar{\eta}^2}{ \bar{\vartheta}[\psymbol{{1}/{4}}{-1/4}]^2}\frac{1}{\bar{\eta}^4} &\big[(\bar{\vartheta_3})^2\bar{\vartheta}[\psymbol{1/4}{-1/4}]^2 +i(\bar{\vartheta_4})^2\bar{\vartheta}[\psymbol{1/4}{-3/4}]^2 -(\bar{\vartheta_2})^2\bar{\vartheta}[\psymbol{{3}/{4}}{-1/4}]^2 + \\
     &i(\bar{\vartheta_1})^2\bar{\vartheta}[\psymbol{{3}/{4}}{-3/4}]^2\big] \times\,\frac{1}{2{\eta}^4}\Theta_{D_4^*}(\tau+1){\mathcal{Z}}_F[0,0]  \,,
  \end{aligned}
\end{equation}
where ${\mathcal{Z}}_F[0,0]$ is the right-moving fermionic partition function, which reads
\begin{equation}
    {\mathcal{Z}}_F[0,0]=\frac{1}{2{\eta}^4}
      [(\vartheta_3)^4-(\vartheta_4)^4-(\vartheta_2)^4-(\vartheta_1)^4]\,.
\end{equation}
Also, in the derivation of the above pieces of the partition function we used the property $\Theta_{D_4^*}(\tau+2)=\Theta_{D_4^*}(\tau)$. Now, by performing a modular $\mathcal{S}$ transformation on $\left({Z}_{T^4}{Z}_F\right)[1,2]$ we obtain
\begin{equation}
    \begin{aligned}
        \left({Z}_{T^4}{Z}_F\right)[2,3]=  \frac{-\bar{\eta}^2}{ \bar{\vartheta}[\psymbol{0}{1/4}]^2}\frac{1}{\bar{\eta}^4} &\big[(\bar{\vartheta_3})^2\bar{\vartheta}[\psymbol{1/2}{-3/4}]^2 +(\bar{\vartheta_4})^2\bar{\vartheta}[\psymbol{1/2}{-1/4}]^2 - (\bar{\vartheta_2})^2\bar{\vartheta}[\psymbol{0}{-3/4}]^2 +\\
        &(\bar{\vartheta_1})^2\bar{\vartheta}[\psymbol{0}{-1/4}]^2\big] \times\,\frac{1}{{\eta}^4}\Theta_{D_4}(\tau){\mathcal{Z}}_F[0,0]\,.
    \end{aligned}
    \label{z23 partition}
\end{equation}
In order to get some other pieces of the partition function effortless, we note that
\begin{equation}
    \left({Z}_{T^4}{Z}_F\right)[0,1]=\left({Z}_{T^4}{Z}_F\right)[0,3]\implies \left({Z}_{T^4}{Z}_F\right)[1,0]=\left({Z}_{T^4}{Z}_F\right)[3,0] \,.
    \label{equalities1}
\end{equation}
Then, it is also implied that
\begin{equation}
\begin{aligned}
    &\left({Z}_{T^4}{Z}_F\right)[1,3]=\left({Z}_{T^4}{Z}_F\right)[3,1]\,,\qquad\left({Z}_{T^4}{Z}_F\right)[1,2]=\left({Z}_{T^4}{Z}_F\right)[3,2]\,,\\
    &\left({Z}_{T^4}{Z}_F\right)[1,1]=\left({Z}_{T^4}{Z}_F\right)[3,3]\,,\qquad\left({Z}_{T^4}{Z}_F\right)[2,3]=\left({Z}_{T^4}{Z}_F\right)[2,1]\,.
    \end{aligned}
    \label{equalities2}
\end{equation}
Note that \eqref{equalities1} and \eqref{equalities2} are in agreement with the properties of the circle partition function \eqref{circle ids} (for this example $p=4$). Also, from the above equations and \eqref{circle ids} we can see that the $k=1$ and $k=3$ sectors and equivalent. We will use all these later in order to check modular invariance easier.

Regarding the remaining pieces of the partition function we find
\begin{equation}
\begin{aligned}
      \left({Z}_{T^4}{Z}_F\right)[0,2]=2\left(\frac{\bar{\eta}}{\bar{\vartheta_2}}\right)^2\frac{1}{\bar{\eta}^4} &[(\bar{\vartheta_3}\bar{\vartheta_4})^2-(\bar{\vartheta_4}\bar{\vartheta_3})^2-(\bar{\vartheta_2}\bar{\vartheta_1})^2-(\bar{\vartheta_1}\bar{\vartheta_2)^2}] \times \\
      & \frac{1}{{\eta}^4}\Theta_{D_4}(\tau){\mathcal{Z}}_F[0,0]\,. \xrightarrow{\mathcal{S}}
       \end{aligned}
\end{equation}
\begin{equation}
\begin{aligned}
    \left({Z}_{T^4} {Z}_F\right)[2,0]= 2\left(\frac{\bar{\eta}}{\bar{\vartheta_4}}\right)^2\frac{1}{\bar{\eta}^4}  &[(\bar{\vartheta_3}\bar{\vartheta_2})^2+(\bar{\vartheta_4}\bar{\vartheta_1})^2-(\bar{\vartheta_2}\bar{\vartheta_3})^2+(\bar{\vartheta_1}\bar{\vartheta_4)^2}] \times \\
    & \frac{1}{2{\eta}^4}\Theta_{D_4^*}(\tau){\mathcal{Z}}_F[0,0]\,.\xrightarrow{\mathcal{T}}
     \end{aligned}
\end{equation}
\begin{equation}
\begin{aligned}
    \left({Z}_{T^4}{Z}_F\right)[2,2]= 2\left(\frac{\bar{\eta}}{\bar{\vartheta_3}}\right)^2\frac{1}{\bar{\eta}^4}  &[(\bar{\vartheta_3}\bar{\vartheta_1})^2+(\bar{\vartheta_4}\bar{\vartheta_2})^2-(\bar{\vartheta_2}\bar{\vartheta_4})^2+(\bar{\vartheta_1}\bar{\vartheta_3)^2}] \times \\
    &\frac{1}{2{\eta}^4}\Theta_{D_4^*}(\tau+1){\mathcal{Z}}_F[0,0]\,.
    \end{aligned}
    \label{z22partition}
\end{equation}
Now we can put all pieces of the partition function together, and verify modular invariance. Under a modular $\mathcal{T}$ transformation we find
\begin{equation}
    \begin{aligned}
   & \qquad \qquad Z[0,1]\,,Z[0,2]\,,Z[2,1]:\text{invariant}\,, \\
   & \qquad \qquad Z[1,0]\to Z[1,3]\to Z[1,2]\to Z[1,1]\to Z[1,0]\,,\\
   & \qquad \qquad Z[2,0]  \xleftrightarrow{} Z[2,2]\,.
    \end{aligned}
    \label{modular T orbit}
\end{equation}
Under a modular $\mathcal{S}$ transformation we find
\begin{equation}
    \begin{aligned}
   & Z[0,1]\xleftrightarrow{}Z[1,0]\,,\quad Z[0,2]\xleftrightarrow{}Z[2,0]\,,\\
   &Z[1,1]\xleftrightarrow{}Z[1,3]\,,\quad Z[1,2]\xleftrightarrow{}Z[2,1]\,,\\
   &Z[2,2]:\text{invariant}\,.
    \end{aligned}
    \label{modular S orbit}
\end{equation}
The transformations \eqref{modular T orbit} and \eqref{modular S orbit} together with \eqref{circle ids}, \eqref{equalities1} and \eqref{equalities2} ensure modular invariance. 

Now, we move on to the spectrum of our model. We start from the untwisted sector, in which the partition function can be expanded as\footnote{Similarly with the $\mathbb{Z}_2$ model of section \ref{sec:N=6}, we omit writing down the the $T^4$ lattice sum because, again, states carrying $T^4$ momenta are not the lowest excited states.}
\begin{equation}
    {Z}[0,l]= (q\bar{q})^{-\frac{1}{2}}\sum_{n,w \in \mathbb{Z}}e^{\frac{\pi i n}{2}l}\, q^{\frac{\alpha'}{4}P_{R}^2(0)}\, (\bar{q})^{\frac{\alpha'}{4}P_{L}^2(0)}\sum_{{r},\tilde{{r}}} q^{\frac{1}{2}{r}^2}\,(\bar{q})^{\frac{1}{2}\tilde{r}^2}e^{\frac{\pi il}{2} (\tilde{r}_3+\tilde{r}_4)}\, \left(1+\cdots\right)\,.
    \label{N=6 massless untwisted4}
\end{equation}
Using \eqref{N=6 massless untwisted4} and \autoref{tablemasslessstates}, it is easy to verify that the spectrum of lightest states in the untwisted sector of the $\mathbb{Z}_4$, $\mathcal{N}=6$ orbifold is exactly the same with the spectrum of the $\mathbb{Z}_2$, $\mathcal{N}=6$ orbifold of section \ref{sec:N=6}. Specifically, the massless states constitute the $\mathcal{N}=6$ gravity multiplet in $5D$ and the massive states 
 fit into a complex (1,2) BPS supermultiplet with mass $\left|1/\mathcal{R}\right|$, as in \eqref{bps reps n=6}.
 
We mention here that we can construct Kaluza-Klein towers along the $S^1$ by adding a trivial phase $e^{(\frac{\pi i l}{2})4\mathbb{Z}}$ to all states. Then, the orbifold untwisted spectrum matches exactly with the Scherk-Schwarz  supergravity one found in \cite{Hull:2020byc} (with the identification $\mathcal{R}=4R$, where $R$ is the radius of the Scherk-Schwarz circle). 

We continue with the spectrum of lightest states in the twisted sectors, and we start from the $k=1$ sector. As before, we expand the partition function in powers of $q\bar{q}$ and focus on the lowest order terms. We have
\begin{equation}
     Z[1,l]=q^{-\frac{1}{2}}(\bar{q})^{-\frac{5}{16}}\,\sum_{{n,w\in \mathbb{Z}}}e^{\frac{\pi i n}{2}l} q^{\frac{\alpha'}{4}P_{R}^2(1)} (\bar{q})^{\frac{\alpha'}{4}P_{L}^2(1)}\,\sum_{{r},\tilde{r}}  q^{\frac{1}{2}{r}^2}(\bar{q})^{\frac{1}{2}(\tilde{r}+\tilde{u})^2} e^{\frac{\pi il}{2}(\tilde{r}_3+\tilde{r}_4+1)}(1+\cdots)\,.
\end{equation}
 The weight vectors for the lightest right-moving states are, again, given in \autoref{tablemasslessstates}. The weight vectors for the lightest left-moving states are the same with the weight vectors of \autoref{k=1 N=6 states}. In the $k=1$ sector, orbifold invariant states have to satisfy $\tilde{r}_3+\tilde{r}_4+1=0$ mod 4. We list below the states that we find in each sector

{\noindent NS-NS sector:}
\begin{equation}
\begin{aligned}
 (0,0,\underline{-1,0}) &\otimes (\underline{\pm 1,0},0,0)= 2 \times (\textbf{2},\textbf{2})\\
     (0,0,\underline{-1,0}) &\otimes  (0,0,\underline{\pm1,0})= 8\times (\textbf{1},\textbf{1})
\end{aligned}
\end{equation}
NS-R sector:
\begin{equation}
    \begin{aligned}
       (0,0,\underline{-1,0})&\otimes\pm (\pm\tfrac{1}{2},\pm\tfrac{1}{2},\tfrac{1}{2},\tfrac{1}{2})=4\times (\textbf{2},\textbf{1})\\
         (0,0,\underline{-1,0})&\otimes\pm(\underline{\tfrac{1}{2},-\tfrac{1}{2}},\tfrac{1}{2},-\tfrac{1}{2})=4\times (\textbf{1},\textbf{2})
    \end{aligned}
\end{equation}
R-NS sector:
\begin{equation}
    \begin{aligned}
        (\pm\tfrac{1}{2},\pm\tfrac{1}{2},-\tfrac{1}{2},-\tfrac{1}{2})&\otimes(\underline{\pm 1,0},0,0)=(\textbf{3},\textbf{2})\oplus (\textbf{1},\textbf{2})\\
       (\pm\tfrac{1}{2},\pm\tfrac{1}{2},-\tfrac{1}{2},-\tfrac{1}{2})&\otimes(0,0,\underline{\pm1,0})=4\times (\textbf{2},\textbf{1})
    \end{aligned}
\end{equation}
R-R sector:
\begin{equation}
    \begin{aligned}
            (\pm\tfrac{1}{2},\pm\tfrac{1}{2},-\tfrac{1}{2},-\tfrac{1}{2}) &\otimes\pm (\pm\tfrac{1}{2},\pm\tfrac{1}{2},\tfrac{1}{2},\tfrac{1}{2})=2\times (\textbf{3},\textbf{1})\oplus 2 \times (\textbf{1},\textbf{1})\\
        (\pm\tfrac{1}{2},\pm\tfrac{1}{2},-\tfrac{1}{2},-\tfrac{1}{2})&\otimes \pm (\underline{\tfrac{1}{2},-\tfrac{1}{2}},\tfrac{1}{2},-\tfrac{1}{2})=2\times (\textbf{2},\textbf{2})
    \end{aligned}
\end{equation}
In total we find 1 gravitino (\textbf{3},\textbf{2}), 2 tensors (\textbf{3},\textbf{1}), 4 vectors (\textbf{2},\textbf{2}), 13 dilatini, $8\times(\textbf{2},\textbf{1})$ and $5\times(\textbf{1},\textbf{2})$, and 10 scalars (\textbf{1},\textbf{1}). All these fields have mass $\left|{\mathcal{R}}/{4\alpha'}\right|$ due to the $\tfrac{1}{4}$-winding on the circle (cf. \eqref{twisted masses}). In the $k=3$ sector we find exactly the same spectrum (in the $k=3$ sector the lightest states are those with winding number $w=-1$). Together, the fields from both the $k=1$ and $k=3$ sectors fit into a complex (1,2) BPS supermultiplet, as in \eqref{bps reps n=6}.

Finally, in the $k=2$ sector we have
\begin{equation}
     Z[2,l]=2  q^{-\frac{1}{2}}(\bar{q})^{-\frac{1}{4}}\,\sum_{{n,w\in \mathbb{Z}}}e^{\frac{\pi i n}{2}l} q^{\frac{\alpha'}{4}P_{R}^2(2)} (\bar{q})^{\frac{\alpha'}{4}P_{L}^2(2)}\,\sum_{{r},\tilde{r}}  q^{\frac{1}{2}{r}^2}(\bar{q})^{\frac{1}{2}(\tilde{r}+2\tilde{u})^2} e^{\frac{\pi i l}{2}(\tilde{r}_3+\tilde{r}_4-1)}(1+\cdots)\,,
\end{equation}
where the overall factor of $2$, can be read of from the relevant pieces of the partition function \eqref{z23 partition} and \eqref{z22partition}, after bosonization (also, recall that $Z[2,1]=Z[2,3]$). Regarding the spectrum of the $k=2$ sector, the weight vectors of the lightest states coincide with those in the $k=1$ sector. However, in the $k=2$ sector states with $\tilde{r}=(0,0,\underline{-1,0})$ or $\tilde{r}=(\pm\frac{1}{2},\pm\frac{1}{2},-\frac{1}{2},-\frac{1}{2})$  survive the orbifold projection only if suitable combination of momentum modes, winding modes and $T^4$ right-moving momenta are added to these states\footnote{I would like to thank Guillaume Bossard for useful correspondence on this point.}. These are higher excited stringy states, which we omit writing down. 

We conclude this section by discussing the low-energy spectrum that is obtained without the shift on the circle, i.e. from the non-freely acting orbifold $T^4/\mathbb{Z}_4 \times S^1$. In this case, there are no massive fields in the untwisted sector; these are projected out of the spectrum instead. In addition to the 6 massless gravitini from the untwisted sector, there are another 2 massless gravitini coming from the $k=1$ and $k=3$ twisted sectors, as now these states do not carry fractional winding numbers. This leads to a supersymmetry enhancement from 24 to 32 supersymmetries. In the $k=2$ sector all states are massive even without the fractional windings. Hence, we see that we simply obtain an $\mathcal{N}=8$ theory in five dimensions. Furthermore, in the de-compactification limit $\mathcal{R}\to \infty$, which corresponds to an asymmetric $T^4/\mathbb{Z}_4$ orbifold compactification, one obtains an $\mathcal{N}=8, D=6$ theory.  

The same result can be obtained from the lowest lying spectrum of our freely acting orbifold in the limit $\mathcal{R}\to \infty$. Recall that the mass of all fields in the untwisted sector is equal to $|1/\mathcal{R}|$. In the de-compactification limit these fields become massless and we obtain 2 additional massless gravitini. On the other hand, the mass of the fields in the twisted sector is proportional to $\mathcal{R}$. These become infinitely massive and decouple. Consequently, we find 8 massless gravitini and we retrieve the $\mathcal{N}=8, D=6$ theory. In fact, in the de-compactification limit all our orbifolds reduce to the $\mathcal{N}=8, D=6$ theory because, as we will also see in the next examples, the masses of all fields in the untwisted sector are proportional to $1/\mathcal{R}$, while in the twisted sectors they are proportional to $\mathcal{R}$ (for large $\mathcal{R}$).

Similar arguments also hold for the $\mathbb{Z}_{2}$ and $\mathbb{Z}_3$, $\mathcal{N}=6$ orbifolds. Finally, it is worth noting that both freely and non-freely acting orbifolds prevent the appearance of an $\mathcal{N}=6, D=6$ theory. Such a theory can be defined classically but in the quantum level is inconsistent since it suffers from gravitational anomalies (for a discussion on the $\mathcal{N}=6, D=6$ supergravity see also \cite{d1998n}).

\subsection{$\mathcal{N}=4$}
\label{sec:N=4 symmetric and asymmetric}
In this section, we discuss orbifolds with $\mathcal{N}=4$ supersymmetry in five dimensions. These models can be realized by symmetric or asymmetric constructions. Here we will present a symmetric $\mathbb{Z}_4$,  $\mathcal{N}=4\,(0,2)$ and an asymmetric  $\mathbb{Z}_2$,  $\mathcal{N}=4\,(1,1)$  orbifold.

\subsection*{A symmetric $\mathbb{Z}_4,$ $\mathcal{N}=4\,(0,2)$ orbifold}
\label{symmetric z4}

As a first example, we consider a non-chiral symmetric $\mathbb{Z}_4$ orbifold (cf. example \ref{non-chiral N=4}) with twist vectors $\tilde{u}=u=(0,0,\tfrac{1}{4},\tfrac{1}{4})$, or equivalently $\vec{m}=(\tfrac{\pi}{2},\tfrac{\pi}{2},0,0)$, breaking half of the left and right-moving supersymmetries. We choose the torus lattice to be the $(A_1)^4$ root lattice with basis vectors $R_1(1,0,0,0)$, $R_1(0,1,0,0)$, $R_2(0,0,1,0)$ and $R_2(0,0,0,1)$, and we set the $B$-field to zero. This $\mathbb{Z}_4$ orbifold acts non-trivially on all toroidal dimensions. Consequently, there will be no invariant sublattices of left or right-moving momenta contributing to ${Z}_{T^4}[k,l]$ (for $k$ or $l\neq 0$). As was discussed in section \ref{sec:Partition function}, the partition functions for the symmetric orbifolds that we consider are always modular invariant and we will not  present them here. 

Regarding the orbifold spectrum, we construct closed string states following the same procedure as in section \ref{sec:N=6}. We begin with the massless spectrum in the untwisted sector where invariant states satisfy $\tilde{r}_3+\tilde{r_4}-r_3-r_4=0$ mod 4. We find the following states

{\noindent NS-NS sector:}
\begin{equation}
    \begin{aligned}
      (\underline{\pm 1,0},0,0) &\otimes (\underline{\pm 1,0},0,0)=\textbf{5}\oplus3\times\textbf{3}\oplus2\times\textbf{1}\\
    \pm[  (0,0,\underline{ 1,0}) &\otimes (0,0,\underline{ 1,0})] = 8\times\textbf{1}
    \end{aligned}
\end{equation}
NS-R sector:
\begin{equation}
    \begin{aligned}
      (\underline{\pm 1,0},0,0) &\otimes \pm(\underline{\tfrac{1}{2},-\tfrac{1}{2}},\tfrac{1}{2},-\tfrac{1}{2})=2\times \textbf{4}\oplus 4\times \textbf{2}\\
      \pm[ (0,0,\underline{ 1,0}) &\otimes (\pm\tfrac{1}{2},\pm\tfrac{1}{2},\tfrac{1}{2},\tfrac{1}{2})]=   4\times \textbf{2}
    \end{aligned}
\end{equation}
R-R sector:
\begin{equation}
    \begin{aligned}
    \pm[ (\pm\tfrac{1}{2},\pm\tfrac{1}{2},\tfrac{1}{2},\tfrac{1}{2})&\otimes (\pm\tfrac{1}{2},\pm\tfrac{1}{2},\tfrac{1}{2},\tfrac{1}{2})] = 2\times \textbf{3} \oplus 2\times \textbf{1}\\
    \pm (\underline{\tfrac{1}{2},-\tfrac{1}{2}},\tfrac{1}{2},-\tfrac{1}{2}) &\otimes \pm (\underline{\tfrac{1}{2},-\tfrac{1}{2}},\tfrac{1}{2},-\tfrac{1}{2}) = 4\times \textbf{3} \oplus 4 \times \textbf{1} 
    \end{aligned}
\end{equation}
The spectrum in the R-NS sector is identical with the one found in the NS-R sector. This is expected because we are treating a symmetric orbifold.  Collecting together our results from the four sectors, we find the graviton, 9 vectors, 4 gravitini, 16 dilatini and 16 scalars. These form the $\mathcal{N}=4$ gravity multiplet, consisting of the graviton, 4 gravitini, 6 vectors, 4 dilatini and 1 scalar, coupled to three vector multiplets, each made up from 1 vector, 4 dilatini and 5 scalars. We continue with the massive states

{\noindent NS-NS sector:}
\begin{equation}
    \begin{aligned}
   \pm[   (\underline{\pm 1,0},0,0;1) &\otimes (0,0,\underline{ 1,0})]= 4\times (\textbf{2},\textbf{2})\\
       \pm (0,0,\underline{1,0};-1)&\otimes  (\underline{\pm 1,0},0,0)= 4 \times (\textbf{2},\textbf{2})\\
  \pm[   (0,0,\underline{1,0};-2)&\otimes (0,0,\underline{-1,0})] = 8\times (\textbf{1},\textbf{1})
    \end{aligned}
\end{equation}
NS-R sector:
\begin{equation}
\begin{aligned}
  \pm[  (\underline{\pm 1,0},0,0;1) &\otimes (\pm\tfrac{1}{2},\pm\tfrac{1}{2},\tfrac{1}{2},\tfrac{1}{2})]= 2\times (\textbf{3},\textbf{2}) \oplus 2\times (\textbf{1},\textbf{2})\\
  \pm[ (0,0,\underline{1,0};-2) &\otimes (\pm\tfrac{1}{2},\pm\tfrac{1}{2},-\tfrac{1}{2},-\tfrac{1}{2})]=4\times (\textbf{2},\textbf{1})\\
   \pm (0,0,\underline{1,0};-1) &\otimes \pm(\underline{\tfrac{1}{2},-\tfrac{1}{2}},\tfrac{1}{2},-\tfrac{1}{2})  =8\times (\textbf{1},\textbf{2})
   \end{aligned}
\end{equation}
R-R sector: 
\begin{equation}
    \begin{aligned}
  \pm[  (\pm\tfrac{1}{2},\pm\tfrac{1}{2},\tfrac{1}{2},\tfrac{1}{2};-2)  &\otimes (\pm\tfrac{1}{2},\pm\tfrac{1}{2},-\tfrac{1}{2},-\tfrac{1}{2})]= 2\times (\textbf{3},\textbf{1}) \oplus 2\times (\textbf{1},\textbf{1})\\
 \pm(\pm\tfrac{1}{2},\pm\tfrac{1}{2},\tfrac{1}{2},\tfrac{1}{2};-1)  &\otimes \pm(\underline{\tfrac{1}{2},-\tfrac{1}{2}},\tfrac{1}{2},-\tfrac{1}{2}) = 4\times (\textbf{2},\textbf{2})\\
\pm(\underline{\tfrac{1}{2},-\tfrac{1}{2}},\tfrac{1}{2},-\tfrac{1}{2}) & \otimes \pm(\pm\tfrac{1}{2},\pm\tfrac{1}{2},\tfrac{1}{2},\tfrac{1}{2};1) = 4\times (\textbf{2},\textbf{2})
  \end{aligned}
\end{equation}
\footnote{In the last line of the R-R sector, for clearer notation we denoted the momentum of the states on the right-movers.}Again, in the R-NS sector we find the same spectrum as in the NS-R sector. In total, we find 16 vectors $(\textbf{2},\textbf{2})$ with mass $|{1}/{\mathcal{R}}|$, 2 tensors $(\textbf{3},\textbf{1})$ and 10 scalars $(\textbf{1},\textbf{1})$ with mass $|{2}/{\mathcal{R}}|$, 4 gravitini $(\textbf{3},\textbf{2})$ and 20 dilatini  $(\textbf{1},\textbf{2})$ with mass $|{1}/{\mathcal{R}}|$ and 8 dilatini $(\textbf{2},\textbf{1})$ with mass $|{2}/{\mathcal{R}}|$ (see the mass formulae \eqref{untwisted masses}). These fields fit into two complex $(0,2)$ spin-$\tfrac{3}{2}$ multiplets of the form $(\textbf{3},\textbf{2}) \oplus 4 \times (\textbf{2},\textbf{2}) \oplus 5\times (\textbf{1},\textbf{2})$ with mass $|1/\mathcal{R}|$, and one complex $(0,2)$ tensor multiplet $(\textbf{3},\textbf{1}) \oplus 4 \times (\textbf{2},\textbf{1}) \oplus 5\times (\textbf{1},\textbf{1})$ with mass $|2/\mathcal{R}|$. Finally, we construct Kaluza-Klein towers by adding a trivial phase $e^{(\frac{\pi i l}{2})4\mathbb{Z}}$ to all states and we identify $\mathcal{R}=4R$. In this way, we can verify that the orbifold untwisted spectrum matches exactly with the Scherk-Schwarz supergravity one found in \cite{Hull:2020byc}. Now, we move on to the twisted sectors and we start our analysis with the $k=1$ sector. The weight vectors for the lightest left and right-moving states are the same, and coincide with those listed in  \autoref{k=1 N=6 states}. Orbifold invariant states satisfy $\tilde{r}_3+\tilde{r_4}-r_3-r_4=0$ mod 4. We list below the states that we find in each sector

{\noindent NS-NS sector:}
\begin{equation}
     (0,0,\underline{-1,0})\otimes   (0,0,\underline{-1,0}) = 4\times (\textbf{1},\textbf{1})
\end{equation}
NS-R sector:
\begin{equation}
     (0,0,\underline{-1,0})\otimes (\pm\tfrac{1}{2},\pm\tfrac{1}{2},-\tfrac{1}{2},-\tfrac{1}{2})= 2\times   (\textbf{2},\textbf{1})
\end{equation}
R-NS sector
\begin{equation}
    (\pm\tfrac{1}{2},\pm\tfrac{1}{2},-\tfrac{1}{2},-\tfrac{1}{2}) \otimes (0,0,\underline{-1,0}) = 2\times   (\textbf{2},\textbf{1})
\end{equation}
R-R sector:
\begin{equation}
     (\pm\tfrac{1}{2},\pm\tfrac{1}{2},-\tfrac{1}{2},-\tfrac{1}{2})\otimes  (\pm\tfrac{1}{2},\pm\tfrac{1}{2},-\tfrac{1}{2},-\tfrac{1}{2}) = (\textbf{3},\textbf{1}) \oplus (\textbf{1},\textbf{1})
\end{equation}
The degeneracy of the above states is 4. In total, we find 4 tensors $(\textbf{3},\textbf{1})$ 16 dilatini $(\textbf{2},\textbf{1})$ and 20 scalars $(\textbf{1},\textbf{1})$. These fields have mass $|{\mathcal{R}}/{4\alpha'}|$, due to the $\tfrac{1}{4}$-winding on the circle (cf. \eqref{twisted masses}), and they fit into two complex $(0,2)$ tensor multiplets of the form $(\textbf{3},\textbf{1}) \oplus 4 \times (\textbf{2},\textbf{1}) \oplus 5\times (\textbf{1},\textbf{1})$. Regarding the $k=2$ twisted sector, the weight vectors for the lightest states coincide with those of the $k=1$ sector. Hence, we find the same states. However, the degeneracy of the states in the $k=2$ twisted sector is 10 and their mass is $|{\mathcal{R}}/{2\alpha'}|$ (cf. \eqref{twisted masses}). Finally, the $k=3$ twisted sector is equivalent to the $k=1$ sector. Note that the lightest states in the $k=3$ sector are those with winding $w=-1$\footnote{The same result can be obtained by considering $k=-1$ instead of $k=3$.}. Concluding, in the twisted sectors of the symmetric $\mathbb{Z}_4$, $\mathcal{N}=4\,(0,2)$ orbifold we find 9 complex tensor multiplets, 4 with mass $|{\mathcal{R}}/{4\alpha'}|$ and 5 with mass $|{\mathcal{R}}/{2\alpha'}|$ .

Finally, we would like to briefly discuss here the spectrum obtained by the corresponding non-freely acting orbifold $T^4/\mathbb{Z}_4\times S^1$. In the untwisted sector the massless spectrum consists of the $\mathcal{N}=4$ gravity multiplet coupled to 3 vector multiplets. In the twisted sectors one finds 18 massless vector multiplets. In total, there exist 27 massless vectors.
The resulting number of vectors is consistent with the anomaly cancellation condition in type IIB $\mathcal{N}=4\,(0,2)$ theory in 6$D$, where 21 massless tensor multiplets are required for the anomalies to cancel \cite{townsend1984new}. Then compactification of the 6$D$ theory on a circle yields the 5$D$ theory with exactly 27 vectors, upon dualizing the tensors into vectors (see also \cite{bonetti2013exploring}). For the $T^4/\mathbb{Z}_{3,6}\times S^1$ orbifolds the result is exactly the same and for the $T^4/\mathbb{Z}_2\times S^1$ orbifold, 11 vectors come from the untwisted sector and 16 from the untwisted sector, giving again a total of 27 vectors\footnote{Type IIB on $K3\times S^1$  also gives 27 vectors in $D=5$.}.

\subsection*{An asymmetric $\mathbb{Z}_2,$ $\mathcal{N}=4\,(1,1)$ orbifold} 
\label{subsec N=4 1,1}

Here, we consider an asymmetric $\mathbb{Z}_2$ orbifold with twist vectors $\tilde{u}=(0,0,0,0)$ and $u=(0,0,0,1)$, i.e. $\vec{m}=(0,\pi,0,-\pi)$. An interesting characteristic of this orbifold is the form of the right-moving twist vector $u$, which generates a $(-1)^{F_R}$ action. As we saw in the example of the fermionic monodromies in section \ref{fermionic monodromies}, this orbifold acts trivially on the torus coordinates, but spacetime fermions do feel the twist. Consequently, the torus bosonic partition is given by
\begin{equation}
    {Z}_{T^4} = \frac{1}{(\eta\,\bar{\eta})^4} \Gamma^{4,4}\,,
\end{equation}
which simply corresponds to a $T^4$ compactification and is invariant under modular transformations. For the remaining pieces of the partition function we find

\begin{equation}
\begin{aligned}
     Z[0,1]=Z_{\mathbb{R}^{1,4}}{Z}_{S^1}[0,1]{Z}_{T^4}\frac{1}{4(\eta\,\bar{\eta})^4} &[(\bar{\vartheta}_3)^4-(\bar{\vartheta}_4)^4-(\bar{\vartheta}_2)^4-(\bar{\vartheta}_1)^4] \,\times\\
&[(\vartheta_3)^4-(\vartheta_4)^4+(\vartheta_2)^4+(\vartheta_1)^4]\,.
\end{aligned}
\end{equation}
\begin{equation}
\begin{aligned}
     Z[1,0]=Z_{\mathbb{R}^{1,4}}{Z}_{S^1}[1,0]{Z}_{T^4}\frac{1}{4(\eta\,\bar{\eta})^4} &[(\bar{\vartheta}_3)^4-(\bar{\vartheta}_4)^4-(\bar{\vartheta}_2)^4-(\bar{\vartheta}_1)^4] \,\times\\
&[(\vartheta_3)^4+(\vartheta_4)^4-(\vartheta_2)^4+(\vartheta_1)^4]\,.
\end{aligned}
\end{equation}
\begin{equation}
\begin{aligned}
     Z[1,1]=Z_{\mathbb{R}^{1,4}}{Z}_{S^1}[1,1]{Z}_{T^4}\frac{1}{4(\eta\,\bar{\eta})^4} &[(\bar{\vartheta}_3)^4-(\bar{\vartheta}_4)^4-(\bar{\vartheta}_2)^4-(\bar{\vartheta}_1)^4] \,\times\\
&[-(\vartheta_3)^4-(\vartheta_4)^4-(\vartheta_2)^4+(\vartheta_1)^4]\,.
\end{aligned}
\end{equation}
Under modular transformations, the above pieces of the partition function transform as in \eqref{explicit modular orbits} and this guarantees modular invariance. Concerning the orbifold spectrum, in the untwisted sector orbifold invariant states satisfy $r_4\in \mathbb{Z}$. This means that all NS-NS and R-NS states survive the orbifold projection and remain massless. On the other hand, all NS-R and R-R states are charged under the orbifold action and survive the orbifold projection only with the addition of $n=\pm 1$ momentum modes. In this case, there is no need to write down explicitly the states that we find in each sector. The massless spectrum consists of the graviton, 11 vectors, 26 scalars, 4 gravitini and 24 dilatini. These fields make up the $\mathcal{N}=4$ gravity multiplet coupled to five vector multiplets. The massive spectrum consists of 8 vectors $(\textbf{2},\textbf{2})$, 8 tensors, $4\times(\textbf{3},\textbf{1})$ and $4\times(\textbf{1},\textbf{3})$, 8 scalars $(\textbf{1},\textbf{1})$, 4 gravitini, $2\times(\textbf{3},\textbf{2})$ and $2\times(\textbf{2},\textbf{3})$ and 20 dilatini,  $10\times(\textbf{1},\textbf{2})$ and $10\times(\textbf{2},\textbf{1})$. All these fields have mass $|{1}/{\mathcal{R}}|$ and form two complex $(1,1)$ spin-$\tfrac{3}{2}$ multiplets in the representations
\begin{equation}
    \begin{aligned}
       &(\textbf{3},\textbf{2})\oplus2\times(\textbf{3},\textbf{1})\oplus2\times(\textbf{2},\textbf{2})\oplus(\textbf{1},\textbf{2})\oplus4\times(\textbf{2},\textbf{1})\oplus2\times(\textbf{1},\textbf{1})\,,\\
       &(\textbf{2},\textbf{3})\oplus2\times(\textbf{1},\textbf{3})\oplus2\times(\textbf{2},\textbf{2})\oplus(\textbf{2},\textbf{1})\oplus4\times(\textbf{1},\textbf{2})\oplus2\times(\textbf{1},\textbf{1})\,.
    \end{aligned}
    \label{n=4 1,1 massive reps}
\end{equation}
Finally, for the construction of the Kaluza-Klein towers one works exactly as in section \ref{sec:N=6}. Once again, we can verify that the untwisted orbifold spectrum matches exactly the one found in \cite{Hull:2020byc} from the Scherk-Schwarz reduction on the level of supergravity. 

Now, consider the $k=1$ twisted sector where we use\footnote{Here there is an additional phase $e^{\pi i l}$ coming from the fermionic partition function \eqref{fermionic infinite sums} which is not cancelled by the torus bosonic partition function due to the form of the twist vector $u$. Also, we omit writing down the $T^4$ Narain lattice.} 
\begin{equation}
     Z[1,l]=(q\,\bar{q})^{-\frac{1}{2}}\,\sum_{ {n,w\in \mathbb{Z}}}e^{ {\pi i n}l} q^{\frac{\alpha'}{4}P_{R}^2(1)} (\bar{q})^{\frac{\alpha'}{4}P_{L}^2(1)}\,\sum_{{r},\tilde{r}}  q^{\frac{1}{2}{(r+u)^2}}(\bar{q})^{\frac{1}{2}\tilde{r}^2} e^{{-\pi}il(2r_4+1)}(1+\cdots)\,.
\end{equation}
Since the orbifold acts trivially on the left-movers, the weight vectors for the lightest left-moving states in the absence of momentum and/or winding modes are given in \autoref{tablemasslessstates}. The weight vectors for the lightest right-moving states are listed in \autoref{k=1 N=4 asymmetric states}. Note here that the state with $r=(0,0,0,-1)$ is tachyonic. However, as we shall demonstrate in what follows, tachyonic states do not survive the orbifold projection, as expected in a supersymmetric model.
\renewcommand{\arraystretch}{2}
\begin{table}[h!]
\centering
 \begin{tabular}{|c|c|c|}
    \hline
    Sector &  r & SO(4) rep \\
    \hline
    \hline
  NS  & $(0,0,0,-1)$ &  $(\textbf{1},\textbf{1})$   \\
  \hline
   \multirow{4}{*}{R}   &$(\pm\frac{1}{2},\pm\frac{1}{2},-\frac{1}{2},-\frac{1}{2})$&$(\textbf{2},\textbf{1})$ \\  
   \cline{2-3}
  & $(\pm\frac{1}{2},\pm\frac{1}{2},\frac{1}{2},-\frac{3}{2})$ & $(\textbf{2},\textbf{1})$\\
   \cline{2-3}
  &  $(\underline{\frac{1}{2},-\frac{1}{2}},\frac{1}{2},-\frac{1}{2})$ & $(\textbf{1},\textbf{2})$\\
   \cline{2-3}
  &  $(\underline{\frac{1}{2},-\frac{1}{2}},-\frac{1}{2},-\frac{3}{2})$ & $(\textbf{1},\textbf{2})$\\
    \hline
    \end{tabular}
\captionsetup{width=.9\linewidth}
\caption{\textit{Here we list the weight vectors of the lightest right-moving states in the $k=1$ twisted sector of the asymmetric $\mathbb{Z}_2$, $\mathcal{N}=4$ orbifold and their representations under the massive little group in 5D.}}
\label{k=1 N=4 asymmetric states}
\end{table}
\renewcommand{\arraystretch}{1}

\noindent Orbifold invariant states satisfy $2r_4+1=0$ mod 2. As it follows from \autoref{k=1 N=4 asymmetric states}, all states in the NS-R and R-R sectors are invariant under the orbifold action. These are 8 vectors $(\textbf{2},\textbf{2})$, 8 tensors, $4\times(\textbf{3},\textbf{1})$ and $4\times(\textbf{1},\textbf{3})$, 8 scalars $(\textbf{1},\textbf{1})$, 4 gravitini, $2\times(\textbf{3},\textbf{2})$, and $2\times(\textbf{2},\textbf{3})$ and 20 dilatini,  $10\times(\textbf{1},\textbf{2})$ and $10\times(\textbf{2},\textbf{1})$. All these fields have mass $|{\mathcal{R}}/{2\alpha'}|$, due to the $\tfrac{1}{2}$-winding on the circle (cf. \eqref{twisted masses}), and form two complex $(1,1)$ spin-$\tfrac{3}{2}$ multiplets, exactly as in \eqref{n=4 1,1 massive reps}. Moving on to the NS-NS and R-NS sectors, we notice that states are neither invariant under the orbifold action nor level-matched. However we can fix both issues either by adding $n=-1,w=0$ or $n=+1, w=-1$ momentum and winding modes respectively on the circle. We find the following states

{\noindent NS-NS sector:}
\begin{equation}
    \begin{aligned}
        (\underline{\pm1,0},0,0;-1,0)&\otimes (0,0,0,-1) = (\textbf{2},\textbf{2})\\
          (\underline{\pm1,0},0,0;1,-1)&\otimes (0,0,0,-1) = (\textbf{2},\textbf{2})\\
          (0,0,\underline{\pm1,0};-1,0)&\otimes (0,0,0,-1) = 4\times (\textbf{1},\textbf{1})\\
          (0,0,\underline{\pm1,0};1,-1)&\otimes (0,0,0,-1) = 4\times (\textbf{1},\textbf{1})
    \end{aligned}
\end{equation}
R-NS sector:
\begin{equation}
    \begin{aligned}
         (\pm(\pm\tfrac{1}{2},\pm\tfrac{1}{2},\tfrac{1}{2},\tfrac{1}{2});-1,0)&\otimes (0,0,0,-1) = 2\times (\textbf{2},\textbf{1})\\
         (\pm(\pm\tfrac{1}{2},\pm\tfrac{1}{2},\tfrac{1}{2},\tfrac{1}{2});1,-1)&\otimes (0,0,0,-1) = 2\times (\textbf{2},\textbf{1})\\
         (\pm(\underline{\tfrac{1}{2},-\tfrac{1}{2}},\tfrac{1}{2},-\tfrac{1}{2});-1,0) &\otimes (0,0,0,-1) = 2\times (\textbf{1},\textbf{2})\\
          (\pm(\underline{\tfrac{1}{2},-\tfrac{1}{2}},\tfrac{1}{2},-\tfrac{1}{2});1,-1) &\otimes (0,0,0,-1) = 2\times (\textbf{1},\textbf{2})
    \end{aligned}
\end{equation}
In total we find 2 vectors $(\textbf{2},\textbf{2})$, 8 scalars $(\textbf{1},\textbf{1})$ and 8 dilatini,  $4\times(\textbf{1},\textbf{2})$ and $4\times(\textbf{2},\textbf{1})$. These fields form one complex $(1,1)$ vector multiplet with mass $ m = \left|{1}/{\mathcal{R}}-{\mathcal{R}}/{2\alpha'}\right|$ (cf. \eqref{twisted masses}). As we have seen so far, states in the twisted sectors are generally massive. However, some states can become massless at special points of the moduli space. In the specific example that we discuss here, this can be achieved at circle radius $\mathcal{R}=\sqrt{2\alpha'}$. In this massless limit the massive complex vector multiplet gives two massless real vector multiplets (this can be seen by using the appropriate representations of the states under the massless little group in five dimensions). This is a generic feature of $\mathcal{N}=4\,(1,1)$ theories, where we break all the left, or right-moving supersymmetries, and we will revisit it when we discuss the moduli spaces of the orbifolds in chapter \ref{chap:mod5}.

\subsection{$\mathcal{N}=2$}
\label{sec:N=2 asymmetric orbifold}
In this section, we discuss orbifolds with $\mathcal{N}=2$ supersymmetry in five dimensions. These are non-chiral asymmetric constructions (cf. example \ref{non-chiral N=2}) which break half of the left-moving and all the right-moving supersymmetries (or the other way around). As an example, consider the $\mathbb{Z}_2$ orbifold with twist vectors $\tilde{u}=(0,0,\tfrac{1}{2},\tfrac{1}{2})$ and $u=(0,0,0,1)$, i.e. $\vec{m}=(\pi,\pi,0,-\pi)$. Regarding the bosonic contribution to the partition function, notice that the twist vector $\tilde{u}$ is the same as in section \ref{sec:N=6}. Moreover, the twist vector $u$ acts trivially on the torus coordinates, exactly as the twist vector of section \ref{sec:N=6}. Therefore, the bosonic partition function of this $\mathcal{N}=2$ model coincides with the partition function of the $\mathcal{N}=6$ model obtained in section \ref{sec:N=6}. However, the fermionic partition function is different because the twist vector $u$ generates a $(-1)^{F_R}$ action. Putting all together we find 
\begin{equation}
    \begin{aligned}
       Z[0,1]=Z_{\mathbb{R}^{1,4}}{Z}_{S^1}[0,1]& \left(\frac{\bar{\eta}}{\bar{\vartheta_2}}\right)^2\frac{1}{\bar{\eta}^4} [(\bar{\vartheta_3}\bar{\vartheta_4})^2-(\bar{\vartheta_4}\bar{\vartheta_3})^2-(\bar{\vartheta_2}\bar{\vartheta_1})^2-(\bar{\vartheta_1}\bar{\vartheta_2)^2}] \,\times\\
      &  \frac{1}{{\eta}^4}\Theta_{D_4}(\tau)\frac{1}{{\eta}^4}
      [(\vartheta_3)^4-(\vartheta_4)^4+(\vartheta_2)^4+(\vartheta_1)^4]\,.
    \end{aligned}
\end{equation}
\begin{equation}
    \begin{aligned}
     Z[1,0]=Z_{\mathbb{R}^{1,4}}{Z}_{S^1}[1,0]& \left(\frac{\bar{\eta}}{\bar{\vartheta_4}}\right)^2\frac{1}{\bar{\eta}^4}  [(\bar{\vartheta_3}\bar{\vartheta_2})^2+(\bar{\vartheta_4}\bar{\vartheta_1})^2-(\bar{\vartheta_2}\bar{\vartheta_3})^2+(\bar{\vartheta_1}\bar{\vartheta_4)^2}] \,\times\\
    & \frac{1}{2{\eta}^4}\Theta_{D_4^*}(\tau)\frac{1}{{\eta}^4}
      [(\vartheta_3)^4+(\vartheta_4)^4-(\vartheta_2)^4+(\vartheta_1)^4]\,.
    \end{aligned}
\end{equation}
\begin{equation}
    \begin{aligned}
     Z[1,1]=Z_{\mathbb{R}^{1,4}}{Z}_{S^1}[1,1]& \left(\frac{\bar{\eta}}{\bar{\vartheta_3}}\right)^2\frac{1}{\bar{\eta}^4}  [(\bar{\vartheta_3}\bar{\vartheta_1})^2+(\bar{\vartheta_4}\bar{\vartheta_2})^2-(\bar{\vartheta_2}\bar{\vartheta_4})^2+(\bar{\vartheta_1}\bar{\vartheta_3)^2}] \,\times\\
    & \frac{1}{2{\eta}^4}\Theta_{D_4^*}(\tau+1)\frac{1}{{\eta}^4}
      [-(\vartheta_3)^4-(\vartheta_4)^4-(\vartheta_2)^4+(\vartheta_1)^4]\,.
    \end{aligned}
\end{equation}
These pieces of the partition function satisfy the transformations \eqref{explicit modular orbits}. This guarantees modular invariance\footnote{As in the $\mathcal{N}=6$ example of section \ref{sec:N=6}, the model based on the $(A_1)^4$ lattice, which is also discussed in \cite{sen1995dual}, does not give a modular invariant partition function.}. Now, let us proceed with the construction of the orbifold spectrum. As usual, we start with the massless states in the untwisted sector, which obey $\tilde{r}_3+\tilde{r}_4-2r_4=0$ mod 2 

{\noindent NS-NS sector:}
\begin{equation}
    \begin{aligned}
       (\underline{\pm 1,0},0,0) &\otimes (\underline{\pm 1,0},0,0)=\textbf{5}\oplus3\times\textbf{3}\oplus2\times\textbf{1}\\
     (\underline{\pm 1,0},0,0) &\otimes (0,0,\underline{\pm 1,0})=4\times\textbf{3}\oplus4\times\textbf{1}
    \end{aligned}
\end{equation}
NS-R sector:
\begin{equation}
    \begin{aligned}
      (0,0,\underline{\pm 1,0}) &\otimes \pm (\pm\tfrac{1}{2},\pm\tfrac{1}{2},\tfrac{1}{2},\tfrac{1}{2})=   8\times \textbf{2}  \\
       (0,0,\underline{\pm 1,0}) &\otimes \pm (\underline{\tfrac{1}{2},-\tfrac{1}{2}},\tfrac{1}{2},-\tfrac{1}{2})=8\times \textbf{2}
    \end{aligned}
\end{equation}
R-NS sector:
\begin{equation}
    \begin{aligned}
    \pm (\underline{\tfrac{1}{2},-\tfrac{1}{2}},\tfrac{1}{2},-\tfrac{1}{2}) & \otimes  (\underline{\pm 1,0},0,0) = 2\times \textbf{4} \oplus 4\times \textbf{2}\\
      \pm(\underline{\tfrac{1}{2},-\tfrac{1}{2}},\tfrac{1}{2},-\tfrac{1}{2}) & \otimes   (0,0,\underline{\pm 1,0}) = 8\times \textbf{2}
    \end{aligned}
\end{equation}
R-R sector:
\begin{equation}
    \begin{aligned}
    \pm (\pm\tfrac{1}{2},\pm\tfrac{1}{2},\tfrac{1}{2},\tfrac{1}{2})&\otimes\pm  (\pm\tfrac{1}{2},\pm\tfrac{1}{2},\tfrac{1}{2},\tfrac{1}{2})=4\times \textbf{3}\oplus 4\times \textbf{1}\\
     \pm (\pm\tfrac{1}{2},\pm\tfrac{1}{2},\tfrac{1}{2},\tfrac{1}{2})&\otimes \pm (\underline{\tfrac{1}{2},-\tfrac{1}{2}},\tfrac{1}{2},-\tfrac{1}{2})=4\times \textbf{3}\oplus 4\times \textbf{1}
    \end{aligned}
\end{equation}
We find the graviton, 15 vectors, 14 scalars, 2 gravitini and 28 dilatini. These fields form the $\mathcal{N}=2$ gravity multiplet consisting of the graviton, 2 gravitini and 1 vector, and 14 vector multiplets each consisting of 1 vector 2 dilatini and 1 scalar. We continue with the construction of the massive spectrum

{\noindent NS-NS sector:}
\begin{equation}
    \begin{aligned}
    \pm(0,0,\underline{1,0};-1) & \otimes (\underline{\pm1,0},0,0)=  4 \times (\textbf{2},\textbf{2})\\
     \pm (0,0,\underline{1,0};-1) & \otimes  (0,0,{\pm 1,0}) = 8 \times (\textbf{1},\textbf{1})\\
     \pm [(0,0,\underline{1,0};1) & \otimes (0,0,0,1) = 4 \times (\textbf{1},\textbf{1})\\
     \pm [(0,0,\underline{1,0};-3) & \otimes (0,0,0,-1) = 4 \times (\textbf{1},\textbf{1})
        \end{aligned}
\end{equation}
NS-R sector:
\begin{equation}
    \begin{aligned}
   \pm[  (\underline{\pm1,0},0,0;1) &\otimes (\pm\tfrac{1}{2},\pm\tfrac{1}{2},\tfrac{1}{2},\tfrac{1}{2})] = 2\times (\textbf{3},\textbf{2}) \oplus 2\times (\textbf{1},\textbf{2})\\
   \pm[   (\underline{\pm1,0},0,0;-1) &\otimes  (\underline{\tfrac{1}{2},-\tfrac{1}{2}},\tfrac{1}{2},-\tfrac{1}{2})] = 2\times (\textbf{2},\textbf{3}) \oplus 2\times (\textbf{2},\textbf{1})
    \end{aligned}
\end{equation}
R-NS sector:
\begin{equation}
    \begin{aligned}
    \pm (\pm\tfrac{1}{2},\pm\tfrac{1}{2},\tfrac{1}{2},\tfrac{1}{2};-1) &\otimes  (\underline{\pm1,0},0,0) = 2 \times (\textbf{3},\textbf{2}) \oplus 2 \times (\textbf{1},\textbf{2})\\
     \pm (\pm\tfrac{1}{2},\pm\tfrac{1}{2},\tfrac{1}{2},\tfrac{1}{2};-1) &\otimes (0,0,\pm1,0) = 4 \times   (\textbf{2},\textbf{1})\\
    \pm [(\pm\tfrac{1}{2},\pm\tfrac{1}{2},\tfrac{1}{2},\tfrac{1}{2};1) &\otimes (0,0,0,1)] = 2 \times (\textbf{2},\textbf{1})\\
     \pm [(\pm\tfrac{1}{2},\pm\tfrac{1}{2},\tfrac{1}{2},\tfrac{1}{2};-3) &\otimes (0,0,0,-1)] = 2 \times (\textbf{2},\textbf{1})
      \end{aligned}
\end{equation}
R-R sector:
\begin{equation}
    \begin{aligned}
    \pm (\underline{\tfrac{1}{2},-\tfrac{1}{2}},\tfrac{1}{2},-\tfrac{1}{2}) & \otimes \pm(\pm\tfrac{1}{2},\pm\tfrac{1}{2},\tfrac{1}{2},\tfrac{1}{2};1)= 4 \times (\textbf{2},\textbf{2})\\
   \pm (\underline{\tfrac{1}{2},-\tfrac{1}{2}},\tfrac{1}{2},-\tfrac{1}{2}) & \otimes \pm(\underline{\tfrac{1}{2},-\tfrac{1}{2}},\tfrac{1}{2},-\tfrac{1}{2};-1) =  4 \times (\textbf{1},\textbf{3}) \oplus  4 \times (\textbf{1},\textbf{1})
    \end{aligned}
\end{equation}
\footnote{In the R-R sector, for clearer notation we denoted the momentum of the states on the right-movers.}In total, we find 6 gravitini, $4\times(\textbf{3},\textbf{2})$ and $2\times(\textbf{2},\textbf{3})$, 4 tensors (\textbf{1},\textbf{3}), 8 vectors (\textbf{2},\textbf{2}), 12 dilatini, $8\times(\textbf{2},\textbf{1})$ and $4\times(\textbf{1},\textbf{2})$, and 16 scalars (\textbf{1},\textbf{1}) with mass $|1/\mathcal{R}|$, and in addition, 2 dilatini (\textbf{2},\textbf{1}) and 4 scalars (\textbf{1},\textbf{1}) with mass $|3/\mathcal{R}|$ (see the mass formulae \eqref{untwisted masses}). These fields fit into the following complex multiplets: Two spin-$\tfrac{3}{2}$ multiplets $(\textbf{3},\textbf{2}) \oplus 2 \times (\textbf{2},\textbf{2}) \oplus (\textbf{1},\textbf{2})$, another multiplet containing a spin-$\tfrac{3}{2}$ particle; $ (\textbf{2},\textbf{3}) \oplus 2 \times (\textbf{1},\textbf{3})$, and four hypermultiplets $(\textbf{2},\textbf{1})\oplus 2 \times (\textbf{1},\textbf{1})$ with mass $|1/\mathcal{R}|$, as well as one hypermultiplet with mass $|3/\mathcal{R}|$. Finally, the construction of the Kaluza-Klein towers proceeds in exactly the same way as in section \ref{sec:N=6}. Once again, we confirm that the orbifold spectrum in the untwisted sector matches exactly the one found in \cite{Hull:2020byc} from the Scherk-Schwarz reduction on the level of supergravity. Now, we move on to the $k=1$ twisted sector where we use\footnote{As in the asymmetric $\mathbb{Z}_2, \mathcal{N}=4$ orbifold, here there is an additional phase $e^{\pi i l}$ coming from the fermionic partition function \eqref{fermionic infinite sums} due to the form of the twist vector $u$. Also, we have included explicitly the sum over the invariant momentum sublattice.} 
\begin{equation}
\begin{aligned}
     Z[1,l]=2q^{-\frac{1}{2}}(\bar{q})^{-\frac{1}{4}}\,&\sum_{P\in I^*}q^{\frac{1}{2}P^2}\sum_{ {n,w\in \mathbb{Z}}}e^{ {\pi i n}l} q^{\frac{\alpha'}{4}P_{R}^2(1)} (\bar{q})^{\frac{\alpha'}{4}P_{L}^2(1)}\times\\
     &\sum_{{r},\tilde{r}}  q^{\frac{1}{2}{(r+u)}^2}(\bar{q})^{\frac{1}{2}(\tilde{r}+\tilde{u})^2}\times 
     e^{{\pi}il(\tilde{r}_3+\tilde{r}_4-2r_4)}(1+\cdots)\,.
     \end{aligned}
\end{equation}
The weight vectors for the lightest states in the absence of momentum and/or winding modes are given in \autoref{k=1 N=2 states}. 
\renewcommand{\arraystretch}{2}
\begin{table}[h!]
\centering
 \begin{tabular}{|c|c|c|c|c|}
    \hline
    Sector &  $\tilde{r}$ & SO(4) rep & r & SO(4) rep \\
    \hline
    \hline
  NS & $(0,0,\underline{-1,0})$ & 2$\,\times\,(\textbf{1},\textbf{1})$ & $(0,0,0,-1)$ &  $(\textbf{1},\textbf{1})$   \\
  \hline
   \multirow{4}{*}{R}   &  \multirow{4}{*}{$(\pm\frac{1}{2},\pm\frac{1}{2},-\frac{1}{2},-\frac{1}{2})$}  & \multirow{4}{*}{$(\textbf{2},\textbf{1})$} &$(\pm\frac{1}{2},\pm\frac{1}{2},-\frac{1}{2},-\frac{1}{2})$&$(\textbf{2},\textbf{1})$ \\  
   \cline{4-5}
  & & & $(\pm\frac{1}{2},\pm\frac{1}{2},\frac{1}{2},-\frac{3}{2})$ & $(\textbf{2},\textbf{1})$\\
   \cline{4-5}
  & & & $(\underline{\frac{1}{2},-\frac{1}{2}},\frac{1}{2},-\frac{1}{2})$ & $(\textbf{1},\textbf{2})$\\
   \cline{4-5}
  & & & $(\underline{\frac{1}{2},-\frac{1}{2}},-\frac{1}{2},-\frac{3}{2})$ & $(\textbf{1},\textbf{2})$\\
    \hline
    \end{tabular}
\captionsetup{width=.9\linewidth}
\caption{\textit{Here we list the weight vectors of the lightest left and right-moving states in the $k=1$ twisted sector of the $\mathbb{Z}_2$, $\mathcal{N}=2$ orbifold and their representations under the massive little group in 5D.}}
\label{k=1 N=2 states}
\end{table}
\renewcommand{\arraystretch}{1}

\noindent Orbifold invariant states satisfy $\tilde{r}_3+\tilde{r}_4-2r_4=0$ mod 2 and they have degeneracy 2. All the states in the NS-R and R-R sectors survive the orbifold projection. In these sectors we find 16 dilatini, $8\times(\textbf{2},\textbf{1})$ and $8\times(\textbf{1},\textbf{2})$, 4 tensors (\textbf{3},\textbf{1}), 4 vectors (\textbf{2},\textbf{2}) and 4 scalars (\textbf{1},\textbf{1}). All these fields have mass $\left|{\mathcal{R}}/{2\alpha'}\right|$, due to the $\tfrac{1}{2}$-winding on the circle (cf. \eqref{twisted masses}), and they fit into two complex tensor multiplets $(\textbf{3},\textbf{1})\oplus 2\times (\textbf{2},\textbf{1})\oplus (\textbf{1},\textbf{1})$ and two complex vector multiplets $(\textbf{2},\textbf{2})\oplus 2\times (\textbf{1},\textbf{2})$. Regarding the NS-NS and R-NS sectors, we observe that, as in the $\mathcal{N}=4\,(1,1)$ case, the state $r=(0,0,0,-1)$ is tachyonic. In these sectors states are not level-matched and do not survive the orbifold projection. However, with the addition of $n=-1,w=0$ or $n=+1, w=-1$ momentum and winding modes respectively on the circle, states become level-matched and survive the orbifold. Thereby, we find the following states

{\noindent NS-NS sector:}
\begin{equation}
\begin{aligned}
    (0,0,\underline{-1,0};-1,0)&\otimes (0,0,0,-1) = 2\times (\textbf{1},\textbf{1})\\
    (0,0,\underline{-1,0};1,-1)&\otimes (0,0,0,-1) = 2\times (\textbf{1},\textbf{1})
    \end{aligned}
\end{equation}
R-NS sector:
\begin{equation}
\begin{aligned}
     (\pm\tfrac{1}{2},\pm\tfrac{1}{2},-\tfrac{1}{2},-\tfrac{1}{2};-1,0)&\otimes (0,0,0,-1)= (\textbf{2},\textbf{1})\\
     (\pm\tfrac{1}{2},\pm\tfrac{1}{2},-\tfrac{1}{2},-\tfrac{1}{2};1,-1)&\otimes (0,0,0,-1)= (\textbf{2},\textbf{1})
     \end{aligned}
\end{equation}
These states come with a degeneracy factor of 2. We find 8 scalars (\textbf{1},\textbf{1}) and 4 dilatini (\textbf{2},\textbf{1}) which form two massive hypermultiplets with mass $|{1}/{\mathcal{R}}-{\mathcal{R}}/{2\alpha'}|$. Note that it is possible to make these hypermultiplets massless by fixing the circle radius at $\mathcal{R}=\sqrt{2\alpha'}$ (CF. \eqref{twisted masses}). We will return to this issue in chapter \ref{chap:mod5}, where we discuss the moduli spaces of the orbifolds.

Finally, the states in the NS-NS and R-NS sectors can survive the orbifold projection and become level-matched by adding  appropriate right-moving $T^4$ momenta, $P\in I^*$. In particular, level-matching dictates $P^2=1$. It can be easily verified that, for this orbifold, states with $P^2=1$ come with a multiplicity of 48. So, we obtain 96 scalars (\textbf{1},\textbf{1}) and 48 dilatini (\textbf{2},\textbf{1}). These fields form 24 hypermultiplets with mass $\left|{\mathcal{R}}/{2\alpha'}\right|$, due to the $\tfrac{1}{2}$-winding on the circle.

\subsection{$\mathcal{N}=0$}
\label{sec:N=0}
In this section, we discuss non-supersymmetric symmetric orbifolds (see also \cite{font2002non,kawazu2004non}). These are models with $\tilde{u}=u=(0,0,u_3,u_4)$ which do not satisfy $\pm u_3\pm u_4 = 0$ mod 2 for any choice of signs, such that all supersymmetries are broken. In general, for the construction of states in all sectors we follow the same procedure as in section \ref{sec:N=6}. The interesting feature of non-supersymmetric orbifolds is the appearance of tachyons in the twisted sectors. As we will demonstrate, if the radius of the orbifold circle is large enough compared to the string scale, there will be no tachyons in the spectrum, see e.g. \cite{Dabholkar:2002sy,Angelantonj:2006ut,Scrucca:2001ni,acharya2021stringy} for more examples. As an example, consider a symmetric $\mathbb{Z}_3$ orbifold breaking all supersymmetry, with twist vectors $\tilde{u}=u=(0,0,0,\tfrac{2}{3})$, i.e. $\vec{m}=(\tfrac{2\pi}{3},\tfrac{2\pi}{3},-\tfrac{2\pi}{3},-\tfrac{2\pi}{3})$. For this orbifold is it more convenient to decompose the $T^4$ as $T^2\times T^2$. In this way we can see that the orbifold action leaves one $T^2$ intact. Therefore, from this $T^2$, there will be an invariant lattice of left and right-moving momenta contributing to ${Z}_{T^4}[k,l]$. As usual in toroidal compactification, this lattice is a 4-dimensional even, self-dual Lorentzian lattice $\Gamma^{2,2}$, with volume equal to 1. As a consequence, there are no inconsistencies regarding the degeneracy number of the twisted states. For the other $T^2$, in order for the orbifold to be well-defined, we choose the $A_2$ root lattice with basis vectors $R_1(\sqrt{2},0)$ and $R_1(-\frac{1}{\sqrt{2}},\sqrt{\frac{3}{2}})$. We set the $B$-field to zero.

We start with the spectrum in the untwisted sector. As we are mostly interested in the twisted sectors, we do not write down explicitly all the states but we simply state the results. Orbifold invariant states satisfy $2(\tilde{r}_4-r_4)=0$ mod 3. The massless spectrum consists of the graviton, 15 vectors, 20 scalars and 8 dilatini. Note that there are no massless gravitini in the spectrum, hence the spectrum is non-supersymmetric. The massive spectrum consists of 8 vectors (\textbf{2},\textbf{2)}, 4 tensors, $2\times(\textbf{3},\textbf{1})$ and $2\times(\textbf{1},\textbf{3})$, and 12 scalars (\textbf{1},\textbf{1)} with mass $|{2}/{\mathcal{R}}|$, 2 scalars (\textbf{1},\textbf{1)} with mass $|{4}/{\mathcal{R}}|$ , and 8 gravitini, $4\times(\textbf{3},\textbf{2})$ and $4\times(\textbf{2},\textbf{3})$, and 32 dilatini, $16\times(\textbf{2},\textbf{1})$ and $16\times(\textbf{1},\textbf{2})$, with mass $|{1}/{\mathcal{R}}|$. In addition, we build the Kaluza-Klein towers by adding a trivial phase $e^{(\frac{2\pi i l}{3})3\mathbb{Z}}$ to all states. Finally, we identify the orbifold radius $\mathcal{R}$, with the Scherk-Schwarz radius $R$, by $\mathcal{R}=3{R}$ and we confirm that the entire untwisted spectrum matches with the one found in \cite{Hull:2020byc} from the Scherk-Schwarz reduction on the level of supergravity.

Let us now discuss the spectrum in the twisted sectors. As usual, we are interested in finding the lightest states in the absence of momentum and winding modes. We start from the $k=1$ twisted sector and we use\footnote{Note that in the $k=1$ sector the expression $\prod_{i}2\sin(\pi \text{\footnotesize{gcd}}(1,l)\tilde{u}_i)$ simply becomes $\prod_{i}2\sin(\pi \tilde{u}_i)$, and similarly for ${u}$. Also, we omit writing down the $T^2$ invariant Narain lattice.} 
\begin{equation}
    {Z}[1,l]= 3(q\bar{q})^{-\frac{7}{18}}\sum_{{n,w\in \mathbb{Z}}}e^{\frac{2\pi i n}{3}l} q^{\frac{\alpha'}{4}P_{R}^2(1)} (\bar{q})^{\frac{\alpha'}{4}P_{L}^2(1)}\,\sum_{r,\tilde{r}} q^{\frac{1}{2}({r}+ u)^2} (\bar{q})^{\frac{1}{2}(\tilde{r}+{u})^2}\,e^{\frac{4 \pi i l}{3}(\tilde{r}_4-r_4)}\, \left(1+\cdots\right)\,.
    \label{k=1 N=0}
\end{equation}
The weight vectors for the lightest states are given in \autoref{k=1 N=0 states}.
\renewcommand{\arraystretch}{2}
\begin{table}[h!]
\centering
 \begin{tabular}{|c|c|c|}
    \hline
    Sector &  $\tilde{r},r$ & SO(4) rep\\
    \hline
    \hline
  NS & $(0,0,0,-1)$ & $(\textbf{1},\textbf{1})$\\
  \hline
  \multirow{2}{*}{R}  & $(\pm\frac{1}{2},\pm\frac{1}{2},-\frac{1}{2},-\frac{1}{2})$ &(\textbf{2},\textbf{1})\\
    \cline{2-3}
     & $(\underline{\frac{1}{2},-\frac{1}{2}},\frac{1}{2},-\frac{1}{2})$ & (\textbf{1},\textbf{2})\\
   \hline
    \end{tabular}
\captionsetup{width=.9\linewidth}
\caption{\textit{Here we list the weight vectors of the lightest states in the $k=1$ twisted sector of the symmetric $\mathbb{Z}_3$, $\mathcal{N}=0$ orbifold and their representations under the massive little group in 5D.}}
\label{k=1 N=0 states}
\end{table}
\renewcommand{\arraystretch}{1}

\noindent Orbifold invariant states satisfy $2(\tilde{r}_4-r_4)=0$ mod 3. We list below the states that we find in each sector (all the states below come with a multiplicity 3, which we will omit writing down explicitly)

{\noindent NS-NS sector:}
\begin{equation}
    (0,0,0,-1) \otimes (0,0,0,-1) = (\textbf{1},\textbf{1})
\end{equation}
This state is a scalar with mass (cf. \eqref{twisted masses})
\begin{equation}
    \alpha' m^2= \frac{\mathcal{R}^2}{9\alpha'}-\frac{4}{3}\,.
\end{equation}
We note that this state is tachyonic if 
\begin{equation}
    {\mathcal{R}}<2\sqrt{3\alpha'}\,.
\end{equation}
Taking the circle radius to be above this tachyon bound ensures that the spectrum is tachyon-free. We move on to the R-R sector:
\begin{equation}
    \begin{aligned}
      (\pm\tfrac{1}{2},\pm\tfrac{1}{2},-\tfrac{1}{2},-\tfrac{1}{2}) & \otimes (\pm\tfrac{1}{2},\pm\tfrac{1}{2},-\tfrac{1}{2},-\tfrac{1}{2}) =  (\textbf{3},\textbf{1}) \oplus (\textbf{1},\textbf{1})\\
      (\pm\tfrac{1}{2},\pm\tfrac{1}{2},-\tfrac{1}{2},-\tfrac{1}{2}) & \otimes (\underline{\tfrac{1}{2},-\tfrac{1}{2}},\tfrac{1}{2},-\tfrac{1}{2}) = (\textbf{2},\textbf{2})\\
      (\underline{\tfrac{1}{2},-\tfrac{1}{2}},\tfrac{1}{2},-\tfrac{1}{2}) & \otimes   (\pm\tfrac{1}{2},\pm\tfrac{1}{2},-\tfrac{1}{2},-\tfrac{1}{2})= (\textbf{2},\textbf{2})\\
      (\underline{\tfrac{1}{2},-\tfrac{1}{2}},\tfrac{1}{2},-\tfrac{1}{2}) & \otimes  (\underline{\tfrac{1}{2},-\tfrac{1}{2}},\tfrac{1}{2},-\tfrac{1}{2})  =  (\textbf{1},\textbf{3}) \oplus (\textbf{1},\textbf{1})
    \end{aligned}
\end{equation}
In this sector we find 2 tensors, 1$\times$(\textbf{3},\textbf{1}) and 1$\times$(\textbf{1},\textbf{3}), 2 vectors (\textbf{2},\textbf{2}) and 2 scalars (\textbf{1},\textbf{1}) with mass $|{\mathcal{R}}/{3\alpha'}|$ (cf. \eqref{twisted masses}). In the NS-R/R-NS sectors, states are not level-matched and do not survive the orbifold projection. However, we can fix both issues by adding $n=+1/-1$ momentum modes on the circle:

{\noindent NS-R sector:}
\begin{equation}
    \begin{aligned}
        (0,0,0,-1;1) &\otimes (\pm\tfrac{1}{2},\pm\tfrac{1}{2},-\tfrac{1}{2},-\tfrac{1}{2}) = (\textbf{2},\textbf{1})\\
        (0,0,0,-1;1) &\otimes  (\underline{\tfrac{1}{2},-\tfrac{1}{2}},\tfrac{1}{2},-\tfrac{1}{2})  = (\textbf{1},\textbf{2})
    \end{aligned}
\end{equation}
R-NS sector:
\begin{equation}
    \begin{aligned}
       (\pm\tfrac{1}{2},\pm\tfrac{1}{2},-\tfrac{1}{2},-\tfrac{1}{2};-1) &\otimes (0,0,0,-1) = (\textbf{2},\textbf{1})\\
       (\underline{\tfrac{1}{2},-\tfrac{1}{2}},\tfrac{1}{2},-\tfrac{1}{2};-1)&\otimes (0,0,0,-1) = (\textbf{1},\textbf{2})
    \end{aligned}
\end{equation}
In these sectors we find 4 dilatini, 2$\times$(\textbf{2},\textbf{1}) and 2$\times$(\textbf{1},\textbf{2}), with mass $|{1}/{\mathcal{R}}-{\mathcal{R}}/{3\alpha'}|$ (cf. \eqref{twisted masses}). We note that the states in the NS-R/R-NS sectors become massless for $\mathcal{R}=\sqrt{3\alpha'}$. However, this value is under the tachyon bound. Finally, the spectrum in the $k=2$ twisted sector is identical to the spectrum in the $k=1$ sector.

\subsection*{Tachyon bounds}
In general, the spectrum of non-supersymmetric orbifolds contains tachyons coming from the twisted sectors. Nevertheless, we can repeat the same analysis as in the $\mathbb{Z}_3$ orbifold discussed above, in order to find the tachyon bounds for other non-supersymmetric, symmetric orbifolds. For each model, there is a critical circle radius $\mathcal{R}_*$ above which the string spectrum becomes tachyon-free. The critical radius can be determined by examining the $k=1$ twisted sector\footnote{As $k$ becomes bigger, the critical radius becomes smaller. The strongest constraint comes from the $k=1$ sector.}. For $\mathbb{Z}_p$ orbifolds with twist vectors of the form $(0,0,u_3,u_4)$ with $|u_3|-|u_4|={n}/{p}$, $n\in \mathbb{Z}$, we find 
\begin{equation}
   \frac{\mathcal{R}_*}{\sqrt{\alpha'}}=\sqrt{2 |n| p}\ .
    \label{critical radious non-susy}
\end{equation}
We list in \autoref{tab breaking all} all the twist vectors $u$ that we used to obtain the result \eqref{critical radious non-susy}. These are all the twist vectors (up to a sign and an exchange $u_3\leftrightarrow u_4$) breaking all  right-moving supersymmetries. For each twist vector we compute the number of chiral fixed points. In the case of symmetric orbifolds, ($u=\tilde{u}$) we also write down examples of root lattices generated by root systems of simple Lie algebras, which represent the appropriate torus lattices admitting the corresponding $\mathbb{Z}_p$ symmetries (for symmetries of root lattices see e.g. \cite{baake2002guide,koca2014quasicrystallography} and references thereof). Furthermore, we evaluate the volume, Vol, of these lattices. The root lattices $A_n,B_n$ and $D_n$ are associated to the groups  $\text{SU}(n+1),\text{SO}(2n+1)$ and $\text{SO}(2n)$ respectively. Note that the lattice generated by the $A_2$ system is the same as the $G_2$ root lattice (hexagonal). Also, $A_1\oplus A_1\cong D_2$ and $B_2$ generate the same (square) lattice. Finally, we use a shorthand notation of the form $(A_1)^2\equiv A_1\oplus A_1$.

\begin{table}[h!]
    \centering
    {\renewcommand*{\arraystretch}{2.5}
    \footnotesize{
    \begin{tabular}{|c|c|c|c|c|}
    \hline
      $\mathbb{Z}_p$ & $u$ & $\chi$ & Lattice & Vol  \\
        \hline
        \hline
       $\mathbb{Z}_2$  &  $(0,0,0,1)$ & - & & \\
       \hline
       $\mathbb{Z}_3$  &  $(0,0,0,\tfrac{2}{3})$ & ${\sqrt{3}}$ & $A_2$ & ${\sqrt{3}}$ \\
       \hline
      \multirow{2}{*}{ $\mathbb{\mathbb{Z}}_4$} &   $(0,0,0,\tfrac{1}{2})$ & 2 &   $(A_1)^2$ & 2\\
      \cline{2-5}
              &   $(0,0,\tfrac{1}{4},\tfrac{3}{4})$ & 2 & $(A_1)^4$ & 4\\
              \hline
               $\mathbb{Z}_5$ &  $(0,0,\frac{1}{5},\frac{3}{5})$  & $\sqrt{5}$ & $A_4$ & ${\sqrt{5}}$\\
               \hline
        \multirow{4}{*}{ $\mathbb{\mathbb{Z}}_6$} &   $(0,0,0,\tfrac{1}{3})$ & $\sqrt{3}$ & $A_2$ & ${\sqrt{3}}$\\
      \cline{2-5}
              & $(0,0,\tfrac{1}{3},\tfrac{2}{3})$ & 3 & $(A_2)^2$ & 3\\
              \cline{2-5}
              & $(0,0,\tfrac{1}{2},\tfrac{1}{6})$ & 2 & $(A_1)^2$ $\oplus$ $A_2$ &2${\sqrt{3}}$ \\
       \cline{2-5}
              &  $(0,0,\tfrac{1}{6},\tfrac{5}{6})$ & 1 & $(A_2)^2$ &3\\
    \hline
    $\mathbb{Z}_8$ & $(0,0,0,\frac{1}{4})$& $\sqrt{2}$ & $(A_1)^2$ & 2\\
    \hline
    \end{tabular}
    \begin{tabular}{|c|c|c|c|c|}
    \hline
      $\mathbb{Z}_p$ & $u$ & $\chi$ & Lattice & Vol  \\
        \hline
        \hline
  \multirow{2}{*}{$\mathbb{Z}_8$} 
  & $(0,0,\frac{1}{2},\frac{1}{4})$ & $2\sqrt{2}$ & $(A_1)^4$ &4\\
  \cline{2-5}
  & $(0,0,\frac{1}{8},\frac{3}{8})$  & $\sqrt{2}$ & $B_4,D_4$ & 2\\
    \hline
  \multirow{2}{*}{$\mathbb{Z}_{10}$} &   $(0,0,\frac{1}{5},\frac{2}{5})$  & $\sqrt{5}$ & $A_4$ & ${\sqrt{5}}$ \\
    \cline{2-5}
   & $(0,0,\frac{1}{10},\frac{3}{10})$  & ${1}$ &$A_4$ & ${\sqrt{5}}$ \\
    \hline
   \multirow{4}{*}{$\mathbb{Z}_{12}$}& $(0,0,0,\frac{1}{6})$ & 1  & $A_2$ & $\sqrt{3}$\\
    \cline{2-5}
    & $(0,0,\frac{1}{2},\frac{1}{3})$ & $2\sqrt{3}$ & $(A_1)^2\oplus A_2$ & $2\sqrt{3}$\\
    \cline{2-5}
    & $(0,0,\frac{1}{3},\frac{1}{6})$ & $\sqrt{3}$ & $(A_2)^2$ & 3\\
    \cline{2-5}
   & $(0,0,\frac{1}{12},\frac{5}{12})$  & 1  & $F_4,D_4$ & $\frac{1}{2},2$\\
   \hline
    \multirow{2}{*}{$\mathbb{Z}_{24}$}& $(0,0,\frac{1}{4},\frac{1}{3})$ & $\sqrt{6}$  & $(A_1)^2 \oplus A_2$ & $2\sqrt{3}$\\
    \cline{2-5}
    & $(0,0,\frac{1}{4},\frac{1}{6})$ & $\sqrt{2}$  & $(A_1)^2\oplus A_2$ & $2\sqrt{3}$\\
    \hline
   \end{tabular} 
    }}
    \caption{\textit{Twist vectors breaking all right-moving supersymmetries.}}
    \label{tab breaking all}
\end{table}
\renewcommand{\arraystretch}{1}

\section{Scherk-Schwarz spectrum}
\label{Supergravity}
In this section we  give details of the relation between the compactification of type IIB string theory on freely acting orbifolds and the Scherk-Schwarz reduction of type IIB supergravity presented in \cite{Hull:2020byc}.

We will explicitly construct the untwisted orbifold sector in terms of oscillator states. As in section \ref{closed string spectrum}, we will only focus on the lowest excited states, i.e the states that are massless without the addition of momentum and/or winding modes. This will manifest the correspondence between freely acting orbifolds and Scherk-Schwarz mechanism. 

As we discussed in section \ref{The orbifold action and partition function}, in the untwisted sector the NS-vacuum is a spacetime scalar and the R-vacuum is a spacetime spinor in all target space dimensions. The NS-vacua are invariant under the orbifold action, while the R-vacua transform as in \eqref{transformation of ramond vacua}. Having these at hand, we can discuss the resulting spectrum.

First, we present in \autoref{tablemasslessstates2} the lightest NS and R-sector states that survive the GSO projection. We write down general states that appear both in a left-moving and in a right-moving version, and we write down the orbifold charges that both of these versions carry. Furthermore, we table the representations of these states under both the massless little group SO$(3)\approx\text{SU}(2)$ and the massive little group SO$(4)\approx\text{SU}(2)\times\text{SU}(2)$ in five dimensions. 

\renewcommand{\arraystretch}{1.4}
\begin{table}[h!]
\centering
{\small
\begin{tabular}{|c|c|c|c|c|c|}
\hline
\;Sector\; & State & $L$ orbifold charge & $R$ orbifold charge & SO$(3)$ rep & SO$(4)$ rep \\ \hline\hline
NS & \;\;${b}^{\hat{\mu}}_{-1/2}\ket{0}$\;\; & $1$ & $1$ & ${\textbf{3}\oplus\textbf{1}}$ & {$({\textbf{2}},{\textbf{2}})$} \\ \cline{2-6}
& ${b}^i_{-1/2}\ket{0}$ & $e^{i(m_1\pm m_3)}$ & $e^{i(m_2\pm m_4)}$ & {$2\times\textbf{1}$} & $2\times({\textbf{1}},{\textbf{1}})$ \\ \cline{2-6}
& $\bar{{b}}^i_{-1/2}\ket{0}$ & $e^{-i(m_1\pm m_3)}$ & $e^{-i(m_2\pm m_4)}$ & {$2\times\textbf{1}$} & $2\times({\textbf{1}},{\textbf{1}})$ \\ \hline\hline
R & $|a_{1,2}\rangle$ & $e^{\pm im_1}$ & $e^{\pm im_2}$ & {$2\times\textbf{2}$} & $2\times({\textbf{2}},{\textbf{1}})$ \\ \cline{2-6}
& $|a_{3,4}\rangle$ & $e^{\pm im_3}$ & $e^{\pm im_4}$ & {$2\times\textbf{2}$} & $2\times({\textbf{1}},{\textbf{2}})$ \\ \hline
\end{tabular}}
\captionsetup{width=0.9\linewidth}
\caption{\textit{Here we write down all states that are massless in the absence of momentum and/or winding modes, including their charges under the orbifold action and their representations under the massless and massive little groups in 5D. We write down general states that appear both left-moving and right-moving. The tildes on the oscillators in the left-moving sector, and the subscripts $L$ and $R$ on the vacua are omitted. The index $i$ on the oscillators takes the values 1 and 2. The $+$ sign in the ${L/R}$ orbifold charge corresponds to $i=1$ and the $-$ sign to $i=2$.}}
\label{tablemasslessstates2}
\end{table}
\renewcommand{\arraystretch}{1}

String states are constructed by tensoring the left and right-moving states from \autoref{tablemasslessstates2}. If a state carries a non-trivial orbifold charge, we compensate for this by adding momentum modes on the circle. In \autoref{huiberttable} we give the spectrum of lowest excited string states, including their orbifold charge and little group representations.

\renewcommand{\arraystretch}{1.7}
\begin{table}[h!]
{\centering
\footnotesize{
\begin{tabular}{|c|c|c|c|c|}
\hline
\;Sector\; & State & Orbifold charge & SO$(3)$ rep & SO$(4)$ rep \\ \hline\hline
NS-NS & \;\;$\tilde{b}^{\hat{\mu}}_{-1/2}\ket{0}_L \otimes {b}^{\hat{\nu}}_{-1/2}\ket{0}_R$\;\; & $1$ & $\textbf{5} \oplus 3\times\textbf{3}\oplus2\times \textbf{1}$ & $(\textbf{3}\oplus\textbf{1},\textbf{3}\oplus\textbf{1})$ \\ \cline{2-5}
& $\tilde{b}^{\hat{\mu}}_{-1/2}\ket{0}_L \otimes {b}^i_{-1/2}\ket{0}_R$ & $e^{i(m_2\pm m_4)}$ & {$2\times{\textbf{3}\oplus2\times\textbf{1}}$} & $2\times({\textbf{2}},{\textbf{2}})$ \\ \cline{2-5}
& $\tilde{b}^{\hat{\mu}}_{-1/2}\ket{0}_L \otimes \bar{{b}}^i_{-1/2}\ket{0}_R$ & $e^{-i(m_2\pm m_4)}$ & {$2\times{\textbf{3}\oplus2\times\textbf{1}}$} & $2\times({\textbf{2}},{\textbf{2}})$ \\ \cline{2-5}
& $\tilde{b}^{i}_{-1/2}\ket{0}_L \otimes {b}^{\hat{\mu}}_{-1/2}\ket{0}_R$ & $e^{i(m_1\pm m_3)}$ & {$2\times{\textbf{3}\oplus2\times\textbf{1}}$} & $2\times({\textbf{2}},{\textbf{2}})$ \\ \cline{2-5}
& $\bar{\tilde{b}}^{i}_{-1/2}\ket{0}_L \otimes {b}^{\hat{\mu}}_{-1/2}\ket{0}_R$ & $e^{-i(m_1\pm m_3)}$ & {$2\times{\textbf{3}\oplus2\times\textbf{1}}$} & $2\times({\textbf{2}},{\textbf{2}})$ \\ \cline{2-5}
& $\tilde{b}^{i}_{-1/2}\ket{0}_L \otimes {b}^j_{-1/2}\ket{0}_R$ & $e^{i(m_1\pm m_3)+i(m_2\pm m_4)}$ & {$4\times\textbf{1}$} & $4\times({\textbf{1}},{\textbf{1}})$ \\ \cline{2-5}
& $\tilde{b}^{i}_{-1/2}\ket{0}_L \otimes \bar{{b}}^j_{-1/2}\ket{0}_R$ & $e^{i(m_1\pm m_3)-i(m_2\pm m_4)}$ & {$4\times\textbf{1}$} & $4\times({\textbf{1}},{\textbf{1}})$ \\ \cline{2-5}
& $\bar{\tilde{b}}^{i}_{-1/2}\ket{0}_L \otimes {b}^j_{-1/2}\ket{0}_R$ & $e^{-i(m_1\pm m_3)+i(m_2\pm m_4)}$ & {$4\times\textbf{1}$} & $4\times({\textbf{1}},{\textbf{1}})$ \\ \cline{2-5}
& $\bar{\tilde{b}}^{i}_{-1/2}\ket{0}_L \otimes \bar{{b}}^j_{-1/2}\ket{0}_R$ & $e^{-i(m_1\pm m_3)-i(m_2\pm m_4)}$ & {$4\times\textbf{1}$} & $4\times({\textbf{1}},{\textbf{1}})$ \\ \hline\hline

R-R & $|a_{1,2}\rangle_L \otimes |a_{1,2}\rangle_R$ & $e^{\pm im_1\pm im_2}$ & {$4\times{\textbf{3}\oplus4\times\textbf{1}}$} & $4\times({\textbf{3}\oplus\textbf{1}},{\textbf{1}})$ \\ \cline{2-5}
& $|a_{1,2}\rangle_L \otimes |a_{3,4}\rangle_R$ & $e^{\pm im_1\pm im_4}$ & {$4\times{\textbf{3}\oplus4\times\textbf{1}}$} & $4\times({\textbf{2}},{\textbf{2}})$ \\ \cline{2-5}
& $|a_{3,4}\rangle_L \otimes |a_{1,2}\rangle_R$ & $e^{\pm im_3\pm im_2}$ & {$4\times{\textbf{3}\oplus4\times\textbf{1}}$} & $4\times({\textbf{2}},{\textbf{2}})$ \\ \cline{2-5}
& $|a_{3,4}\rangle_L \otimes |a_{3,4}\rangle_R$ & $e^{\pm im_3\pm im_4}$ & {$4\times{\textbf{3}\oplus4\times\textbf{1}}$} & $4\times({{\textbf{1},\textbf{3}\oplus\textbf{1}}})$ \\ \hline\hline

NS-R & $\tilde{b}^{\hat{\mu}}_{-1/2}\ket{0}_L \otimes |a_{1,2}\rangle_R$ & $e^{\pm im_2}$ & {$2\times{\textbf{4}\oplus4\times\textbf{2}}$} & $2\times({\textbf{3}\oplus\textbf{1}},{\textbf{2}})$ \\ \cline{2-5}
& $\tilde{b}^{\hat{\mu}}_{-1/2}\ket{0}_L \otimes |a_{3,4}\rangle_R$ & $e^{\pm im_4}$ & {$2\times{\textbf{4}\oplus4\times\textbf{2}}$} & $2\times({\textbf{2}},{\textbf{3}\oplus\textbf{1}})$ \\ \cline{2-5}
& $\tilde{b}^{i}_{-1/2}\ket{0}_L \otimes |a_{1,2}\rangle_R$ & $e^{i(m_1\pm m_3)\pm im_2}$ & {$4\times\textbf{2}$} & $4\times({\textbf{2}},{\textbf{1}})$ \\ \cline{2-5}
& $\tilde{b}^{i}_{-1/2}\ket{0}_L \otimes |a_{3,4}\rangle_R$ & $e^{i(m_1\pm m_3)\pm im_4}$ & {$4\times\textbf{2}$} & $4\times({\textbf{1}},{\textbf{2}})$ \\ \cline{2-5}
& $\bar{\tilde{b}}^{i}_{-1/2}\ket{0}_L \otimes |a_{1,2}\rangle_R$ & $e^{-i(m_1\pm m_3)\pm im_2}$ & $4\times\textbf{2}$ & $4\times({\textbf{2}},{\textbf{1}})$ \\ \cline{2-5}
& $\bar{\tilde{b}}^{i}_{-1/2}\ket{0}_L \otimes |a_{3,4}\rangle_R$ & $e^{-i(m_1\pm m_3)\pm im_4}$ & {$4\times\textbf{2}$} & $4\times({\textbf{1}},{\textbf{2}})$ \\ \hline\hline

R-NS & $|a_{1,2}\rangle_L \otimes {b}^{\hat{\mu}}_{-1/2}\ket{0}_R$ & $e^{\pm im_1}$ & {$2\times{\textbf{4}\oplus4\times\textbf{2}}$} & $2\times({\textbf{3}\oplus\textbf{1}},{\textbf{2}})$ \\ \cline{2-5}
& $|a_{3,4}\rangle_L \otimes {b}^{\hat{\mu}}_{-1/2}\ket{0}_R$ & $e^{\pm im_3}$ & {$2\times{\textbf{4}\oplus4\times\textbf{2}}$} & $2\times({\textbf{2}},{\textbf{3}\oplus\textbf{1}})$ \\ \cline{2-5}
& $|a_{1,2}\rangle_L \otimes {b}^i_{-1/2}\ket{0}_R$ & $e^{\pm im_1+i(m_2\pm m_4)}$ & {$4\times\textbf{2}$} & $4\times({\textbf{2}},{\textbf{1}})$ \\ \cline{2-5}
& $|a_{3,4}\rangle_L \otimes {b}^i_{-1/2}\ket{0}_R$ & $e^{\pm im_3+i(m_2\pm m_4)}$ & {$4\times\textbf{2}$} & $4\times({\textbf{1}},{\textbf{2}})$ \\ \cline{2-5}
& $|a_{1,2}\rangle_L \otimes \bar{{b}}^i_{-1/2}\ket{0}_R$ & $e^{\pm im_1-i(m_2\pm m_4)}$ & {$4\times\textbf{2}$} & $4\times({\textbf{2}},{\textbf{1}})$ \\ \cline{2-5}
& $|a_{3,4}\rangle_L \otimes \bar{{b}}^i_{-1/2}\ket{0}_R$ & $e^{\pm im_3-i(m_2\pm m_4)}$ & {$4\times\textbf{2}$} & $4\times({\textbf{1}},{\textbf{2}})$ \\ \hline
\end{tabular}}}
\captionsetup{width=0.9\linewidth}
\caption{\textit{The spectrum of lowest excited string states including their orbifold charge and representations under the massless little group} SO$(3)\approx\text{SU}(2)$ \textit{and the massive little group} SO$(4)\approx\text{SU}(2)\times\text{SU}(2)$ \textit{in five dimensions.}}
\label{huiberttable}
\end{table}
\renewcommand{\arraystretch}{1}

We can now find the field content of our orbifold construction from \autoref{huiberttable}. As an example, take the string state $|a_{1}\rangle_L \otimes {b}^{\hat{\mu}}_{-1/2}\ket{0}_R$, which has orbifold charge $e^{im_1}$. First recall that we can always write $m_1 = {2\pi N_1}/{p}$ where $N_1$ is an integer and $p$ is the rank of the orbifold. To make the state invariant under the orbifold action, we add momentum along the $S^1$. States with momentum on the circle obtain a phase $e^{2\pi i n / p}$ with $n$ the number of modes. If we now choose $n=-N_1$, this phase becomes $e^{-2\pi i N_1 / p}$ which cancels exactly against the phase that the string state had before the addition of momentum. In other words, the state $|a_{1};-N_1,0\rangle_L \otimes {b}^{\hat{\mu}}_{-1/2}\ket{0}_R$ is invariant under the orbifold action and therefore survives in the spectrum. Here we use the notational convention to denote the momentum and winding numbers on the $S^1$ as $\ket{\;\;\;\:;n,w}$ on the left-moving vacuum.

At this point, we would like to mention that for the states in \autoref{huiberttable} it is always possible to find an integer-valued momentum number that cancels the phase due to the orbifold action. All mass parameters can be written as $m_i = 2\pi N_i/p$ with $N_i\in\mathbb{Z}$; any sum or difference of mass parameters can thus also be written as $2\pi/p$ times an integer. If we take this integer with the sign flipped as the momentum number, the total phase cancels.

Next, we can use the little group representations in \autoref{huiberttable} to determine what kind of fields the spectrum consists of. We return to the example state $|a_{1};-N_1,0\rangle_L \otimes {b}^{\hat{\mu}}_{-1/2}\ket{0}_R$. Due to the addition of momentum, the state has become massive with mass $|{N_1}/{\mathcal{R}}|$. From the table we then read off the representation as $({\textbf{3},\textbf{2})\oplus(\textbf{1}},{\textbf{2}})$, i.e. it corresponds to a massive gravitino and a massive dilatino.

We can rewrite the mass $|N_1/\mathcal{R}|$ slightly differently in order to make contact with the Scherk-Schwarz supergravity spectrum. We know that $N_1 = p\,m_1/2\pi$, and we know that the radius of the orbifold circle $\mathcal{R}$ and the radius of the Scherk-Schwarz circle $R$ are related by $\mathcal{R}=pR$. The mass of the state is therefore equal to $|m_1/2\pi R|$. Masses of this form are precisely what was found in \cite{Hull:2020byc}. For completeness, we provide in \autoref{Table: SS reduction table} the masses of the supergravity fields (see also section 3.3 of \cite{Hull:2020byc}).
\begin{table}[h]
    \centering
     {\renewcommand*{\arraystretch}{1.5}
    \begin{tabular}{c|c|c}
    \hline
    Fields & Representation & $|\mu(m_i)|$ \\
    \hline
    \hline
    Scalars & $(\textbf{5},\textbf{5})$ & $|\pm m_1 \pm m_2 \pm m_3 \pm m_4|$   \\
    &  &  $|\pm m_1 \pm m_2 |$ \\ 
     &  &  $|\pm m_3 \pm m_4 |$\\
    &  &  0\\
    \hline
    Vectors & $(\textbf{4},\textbf{4})$ & $|\pm m_{1,2} \pm m_{3,4}|$   \\
    \hline
     Tensors  & $(\textbf{5},\textbf{1})$ & $|\pm m_1 \pm m_2|, 0$   \\
     & $(\textbf{1},\textbf{5})$ & $|\pm m_3 \pm m_4|, 0$   \\
     \hline
      Gravitini  & $(\textbf{4},\textbf{1})$ & $|\pm m_{1,2}|$   \\
     & $(\textbf{1},\textbf{4})$ & $|\pm  m_{3,4}|$   \\
     \hline
     Dilatini & $(\textbf{5},\textbf{4})$ & $|\pm m_1 \pm m_2 \pm m_{3,4}|$   \\
     & &  $|\pm m_{3,4}|$\\
     & $(\textbf{4},\textbf{5})$ & $|\pm m_{1,2}\pm m_3 \pm m_4|$   \\
     & &  $|\pm m_{1,2}|$
     \end{tabular}}
     \caption{\textit{Spectrum of Scherk-Schwarz supergravity. This table is taken from} \cite{Hull:2020byc}; \textit{see also} \cite{deWit:2002vt}. \textit{The mass of each field is $|\mu(m_i)|/2\pi R$. The notation $m_{i,j}$ indicates that both indices $i$ and $j$ occur. There is no correlation between the $\pm$ signs and the $ij$
indices, so that e.g. ($\pm m_1\pm m_2$) denotes 4 different combinations of mass parameters,
and ($\pm m_{1,2}\pm m_{3,4}$) denotes 16 different combinations. For example, the 5 tensors in
the} $(\textbf{5},\textbf{1})$ \textit{representation consist of two with mass $|m_1 + m_2|/2\pi R$, two with mass $|m_1 - m_2|/2\pi R$ and one with mass $0$. }}
    \label{Table: SS reduction table}
\end{table}

\noindent Some further care must be taken when comparing the representation of the monodromy matrix acting on the world-sheet fields or on the supergravity fields. For instance, the two gravitini from the NS-R sector (the $(\textbf{3},\textbf{2})$ and the $(\textbf{2},\textbf{3})$ in \autoref{huiberttable}) only pick up a Spin(4)$_R$ monodromy (with mass parameters $m_2, m_4$), but their spacetime chirality is opposite in six dimensions. Similarly the ones from the R-NS sector transform only under Spin(4)$_L$ and again have opposite $6D$ chirality. As supergravity fields, the $6D$ gravitini sit in the $(\textbf{4},\textbf{1})$ and $(\textbf{1},\textbf{4})$ representations of the R-symmetry $\text{USp}(4)_L\times\text{USp}(4)_R$, where $L,R$ now indicates the $6D$ chirality. Therefore, the monodromy matrix appearing in \cite{Hull:2020byc} (see e.g. eq. (3.39) in that paper) has grouped together $m_1$ and $m_2$, and similarly $m_3$ and $m_4$.

This systematic approach can be used to construct the entire field content coming from the lowest excited string states. Each of the states in \autoref{huiberttable} gives fields whose mass can be read off from their orbifold charge (it is always the absolute value of this linear combination of $m_i$'s times $p/2\pi \mathcal{R}$). The type of fields that this state gives can then be found from its massless or massive little group representation, depending on whether the aforementioned mass is zero (mod $2\pi$) or not. In this way, the entire supergravity spectrum from \cite{Hull:2020byc} can be reproduced. It is important to note however that the Scherk-Schwarz reduction can easily be carried out for any monodromy in Spin(5,5), but only for choices in Spin(4,4) we can compare to the orbifold picture. 

Moreover, recall that we can build the Kaluza-Klein towers on the circle by adding $p\,\mathbb{Z}$ momentum modes to the ones that were added following the procedure above. This addition doesn't change the orbifold charge, so all of these states survive as well. The masses shift by $p\,\mathbb{Z}/\mathcal{R} = \mathbb{Z}/R$, e.g. the masses of the KK-tower on our example state become $|m_1/2\pi R+\mathbb{Z}/R|$. Again, this agrees with the supergravity calculation.

\section{Effective potential and supertraces}
\label{sec:bad examples}

Scherk-Schwarz reductions to five spacetime dimensions yield $\mathcal{N}=8$ gauged supergravity in $D=5$ with vacua that spontaneously break supersymmetry. The $\mathcal{N}=8$ supergravity multiplet in five dimensions contains as bosonic fields a graviton, 27 vector fields and 42 real scalars. The gauge group was discussed in detail in \cite{Hull:2020byc}, where it was shown how the structure constants of the gauge algebra are determined by the twist matrix. In the class of twists we considered in this thesis (with the twist matrix (conjugate to) an element in the R-symmetry group  $\text{Spin}(5)_L\times \text{Spin}(5)_R$), the potential and the structure constants depend on four mass parameters $m_i$. The potential can be written as (see e.g. eq. (3.49) in \cite{Hull:2020byc})
\begin{equation}
    V(\mathcal{H})=\frac{1}{4}e^{-{\sqrt{8/3}}\phi_5}\text{Tr}\Big[M^2+M^T\mathcal{H}^{-1}M\mathcal{H}\Big]\ ,
\end{equation}
where $\phi_5$ is the KK scalar coming from the metric in six dimensions, $\mathcal{H}\in \text{Spin}(5,5)$ parametrises the 42 scalar fields, and $M\in \mathfrak{so}(5,5)$ is the mass matrix containing the mass parameters $m_i$ (see \cite{Hull:2020byc} for more details).

For this potential to belong to the string landscape, in particular the landscape of freely acting orbifolds, a number of conditions need to be satisfied. First of all, the monodromy matrix  should be inside the T-duality group, otherwise there is no obvious CFT description on the worldsheet in terms of an orbifold, as we discussed in detail in section \ref{sec:Orbifold constructions}.

Secondly, in supergravity the mass parameters $m_i$ are continuous parameters, while for a well defined orbifold the mass parameters must be quantized (see section \ref{sec:Orbifold constructions}) and possible accidental massless modes arise when some of the masses add up to a multiple of $2\pi$. In the case of symmetric orbifolds the quantization of the mass parameters is the only constraint for a consistent uplift, since modular invariance of the partition function is ensured. For asymmetric orbifolds, however, modular invariance is not guaranteed and this   restricts the possible supergravity theories that can be consistently uplifted to string theory (as freely acting orbifolds). Furthermore, even if modular invariance is achieved, one should carefully examine the degeneracy of states in the twisted sectors. As we have already mentioned, for a sensible theory this degeneracy must be an integer number.

\subsection*{Supertraces}
\label{n=0 and supertraces}

The Scherk-Schwarz reductions we considered generate positive definite scalar potentials at the classical level. In the presence of partial supersymmetry, the corresponding Minkowski vacua are perturbatively stable, and hence the cosmological constant vanishes. When all of the supersymmetries are spontaneously broken, a one-loop cosmological constant, which is determined by the minimum of the one-loop effective potential, may be non-vanishing. 

The effective potential can be expressed in terms of various {supertraces}, as it was first shown in \cite{coleman1973radiative}. Supertraces are defined as weighted sums over the masses of all fields in the spectrum of the theory, that is
\begin{equation}
  \text{Str}M^{2\beta}=   N_{\phi}(M_{\phi})^{2\beta}-2N_{\chi}(M_{\chi})^{2\beta}+3N_{B_{\mu \nu}}(M_{B_{\mu \nu}})^{2\beta}
    +4N_{A_{\mu}}(M_{A_{\mu}})^{2\beta}-6N_{\psi_{\mu}}(M_{\psi_{\mu}})^{2\beta},
    \label{explicit}
\end{equation}
where $\beta > 0$ is an integer\footnote{$\text{Str}M^0$ is equal to bosonic minus fermionic degrees of freedom and is always zero.} and we denote by $N_{\text{field}}$ the number of fields with mass $M_{\text{field}}$. Each field is multiplied by the corresponding (massive) degrees of freedom in five dimensions and fermion fields appear with a minus sign.

Regarding supersymmetric theories, all supertraces vanish on Minkowski vacua \cite{zumino1975supersymmetry}. This is a result of bose-fermi degeneracy. In supersymmetric theories all fields fit in supermultiplets and as a consequence, all supertraces are identically zero. For non-supersymmetric theories the situation is more complicated, due to the absence of bose-fermi degeneracy. However, it was noticed in \cite{Cremmer:1979uq,ferrara1979mass} that for Scherk-Schwarz reductions of $\mathcal{N}=8$ supergravities with fully broken supersymmetry
    \begin{equation}
   \text{Str}M^2=\text{Str}M^4=\text{Str}M^6=0\,,\qquad \text{Str}M^8\neq 0\,.
   \label{first supertraces}
\end{equation}
The same result was later obtained in \cite{Dall'Agata:2012cp} for any gauging of $\mathcal{N}=8$ supergravity.

Here we also perform the supertrace calculation for an $\mathcal{N}=0$ supergravity theory with all mass parameters non-zero. Our result is consistent with \eqref{first supertraces}. Furthermore, we explicitly find the value of $\text{Str}M^8$ to be
\begin{equation}
    \text{Str}M^8=40320\left(m_1\,m_2\, m_3\, m_4\right)^2\,,
    \label{str8}
\end{equation}
which is positive definite. As shown in \cite{Dall'Agata:2012cp}, $\text{Str}M^2=\text{Str}M^4=\text{Str}M^6=0$ and $\text{Str}M^8>0$ imply that the one-loop effective potential is negative definite.

\clearpage

\thispagestyle{empty}
\chapter{Moduli spaces }
\label{chap:mod5}

In chapters \ref{chap:fao} and \ref{chap:spectrum}, we discussed the construction of freely acting (asymmetric) orbifolds, and we presented models preserving $\mathcal{N}=6,4,2$ or $0$ supersymmetry in five dimensions. The resulting theories contain (among other fields) massless scalar fields, which parametrise the moduli spaces of the orbifolds. The moduli spaces are the main subject of this chapter.

First, we present in section \ref{modulispace5d} the moduli space of the theory before orbifolding. Then, in section \ref{24or16} we discuss the classical moduli spaces of theories preserving 24 or 16 supersymmetries in five dimensions. This amount of supersymmetry fixes the form of the moduli space and the classical result holds at the quantum level as well. In section \ref{8sup} we determine the classical moduli spaces for theories preserving 8 supersymmetries in five dimensions. For $\mathcal{N}=2$, $D=5$ theories the moduli space is not fixed by supersymmetry and the classical result can be modified at the quantum level. However, in section \ref{quantasp} we show that for the models studied in this chapter no quantum corrections to the metric on the moduli space arise.

\section{Type IIB on $T^5$}
\label{modulispace5d}

Consider first type IIB string theory compactified on $T^5$, before orbifolding. The low-energy  theory is $D=5$, $\mathcal{N}=8$ supergravity with rigid duality symmetry $G=\text{E}_{6(6)}$ and local R-symmetry $\hat{H}=\text{Sp}(4)$ (sometimes referred to as $\text{USp}(8)$).
The massless spectrum of the theory contains 42 scalar fields, which parametrise the moduli space
\begin{equation}
    \mathcal{M} =\frac{G}{H} = \frac{\text{E}_{6(6)}}{\text{Sp}(4)/\mathbb{Z}_2}\,.
\end{equation}
Note that $H=\text{Sp}(4)/\mathbb{Z}_2$ is the maximal compact subgroup of $\text{E}_{6(6)}$. In the string theory, the global $G=\text{E}_{6(6)}$ is broken to its discrete subgroup $G(\mathbb{Z})=\text{E}_{6(6)}(\mathbb{Z})$, forming the U-duality group \cite{Hull:1994ys}. The T-duality subgroup of this is  Spin$(5,5;\mathbb{Z})$.

The orbifolds we are considering are each specified by  a twist matrix which is an element  
$\hat{\cal M}\in  \hat{H}= \text{Sp}(4)$
that projects to an element
$ {\cal M}\in \text{Sp}(4)/\mathbb{Z}_2$ which is in the discrete U-duality subgroup of
$  \text{E}_{6(6)}(\mathbb{Z})$, i.e.\ 
$\mathcal{M}\in  \text{E}_{6(6)}(\mathbb{Z})\cap \text{Sp}(4)/\mathbb{Z}_2$. We require that the 
twist is in the 6-dimensional T-duality group arising from compactifying to 6 dimensions on $T^4$. Then
${\cal M}\in$ Spin$(4,4;\mathbb{Z})$ and
$\hat{\cal M}\in$ Spin(4)$\times $Spin(4), so that the twist is a T-duality transformation on $T^4$ combined with an R-symmetry twist (for more details we refer to chapter \ref{chap:fao})
\begin{equation}
    ({\cal M},\hat {\cal M})\in
{\rm Spin}(4,4;\mathbb{Z})
\,
\times \,
[{\rm Spin}(4)\times {\rm Spin}(4) ]
\subset 
\text{E}_{(6)6}(\mathbb{Z})\times \text{Sp}(4)\ .
\end{equation}
In the supergravity theory, the orbifold corresponds to a
Scherk-Schwarz reduction and in string theory is a compactification with duality twist
\cite{Dabholkar:2002sy,Gkountoumis:2023fym}.
In the supergravity theory
the twist 
generates a $\text{U}(1)$ subgroup of 
$\text{E}_{6(6)}\times \text{Sp}(4)$.
Let the subgroup of $\text{E}_{6(6)}\times \text{Sp}(4)$ that commutes with this $\text{U}(1)$ be
\begin{equation}
\mathcal{K}\times \hat{\mathcal{K}} \subset \text{E}_{6(6)}\times \text{Sp}(4)\ .
\end{equation}
Then the twist will break the supergravity
duality symmetry 
and  R-symmetry
but a subgroup
$\mathcal{K}\subset \text{E}_{6(6)}$
of the duality symmetry and a subgroup
$\hat{\mathcal{K}}\subset
\text{Sp}(4)
$
will remain as symmetries.
There will be massless scalar fields
in the coset space
$\mathcal{K}/\bar {\mathcal{K}}$ where 
$\bar{\mathcal{K}}$ is the  compact subgroup of $\mathcal{K}$:
$\bar {\mathcal{K}}=\mathcal{K}\cap \text{Sp}(4)/\mathbb{Z}_2
$.
In string theory, a discrete subgroup
$\mathcal{K}(\mathbb{Z})\subset \text{E}_{6(6)}(\mathbb{Z})$ of the U-duality is expected to remain as a duality symmetry and a local $\hat{\mathcal{K}}\subset \text{Sp}(4)$ as an R-symmetry.

The orbifold twist is accompanied by a shift that makes the orbifold freely acting. Recall that the shift along the circle coordinate can be represented by a shift vector $v$ (cf. \eqref{shiftvector1}). In the quantum theory, the T-duality group associated with $S^1$ is $\mathbb{Z}_2$. In the lattice basis, this $\mathbb{Z}_2$ group acts as a reflection. In particular, it acts on the shift vector as $v\to -v$. As we have already discussed in chapter \ref{chap:fao} (see discussion around \eqref{pvector1,1}), $v$ and $-v$ define the same orbifold. Hence, the presence of the shift vector does not break the aforementioned $\mathbb{Z}_2$ subgroup. This will play an important role in section \ref{quantasp}.

Finally, note that this discussion of duality symmetries applies to the untwisted orbifold sector, and we discuss the subgroup of the original duality that is a symmetry of the untwisted sector.  However, the orbifold introduces twisted sectors, and there can be new duality symmetries relating the untwisted and twisted sectors that do not directly arise from the duality symmetry of the theory before the orbifold.
For example, consider type IIA string theory compactified on $T^4$, with U-duality symmetry ${\rm Spin}(5,5;\mathbb{Z})$. Consider now the $\mathbb{Z}_2$ orbifold of this which gives a special point in the ${K}3$ moduli space. The untwisted sector is invariant under an $\text{SO}(4,4;\mathbb{Z})$ subgroup of the duality group, but the full theory in fact has a duality symmetry $\text{SO}(4,20;\mathbb{Z})$. Note that $\text{SO}(4,20)$ is not a subgroup of ${\rm Spin}(5,5)$ and the extra duality symmetries include ones mixing untwisted and twisted sectors. Whether or not there are extra symmetries of this kind depends on the model.
We will see below that some of our orbifold examples have this kind of duality enhancement, at special points in the moduli space.

\section{Theories with 24 or 16 supersymmetries}
\label{24or16}
Here, we discuss the moduli spaces of theories preserving 24 or 16 supersymmetries in five dimensions. This amount of supersymmetry fixes the form of the moduli space and no quantum corrections to the metric on the moduli space can arise. Some of the examples of the next two subsections have already been discussed in chapter \ref{chap:spectrum}, and some examples are new. Recall that the untwisted spectrum of our orbifolds can be easily determined using \autoref{huiberttable}. This is sufficient for the purposes of this section, in which we focus on the determination of the moduli spaces, since for freely acting orbifolds all fields coming from the twisted sectors are massive for generic values of the circle radius.

\subsection{$\mathcal{N}=6$}

There is a unique $\mathcal{N}=6$, $D=5$ supergravity theory, which can be realized by asymmetric orbifolds, as the $\mathbb{Z}_2$ and  $\mathbb{Z}_4$ orbifolds that we constructed in chapter \ref{chap:spectrum} (see section \ref{n=6orbis}). The massless spectrum of this theory constitutes the $\mathcal{N}=6, D=5$ supergravity multiplet, which constains the graviton, 6 gravitini, 15 vectors  20 dilatini and 14 scalars \cite{Cremmer:1980gs}. The U-duality group is $\mathcal{K}=\text{SU}^*(6)$ and the R-symmetry group is $\hat{\mathcal{K}}=\text{Sp}(3)$. The 14 scalars parametrise the moduli space
\begin{equation}
{\cal M}_{\mathcal{N}=6}=\frac{\text{SU}^*(6)}{\text{Sp}(3)}\ .
\end{equation}
6 of the 14 massless scalars descend from the NS-NS sector and 8 come from the R-R sector. 28 scalars have become massive in the Scherk-Schwarz reduction. 

Similarly, one can construct the moduli space of the $\mathcal{N}=6$, $\mathbb{Z}_3$ orbifold,  based on the lattice $\Gamma^{4,4}(A_2\oplus A_2)\oplus\Gamma^{1,1}$, with mass parameters $\vec{m}=(\tfrac{2\pi}{3},0,0,0)$, corresponding to $\tilde u=(0,0,\frac{1}{3},\frac{1}{3})$ and $u=(0,0,0,0)$. It has the same massless spectrum as the $p=2,4$ orbifolds and hence the same moduli space, dictated by $\mathcal{N}=6$ supersymmetry. Asymmetric orbifolds with $\mathcal{N}=6$ supersymmetry have also been constructed in \cite{Ferrara:1989nm} and more recently in \cite{bianchi2022perturbative,Bianchi:2008cj,Bianchi:2010aw}.

\subsection{$\mathcal{N}=4$}

The massless spectrum of theories preserving 16 supersymmetries in five dimensions consists of the $\mathcal{N}=4$, $D=5$ gravity multiplet, containing 1 scalar, coupled to $n$ vector multiplets, each containing 5 scalars. The U-duality group is $\mathcal{K}=\text{SO}(1,1)\times \text{SO}(5,n)$ \cite{Andrianopoli:2002aq} and the R-symmetry group is $\hat{\mathcal{K}}=\text{Sp}(2)\cong \text{Spin}(5)$ \cite{Seiberg:1997ax}. The moduli space is fixed by supersymmetry. It is completely determined by the number $n$ of vector multiplets and is given by \cite{Awada:1985ep}
\begin{equation}
{\cal M}_{\mathcal{N}=4}=\mathbb{R}^+\times \frac{\text{SO}(5,n)}{\text{SO}(5)\times \text{SO}(n)}\,,
\end{equation}
 where $\mathbb{R}^+=\text{SO}(1,1)/
\mathbb{Z}_2$.

\subsection*{Models with one vector multiplet}
In general, for the $\mathcal{N}=4, D=5$ theories, if there are no accidental massless modes\footnote{Recall that accidental massless modes arise when some (or all) of the mass parameters add up to a multiple of $2\pi$.}, the moduli space is given by
\begin{equation}
{\cal M}=\mathbb{R}^+\times \frac{\text{SO}(5,1)}{\text{SO}(5)}\ .
\label{mv1v}
\end{equation}
We can obtain such theories by using freely acting asymmetric orbifolds. For instance, consider the asymmetric $\mathbb{Z}_6\,(0,2)$  orbifold, based on the lattice $\Gamma^{4,4}(A_2\oplus A_2)\oplus\Gamma^{1,1}$, with mass parameters $\vec{m}=(\tfrac{\pi}{3},\pi,0,0)$, corresponding to $\tilde u=(0,0,\frac{1}{6},\frac{1}{6})$ and $u=(0,0,\frac{1}{2},\frac{1}{2})$. As we can see from \autoref{huiberttable}, there are 6 massless scalars, 2 from the NS-NS and 4 from the R-R sector. There is only one vector multiplet. 3 vectors come from the NS-NS sector, and 4 from the R-R sector. The 6 scalars parametrise the moduli space \eqref{mv1v}.

Another interesting model, based on the same lattice, is the asymmetric $\mathbb{Z}_6\,(1,1)$ orbifold with $\vec{m}=(\tfrac{\pi}{3},0,\pi,0)$, corresponding to $\tilde{u}=\left(0,0,\tfrac{2}{3},-\tfrac{1}{3}\right)$ and $u=(0,0,0,0)$ . It has the same $D=5$ massless field content and moduli space, but the ten-dimensional origin is different, as all the bosonic fields, both scalars and vectors, come from the NS-NS sector, and the theory contains no massless R-R fields. Finally, another orbifold with moduli space \eqref{mv1v} is the the asymmetric $\mathbb{Z}_4\,(1,1)$ orbifold, based on the $\Gamma^{4,4}(D_4)\oplus\Gamma^{1,1}$ lattice, with mass parameters $\vec{m}=(\pi,0,-\frac{\pi}{2},0)$ or in terms of twist vectors $\tilde{u}=\left(0,0,\tfrac{1}{4},\tfrac{3}{4}\right)$ and $u=(0,0,0,0)$. Modular invariance of the above models can be easily verified using \eqref{modular conditions on twist vectors} and \eqref{modcond2}.

\subsection*{Models with three vector multiplets}
As an example, consider the $\mathbb{Z}_4\, (0,2)$ symmetric orbifold model of section \ref{symmetric z4}. We saw that the massless spectrum of this orbifold consisted of the $D=5$, $\mathcal{N}=4$ supergravity multiplet, containing 1 real scalar and 6 vectors, together with three massless vector multiplets, each containing 5 scalars. Out of the 15 scalars from the vector multiplets, 9 come from the NS-NS sector, and are remaining geometric moduli, and 6 from the R-R sector. The total scalar manifold with the $16=1+15$ scalars including those of the three vector multiplets is
\begin{equation}
{\cal M}=\mathbb{R}^+\times \frac{\text{SO}(5,3)}{\text{SO}(5)\times \text{SO}(3)}\ .
\end{equation}
One gets the same moduli space for the symmetric $\mathbb{Z}_3$ and $\mathbb{Z}_6$ orbifolds\footnote{One choice of lattice for these two models could be two copies of the $A_2$ lattice that we used in the example of section \ref{sec:N=0}.}. For these models we get two accidental massless vector multiplets. Similar models can also be obtained by using asymmetric orbifolds.

\subsection*{Models with five vector multiplets}
Here, we consider the symmetric $\mathbb{Z}_2\,(0,2)$  orbifold, based on two copies of the $D_2$ root lattice, with $\vec{m}=(\pi,\pi,0,0)$, corresponding to $\tilde{u}=u=(0,0,\frac{1}{2},\frac{1}{2})$. We find from \autoref{huiberttable} that there are $26=1+25$ massless scalars, parameterising the moduli space
\begin{equation}\label{nV=5}
{\cal M}=\mathbb{R}^+\times \frac{\text{SO}(5,5)}{\text{SO}(5)\times \text{SO}(5)}\ .
\end{equation}
For this $\mathbb{Z}_2$ orbifold, 3 vectors come from the NS-NS sector, and 8 from the R-R sector. As the gravity multiplet contains 6 vectors, we are left over with 5 vectors to form vector multiplets indeed. Four of these multiplets are comprised by accidental massless modes.

There is another way to get the moduli space \eqref{nV=5}, namely from the asymmetric $\mathbb{Z}_2\,(1,1)$ orbifold that we discussed in section \ref{subsec N=4 1,1}. The bosonic massless field content is the same, except that the fields have a different ten-dimensional origin. In this case, all the vectors and scalars come from the NS-NS sectors. Furthermore, for the asymmetric $\mathbb{Z}_2$ orbifold, we saw that for a certain value of the orbifold circle radius, we can obtain two extra massless vector multiplets coming from the twisted sector. In this case, the moduli space becomes
\begin{equation}\label{nV=7}
{\cal M}=\mathbb{R}^+\times \frac{\text{SO}(5,7)}{\text{SO}(5)\times \text{SO}(7)}\ .
\end{equation}

\subsection*{Gauge group}

The $\mathcal{N}=4$ orbifolds we discussed yield an odd number $n$ of massless vector multiplets, with $n\leq 21$ (all massless tensors are dualized into vectors), and the corresponding unbroken gauge group (at generic points in the moduli space) is U$(1)^n$. First of all, this result agrees with the upper bound on the rank of the gauge group $r_G=26-D$ given in \cite{montero2021cobordism,kim2020four}. Secondly, it should be pointed out that in all the examples we gave $n\in 2\mathbb{Z}+1$. Furthermore, it is easy to derive from \autoref{huiberttable} that all orbifolds give an odd number of untwisted vector multiplets. Regarding the twisted sectors, we saw that in $\mathcal{N}=4\,(1,1)$ theories (where we break all the left, or right-moving supersymmetries) it is possible to tune the circle radius such that some of the twisted vector multiplets become massless. In this massless limit each complex vector multiplet will give two real vector multiplets. As a result, regardless of the specific orbifold, we can only get an even number of real vector multiplets coming from the twisted sectors. Therefore, combining both untwisted and twisted sectors we can see that $n\in 2\mathbb{Z}+1$ is a generic feature of our orbifolds.

The appearance of only an odd number of vector multiplets seems to be a characteristic of a broader class of string constructions with 16 supersymmetries in $5D$. Theories with 16 supersymmetries in $D>6$ have been studied extensively (see e.g. \cite{de2001triples,bedroya2022compactness} and references therein). In addition, $\mathcal{N}=(1,1)$ theories in $6D$ were recently classified in \cite{fraiman2023unifying}.
Dimensional reduction of the various higher dimensional theories given in these references always yields an odd number of vector multiplets in $5D$, which supports our findings\footnote{Similar results were found in the context of four dimensional asymmetric orbifolds in \cite{Blumenhagen:2016rof}. In four dimensions, the number of vector multiplets in $\mathcal{N}=4$ was always found to be even, consistent with odd number in five dimensions.}. Consequently, cases with $n \in 2\mathbb{Z}$ appear not to be part of the string landscape of $\mathcal{N}=4, D=5$ Minkowski vacua\footnote{
In \cite{dabholkar1999string}, an asymmetric orbifold was proposed which appeared to give pure supergravity with no vector multiplets. However, the monodromy in this example involves a single T-duality and so the construction is not a quotient by a symmetry of the IIB string theory: the bosonic action of the monodromy is in  $\text{O}(d,d)$, not  in $\text{SO}(d,d)$. Such a quotient seems problematic and is outside the class of constructions we discuss here.}.
They do however seem to appear as moduli spaces in AdS$_5$ vacua in the context of holography, see e.g. section 7 of \cite{Freedman:1999gp} (and earlier work obtained in \cite{Khavaev:1998fb}) for the case of two vector multiplets, $n=2$.

\section{Theories with 8 supersymmetries}
\label{8sup}

As we have already mentioned above, for theories preserving 8 supersymmetries in five dimensions, which is the minimum amount of supersymmetry in $D=5$, the moduli space is not fixed by supersymmetry and quantum corrections may arise. As an example, consider the compactification of M-theory  on a Calabi-Yau threefold (CY$_3$), which gives a five dimensional theory with eight supersymmetries, that is  $\mathcal{N}=2$ supersymmetry in $D=5$. The spectrum contains $h_{1,2}+1$ massless hypermultiplets and $h_{1,1}-1$ vector multiplets.  Generically, the hypermultiplet moduli space receives quantum corrections coming from M2 and M5-brane instantons \cite{Becker:1995kb}. These deform the metric on the classical moduli space to the full quantum corrected metric on the hypermultiplet moduli space. The vector multiplet moduli space is determined by the intersection numbers of the CY$_3$ \cite{Cadavid:1995bk}. It was described in \cite{witten:1996qb} and exhibits a rich structure of topology-changing transitions.

\subsection{Analysis of the models}
\label{OSS}

Here, we will discuss models preserving $\mathcal{N}=2$ supersymmetry in $D=5$, based on the freely acting asymmetric orbifolds presented in chapters \ref{chap:fao} and \ref{chap:spectrum}. Particular emphasis will be given to orbifolds for which
 the duality group of the effective supergravity theory is a subgroup of Spin$(5,5)\times \text{SO}(1,1)$;  this is a maximal subgroup of the U-duality group $\text{E}_{6(6)}$.\footnote{Other orbifolds can give rise to  duality symmetries that are subgroups of other maximal subgroups of $\text{E}_{6(6)}$.}
 Here, Spin$(5,5)$ arises from the 5-dimensional T-duality group and \text{SO}$(1,1)$ from the 10-dimensional SL(2,$\mathbb{R}$) symmetry. Of course, as we have already discussed, the twist, being in the T-duality group of $T^5$, will further break this group. Then the U-duality group of the effective theory will be given by
\begin{equation}
\mathcal{K}=\text{SO}(1,1)\times \mathcal{C} \subset
\text{SO}(1,1)\times
{\rm Spin}(5,5)\,.
\label{ K = so(1,1) x c}
\end{equation}
Here $\mathcal{C}$ is the subgroup of
${\rm Spin}(5,5)$ that commutes with the twist. In the quantum theory we expect the group $\mathcal{C}$ to be broken to its discrete subgroup $\mathcal{C}(\mathbb{Z})$, forming the T-duality group of the orbifold. Moreover, we focus on orbifolds that break the supersymmetry to $\mathcal{N}=2$, so that the orbifold R-symmetry $\hat{\mathcal{K}}$ contains an $\text{SU}(2)$ which is the R-symmetry of the effective $\mathcal{N}=2$ theory that
constitutes the massless sector.

Now, the compact subgroup of $\mathcal{K}$ is
\begin{equation}
\bar {\mathcal{K}}=\mathcal{K}\cap \text{Sp}(4)/\mathbb{Z}_2=\mathbb{Z}_2\times
\bar {\mathcal{C}}\,,
\end{equation}
where
$\bar {\mathcal{C}}=\mathcal{C}\cap 
[{\rm Spin}(5)\times {\rm Spin}(5)]
$.
Then, the moduli space is
\begin{equation}
\mathcal{M}_{\mathcal{O}}=\mathcal{K}/\bar {\mathcal{K}} = \mathbb{R}^+
\times \mathcal{C}/\bar {\mathcal{C}}
\subset  \mathbb{R}^+
\times
\frac{
{\rm Spin}(5,5)}
{
{\rm Spin}(5)\times {\rm Spin}(5)
}\,,
\end{equation}
 where $\mathbb{R}^+=\text{SO}(1,1)/
\mathbb{Z}_2$. 

We will construct models based on asymmetric orbifolds for which the massless spectrum consists purely of NS-NS fields coming only from the untwisted sector. In particular, all fields in the R-R sector, as well as all states in the twisted sectors will be massive. In this case the duality group of the supergravity theory can be at most Spin$(5,5)\times \text{SO}(1,1)$.
The duality group Spin$(5,5)$
only acts on the NS-NS fields
through SO$(5,5)$.
Since we have only massless NS-NS scalars, the   moduli space of the classical effective theory can be at most 
\begin{equation}
\mathcal{M}_{\text{NS}}=\mathbb{R}^+\times \frac{\text{SO}(5,5)}{\text{SO}(5)\times \text{SO}(5)}\,,
\end{equation}
where the $\mathbb{R}^+$ factor is parameterised in terms of the five-dimensional dilaton $\phi_5$ by $\lambda_5=\langle e^{\phi_5}\rangle$,
and the remaining coset is parameterised by the 25 scalars that are the NS-NS moduli  of $T^5$. 
For the orbifold  some of these 25 scalars will become massive and will no longer be moduli of the orbifolded theory. Hence, the 
coset space will
be reduced to a subspace
 parameterised only by those scalars that are invariant under the twist, and as a consequence remain massless and are moduli of the orbifolded theory. On the other hand, the dilaton is inert under the orbifold action and will be a modulus in all of our constructions. The five-dimensional dilaton is T-duality invariant and parametrizes the space $\mathbb{R}^+$ preserved by  $\mathcal{C}$.
 Having said all these, we conclude that the   classical moduli space of the  orbifold   
 $\mathcal{M}_{\mathcal{O}}$ will be exactly the subspace of $\mathcal{M}_{\text{NS}}$ that is compatible with the twist, that is 
 \begin{equation}
\mathcal{M}_{\mathcal{O}}=\mathcal{K}/\bar {\mathcal{K}} = \mathbb{R}^+\times \mathcal{C}/\bar {\mathcal{C}}\,.
\end{equation}
For the quantum theory we expect the U-duality group $\text{SO}(1,1)\times
\mathcal{C}$ to be broken to is discrete subgroup $\mathbb{Z}_2\times\mathcal{C}(\mathbb{Z})$. As we mentioned before, the classical SO(1,1) symmetry is inherited from the type IIB SL(2,$\mathbb{R}$) symmetry in ten dimensions. There, it acts as $\tau \to a^2 \tau$, so that it shifts the (ten-dimensional) dilaton $\Phi$ and rescales the axion $C_0$.\footnote{Recall that $\tau\to\frac{a\tau+b}{c\tau+d}$, where $\tau=C_0+ie^{-\Phi}$.} In the quantum theory, SO(1,1) is broken to $\mathbb{Z}_2$, corresponding to the integer values $a=\pm 1$. This $\mathbb{Z}_2$ leaves the dilaton invariant but acts on the NS-NS and R-R two-forms as $B_2 \to -B_2$ and $C_2\to -C_2$. Upon reducing to five dimension, this $\mathbb{Z}_2$ is preserved since it commutes with the twist, but it does not act on the five-dimensional dilaton. The quantum symmetries and duality groups will play an important role in showing the absence of quantum corrections to the moduli spaces, one of the main results of this chapter.

For any string compactification to 5 dimensions that preserves $\mathcal{N}=2$ supersymmetry there is a low-energy effective field theory consisting of $\mathcal{N}=2$ supergravity coupled to hypermultiplets and vector multiplets.
The moduli are the massless scalar fields in the hypermultiplets and vector multiplets, and are governed by non-linear sigma models 
on the hypermultiplet moduli space and vector multiplet moduli space, respectively.

For models such as the ones we are considering here in which the scalar giving the string coupling constant is in a vector multiplet, the hypermultiplet moduli space metric must be independent of the string coupling and so must be given by the classical result with no quantum corrections.
However, in general the vector multiplet moduli space metric may have quantum corrections. 

For our models the supersymmetry is spontaneously broken rather than explicitly broken, and, at least at the classical level, there is a  truncation to an $\mathcal{N}=8, D=5$ supergravity theory.
As we shall see, this means that there is a hidden $\mathcal{N}=8$ local supersymmetry that  relates the hypermultiplets and vector multiplets. Let us discuss this issue in more detail now.

As we have explained in chapter \ref{chap:fao}, the  orbifold construction can be viewed as a compactification with duality twist of the kind introduced in \cite{Dabholkar:2002sy}. Type IIB string theory compactified on $T^4$ gives a 6-dimensional theory with U-duality group $\text{Spin}(5,5,\mathbb{Z})$ and
scalars taking values in the moduli space of IIB on $T^4$, which is $\text{Spin}(5,5)/\left[\text{Spin}(5)\times \text{Spin}(5)\right]$
quotiented by the U-duality group \cite{Hull:1994ys}.
This $6D$ theory is then
compactified on a further circle with a U-duality twist, i.e.\ with a  $\text{Spin}(5,5,\mathbb{Z})$ monodromy M on the circle.
At a point in the moduli space that is fixed under the action of M, the construction becomes an orbifold by the symmetry M, which is a symmetry of the IIB theory on  $T^4$ at that point in moduli space, combined with a shift on the extra circle \cite{Dabholkar:2002sy}.  For the models considered here the monodromy is a T-duality transformation, i.e.\ $\text{M}=\mathcal{M}$, so that this construction gives a freely acting  orbifold. This viewpoint is useful as it enables us to formulate the construction without restricting to a particular point in moduli space.

The toroidal compactification of IIB supergravity to 6 dimensions has a consistent truncation to 6-dimensional $\mathcal{N}=8$ supergravity, and this is also a consistent truncation of the full string theory compactified on $T^4$. Then the string theory compactification with duality twist has a consistent truncation to  the Scherk-Schwarz compactification of
$\mathcal{N}=8$ supergravity in 6 dimensions on $S^1$.
This compactification 
gives a gauged $\mathcal{N}=8$ supergravity in 5 dimensions
coupled to
an infinite number of massive
$\mathcal{N}=8$ supermultiplets arising from the twisted circle reduction.
This $\mathcal{N}=8$ theory
in turn has a consistent truncation to
 give an $\mathcal{N}=2$ supergravity coupled to supermatter in 5 dimensions, resulting from  the massless fields that are invariant under the twist \cite{Duff:1985jd}. 
This is in accord with the arguments of \cite{Lee:2014mla}.

Of particular interest are the cases in which there is in fact a consistent truncation to a
 gauged $\mathcal{N}=8$ supergravity. We believe this to be the case for all of our examples here, but we will not assume this. Instead, we shall show that our results are consistent with this.

 In some cases there may be additional \say{accidental} massless scalars of the type discussed in \cite{Hull:2020byc} (cf. section 3.9).  Such accidental modes arise when some (or all) of the mass parameters add up to $2\pi n$, $n \in \mathbb{Z}^*$.
 Then some of the massive $\mathcal{N}=8$ supermultiplets arising from the twisted circle reduction 
 contain massless scalars. In particular, these $\mathcal{N}=8$ supermultiplets
 decompose into $\mathcal{N}=2$ supermultiplets of different masses and in the accidental case some of these $\mathcal{N}=2$ supermultiplets turn out to be massless.
 For the rest of this section, we restrict ourselves to the cases in which there are no accidental massless scalars of this type.

The scalars of the  $\mathcal{N}=8$ supergravity multiplet
include the moduli of the $T^4$
and
have a scalar potential.
The scalar potential   has a minimum at which the potential vanishes and the location of the minimum in the scalar target space fixes the $T^4$ moduli to be at the point where the $T^4$ CFT is invariant under the $\mathbb{Z}_p$ symmetry that we use in the orbifold. At the  minimum there is a Minkowski vacuum  in which
the $\mathcal{N}=8$ supersymmetry is spontaneously broken. For the models we consider here, it is broken  to $\mathcal{N}=2$ and then the $\mathcal{N}=8$ supergravity multiplet can be decomposed into  $\mathcal{N}=2$ multiplets.
The massless $\mathcal{N}=2$  multiplets give $\mathcal{N}=2$ supergravity coupled to hypermultiplets and vector multiplets. In addition, there are massive multiplets including $\mathcal{N}=2$ gravitino multiplets that contain 6 massive gravitini. If the mass parameters $m_i$ in the twist are set to zero, the massive multiplets all become massless and ungauged $\mathcal{N}=8, D=5$ supergravity is recovered.

The scalars of the ungauged
$\mathcal{N}=8$ supergravity in 5 dimensions take values in the coset space $\text{E}_{6(6)}/\left[\text{Sp}(4)/\mathbb{Z}_2\right]
$. Our construction results in some of these scalars being eaten by vector fields which become massive and some gaining masses through the scalar potential.
The remaining scalars which are massless are the moduli. They correspond to flat directions of the potential and   lie in a subspace of
$\text{E}_{6(6)}/\left[\text{Sp}(4)/\mathbb{Z}_2\right]
$.
As we argued above, these remaining scalars 
in fact take values in the classical moduli space
\begin{equation}
\mathcal{M}_{\mathcal{O}}=\mathcal{K}/\bar {\mathcal{K}} = \mathbb{R}^+\times \mathcal{C}/\bar {\mathcal{C}}\,.
\end{equation}
For compactifications with 
$\mathcal{N}=2$ supersymmetry this space must factorise into a vector multiplet moduli space and 
a hypermultiplet moduli space
\begin{equation}
\mathcal{M}_{\mathcal{O}}=
{\cal M}_V \times {\cal M}_H\,,
\end{equation}
so that $\mathcal{C}$ 
must have a factorisation
\begin{equation}
\mathcal{C}=\mathcal{C}_V
\times
\mathcal{C}_H\,,
\end{equation}
with a corresponding 
factorisation of
$\bar {\mathcal{C}}$
\begin{equation}
\bar {\mathcal{C}}=\bar {\mathcal{C}}_V
\times
\bar {\mathcal{C}}_H\,,
\end{equation}
so that
\begin{equation}
{\cal M}_V 
=\mathbb{R}^+\times 
{\mathcal{C}}_V/\bar {\mathcal{C}}_V\qquad\text{and} \qquad {\cal M}_H 
=
{\mathcal{C}}_H/\bar {\mathcal{C}}_H\,.
\label{factorization of C}
\end{equation}
The metric on the moduli space 
$\mathcal{M}_{\mathcal{O}}=
{\cal M}_V \times {\cal M}_H$
is the restriction of the  
canonical coset metric
on
the $\text{E}_{6(6)}/\left[\text{Sp}(4)/\mathbb{Z}_2\right]
$ coset space to this subspace.

In the quantum theory,  the set of moduli remains the same (since we are restricting to the cases with no extra accidental massless scalars) but the  metric 
for the vector moduli space ${\cal M}_V$
could in principle receive quantum corrections.
However, 
$\mathcal{N}=8$ local supersymmetry should be preserved in the quantum theory and this is highly restrictive.
The quantum metric must also be a
restriction of the
coset metric
on
the $\text{E}_{6(6)}/\left[\text{Sp}(4)/\mathbb{Z}_2\right]$.
One way in which this could happen that is consistent with the symmetries of the system  would be if 
${\mathcal{C}}$ were deformed to a conjugate subgroup
\begin{equation}
{\mathcal{C}}'=g\,{\mathcal{C}}g^{-1}\,,
\end{equation}
where
$g\equiv g(\phi_5) $
is a dilaton-dependent element of $\text{E}_{6(6)}$. This transformation would need to preserve
${\cal M}_H $ as this receives no quantum corrections, so 
$g\,{\mathcal{C}}_Hg^{-1}={\mathcal{C}}_H$.
However, such a deformation would also lead to a deformation of the twist ${\cal M}$
 and, since this deformation depends continuously on the dilaton, will be inconsistent with the 
 restriction that
${\cal M}\in$ Spin$(4,4;\mathbb{Z})$.

It seems that any deformation would be ruled out in this way.
This would then imply that the metric on the vector multiplet moduli space can receive no quantum corrections. Later, we shall give independent arguments that this is the case, using more details of the models under consideration.
These arguments also extend to the case with accidental massless scalars.
This then provides further evidence supporting the discussion in this section.

In the next, we will discuss models in which there are no massless fields coming from the R-R sector, and all fields from the twisted sectors are massive. Such models can be realised via freely acting asymmetric orbifolds. If there are no massless R-R scalars, the moduli space of both vector multiplets and hypermultiplets is entirely spanned by NS-NS scalars. Since there are no massless R-R vectors, there are no D-branes carrying R-R charges that can correct the metric on the moduli space, but there are presumably non-supersymmetric D-branes. Then, the classical moduli space can be determined as described above. Note that if the orbifold spectrum contains massless states coming from the twisted sectors, the moduli space will be enhanced. However, our orbifolds are freely acting and all the states coming from the twisted sectors are massive for generic values of the circle radius. In all the examples we discuss here, we will take the radius of the circle to be large enough compared with the string scale to ensure that there will be no massless states coming from the twisted sectors (see also \cite{Baykara:2023plc}). 

Finally, notice that our analysis of the moduli space is exact in $\alpha'$. This is because we have determined the T-duality group $\mathcal{C}$ exactly. Any quantum correction in $\alpha'$ should come with a dimensionful parameter, such as e.g.\ $L/\sqrt{\alpha'}$ for any radius $L$ on the $T^5$. Such corrections would not be compatible with the symmetry $\mathcal{C}$ acting on the moduli space $\mathcal{C}/\mathcal{C}_H$. As in the case of higher extended supersymmetry, $\mathcal{N}\geq 4$, $\sqrt{\alpha'}$ corrections will however contribute to higher derivative terms in the effective action.

\subsection{Determination of the moduli space}

In order to determine the moduli space of the orbifold, we need to find the group $\mathcal{C}\subset \text{Spin}(5,5)$ that commutes with the twist (this issue was first addressed in \cite{Spalinski:1991yw,Spalinski:1991vd}; for some other references see e.g.\ \cite{ferrara1992moduli,Bailin:1993ri,Erler:1992av,{cardoso1994moduli}}). For models with only NS-NS scalars, it is sufficient to work with the subgroup $\text{SO}(5,5)\subset \text{Spin}(5,5)$, as explained in section \ref{OSS}. A general element $\hat{h}\in \text{SO}(5,5)$ is a $10\times 10$ matrix
\begin{equation} \hat{h}=
   \begin{pmatrix}
\hat{a}&\hat{b}\\
\hat{c}&\hat{d}
\end{pmatrix}
\label{general SO(5,5) element}
\end{equation}
that satisfies
\begin{equation}
    \hat{h}^t\, \eta\,\hat{h} = \eta\,,  \qquad \eta=\begin{pmatrix}
    {1}_5 & 0\\
0& -1_5 
\end{pmatrix}\,, 
\label{eta frame}
\end{equation}
and $\text{det}(\hat{h})=1$. Here $\hat{a},\hat{b},\hat{c},\hat{d}$ are $5\times 5$ matrices, $t$ denotes the transpose matrix and $1_5$ is the $5\times 5$ unit matrix. From \eqref{general SO(5,5) element}, \eqref{eta frame} we see that the matrices $\hat{a},\hat{b},\hat{c},\hat{d}$ should satisfy the constraints
\begin{equation}
    \hat{a}^t\hat{a}- \hat{c}^t\hat{c}=1_5\,,\qquad   \hat{a}^t\hat{b}- \hat{c}^t\hat{d}=0_5\,,\qquad \hat{d}^t\hat{d}- \hat{b}^t\hat{b}=1_5\,.
    \label{matrices a,b,c,d constraints}
\end{equation}
We will also express the orbifold action in terms of an $\text{SO}(5,5)$ matrix. We embed $\mathcal{M}_{\theta}=(\mathcal{N}_L,\mathcal{N}_R) \in \text{SO}(4)_L\times \text{SO}(4)_R \subset \text{SO}(4,4)\,$ in $\text{SO}(5,5)$ as follows
\begin{equation}
    \hat{\mathcal{M}}_{\theta}=(\hat{\mathcal{N}}_L,\hat{\mathcal{N}}_R) \in \text{SO}(5)_L\times \text{SO}(5)_R \subset \text{SO}(5,5)\,,
    \label{general twist element}
\end{equation}
with
\begin{equation}\hat{\mathcal{N}}_L=
     \begin{pmatrix}
\mathcal{N}_L&0\\
0&1
\end{pmatrix}\,, \qquad \hat{\mathcal{N}}_R=
     \begin{pmatrix}
\mathcal{N}_R&0\\
0&1
\end{pmatrix}\,.
\label{general twist embedded in SO(5,5)}
\end{equation}
Now, in order to find the matrix $\hat{h}$ that commutes with the twist, we simply need to solve the matrix equation
\begin{equation}
    \left[\hat{\mathcal{M}}_{\theta},\hat{h}\right]=0\,.
    \label{commutant}
\end{equation}
This gives the following set of equations
\begin{equation}
  \left[\hat{\mathcal{N}}_L,\hat{a}\right]=0\,,\qquad  \hat{\mathcal{N}}_L\,\hat{b}=\hat{b}\,\hat{\mathcal{N}}_R\,,\qquad  \hat{\mathcal{N}}_R\,\hat{c}=\hat{c}\,\hat{\mathcal{N}}_L\,,\qquad \left[\hat{\mathcal{N}}_R,\hat{d}\right]=0\,.  
  \label{general commutation relations for h,g}
\end{equation}
In practise, in order to solve \eqref{commutant}, we start from a given twist $\hat{\mathcal{M}}_{\theta}\in \text{SO}(5,5)$ and a general matrix $\hat{h}=\begin{psmallmatrix}
\hat{a}&\hat{b}\\
\hat{c}&\hat{d}
\end{psmallmatrix}\in \text{GL}(10;\mathbb{R})$. Then we specify the form of the sub-matrices $\hat{a},\hat{b},\hat{c},\hat{d}$ satisfying \eqref{general commutation relations for h,g}. Afterwards, we demand that $\hat{h}\in \text{SO}(5,5)$ by imposing the conditions \eqref{matrices a,b,c,d constraints} and det$(\hat{h})=1$. This yields the group ${\mathcal{C}}$ that commutes with the twist. Then, as we discussed in section \ref{OSS}, the moduli space will be $\mathbb{R}^+\times \mathcal{C}/\bar {\mathcal{C}}$, where $\bar {\mathcal{C}}$ is the compact subgroup of ${\mathcal{C}}$. 

In what follows, we will apply this general analysis to specific examples with $\mathcal{N}=2$ supersymmetry in $D=5$. In order to construct such models we need to turn on three mass parameters. To start with, we choose $m_1,m_2,m_3\neq 0$ (mod $2\pi$) and $m_4=0$. Then, as can be read off from \autoref{huiberttable}, models without massless R-R states should satisfy the following conditions
\begin{equation}
    \begin{aligned}
        &\pm m_1 \pm m_2 \neq 0\quad \text{mod} \quad 2\pi\,,\\
         &\pm m_3 \pm m_2 \neq 0\quad \text{mod} \quad 2\pi\,,
    \end{aligned}
    \label{mass parameters for no R-R fields}
\end{equation}
for any choice of signs. We could have also chosen another vanishing mass parameter instead of $m_4$, e.g.\ $m_1=0$. This change would lead to dual models; these will be discussed in detail in section \ref{dual pairs}. For now, we  work with $m_4=0$. Using the results of chapter \ref{chap:spectrum}, we can provide a few concrete examples of models with no R-R massless states. In the next subsections, we will present such examples with $0,1$ or $2$ hypermultiplets, which involve orbifolds of rank $4,6$ and $12$. We mention here that we did not find any models of rank $2$ and $3$ without massless R-R states. Also, some of the models below were also discussed in  \cite{Baykara:2023plc}.

\subsection{Models with no hypermultiplets}
\label{modelsno}

In this subsection, we focus on models with no hypermultiplets, $n_H=0$, so there will only be vector multiplets. These are quite special in the landscape of string theories, as such models cannot be obtained from geometric compactifications such as M-theory on a CY$_3$. Using \autoref{huiberttable}, we see that such models can be obtained if 
\begin{equation}\label{nohyper}
    \pm m_1 \pm m_2 \pm m_3\neq 0\quad \text{mod} \quad 2\pi\,,
\end{equation}
for any choice of signs. We mention here that each choice of signs that does not satisfy the above condition results in two  massless scalars and one massless dilatino. Such choices are two-fold degenerate, so we obtain one massless hypermultiplet  for each choice of signs for which $\pm m_1 \pm m_2 \pm m_3$ vanishes.

\subsection*{$\mathbb{Z}_{12}$ with $n_V=2$ and $n_H=0$}
\label{z12 with nv=2 and nh=0}

The first example we consider is the orbifold with the following mass parameters
\begin{equation}
    m_1=\frac{2\pi}{3}\ , \qquad m_2= \frac{\pi}{2}\ ,\qquad m_3=\frac{\pi}{3} \ ,\qquad m_4=0\ .
\end{equation}
The corresponding twist vectors are
\begin{equation}
    \tilde{u}=\left(\frac{1}{2},\frac{1}{6}\right)\ ,\qquad u=\left(\frac{1}{4},\frac{1}{4}\right)\,,
    \label{twist vectors for primary example}
\end{equation}
which satisfy \eqref{modular conditions on twist vectors} (for this model $p=12$). These vectors give rise to the twist matrices
\begin{equation}
    \mathcal{N}_L=\begin{pmatrix}
        R(\pi)&0\\
        0&R(\frac{\pi}{3})
    \end{pmatrix}\,,\qquad \mathcal{N}_R=\begin{pmatrix}
        R(\frac{\pi}{2})&0\\
        0&R(\frac{\pi}{2})
    \end{pmatrix}\,.
    \label{twist matrices for primary example}
\end{equation}
This is one of the models constructed in \cite{Baykara:2023plc}. The massless spectrum of this orbifold consists of the $\mathcal{N}=2, D=5$ gravity multiplet and two vector multiplets.
The appropriate lattice to begin with is
\begin{equation}
    \Gamma^{5,5}=\Gamma^{4,4}(D_4)\oplus\Gamma^{1,1}\ .
    \label{lattice d4 + gamma1,1}
\end{equation}
Given the twist matrices \eqref{twist matrices for primary example} and using the properties of the $D_4$ lattice\footnote{One such useful property is that for $p_L-p_R \in \Lambda_R(D_4)$, $p_L^2+p_R^2\in 2\mathbb{Z}$.}, it can be verified that \eqref{modcond2} is satisfied. Note that since the orbifold acts non-trivially in all torus directions, the invariant sublattice is simply $\Gamma^{1,1}$.  

Let us now proceed with the calculation of the commutant. In order to find the matrix $\hat{h}=\begin{psmallmatrix}\hat{a} & \hat{b} \\ \hat{c} & \hat{d} \end{psmallmatrix}$, we need to solve \eqref{general commutation relations for h,g} for the twist specified by \eqref{twist matrices for primary example}. We find
\begin{equation}
    \hat{a}=\begin{pmatrix}
           a_{4\times 4} &0_{4\times 1}\\
            0_{1\times 4} &  a_{1\times 1}
        \end{pmatrix}\,,\quad \hat{b}=\begin{pmatrix}
           0_{4\times 4} &0_{4\times 1}\\
            0_{1\times 4} &  b_{1\times 1} 
        \end{pmatrix}\,,\quad \hat{c}=\begin{pmatrix}
           0_{4\times 4} &0_{4\times 1}\\
            0_{1\times 4} &  c_{1\times 1}
        \end{pmatrix}\,,\quad  \hat{d}=\begin{pmatrix}
           d_{4\times 4} &0_{4\times 1}\\
            0_{1\times 4} &  d_{1\times 1}
        \end{pmatrix}\,,
\end{equation}
where $a_{1\times 1},b_{1\times 1},c_{1\times 1},d_{1\times 1}$ are arbitrary numbers, $a_{4\times 4}$ takes the form
\begin{equation}a_{4\times 4}=
    \begin{pmatrix}
    a^1_{2\times 2} & 0_{2\times 2}\\
           0_{2\times 2} & a^2_{2\times 2}\\
\end{pmatrix}\,,\quad \text{with} \quad
a^2_{2\times 2}=
    \begin{pmatrix}
      a & \tilde{a}\\
      -\tilde{a} & a
    \end{pmatrix}\,,\quad\text{and}\quad  a^1_{2\times 2}\quad \text{arbitrary}\,,
    \label{matrix a form commuting}
\end{equation}
and $d_{4\times 4}$ is of the following form
\begin{equation}d_{4\times 4}=
\begin{pmatrix}
    d^1_{2\times 2} & d^2_{2\times 2}\\
            d^3_{2\times 2} & d^4_{2\times 2}\\
\end{pmatrix}\,,\quad \text{with} \quad
d^i_{2\times 2}=
    \begin{pmatrix}
      d_i & \tilde{d}_i\\
      -\tilde{d}_i & d_i
    \end{pmatrix}\,,\quad i=1,\ldots4\,.
    \label{d SO(4) subgroup}
\end{equation}
It will be useful to rearrange some columns and rows of $\hat{h}$ in order to bring $d_{4\times 4}$ to the form\footnote{This rearrangement doesn't affect the other sub-matrices in $\hat{h}$. Also, a permutation of rows and columns is an automorphism of $\text{GL}(10;\mathbb{R})$.} (see also \cite{cardoso1994moduli} for more details)
\begin{equation}
      \begin{pmatrix}
      d_1&d_2&\tilde{d}_1&\tilde{d}_2\\
      d_3&d_4&\tilde{d}_3&\tilde{d}_4\\
       -\tilde{d}_1&-\tilde{d}_2&{d}_1&{d}_2\\
       -\tilde{d}_3&-\tilde{d}_4&{d}_3&{d}_4
    \end{pmatrix} \,.
\end{equation}
Now, we can use the isomorphism
\begin{equation}
    \begin{pmatrix}
      x & \tilde{x}\\
      -\tilde{x} & x
    \end{pmatrix}\,\in \text{GL}(2n;\mathbb{R})\, \cong \, z =x+i\tilde{x} \in   \text{GL}(n;\mathbb{C})\,,
    \label{GL isomorphism}
\end{equation}
where $x$ and $\tilde x$ are real $n\times n$ matrices, in order to represent $d_{4\times 4}$ as
\begin{equation}
 \begin{pmatrix}
      d_1&d_2&\tilde{d}_1&\tilde{d}_2\\
      d_3&d_4&\tilde{d}_3&\tilde{d}_4\\
       -\tilde{d}_1&-\tilde{d}_2&{d}_1&{d}_2\\
       -\tilde{d}_3&-\tilde{d}_4&{d}_3&{d}_4
    \end{pmatrix} \xrightarrow{\eqref{GL isomorphism}}  \begin{pmatrix}
      d_1+i\tilde{d}_1 &  d_2+i\tilde{d}_2\\
       d_3+i\tilde{d}_3 &  d_4+i\tilde{d}_4
    \end{pmatrix}= \begin{pmatrix}
      z_1 & z_2\\
      z_3 & z_4
    \end{pmatrix} \in   \text{GL}(2;\mathbb{C})\,.
    \label{d4 matrix complex rep}
\end{equation}
So far, we have found the form of a generic matrix $\hat{h}\in \text{GL}(10;\mathbb{R})$ that commutes with the particular twist specified by \eqref{twist matrices for primary example}. Now, we demand that $\hat{h}\in \text{SO}(5,5)$. First, by imposing the constraints \eqref{matrices a,b,c,d constraints} we find
    \begin{align}
       & {a}_{4\times 4}^t{a}_{4\times 4}=1_{4\times 4}\,,\label{aal1}\\
       &{a}_{1\times 1}^t{a}_{1\times 1}-c_{1\times 1}^tc_{1\times1}=1_{1\times 1}\,,\label{aal2}\\
       & {a}_{1\times 1}^tb_{1\times 1}-c_{1\times 1}^td_{1\times 1}=0_{1\times 1}\,,\label{aal3}\\
       &d_{1\times 1}^td_{1\times 1}-b_{1\times 1}^tb_{1\times 1}=1_{1\times1}\,,\label{aal4}\\
       &d_{4\times 4}^td_{4\times 4}=1_{4\times4}\,.\label{aal5}
          \end{align}
From \eqref{aal1} and \eqref{matrix a form commuting} we see that $a^i_{2\times 2}\in \text{O}(2)$, for $i=1,2$. From \eqref{aal2}-\eqref{aal4} it follows that $a_{1\times 1},b_{1\times 1},c_{1\times 1}$ and $d_{1\times 1}$ form an $\text{O}(1,1)$ group. Also, \eqref{aal5} together with \eqref{d4 matrix complex rep} yield $d_{4\times 4}\in \text{U}(2)\subset \text{O}(4)$. Finally, by demanding det$(\hat{h})=1$ we conclude that the group commuting with the twist is
\begin{equation}
   \mathcal{C}= \text{SO}(1,1)\times \text{SO}(2)^2 \times \text{SU}(2)\times \text{U}(1)\,.
\end{equation}
As explained in section \ref{OSS}, the moduli space is given by
\begin{equation}
    \mathbb{R}^+\times \mathcal{C}/ \bar {\mathcal{C}}\,.
\end{equation}
Hence, we find 
\begin{equation}\label{modspace2}
    {\cal M}_V=\mathbb{R}^+\times \mathbb{R}^+\ ,
\end{equation}
where we recall that $\mathbb{R}^+=\text{SO}(1,1)/\mathbb{Z}_2$. This moduli space is parameterised by the string coupling and the radius of the circle $\mathcal{R}$. Also, the  isometry group of the moduli space is $\text{SO}(1,1)\times \text{SO}(1,1)$. 

In terms of real special geometry \cite{Gunaydin:1983bi,Gunaydin:1983rk}, the moduli space \eqref{modspace2} can be described by the cubic polynomial
\begin{equation}
    C(h)=d_{ABC}h^Ah^Bh^C\ ,
\end{equation}
with the only two non-vanishing $d$-symbols being\footnote{The overall normalization of the $d$-symbols is not fixed by the arguments here, but fixed by the conventions used in e.g.\ \cite{deWit:1991nm}, see eqn. (1.13) of that reference. If we keep the normalization arbitrary, but with $d_{133}=-d_{122}$, then the metric in \eqref{classmetric} is multiplied by $d_{122}$ which has to be taken positive for positive definiteness of the metric.}
\begin{equation}\label{d-symbols1}
    d_{122}=1\ ,\qquad d_{133}=-1\ .
\end{equation}
This gives
\begin{equation}
    C(h)=3h^1\Big((h^2)^2-(h^3)^2\Big)\,.
    \label{cubic pol for ex1}
\end{equation}
This is a special example of a homogeneous real special geometry. It falls into the classification of $d$-symbols given in \cite{deWit:1991nm, deWit:1992wf} corresponding to homogeneous real special geometries. The moduli spaces of real special geometry are found by setting $C(h)=1$. For the cubic polynomial \eqref{cubic pol for ex1}, this equation can be solved  by the parametrisation
\begin{equation}\label{repsh}
    h^1=\frac{1}{3a^2}e^{2\varphi/3}\ ,\qquad h^2= a\,e^{-\varphi/3}\cosh{(\sigma/\sqrt{3})}\ ,\qquad h^3=a\,e^{-\varphi/3}\sinh{(\sigma/\sqrt{3})}   \ ,
\end{equation}
for any real constant $a$. The induced metric can then be computed from
\begin{equation}\label{classmetric}
    {\rm d}s^2=-9\,d_{ABC}h^A{\rm d}h^B{\rm d}h^C= {\rm d}\varphi^2+{\rm d}\sigma^2   \ ,
\end{equation}
which is the canonical, flat metric on \eqref{modspace2}. The normalization is chosen such that later we can identify the expectation value of $e^{\varphi}$ with the (five-dimensional) string coupling and the expectation value of $e^{2\sigma/\sqrt{3}}$
with 
$\mathcal{R}/\sqrt{\alpha'}$ where $\mathcal{R}$ is the radius of the circle.

Now, the duality group of the effective theory, $\mathcal{K}=\text{SO}(1,1)\times \mathcal{C}$, contains $\text{SO}(1,1)\times \text{SO}(1,1)$. One $\text{SO}(1,1)$ isometry group acts linearly on $(h^2,h^3)$ and leaves the cubic polynomial invariant. The part connected to the identity corresponds to a shift in $\sigma$, so is  a rescaling of the radius $\mathcal{R}$. The $\text{SO}(1,1)$ element $-1_{2\times 2}$ is realised by $a\to -a$, i.e.\ $(h^2,h^3)\to -(h^2,h^3)$, leaving the radius invariant. The other continuous $\text{SO}(1,1)$ isometry is a scale transformation generating  a shift in the dilaton, see also section \ref{OSS}. This acts as 
$h^1\to e^{-2b}h^1, h^2\to e^b h^2, h^3\to e^b h^3$. This does not act on the radius $\mathcal{R}$ and therefore the two $\text{SO}(1,1)$ factors commute. In the quantum theory we expect that both $\text{SO}(1,1)$ factors are broken to $\mathbb{Z}_2$ subgroups under which the dilaton is invariant (see section \ref{OSS}), while $(h^2,h^3)\to -(h^2,h^3)$. We discuss this further later in section \ref{absence}.

\subsection*{$\mathbb{Z}_{12}$ with $n_V=4$ and $n_H=0$}
\label{Z12 example with zero hypers}

As a second example, let us consider the orbifold with mass parameters 
\begin{equation}
    m_1=\frac{2\pi}{3}\ , \qquad m_2= \frac{\pi}{2}\ ,\qquad m_3=\frac{2\pi}{3} \ ,\qquad m_4=0\ ,
\end{equation}
or, in terms of the twist vectors, 
\begin{equation}
    \tilde{u}=\left(\frac{2}{3},0\right)\ ,\qquad u=\left(\frac{1}{4},\frac{1}{4}\right)\ .
    \label{twist vectors z12 4 vm}
\end{equation}
This model was also discussed in \cite{Baykara:2023plc}, and the appropriate lattice  is \begin{equation}
    \Gamma^{5,5}=\Gamma^{4,4}(D_4)\oplus\Gamma^{1,1}\ .
\end{equation}
The conditions for modular invariance can be checked similarly with the previous example. The massless spectrum of this orbifold consists of the $\mathcal{N}=2, D=5$ gravity multiplet and four vector multiplets.

To find the invariant sublattice $I\subset \Gamma^{4,4}(D_4)$  we work in two steps. First, we see from the twist vector $u$ in \eqref{twist vectors z12 4 vm} that $p_R={0}$, which, combined with the condition $p_L-p_R \in \Lambda_R({D_4})$, implies that $p_L\in \Lambda_R({D_4})$, that is $p_L=(p_L^1,p_L^2,p_L^3,p_L^4)$, with $p_L^i\in \mathbb{Z}$ and $\sum_i p_L^i \in 2\mathbb{Z}$, for $i=1,\ldots,4$.\footnote{In general $\Lambda_R({D_n})$ is the lattice of vectors with integer entries, summing up to an even number.} Secondly, from the twist vector $\tilde{u}$ in \eqref{twist vectors z12 4 vm} we see that not all components of $p_L$ are invariant under the orbifold action, which indicates that the invariant lattice is spanned by the vectors $(p_L,0)=(0,0,p_L^3,p_L^4,{0})$, with $p_L^3,p_L^4\in \mathbb{Z}$ and $p_L^3+p_L^4\in2\mathbb{Z}$, namely $I=\Lambda_R({D_2})$. Finally, recall that the orbifold leaves $\Gamma^{1,1}$ invariant, and hence the complete orbifold invariant lattice is
\begin{equation}
    \hat{I}=I \oplus\Gamma^{1,1}\ .
\end{equation}
We now move on to the calculation of $\mathcal{C}$. For the twist specified by \eqref{twist vectors z12 4 vm} the equations \eqref{general commutation relations for h,g} can be solved by the following matrices
\begin{equation}
     \hat{a}=\begin{pmatrix}
           a_{2\times 2} &0_{2\times 3}\\
            0_{3\times 2} &  a_{3\times 3}
            \end{pmatrix}\,,\quad\hat{b}=\begin{pmatrix}
           0_{4\times 4} &0_{2\times 1}\\
            0_{1\times 4} &  b_{3\times 1} \end{pmatrix}\,,\quad\hat{c}=\begin{pmatrix}
           0_{4\times 4} &0_{4\times 1}\\
            0_{1\times 2} &  c_{1\times 3} \end{pmatrix}\,,\quad  \hat{d}=\begin{pmatrix}
           d_{4\times 4} &0_{4\times 1}\\
            0_{1\times 4} &  d_{1\times 1}
        \end{pmatrix}\,,
        \label{z12 matrices a,b,c,d}
\end{equation}
where $a_{3\times 3},b_{3\times 1},c_{1\times 3}, d_{1\times 1}$ are arbitrary, $a_{2\times 2}$ takes the same form as the matrix $a_{2\times 2}^2$ in \eqref{matrix a form commuting} and $d_{4\times 4}$ is the same as in \eqref{d SO(4) subgroup}. As in the previous example, we use the isomorphism \eqref{GL isomorphism} in order to rewrite $d_{4\times 4}$ as in \eqref{d4 matrix complex rep}. Then, imposing the constraints \eqref{matrices a,b,c,d constraints} yields
 \begin{align}
       & {a}_{2\times 2}^t{a}_{2\times 2}=1_{2\times 2}\,,\label{aaal1}\\
       &{a}_{3\times 3}^t{a}_{3\times 3}-c_{3\times 1}^tc_{1\times3}=1_{3\times 3}\,,\label{aaal2}\\
       & {a}_{3\times 3}^tb_{3\times 1}-c_{3\times 1}^td_{1\times 1}=0_{3\times 1}\,,\label{aaal3}\\
       &d_{1\times 1}^td_{1\times 1}-b_{1\times 3}^tb_{3\times 1}=1_{1\times1}\,,\label{aaal4}\\
       &d_{4\times 4}^td_{4\times 4}=1_{4\times4}\,.\label{aaal5}
          \end{align}
From \eqref{aaal1} and  \eqref{matrix a form commuting} we see that $a_{2\times 2}\in \text{O}(2)$. From \eqref{aaal2}-\eqref{aaal4} it follows that $a_{3\times 3},b_{3\times 1},c_{1\times 3}$ and $d_{1\times 1}$ form an $\text{O}(3,1)$ group. Also, \eqref{aaal5} together with \eqref{d4 matrix complex rep} yield $d_{4\times 4}\in \text{U}(2)\subset \text{O}(4)$. Finally, imposing $\text{det}(\hat{h})=1$ yields
\begin{equation}
   \mathcal{C}=\text{SO}(3,1)\times \text{SO}(2) \times \text{SU}(2)\times \text{U}(1)\,.
\end{equation}
Now, the moduli space can be easily determined by the analysis of section \ref{OSS}. We find
\begin{equation}
     {\cal M}_V=\mathbb{R}^+\times \frac{\text{SO}(3,1)}{\text{SO}(3)}\ ,
\end{equation}
and we recall that $\mathbb{R}^+$ is parameterised by the string coupling.

So far, we have constructed a seemingly consistent string theory, with a modular invariant partition function and a moduli space consistent with $\mathcal{N}=2, D=5$ local supersymmetry.
However, this model seems to fail at satisfying the integrality condition that is related to the degeneracies of states in the twisted sectors. In particular, the degeneracy of the ground state in the twisted sectors $D(k)$ is given by the number of \say{chiral} fixed points $\tilde{\chi}\cdot\chi$ divided by the volume of the invariant sublattice $\text{Vol}(\hat{I})$, and for a consistent physical theory $D(k)$ should be an integer \cite{Narain:1986qm,narain1991asymmetric}. Now, we can compute for example $D(k=1)$ for the $\mathbb{Z}_{12}$ orbifold discussed above. We find $\tilde{\chi}\cdot\chi=2 \sin \left(\frac{2\pi}{3}\right) 4\sin^2\left(\frac{\pi}{4}\right)= 2\sqrt{3}$ and $\text{Vol}(\hat{I})=2$, which implies $D(k=1)=\sqrt{3}$.

On the basis of this, such a model would usually be discarded. However, in section \ref{dual pairs} we will construct a dual model (see the first model in \autoref{Table of dual pairs with R-R field}), which does not suffer from the same problem. So, it would be interesting to see if any physical interpretation  could be given to the non-integral degeneracy, or if the dual model should also be discarded, presumably due to a non-perturbative inconsistency.

\subsection*{$\mathbb{Z}_6$ with $n_V=6$ and $n_H=0$}
\label{{Z}_6 with n_V=6 and n_H=0}

As a final example with no hypermultiplets, consider the orbifold with mass parameters
\begin{equation}
    m_1=\pi\ , \qquad m_2= \frac{2\pi}{3}\ ,\qquad m_3=\pi \ ,\qquad m_4=0\ .
\end{equation}
The corresponding twist vectors are
\begin{equation}
    \tilde{u}=\left(1,0\right)\,,\qquad u=\left(\frac{1}{3},\frac{1}{3}\right)\,,
    \label{twist vectors model 1}
\end{equation}
satisfying the condition \eqref{modular conditions on twist vectors} (this model was also studied in \cite{Baykara:2023plc}). The massless spectrum of this orbifold consists of the $\mathcal{N}=2, D=5$ gravity multiplet and six vector multiplets, as can be read off from \autoref{huiberttable}. Notice that this is an example in which accidental massless modes arise, as $m_1+m_3=2\pi$.

The choice of lattice (which can be easily checked that satisfies \eqref{modcond2}) is
\begin{equation}
    \Gamma^{5,5}=\Gamma^{4,4}(A_2\oplus A_2)\oplus\Gamma^{1,1}\ .
    \label{Narain lattice A2+A2}
\end{equation}
The sublattice $I\subset \Gamma^{4,4}(A_2\oplus A_2)$ that is invariant under the twist is
\begin{equation}
    I=\Lambda_R(A_2\oplus A_2)\,.
    \label{invarian sublattice for zero hyper model I}
\end{equation}
This can be understood by noting that the invariant lattice is spanned by the vectors $(p_L,0)$, which combined with the condition ${p}_L-{p}_R\in \Lambda_R(A_2\oplus A_2)$ yields ${p}_L\in \Lambda_R(A_2\oplus A_2)\subset \Lambda_W (A_2\oplus A_2)$. Recall that the orbifold acts as a shift on $\Gamma^{1,1}$ and leaves this lattice invariant. So, the complete orbifold invariant lattice is
\begin{equation}
    \hat{I}=\Lambda_R(A_2\oplus A_2) \oplus\Gamma^{1,1}\ .
\end{equation}
Let us now proceed with the calculation of $\mathcal{C}$. Given the twist vectors \eqref{twist vectors model 1}, we find that the equations \eqref{general commutation relations for h,g} can be solved by the matrices
\begin{equation}
 \hat{a}: \text{arbitrary}\,\qquad   \hat{b}=\begin{pmatrix}
        0_{5\times 4} & b_{5\times 1}\end{pmatrix}\,,\qquad \hat{c}= \begin{pmatrix}
            0_{4\times 5}\\
            c_{1\times 5}
        \end{pmatrix}\,,\qquad \hat{d}=\begin{pmatrix}
           d_{4\times 4} &0_{4\times 1}\\
            0_{1\times 4} &  d_{1\times 1}
        \end{pmatrix}\,,
\end{equation}
where $b_{5\times 1}$, $c_{1\times 5}$ and $d_{1\times 1}$ are arbitrary matrices, while $d_{4\times 4}$ is of the form \eqref{d SO(4) subgroup}. Once again, we employ the isomorphism \eqref{GL isomorphism} in order to bring  $d_{4\times 4}$ to the form \eqref{d4 matrix complex rep}. Now, by imposing the constraints \eqref{matrices a,b,c,d constraints} we find
    \begin{align}
       & \hat{a}_{5\times 5}^t\hat{a}_{5\times 5}-c_{5\times 1}^tc_{1\times 5}=1_{5\times 5}\,,\label{al1}\\
       & \hat{a}_{5\times 5}^tb_{5\times 1}-c_{5\times 1}^td_{1\times 1}=0_{5\times 1}\,,\label{al2}\\
       &d_{1\times 1}^td_{1\times 1}-b_{1\times 5}^tb_{5\times 1}=1_{1\times1}\,,\label{al3}\\
       &d_{4\times 4}^td_{4\times 4}=1_{4\times4}\,.\label{al4}
          \end{align}
From \eqref{al1}-\eqref{al3} it immediately follows that the matrices $\hat{a}_{5\times 5}$, $b_{5\times 1}$, $c_{1\times 5}$ and $d_{1\times 1}$ form an $\text{O}(5,1)$ group. Also, \eqref{al4} together with \eqref{d4 matrix complex rep} imply that $d_{4\times 4}\in \text{U}(2)\subset \text{O}(4)$. By imposing det$(\hat{h})=1$ we conclude that 
\begin{equation}\label{dualitygroup1}
   \mathcal{C}=\text{SO}(5,1)\times \text{SU}(2) \times \text{U}(1)\,,
\end{equation}
which, according to section \ref{OSS}, yields
\begin{equation}\label{modspace1}
    {\cal M}_V=\mathbb{R}^+\times \frac{\text{SO}(5,1)}{\text{SO}(5)}\,.
\end{equation}
Once again, the $\mathbb{R}^+$ factor is parameterised by the (five-dimensional) string coupling.

\subsection{A model with one hypermultiplet}
\label{z6 with 1 hyper 4 vectors}
Next, we focus on models with one hypermultiplet. Surprisingly, we could find only one such model, which, besides the one hypermultiplet, also has four vector multiplets. It is a $\mathbb{Z}_6$
orbifold specified by the mass parameters
\begin{equation}
    m_1=\frac{\pi}{3}\ , \qquad m_2=\frac{2\pi}{3}\ ,\qquad m_3=\frac{\pi}{3}\ , \qquad m_4=0\ ,
\end{equation}
corresponding to the twist vectors 
\begin{equation}
    \tilde{u}=\left(\frac{1}{3},0\right)\ ,\qquad u=\left(\frac{1}{3},\frac{1}{3}\right)\ .
\end{equation}
This is a $\mathbb{Z}_3$ rotation on the bosons, but it acts as a $\mathbb{Z}_6$ on the R-vacua\footnote{From \autoref{huiberttable} we can see that orbifold charge of states in the R-sector is of the form $e^{\pm im_i}$.}. 

The massless spectrum of this model is given by $\mathcal{N}=2$ supergravity coupled to four vector multiplets and one hypermultiplet. The lattice we start with is $\Gamma^{4,4}(A_2\oplus A_2)\oplus\Gamma^{1,1}$. It can be easily verified that the twist vectors satisfy the conditions \eqref{modular conditions on twist vectors} and also
\begin{equation}
    p\mathcal{M}_{\theta}^{\,3}p\equiv p_L\mathcal{N}_L^{\,3}p_L - p_R\mathcal{N}_R^{\,3}p_R=p_L^2-p_R^2 \in 2\mathbb{Z}\ ,
\end{equation}
and so, modular invariance is ensured. The sublattice $I\subset \Gamma^{4,4}(A_2\oplus A_2)$ that is invariant under the twist is
\begin{equation}
    I=\Lambda_R(A_2)\,.
    \label{invarian sublattice for one hyper model}
\end{equation}
This can be understood by first writing $(p_L,p_R)=(p_L^1,p_L^2,p_R^1,p_R^2)$\footnote{Here the labels $1,2$ correspond to the complex torus coordinates $W^{1,2}$.} and then noting that the invariant lattice is spanned by the vectors $(0,p_L^2,0,0)$. This result combined with the condition ${p}_L-{p}_R\in \Lambda_R(A_2\oplus A_2)$, which is equivalent to ${p}_L^i-{p}_R^i\in \Lambda_R(A_2), i=1,2$, yields ${p}_L^2\in \Lambda_R(A_2)\subset \Lambda_W (A_2)$. Recall that the orbifold acts as a shift on $\Gamma^{1,1}$ and leaves this lattice invariant. So, the complete orbifold invariant lattice is
\begin{equation}
    \hat{I}=I \oplus\Gamma^{1,1}\ .
\end{equation}
Let us now focus on the calculation of $\mathcal{C}$. By comparing this model with the one presented in section \ref{Z12 example with zero hypers} we can immediately see that the matrices $\hat{a},\hat{d}$ should have exactly the same form as in \eqref{z12 matrices a,b,c,d}. For the matrices $\hat{b},\hat{c}$ we find
\begin{equation}
    \hat{b}=\begin{pmatrix}
           b^1_{2\times 2} &b^2_{2\times 2} & 0_{2\times 1}\\
            0_{3\times 2} &  0_{3\times 2}  &b_{3\times 1} \end{pmatrix}\,,\qquad\hat{c}=\begin{pmatrix}
            c^1_{2\times 2} &0_{2\times 3}\\
            c^2_{2\times 2} &0_{2\times 3}\\
            0_{1\times 2} &  c_{1\times 3}  \end{pmatrix}\,,
\end{equation}
where the matrices $ b^i_{2\times 2}, c^i_{2\times 2}, i=1,2$ are of the same form as the matrix $a^2_{2\times 2}$ in \eqref{matrix a form commuting}, while $b_{3\times 1},c_{1\times 3}$ are arbitrary. Specifically, the matrix $\hat{h}$ that commutes with the twist reads
\begin{equation}
  \hat{h}=  \begin{pmatrix}
        a & \tilde{a} & 0 &0 &0 &b_1&\tilde{b}_1&b_2&\tilde{b}_2&0\\
         -\tilde{a}&a & 0 &0 &0 &-\tilde{b}_1&b_1&-\tilde{b}_2&b_2&0\\ 
         0&0&a_1&a_2&a_3&0&0&0&0&b_3\\
         0&0&a_4&a_5&a_6&0&0&0&0&b_4\\
         0&0&a_7&a_8&a_9&0&0&0&0&b_5\\
          c_1 & \tilde{c}_1 & 0 &0 &0 &d_1&\tilde{d}_1&d_2&\tilde{d}_2&0\\
           -\tilde{c}_1&c_1 & 0 &0 &0 &-\tilde{d}_1&d_1&-\tilde{d}_2&d_2&0\\
            c_2 & \tilde{c}_2 & 0 &0 &0 &d_3&\tilde{d}_3&d_4&\tilde{d}_4&0\\
           -\tilde{c}_2&c_2 & 0 &0 &0 &-\tilde{d}_3&d_3&-\tilde{d}_4&d_4&0\\
           0&0&c_3&c_4&c_5&0&0&0&0&d_5
    \end{pmatrix}\,.
\end{equation}
By rearranging some columns and rows of $\hat{h}$ we obtain
\begin{equation}
  \hat{h}= \begin{pmatrix}
        a & b_1 & 0 &0 &0 &b_2&\tilde{a}&\tilde{b}_1&\tilde{b}_2&0\\
         c_1&d_1 & 0 &0 &0 &d_2&\tilde{c}_1&\tilde{d}_1&\tilde{d}_2&0\\ 
         0&0&a_1&a_2&a_3&0&0&0&0&b_3\\
         0&0&a_4&a_5&a_6&0&0&0&0&b_4\\
         0&0&a_7&a_8&a_9&0&0&0&0&b_5\\
          c_2 & {d}_3 & 0 &0 &0 &d_4&\tilde{c}_2&\tilde{d}_3&\tilde{d}_4&0\\
           -\tilde{a}&-\tilde{b}_1 & 0 &0 &0 &-\tilde{b}_2&a&b_1&b_2&0\\
            -\tilde{c}_1& -\tilde{d}_1 & 0 &0 &0 &-\tilde{d}_2&c_1&d_1&{d}_2&0\\
           -\tilde{c}_2&-\tilde{d}_3 & 0 &0 &0 &-\tilde{d}_4&c_2&{d}_3&d_4&0\\
           0&0&c_3&c_4&c_5&0&0&0&0&d_5
    \end{pmatrix}\,.
\end{equation}
In this form, we can see by using the isomorphism \eqref{GL isomorphism} and imposing the constraints \eqref{matrices a,b,c,d constraints} that the matrices $a^1_{2\times 2}, b^i_{2\times 2}, c^i_{2\times 2} $ (for $i=1,2$) and $d_{4\times 4}$ form a $\text{U}(1,2)$ subgroup of $\text{O}(2,4)$. Also, it follows immediately that the matrices $a_{3\times 3},b_{3\times 1},c_{1\times 3}, d_{1\times 1}$ form an $\text{O}(3,1)$ group. Finally, by demanding $\text{det}(\hat{h})=1$, we find 
 \begin{equation}
     \mathcal{C}=\text{SO}(3,1)\times \text{SU}(1,2)\times \text{U}(1)\ .
      \label{T-orbi}
 \end{equation}  
The constraints from supergravity and the number of vector multiplets and hypermultiplets, $n_V=4$ and $n_H=1$ respectively, fix the complete answer for the moduli space. It is given by
\begin{equation}
    {\cal M}_{\mathcal{O}}={\cal M}_V\times {\cal M}_H\ ,
\end{equation}
where the vector multiplet scalars live in 
\begin{equation}
    {\cal M}_V=\mathbb{R}^+\times \frac{\text{SO}(3,1)}{\text{SO}(3)}\ ,
\end{equation}
and the hypermultiplet scalars on the quaternion-K{\"a}hler manifold 
\begin{equation}
{\cal M}_H=\frac{\text{SU}(1,2)}{\text{U}(2)}\ .
\end{equation}
This space is sometimes called the universal hypermultiplet moduli space. It is important to stress that the metric on this space does not receive quantum corrections as the dilaton sits in a vector multiplet.

\subsection{Models with two hypermultiplets}
\label{models2}

\subsection*{$\mathbb{Z}_6$ with $n_V=2$ and $n_H=2$}
\label{z6 orbifold with nv=vh=2}

The first example of this section is the orbifold with mass parameters given by  
\begin{equation}
    m_1=\pi \ ,\qquad m_2=\frac{\pi}{3}\ , \qquad m_3=\frac{2\pi}{3}\ ,\qquad m_4=0\ .
\end{equation}
The corresponding twist vectors are
\begin{equation}
    \tilde{u}=\left(\frac{5}{6},\frac{1}{6}\right)\ ,\qquad u=\left(\frac{1}{6},\frac{1}{6}\right)\ .
    \label{z6 twist class}
\end{equation}
Once again, we choose the lattice $\Gamma^{4,4}(A_2\oplus A_2)\oplus \Gamma^{1,1}$ and we verify that the twist vectors satisfy the conditions \eqref{modular conditions on twist vectors}. Also,
\begin{equation}
    p\mathcal{M}_{\theta}^{\,3}p\equiv p_L\mathcal{N}_L^{\,3}p_L - p_R\mathcal{N}_R^{\,3}p_R=-(p_L^2-p_R^2) \in 2\mathbb{Z}\ .
\end{equation}
So, modular invariance is guaranteed. Notice that only $\Gamma^{1,1}$ is left invariant under the twist.

Regarding the massless spectrum of this model, except for the gravity multiplet one finds two vector multiplets and two hypermultiplets.  One hypermultiplet consists of accidental massless modes, because $m_1+m_2+m_3=2\pi$. The particular twist specified by \eqref{z6 twist class} falls into the class of twists studied in \cite{cardoso1994moduli}, and $\mathcal{C}$ can be easily determined using the techniques developed in the previous examples. The result is 
\begin{equation}
\mathcal{C}=\text{SO}(1,1)\times \text{SU}(2,2)\times \text{U}(1)\ .
\label{comm5dcase}
\end{equation}
This fixes the moduli space to be
\begin{equation}
    {\cal M}_{\mathcal{O}}={\cal M}_V\times {\cal M}_H\ ,
\end{equation}
where the vector multiplet scalars live in 
\begin{equation}
    {\cal M}_V=\mathbb{R}^+\times \mathbb{R}^+\ .
\end{equation}
This space is parameterised by the string coupling and the radius of the circle and is the same as that in the example discussed in \ref{z12 with nv=2 and nh=0}. The hypermultiplet scalars form the quaternion-K{\"a}hler manifold of real dimension 8 (with $\text{SU}(2,2)\simeq \text{SO}(4,2)$),
\begin{equation}
{\cal M}_H=\frac{\text{SU}(2,2)}{\text{SU}(2)\times \text{SU}(2)\times \text{U}(1)}\simeq \frac{\text{SO}(4,2)}{\text{SO}(4)\times \text{SO}(2)}\ .
\end{equation}

\subsection*{$\mathbb{Z}_4$ with $n_V=4$ and $n_H=2$}

As a second example, we consider the orbifold with mass parameters given by 
\begin{equation}
    m_1=\frac{\pi}{2} \ ,\qquad m_2=\pi\ ,\qquad m_3=\frac{\pi}{2}\ ,\qquad m_4=0\ .
\end{equation}
The corresponding twist vectors are
\begin{equation}
    \tilde{u}=\left(\frac{1}{2},0\right)\ ,\qquad u=\left(\frac{1}{2},\frac{1}{2}\right)\ ,
    \label{z4 twist vectors}
\end{equation}
which satisfy the conditions \eqref{modular conditions on twist vectors}. The massless spectrum of this model consists of the  gravity multiplet coupled to two hypermultiplets and four vector multiplets. One hypermultiplet is again made up by accidental massless modes, as $m_1+m_2+m_3=2\pi$. The lattice we choose is\footnote{Here we mention the isomorphism $D_2\cong A_1\oplus A_1$.} $\Gamma^{4,4}(D_2\oplus D_2)\oplus \Gamma^{1,1}$, and  we check 
\begin{equation}
    p\mathcal{M}_{\theta}^{\,2}p\equiv p_L\mathcal{N}_L^{\,2}p_L - p_R\mathcal{N}_R^{\,2}p_R=p_L^2-p_R^2 \in 2\mathbb{Z}\ .
\end{equation}
Hence, modular invariance is ensured. Following similar arguments as in section \ref{z6 with 1 hyper 4 vectors}, we find that the invariant lattice is
\begin{equation}
    \hat{I} = \Lambda_R(D_2)\oplus \Gamma^{1,1}\,.
\end{equation}
Given the simple form of the twist vectors \eqref{z4 twist vectors} and the analysis of the previous sections, it is easy to see that $\hat{h}$ takes the form
\begin{equation}
    \hat{a}=\begin{pmatrix}
           a_{2\times 2} &0_{2\times 3}\\
            0_{3\times 2} &  a_{3\times 3}
        \end{pmatrix},  \hat{b}=\begin{pmatrix}
           b^1_{2\times 2} &b^2_{2\times 2} & 0_{2\times 1}\\
            0_{3\times 2} &  0_{3\times 2}  &b_{3\times 1} \end{pmatrix},\hat{c}=\begin{pmatrix}
            c^1_{2\times 2} &0_{2\times 3}\\
            c^2_{2\times 2} &0_{2\times 3}\\
            0_{1\times 2} &  c_{1\times 3}  \end{pmatrix},  \hat{d}=\begin{pmatrix}
           d_{4\times 4} &0_{4\times 1}\\
            0_{1\times 4} &  d_{1\times 1}
        \end{pmatrix}\,.
\end{equation}
Imposing the constraints \eqref{matrices a,b,c,d constraints} and $\text{det}(\hat{h})=1$ yields
\begin{equation}
  \mathcal{C}=\text{S}\left[\text{O}(3,1)\times \text{O}(2,4)\right]\ .
\end{equation}
The vector multiplet and hypermultiplet moduli spaces are
\begin{equation}
    {\cal M}_V=\mathbb{R}^+\times \frac{\text{SO}(3,1)}{\text{SO}(3)}\ ,\qquad {\cal M}_H=\frac{\text{SO}(4,2)}{\text{SO}(4)\times \text{SO}(2)}\ .
   \end{equation}

\subsection*{Concluding remarks}

We conclude this section with some general remarks about the (classical) moduli spaces in theories with $\mathcal{N}=2$ supersymmetry in $D=5$. First of all, the models presented in this section give vector multiplet moduli spaces that  are of the general form 
\begin{equation}\label{VMMod}
    {\cal M}_V=\mathbb{R}^+\times \frac{\text{SO}(n_V-1,1)}{\text{SO}(n_V-1)}\ ,
\end{equation}
consistent with the constraints from real special geometry in $\mathcal{N}=2,D=5$, with $n_V$ the number of vector multiplets. Indeed, when the moduli space factorises, it must be of the form \eqref{VMMod} 
\cite{Gunaydin:1983bi,Gunaydin:1983rk}.
The cubic polynomial of real special geometry is then of the factorised form 
\begin{equation}
    C(h)=h^1 Q(h^I)\ ,\qquad Q(h^I)=h^I\bar{\eta}_{IJ}h^J\ ,
\end{equation}
with $I=(i,2)\,,i=3,...,n_V+1$ and where $\bar{\eta}$ is a quadratic form of (mostly minus) signature $(n_V-1,1)$, making the SO$(n_V-1,1)$ isometry manifest. The factorization of $\mathcal{M}_V$ is a special property and not the most general $\mathcal{N}=2, D=5$ supergravity coupling of vector multiplets. It is a consequence of the underlying $\mathcal{N}=8$ supersymmetry which is spontaneously broken, and the fact that $\text{E}_{6(6)}$ has a maximal subgroup which factorises as $\text{SO}(1,1) \times \text{Spin}(5,5)$, as explained in section \ref{OSS}. 

Hypermultiplet scalars coupled to supergravity with eight supersymmetries parametrise (non-compact) quaterion-K\"ahler (QK) manifolds. The cases studied here correspond to homogeneous spaces $G/H$. These are classified, and the ones appearing in this section belong to the series
\begin{equation}
    X(n_H)=\frac{\text{SU}(n_H,2)}{\text{SU}(n_H)\times \text{U}(2)} \ , \qquad Y(n_H)=\frac{\text{SO}(n_H,4)}{\text{SO}(n_H)\times \text{SO}(4)}\ ,
\end{equation}
with $n_H$ the number of hypermultiplets. The examples we discussed in this section correspond to $X(1)$ and $X(2)$. Note that $X(2)\simeq Y(2)$ up to double covers that for our purposes are not relevant and therefore, not explicitly written. 

Moduli spaces that are not of the form \eqref{VMMod} can also be obtained by freely acting asymmetric orbifolds. As an example, consider the $\mathbb{Z}_2$ orbifold discussed in section \ref{sec:N=2 asymmetric orbifold}. The massless spectrum of this orbifold consists of the $\mathcal{N}=2$, $D=5$ gravity multiplet, coupled to 14 vector multiplets, and no hypermultiplets. In total there are 14 real scalars, one per each massless vector multiplet, and the moduli space is 
\begin{equation}
{\cal M}=\frac{\text{SU}^*(6)}{\text{USp}(6)}\ .
\end{equation}
This is actually one of the magic square supergravities of \cite{Gunaydin:1983rk}.

Another interesting example is the $\mathbb{Z}_4$ orbifold, based on the $D_4$ lattice, with $\vec{m}=(\tfrac{\pi}{2},\tfrac{\pi}{2},\tfrac{\pi}{2},0)$, corresponding to $\tilde{u}=(0,0,\frac{1}{2},0)$ and $u=(0,0,\frac{1}{4},\frac{1}{4})$.  This orbifold gives a massless spectrum consisting of eight vector multiplets coupled to supergravity (cf. \autoref{huiberttable}), and no hypermultiplets. The most likely candidate for the (classical) moduli space is another of the magic square supergravities,
\begin{equation}
    {\cal M}_{\mathcal{N}=2}^{\mathbb{Z}_4}=\frac{\text{SL}(3,\mathbb{C})}{\text{SU}(3)}\ ,
\end{equation}
which is known to be a truncation of $\mathcal{N}=8, D=5$ supergravity if we break the $\mathcal{N}=8$ supermultiplet into $\mathcal{N}=2$ supermultiplets and truncate the spin-$\tfrac{3}{2}$ and matter multiplets  \cite{Gunaydin:1983rk}. In Scherk-Schwarz, we do not truncate these multiplets; instead, these become lifted and massive.  Similar models of hyper-free superstrings were constructed in \cite{Dolivet:2007sz}. It is interesting that we can produce two out of the four magical supergravity theories of \cite{Gunaydin:1983rk,Gunaydin:1983bi}, but apparently not the other two (with 5 and 26 vector multiplets). 

What all of our models seem to have in common is that the total number of vectors is always odd. In $\mathcal{N}=2$ theories in $D=5$, there is one vector in the gravity multiplet, and we find always an even number of vector multiplets. It would be nice to relate these observations to the work in higher dimensions in $D\geq 7$ and the string lamppost principle \cite{montero2021cobordism}.

\section{Quantum aspects}
\label{quantasp}

So far, we have determined the classical moduli spaces of theories preserving 8 supersymmetries in five dimensions. In all the examples that we presented the dilaton lies in a vector multiplet, hence the classical vector multiplet moduli spaces of section \eqref{8sup} can be modified my quantum corrections. However, in this section we will argue that, for our models, the metric on the vector multiplet moduli space receives no quantum corrections.

In order to achieve this we will first construct dual pairs of orbifold models. These are pairs of orbifolds related by a strong-weak coupling duality. By combining these dual pairs with the duality group of the orbifold, which we have explicitly computed in the previous section, we  will show the absence of quantum corrections.

\subsection{Dual pairs}
\label{dual pairs}
In this section we construct dual pairs \`a la Sen and Vafa \cite{sen1995dual}. We will briefly discuss the procedure of constructing dual pairs here; for more details we refer to \cite{sen1995dual}. The starting point is type II string theory compactified on $T^4$, with moduli space
$\mathcal{M}_{T^4}=\text{Spin}(5,5)/\left[\text{Spin}(5)\times \text{Spin}(5)\right]$.
The U-duality group of the theory is $\text{Spin}(5,5;\mathbb{Z})$ which contains the T-duality group  $\text{Spin}(4,4;\mathbb{Z})$. Take two elements $g,g'\in \text{Spin}(4,4;\mathbb{Z})$ of order $p$ that are conjugate in $\text{Spin}(5,5;\mathbb{Z})$ but not in $\text{Spin}(4,4;\mathbb{Z})$, that is
\begin{equation}
    \hat{g} \,g\,\hat{g}^{-1} = g'\,,\quad \hat{g}\,\in \text{Spin}(5,5;\mathbb{Z})\,.
    \label{general duality element}
\end{equation}
Consider a point in the moduli space $m\in \mathcal{M}_{T^4}$
that is invariant under the action of $g$, so that $g \cdot m=m$. Then the point $m'=\hat g \cdot m\in \mathcal{M}_{T^4}$ is invariant under the action of $g'$. 

Next, compactify on a  further circle $S^1$. At the point in the moduli space $m\in \mathcal{M}_{T^4}$
we can orbifold by the action of $g$ combined with a shift by $2\pi \mathcal{R}/p$ on the circle and at
the point in the moduli space $m'\in \mathcal{M}_{T^4}$
we can orbifold by the action of $g'$ combined with a shift by $2\pi \mathcal{R}'/p$ on the  circle.
The two orbifold theories should then be equivalent non-perturbatively
\cite{Vafa:1995gm,sen1995dual}, with a relation between $\mathcal{R}$ and $\mathcal{R}'$ that we specify below.

In the context of type IIB string theory, the particular element $\hat{g}$ chosen in \cite{sen1995dual} can be expressed in terms of the SL$(2;\mathbb{Z}$) S-duality element $s$ that inverts the ten-dimensional axion-dilaton field, and the SO$(4,4;\mathbb{Z}$) T-duality element $t_{67}\cdot t_{89}$, 
where $t_{ij}$ inverts the
 K{\"a}hler modulus of the $i-j$ plane, so that
\begin{equation}
    \hat{g} = s\cdot t_{67}\cdot t_{89} \cdot s^{-1}\,.
    \label{duality element omega}
\end{equation}
If the R-R scalars are set to zero, this $\hat{g}$ acts on the six-dimensional dilaton $\phi_6$ and the NS-NS three-form $H_3$ exactly as string-string duality acts on these fields in six dimensions \cite{sen1995dual}, that is (see also \cite{Duff:1994zt,Witten:1995ex})
\begin{equation}
    \phi_6'=-\phi_6\,,\qquad H_3' = e^{-2\phi_6}*H_3\,.
\end{equation}This induces the following transformation on the 6-dimensional metric in string frame (in Einstein frame, the metric is invariant)
\begin{equation}
    G_{\hat{\mu}\hat{\nu}}'= e^{-2\phi_6}G_{\hat{\mu}\hat{\nu}}\,, \qquad \hat{\mu},\hat{\nu}=0,\ldots,5\,.
\end{equation}
Also, it follows that under string-string duality the strong coupling regime of one theory corresponds to the weak coupling regime of the dual theory
as 
\begin{equation}
   \phi_6'=-\phi_6\implies \lambda_6'=\frac{1}{\lambda _6}\,,
\end{equation}
 where $\lambda _6=\langle e^{\phi_6}\rangle$ and similarly for $\lambda_6'$.
Moreover, we can determine how the duality works upon further compactification to five dimensions \cite{Witten:1995ex}; this will be essential for our discussion later in this section. We have\footnote{For convenience, we set $\alpha '=1$ from now on.}
\begin{equation}
    G_{55}'= e^{-2\phi_6}G_{55}\implies\mathcal{R}'=\frac{\mathcal{R}}{\lambda_6}\,.
\end{equation}
We can rewrite the above equations in terms of the $5D$ string coupling constant $\lambda_5$, using $\lambda_6=\sqrt{\mathcal{R}}\lambda_5$ and similarly for $\lambda_6'$. We obtain
\begin{equation}
    \lambda_5'=\frac{1}{\sqrt{\lambda_5}\mathcal{R}^{{3}/{4}}}\,,\qquad \mathcal{R}'=\frac{\sqrt{\mathcal{R}}}{\lambda_5}\,.
    \label{5d string-string duality}
\end{equation}
We can also perform a T-duality transformation 
taking the circle of radius $\mathcal{R}'$ to the T-dual circle of radius $\widetilde{\mathcal{R}}=1/\mathcal{R}'
$
under which the five-dimensional string coupling remains invariant, $ \lambda_5'=\widetilde{\lambda}_5$,
so that $\lambda'_5=\lambda'_6/\sqrt{\mathcal{R}'}
=\widetilde\lambda_6/\sqrt{\widetilde{\mathcal{R}}}$ 
implies
\begin{equation}
   \lambda_6'=\frac{\widetilde{\lambda}_6}{\widetilde{\mathcal{R}}}\,.
\end{equation}
After the T-duality transformation, \eqref{5d string-string duality} becomes
\begin{equation}
     \widetilde{\lambda}_5=\frac{1}{\sqrt{\lambda_5}\mathcal{R}^{{3}/{4}}}\,,\qquad \widetilde{\mathcal{R}}=\frac{\lambda_5}{\sqrt{\mathcal{R}}}\,.
     \label{T-dual string-string duality}
\end{equation}
Finally, we can rewrite \eqref{T-dual string-string duality} in terms of the six-dimensional couplings as
\begin{equation}
     \widetilde{\lambda}_6=\frac{1}{\mathcal{R}}\,,\qquad \widetilde{\mathcal{R}}=\frac{\lambda_6}{{\mathcal{R}}}\,.
\end{equation}
For our purposes, we start from type IIB theory and, following the procedure presented above (cf.\,\eqref{general duality element}-\eqref{duality element omega}), we construct a dual pair of type IIB orbifolds, $\mathcal{O}_{\text{IIB}}$ and $\mathcal{O}_{\text{IIB}}'$. Then we apply a T-duality transformation on $\mathcal{O}_{\text{IIB}}'$ and we obtain a T-dual type IIA orbifold $\widetilde{\mathcal{O}}_{\text{IIA}}$. Note that the duality between $\mathcal{O}_{\text{IIB}}$ and $\mathcal{O}_{\text{IIB}}'$ is non-perturbative, while the duality between $\mathcal{O}_{\text{IIB}}'$ and $\widetilde{\mathcal{O}}_{\text{IIA}}$ is perturbative, and the models $\mathcal{O}_{\text{IIB}}'$ and $\widetilde{\mathcal{O}}_{\text{IIA}}$ have the same perturbative spectrum and moduli space. Now, according to \eqref{T-dual string-string duality}, the strong coupling regime of the $\mathcal{O}_{\text{IIB}}$ model at fixed radius corresponds to weak coupling and large radius of the dual $\widetilde{\mathcal{O}}_{\text{IIA}}$ model.

Consider the $\mathbb{R}^+\times \mathbb{R}^+$ subspace of the moduli space parameterised by $\lambda_5,\mathcal{R}$.
It can equivalently be viewed as $\mathbb{R}^2$ 
 with coordinates $\phi_5,\sigma$ where
$\lambda_5=\langle e^{\phi_5}\rangle$ and $\mathcal{R}=e^{2\sigma/{\sqrt{3}}}$.
In the examples we discuss in this chapter, the classical moduli space metric for this subspace is 
\begin{equation}\label{isom-metrica}
    {\rm d}s^2=
    {\rm d}\phi_5^2+{\rm d}\sigma^2\ .
\end{equation}
The U-duality
between $\mathcal{O}_{\text{IIB}}'$ and $\mathcal{O}_{\text{IIB}}$
then gives
\begin{equation}\label{dictionaryOprimeO}
    \phi_5'=-\frac{1}{2}\phi_5-\frac{\sqrt{3}}{2}\sigma\ ,\qquad \sigma'=-\frac{\sqrt{3}}{2}\phi_5+\frac{1}{2}\sigma\ ,
\end{equation}
while the T-duality
between $\mathcal{O}_{\text{IIB}}'$ and $\widetilde{\mathcal{O}}_{\text{IIA}}$ gives
\begin{equation}\label{dictionaryOtildeO}
    \tilde{\phi}_5=-\frac{1}{2}\phi_5-\frac{\sqrt{3}}{2}\sigma\ ,\qquad \tilde{\sigma}=\frac{\sqrt{3}}{2}\phi_5-\frac{1}{2}\sigma\ ,
\end{equation}
Notice that this is a rotation through an angle $2\pi/3$ and, since it is a rotation, it is an isometry of the flat metric (\ref{isom-metrica}), taking it to
\begin{equation}\label{isom-metric}
  {\rm d}s^2=  {\rm d}\tilde{\phi}_5^2+{\rm d}\tilde{\sigma}^2\ .
\end{equation}
This fact will be important in our arguments showing the absence of quantum corrections to the moduli spaces. For instance, in the two-moduli example of section \ref{z12 with nv=2 and nh=0}, the classical moduli space
 $\mathbb{R}^+\times \mathbb{R}^+$ has a
metric that is precisely the flat metric (\ref{isom-metrica}), and hence is unchanged by the strong-weak duality. We will return to this in the examples below and in the next section.

Finally, we mention here that the five-dimensional duality proposed in \cite{Witten:1995ex} was between type IIB string theory on $\mathbb{R}^{1,4}\times S^1 \times {K}3$ and heterotic string theory on $\mathbb{R}^{1,4}\times T^5$, and it was argued there that the strong coupling regime of the heterotic string compactification corresponds to the type IIB compactification at large radius. This duality was derived from string-string duality in six dimensions, which relates the heterotic string on $\mathbb{R}^{1,5}\times T^4$ with the type IIA string on $\mathbb{R}^{1,5}\times {K}3$. 

\subsection{Explicit examples}

In this section, we present explicit examples of dual type II pairs with $\mathcal{N}=2$ supersymmetry in $5D$. In practice, for the construction of a dual pair we start with an orbifold twist $\mathcal{M}_{\theta}$ specified by the twist vectors $\tilde{u},u$, that has the form
\begin{equation}
    \mathcal{M}_{\theta}= \text{diag}\left(R(2\pi\tilde{u}_3),R(2\pi\tilde{u}_4),R(2\pi u_3),R(2\pi u_4)\right)\,,
\end{equation}
according to our analysis in chapter \ref{chap:fao} (cf. \eqref{standarformM}, \eqref{theta's}, \eqref{u's}). We will refer to the orbifold specified by $\mathcal{M}_{\theta}$ as the \say{initial} model. Then, the action of $\hat g$ on the twist vectors is simply given by (see \cite{sen1995dual} for details)
\begin{equation}
    \begin{pmatrix}
        \tilde{u}_3'\\
        \tilde{u}_4'\\
        u_3'\\
        u_4'
    \end{pmatrix} =\frac{1}{2} \begin{pmatrix}
        1&-1&1&-1\\
        -1&1&1&-1\\
        1&1&1&1\\
        -1&-1&1&1
    \end{pmatrix}\,\begin{pmatrix}
        \tilde{u}_3\\
        \tilde{u}_4\\
        u_3\\
        u_4
    \end{pmatrix}\,.
    \label{matrix for dual pairs}
\end{equation}
This induces a duality transformation on the mass parameters, and using \eqref{u's} we find
\begin{equation}
    m_1'=m_4\ ,\qquad m_2'=m_2\ ,\qquad m_3'=m_3\ ,\qquad m_4'=m_1\ ,
\end{equation}
so it simply amounts to an exchange $m_1\leftrightarrow m_4$. 
For example, starting from a twist with $m_4=0$
\begin{equation}
    \mathcal{M}_{\theta} = \text{diag}\left(R(m_1+m_3),R(m_1-m_3),R(m_2),R(m_2)\right)\,,
\end{equation}
the twist of the dual model is 
\begin{equation}
    \mathcal{M}_{\theta}' = \text{diag}\left(R(m_3),R(-m_3),R(m_2+m_1),R(m_2-m_1)\right)\,.
\end{equation}
We mention here that it is possible to start with a model for which the only massless fields are in the NS-NS sector and under the duality transformation end up with a model with both NS-NS and R-R massless fields.

\subsection*{Dual pairs without R-R fields}
\label{dual pairs without R-R fields}
Here, we work out two concrete examples of $\mathcal{N}=2$ dual pairs in $5D$ where the massless spectra of both the initial and the dual theory consist purely of NS-NS states. This has the advantage that the moduli space of the dual theory can be found using the same techniques as for the examples of sections \ref{modelsno}-\ref{models2}, namely by finding the commutant with the T-duality group. We could find only two such models.

\subsection*{Dual pair I:}
\label{dual pair I}
We start from the $\mathbb{Z}_{12}$ orbifold presented in section \ref{z12 with nv=2 and nh=0} with $n_V=2$ and $n_H=0$. The mass parameters and twist vectors of this model are 
\begin{equation}
     m_1=\frac{2\pi}{3}\ , \qquad m_2= \frac{\pi}{2}\ ,\qquad m_3=\frac{\pi}{3} \ ,\qquad m_4=0\,,
\end{equation}
\begin{equation}
     \tilde{u}=\left(\frac{1}{2},\frac{1}{6}\right)\ ,\qquad u=\left(\frac{1}{4},\frac{1}{4}\right)\,,
\end{equation}
and its moduli space was found to be
\begin{equation}
     {\cal M}_{\mathcal{O}}=\mathbb{R}^+\times \mathbb{R}^+\ .
     \label{moduli space dual pair i}
\end{equation}
The two scalars parametrizing this moduli space are the string coupling $\lambda_5=\langle e^{\varphi}\rangle$ and the radius of the circle $\mathcal{R}
=e^{2\sigma/\sqrt{3}}$ (recall that we are setting $\alpha'=1$), and the classical metric on ${\cal M}_{\mathcal{O}}$ is the flat metric \eqref{classmetric}.

Regarding the dual model, we can calculate the mass parameters and twist vectors using \eqref{matrix for dual pairs} and \eqref{u's}. We find
\begin{equation}
     m_1'=0\ , \qquad m_2'= \frac{\pi}{2}\ ,\qquad m_3'=\frac{\pi}{3} \ ,\qquad m_4'=\frac{2\pi}{3}\,,
\end{equation}
\begin{equation}
     \tilde{u}'=\left(\frac{1}{6},-\frac{1}{6}\right)\ ,\qquad u'=\left(\frac{7}{12},-\frac{1}{12}\right)\,.
\end{equation}
This is again a $\mathbb{Z}_{12}$ orbifold with $n_V=2$ and $n_H=0$, which we denote by $\mathcal{O}'$, and all the massless fields come from the NS-NS sector. So, we can determine the moduli space $\mathcal{M}_{\mathcal{O}'}$ of this dual model using the techniques developed in sections \ref{OSS}-\ref{models2}. We find
\begin{equation}
     {\cal M}_{\mathcal{O}'}=\mathbb{R}^+\times \mathbb{R}^+\ .
     \label{moduli space dual pair i dual}
\end{equation}
Classically, in the dual variables, the metric is again the flat metric. The moduli space is parameterised by the string coupling $\lambda_5'=\langle e^{\varphi'}\rangle$ and the radius of the circle $\mathcal{R}'
=e^{2\sigma'/\sqrt{3}}$. 
T-duality takes the circle radius to  $\widetilde{\mathcal{R}}=1/\mathcal{R}'
$
 and gives the T-dual IIA orbifold model, denoted by $\widetilde{\mathcal{O}}$. Since it is obtained by a T-duality of the $\mathcal{O}'$ orbifold, we have that $\tilde\sigma=-\sigma'$ and so the moduli space is again given by
\begin{equation}
{\cal{M}}_{\widetilde{\mathcal{O}}}=\mathbb{R}^+\times \mathbb{R}^+\ ,
\end{equation}
parameterised by $\widetilde{\mathcal{R}}$ and $\widetilde{\lambda}_5$. The classical metric is again the flat one, as it is obtained from the T-dual of the $\mathcal{O}'$ theory. 

We  now consider quantum corrections to the metric on the moduli space \eqref{moduli space dual pair i}.
Consider first the initial orbifold ${\mathcal{O}}$.
For the classical theory, the moduli space metric is (\ref{isom-metrica}) and is independent of the dilaton $\phi_5$.
Quantum effects can in principle deform this moduli space metric and lead to dilaton-dependence.
Now, as we can see from \eqref{T-dual string-string duality}, 
the theory $ {\cal{O}}$ at  strong coupling (large $\lambda_5$) and fixed radius $\mathcal{R}$ is given by the dual theory $\widetilde{\cal{O}}$ at
weak coupling (in $\tilde{\lambda}_5$). However, we have seen that the classical moduli space metric for the dual
theory $\widetilde{\cal{O}}$ is again the flat metric
(\ref{isom-metric}). The classical theory $\widetilde{\cal{O}}$ is the strong-coupling limit of the initial theory $ {\cal{O}}$, so we see that the strong-coupling limit of the moduli space metric on $ {\cal{O}}$
is the same as for the classical theory, greatly limiting the form any quantum corrections to the metric can take.
In the next section, we present a stronger version of this argument in which we parametrise the possible quantum corrections and then  use duality to argue that no quantum corrections in fact arise.

\subsection*{Dual pair II:}
\label{dual pair II}
Here, we present another example where the massless spectra of both the initial and the dual model consist purely of NS-NS fields. The initial model is specified by the following mass parameters and twist vectors
\begin{equation}
     m_1=\pi\ , \qquad m_2= \frac{\pi}{3}\ ,\qquad m_3=\frac{2\pi}{3} \ ,\qquad m_4=0\,,
\end{equation}
\begin{equation}
     \tilde{u}=\left(\frac{5}{6},\frac{1}{6}\right)\ ,\qquad u=\left(\frac{1}{6},\frac{1}{6}\right)\,,
\end{equation}
This is a $\mathbb{Z}_6$ orbifold with $n_V=n_H=2$ discussed in section \ref{z6 orbifold with nv=vh=2}. The moduli space of this model is
\begin{equation}
{\cal M}_{\mathcal{O}}={\cal M}_V\times{\cal M}_H=\mathbb{R}^+\times \mathbb{R}^+\times \frac{\text{SO}(4,2)}{\text{SO}(4)\times \text{SO}(2)}\ ,
\label{moduli space dual pair II}
\end{equation}
where ${\cal M}_V=\mathbb{R}^+\times \mathbb{R}^+$ is parameterised by the string coupling and the radius of the circle. For the dual model we find (cf.\,\eqref{matrix for dual pairs},\eqref{u's})
\begin{equation}
     m_1'=0\ , \qquad m_2'= \frac{\pi}{3}\ ,\qquad m_3'=\frac{2\pi}{3} \ ,\qquad m_4'=\pi\,,
\end{equation}
\begin{equation}
     \tilde{u}'=\left(\frac{1}{3},-\frac{1}{3}\right)\ ,\qquad u'=\left(\frac{2}{3},-\frac{1}{3}\right)\,.
\end{equation}
This is a $\mathbb{Z}_6$ orbifold with $n_V=n_H=2$, with only NS-NS massless fields. Once again, we can specify classically $\mathcal{M}_{\mathcal{O}}'$ using the methods of sections \ref{OSS}-\ref{models2}. We find that 
\begin{equation}
  \mathcal{M}_{\mathcal{O}}' = \mathbb{R}^+\times \mathbb{R}^+\times \frac{\text{SO}(4,2)}{\text{SO}(4)\times \text{SO}(2)}\,.
  \label{dual pair II dual moduli space}
\end{equation}
In section \ref{absence} we 
 argue that the classical moduli space metric on \eqref{moduli space dual pair II} is exact and receives no quantum corrections.

\subsection*{Dual pairs with R-R fields}
\label{dual models with R-R fields}
As we mentioned in the beginning of this section, it is possible to start with a model without massless R-R fields and, after applying the transformation \eqref{matrix for dual pairs}, obtain a dual model with massless R-R fields. We collect such dual pairs in \autoref{Table of dual pairs with R-R field}. The moduli space of the initial models was specified in sections \ref{modelsno}-\ref{models2}. Regarding the moduli space of the dual models, we could not use the techniques of sections \ref{OSS}-\ref{models2} because the spectra of those models contain R-R massless states. However, classically the moduli spaces of the dual models (with R-R massless scalars) can be determined by the constraints of spontaneously broken $\mathcal{N}=8 \to \mathcal{N}=2, D=5$ supergravity, and the classical vector multiplet moduli spaces are again of the form \eqref{VMMod}. In the next section, we will present evidence for the absence of quantum corrections to the vector multiplet moduli spaces of these dual pairs as well.

\begin{table}[ht!]
\centering
{\renewcommand*{\arraystretch}{2.6}
{\small
	\begin{tabular}{|c|c|c|c|c|c|}
 \hline
  $\mathbb{Z}_p$	&  $(m_1,m_2,m_3,m_4)$ & $\tilde{u},u$ & $(m_1',m_2',m_3',m_4')$ & $\tilde{u}',u'$   & $(n_V,n_H)$ \\ \hline
 $\mathbb{Z}_{12}$ & $(\tfrac{2\pi}{3},\tfrac{\pi}{2},\tfrac{2\pi}{3},0)$ & $(\tfrac{2}{3},0),(\tfrac{1}{4},\tfrac{1}{4})$ & $(0,\tfrac{\pi}{2},\tfrac{2\pi}{3},\tfrac{2\pi}{3})$ & $(\tfrac{1}{3},-\tfrac{1}{3}),(\tfrac{7}{12},-\tfrac{1}{12})$   & $(4,0)$ \\ \hline
   $\mathbb{Z}_6$ & $(\pi,\tfrac{2\pi}{3},\pi,0)$ & $(1,0),(\tfrac{1}{3},\tfrac{1}{3})$ & $(0,\tfrac{2\pi}{3},\pi,\pi)$ & $(\tfrac{1}{2},-\tfrac{1}{2}),(\tfrac{5}{6},-\tfrac{1}{6})$  & $(6,0)$\\ \hline
 $\mathbb{Z}_6$ & $(\tfrac{\pi}{3},\tfrac{2\pi}{3},\tfrac{\pi}{3},0)$ & $(\tfrac{1}{3},0),(\tfrac{1}{3},\tfrac{1}{3})$ &  $(0,\tfrac{2\pi}{3},\tfrac{\pi}{3},\tfrac{\pi}{3})$  &  $(\tfrac{1}{6},-\tfrac{1}{6}),(\tfrac{1}{2},\tfrac{1}{6})$  &$(4,1)$\\ \hline 
  ${\mathbb{Z}_4}$ & $(\tfrac{\pi}{2},\pi,\tfrac{\pi}{2},0)$ & $(\tfrac{1}{2},0),(\tfrac{1}{2},\tfrac{1}{2})$ & $(0,\pi,\tfrac{\pi}{2},\tfrac{\pi}{2})$ & $(\tfrac{1}{4},-\tfrac{1}{4}),(\tfrac{3}{4},\tfrac{1}{4})$ & $(4,2)$ \\ \hline
     	\end{tabular}}}
        \caption{\textit{Table of $\mathcal{N}=2$ dual pairs in $5D$. First, we write down the mass parameters and twist vectors of the initial models, which have no R-R massless states. Then we write down the mass parameters and the twist vectors of the dual models, which do have R-R massless states. In the last column we write down the number of vector multiplets $n_V$ and hypermultiplets $n_H$, which is of course the same both for the initial model and its dual.}}
        \label{Table of dual pairs with R-R field}
\end{table}

\subsection{Absence of quantum corrections to the vector multiplet moduli space}\label{absence}

In this section, we argue
that there are no quantum corrections to the vector multiplet moduli space of the models we have constructed, using the properties of dual pairs discussed in the previous section. This furthermore leads to predictions about the absence of certain Chern-Simons couplings in the effective action. 

\subsection*{Dual NS-NS pairs with $n_V=2$}

In this subsection we discuss  the possibility of quantum corrections for the two examples presented in sections \ref{z12 with nv=2 and nh=0} and \ref{z6 orbifold with nv=vh=2}. These two examples only differ in the number of hypermultiplets, but it is already known that the hypermultiplet moduli space metric does not receive quantum corrections. Consequently, we  focus on the vector multiplet moduli space metric. For both models, the classical vector multiplet moduli space metric is given by the flat metric (\ref{isom-metrica}) on $\mathbb{R}^2$. In $\mathcal{N}=2, D=5$ supergravity, the vector multiplet moduli space metric is completely characterised by the $d$-symbols, which are real numbers that do not depend on moduli such as the string coupling.
However, as we shall see below, it is in principle possible that the 
$d$-symbols for the quantum theory can jump and are different from those of the classical theory. Such changes to the $d$-symbols can be computed from loop effects that induce transitions in the Chern-Simons couplings, see e.g.\ \cite{witten:1996qb}.

We first notice that the classical values for the $d$-symbols \eqref{d-symbols1} can also be determined from the Scherk-Schwarz reduction from six to five dimensions. Indeed, as was shown in \cite{Hull:2020byc}, the $d$-symbols \eqref{d-symbols1} follow from the reduction of the $6D$ self-dual and anti-self-dual tensors that remain massless under the twist. It was shown explicitly in \cite{Hull:2020byc} that they generate the $5D$ Chern-Simons couplings $d_{ABC}A^A\wedge F^B \wedge F^C$, with $d_{122}=-d_{133}=1$, where $A^{1}$ corresponds to the graviphoton, i.e.\ the Kaluza-Klein vector coming from the metric. Furthermore, classically, $d_{111}=0$, since no Chern-Simons term appears with only the graviphoton. At the quantum level, however, $d_{111}$ can be generated from  one-loop effects by integrating out charged matter, as was also discussed  in \cite{Hull:2020byc} (see e.g.\ \cite{Antoniadis:1995vz,Bonetti:2013cza} for some further literature on loop corrections to Chern-Simons terms). 

The geometry of the scalar manifold could then be corrected by a term proportional to $d_{111}$. We now claim that the only possible quantum correction to the cubic polynomial is
\begin{equation}\label{pertC1}
    C(h)=d_{ABC}h^Ah^Bh^C=3h^1\Big((h^2)^2-(h^3)^2\Big)+d_{111}(h^1)^3\, .
\end{equation}
This
consists of the classical term proportional to $h^1$ and   perturbative corrections proportional to $(h^1)^3$.
Such a term, proportional to $(h^1)^3$, cannot be absorbed into a field redefinition, so it would be a genuine deformation. 
Notice that it furthermore breaks the continuous $\text{SO}(1,1)$ dilatonic scale transformation but not its quantum discrete subgroup $\mathbb{Z}_2$, which acts trivially on the dilaton. 

There is another argument why the $(h^1)^3$ term is the only possible quantum correction. Any other cubic term can not be added, since it is either of the same order in $h^1$, i.e.\ a classical term, or it is forbidden by the quantum symmetry $\mathbb{Z}_2$, which sends $h^{2,3}\to -h^{2,3}$ (see the discussion in the end of section \ref{z12 with nv=2 and nh=0}). In particular, a term quadratic in $h^1$ should be accompanied by linear terms in $h^{2,3}$ but these are forbidden by the aforementioned $\mathbb{Z}_2$. This is consistent with the fact that at one-loop, only Chern-Simons terms with external graviphotons can be generated. It is furthermore also consistent with the fact that there are no $\alpha'$ corrections, as the radius sits in the variables $h^2$ and $h^3$ and any correction in $\alpha'$ must come in the combination $\mathcal{R}/\sqrt{\alpha'}$.

The moduli space metric following from (\ref{pertC1}) is 
found as follows. In order to solve $C(h)=1$ in \eqref{pertC1}, we parametrise $h^2$ and $h^3$ as in \eqref{repsh}, and solve for $h^1$ perturbatively in powers of the string coupling, 
\begin{equation}
\label{changex}
h^1=\frac{1}{3a^2}e^{2\varphi/3}\Big(1-\frac{d_{111}}{27a^6}\,e^{2\varphi}+\cdots\Big)\ .
\end{equation}
Plugging this back into the metric ${\rm d}s^2=-9\,d_{ABC}h^A{\rm d}h^B{\rm d}h^C$ gives the perturbative quantum corrections to the metric. 
With the identification $\phi_5=\varphi$, this gives a power series in $e^{2\phi_5}$.
The resulting metric is then of the form
\begin{equation}\label{isom-metricb}
    {\rm d}s^2=
    \left( \delta_{ab}+  h_{ab} (e^{2\phi_5})\right)
    {\rm d}x^a {\rm d}x^b\ ,
\end{equation}
where $x^a=(\phi_5,\sigma)$. This is the classical metric (\ref{isom-metrica}) plus a 
correction $h_{ab} (e^{2\phi_5})$ comprising the quantum corrections which form a power series  in $e^{2\phi_5}$.
In this case, there are no non-perturbative contributions (we could in principle at this stage consider the possibility of non-perturbative contributions to $h_{ab} (e^{2\phi_5})$; our argument below would then confirm that  there are no non-perturbative contributions).

In section \ref{dual pair I} we constructed the dual model, and we found  that the classical vector multiplet moduli space metric of the dual theory is again  flat. The further T-duality discussed in section \ref{dual pair I} then gives another dual model with a flat moduli space metric (\ref{isom-metric}). On the other hand, the metric for the moduli space of the dual theory can  be obtained from (\ref{isom-metricb}) by
the coordinate transformation (\ref{dictionaryOtildeO}).
The classical limit of the resulting dual metric is obtained by setting $\tilde{\phi}_5=0$ and the components of this metric have non-trivial dependence on
$\tilde{\sigma}$ if $h_{ab}\ne 0$. 
Explicitly, transforming to the tilde coordinates and setting $\tilde{\phi}_5=0$ amounts to transforming $\phi_5$ to $\sqrt 3 \tilde\sigma$. For example, if $h_{ab}$ is independent of $\sigma$,
then the classical metric of the dual moduli space is\footnote{If $h_{ab}$ depends explicitly on $\sigma$, then there would be a similar but more complicated formula.}
\begin{equation}\label{isom-metricdual}
    {\rm d}s^2=
    \left( \delta_{ab}+ 
   \tilde  h_{ab} (e^{2\sqrt 3 \tilde\sigma})
     \right)
    {\rm d}\tilde x^a {\rm d}\tilde x^b\ .
\end{equation}
However,  the classical moduli space metric for the dual theory is (\ref{isom-metric}) and so   these can only agree if $h_{ab}= 0$.
Therefore, we conclude that the classical moduli space is exact, and no quantum corrections arise.

The absence of quantum corrections leads to the prediction that in the quantum theory, $d_{111}$ vanishes. This means that the result of integrating out all $A^1$-charged states in string theory cancels out; in other words, no Chern-Simons terms in the Kaluza-Klein vector are induced by quantum corrections. This is a non-trivial prediction, which would be interesting to be verified by an explicit string theory calculation. Such a calculation could involve integrating out non-perturbative states, as the prediction relies on U-duality. In fact, one can verify that integrating out only low-mass charged states, as in \cite{Hull:2020byc}, does not lead to vanishing Chern-Simons terms in general.

\subsection*{Dual pairs with R-R fields}

The classical values for the $d$-symbols in all the examples discussed in this chapter are of the form
\begin{equation}\label{dsymb}
d_{122}=1\ ,\qquad d_{1ij}=-\delta_{ij}\ ,\qquad i=3,...,n_{V+1}\ ,
\end{equation}
and correspond to factorizable moduli spaces \eqref{VMMod}. They are homogeneous spaces and fall in the classification of \cite{deWit:1991nm,deWit:1992wf}\footnote{\label{basisL(0,P)}The parametrization of the $d$-symbols given in \eqref{dsymb} is the one from \cite{Gunaydin:1983rk}. It is related to the parametrization given in \cite{deWit:1991nm} by a basis transformation. The moduli spaces \eqref{VMMod} are denoted by $L(0,P)$ in \cite{deWit:1991nm,deWit:1992wf} and are a special case of the spaces $L(0,P,\dot P)$ with $\dot P=0$. They are all homogeneous spaces and for $\dot P=0$ become symmetric spaces that furthermore factorise as in \eqref{VMMod}, see equation (4.10) in \cite{deWit:1991nm}.}. The isometry group $\text{SO}(n_V-1,1)\subset \mathcal{K}$ acts linearly on $h^I=(h^i,h^2)$. In the quantum theory, this group is contained in the U-duality group and should be taken over the integers. For an even number of vector multiplets, it contains the element $-1_{n_V\times n_V}$ which sends $h^I\to - h^I$. The vector in the direction of $h^1$ corresponds to the graviphoton, and classically, $d_{111}=0$ (recall that no Chern-Simons term appears with only the graviphoton). As in the previous examples, at the quantum level  $d_{111}$ could be generated from a one-loop induced Chern-Simons term.
The metric on the moduli space could then be corrected with  terms proportional to $d_{111}$. The only possible quantum correction to the cubic polynomial is
\begin{equation}
    C(h)=d_{ABC}h^Ah^Bh^C=3h^1Q(h^I)+d_{111}(h^1)^3\ .
\end{equation}
No other cubic term can   be added, since it would be either of the same order in $h^1$, i.e.\ the classical term, or it would be forbidden by the quantum symmetry under which   $h^{I}\to -h^{I}$. For this argument to work, it is important that there is only an even number of vector multiplets, otherwise $-1_{n_V\times n_V}$ is not an element of SO($n_V-1,1$). In all our models $n_V$ turns out to be even.

The next step is to argue that  
$d_{111}$ actually vanishes. This means that the result of integrating out all $A^1$-charged states in string theory is zero due to cancellations. The argument for the absence of quantum corrections uses again the dual pairs presented in section \ref{dual models with R-R fields}. The dual theories of the $n_V=4,6$ models are presented in the top two lines of \autoref{Table of dual pairs with R-R field}.
Duality enables us to determine the moduli space of one theory at strong coupling by going to weak coupling and large radius in a dual theory, as was explained in section \ref{dual pairs}. Classically, the vector multiplet moduli spaces for theories with $n_V>1$  are of the form \eqref{VMMod}. For theories without R-R scalars we have explicitly derived this result. For the dual theories, which do have R-R scalars, we have not shown this explicitly. However, as we discussed in section \ref{OSS}, the classical moduli space must be a subspace of $\text{E}_{6(6)}/\left[\text{Sp}(4)/\mathbb{Z}_2\right]$. This is again a homogeneous space and, combined with the constraints from $\mathcal{N}=2$ supergravity in $5D$ for $n_V=2,4,6$,  fixes classically the moduli space to be of the form \eqref{VMMod}\footnote{For higher (even) numbers of vector multiplets, there exist the special magic supergravities with moduli spaces which do not fit in \eqref{VMMod}. For example, the moduli spaces $\text{SL}(3,\mathbb{C})/\text{SU}(3)$ and $\text{SU}^*(6)/\text{USp}(6)$ can be obtained for $n_V=8$ and 14, respectively. These, however, always have massless R-R fields and are self-dual (in the sense of duality presented in section \ref{dual pairs}), but are worthwhile studying too.}.

Now, start with a theory $\mathcal{O}_{\text{IIB}}$ with no R-R scalars with classical vector multiplet moduli space $\mathcal{M}_V$, as given in \eqref{VMMod} and equipped with the canonical coset metric with the isometries inherited from $\mathcal{M}_V$ as a homogeneous space. The dual theory $\mathcal{O}_{\text{IIB}}'$ (with R-R scalars) has the same number of vector multiplets and by the arguments given above, it has the same classical vector multiplet moduli space and coset metric.  The same holds for the T-dual theory $\widetilde{\mathcal{O}}_{\text{IIA}}$. In order to examine if the moduli space metric on $\mathcal{M}_V$ is deformed by quantum corrections at strong coupling, in particular by the presence of a $d_{111}(h^1)^3$ term, we can simply go to the dual theory $\widetilde{\mathcal{O}}_{\text{IIA}}$ at weak coupling in $\widetilde{\lambda}_5$ and large radius $\widetilde{\mathcal{R}}$ and study the moduli space there. But this is a regime where the classical answer can be trusted. Since the classical moduli space of the $\widetilde{\mathcal{O}}_{\text{IIA}}$ theory is again homogeneous, with the same isometries as in the classical $\mathcal{O}_{\text{IIB}}$ theory, the coset metrics must be the same. But since one is the strong coupling version of the other, there cannot be any quantum corrections. 

Another way of saying this is that the real special geometry with the $d_{111}(h^1)^3$ term included does not fall into the classification of homogeneous spaces \cite{deWit:1991nm}, as one can check using (\ref{changex}). This would imply that the quantum moduli space would not be homogeneous.  But the dual theory describing the strong coupling regime has  in fact a homogeneous moduli space. Consequently, $d_{111}$ must be zero\footnote{In a theory with the $d_{111}(h^1)^3$ term included, at strong string coupling, the $d_{111}(h^1)^3$ term would dominate over the terms of order $h^1$. A theory with a cubic form $C(h)$ proportionally to $(h^1)^3$ only, would describe a theory with no vector multiplets, which would then be   supergravity coupled to hypermultiplets, and this is in contradiction with duality.}.

\clearpage
\thispagestyle{empty}

\chapter{Orbifolds and swampland}
\label{chap:swampland}

In this chapter, we will discuss some of the ideas of the swampland program
\cite{vafa2005string} (see \cite{Palti:2019pca,Agmon:2022thq} for a review and lecture notes) in the context of freely acting asymmetric orbifolds. Asymmetric orbifolds constitute an interesting region in the string landscape of non-geometric string compactifications, and their corresponding effective supergravity theories provide highly non-trivial examples in which the various conjectures of the swampland program can be tested.

Here, we will study orbifolds that have target spaces of the form $\mathbb{R}^{1,3}\times  T^2\times T^4$. The only difference with the orbifolds studied in the previous chapters is that we have replaced the circle on which the orbifold acts as a shift, with a two-torus. As we will explain in detail in the next, the replacement of the circle with a two-torus will provide a richer and more interesting structure on the moduli space of the orbifolded theory.

First, in section \ref{torpartition} we present the partition function of the two-torus on which the orbifold acts as a shift. In section \ref{duality groups of orbifolded theory}, we discuss the S and T-duality groups of the  orbifolded string theory, as well as those of the effective supergravity theory. Then, in section \ref{z12 with nv=3 and nh=0} we construct an $STU$-like model preserving 8 supersymmetries in four dimensions by using an asymmetric freely acting $\mathbb{Z}_6$ orbifold, and we analyse the spectrum of lightest states in the untwisted and twisted orbifold sectors. In section \ref{moduli and swampland}, we discuss the classical moduli space of our orbifold model and we carefully determine all special points and lines in the moduli space where generically massive states can become massless. This is essential for investigating if various swampland conjectures are valid; here we will mostly focus on the swampland distance conjecture \cite{Ooguri:2006in}.

\section{Toroidal partition function}
\label{torpartition}

First of all, we recall that the orbifold partition function takes the general form (cf. section \ref{sec:Partition function})
\begin{equation}
    Z(\tau,\bar \tau)=\frac{1}{p}\sum_{k,l=0}^{p-1}{Z}[k,l](\tau,\bar \tau)\ ,
    \label{generic orbi par 4d}
\end{equation}
where $p$ is the orbifold rank, $k$ characterizes the various sectors and $l$ implements the orbifold projection in each sector. For the models that we discuss in this chapter the partition function further factorizes as 
\begin{equation}
    {Z}[k,l]= {Z}_{\mathbb{R}^{1,3}}\,  {Z}_{T^2}[k,l]  {Z}_{T^4}[k,l]  {Z}_F[k,l]\,.
    \label{partition function 4d models}
\end{equation}
Here, ${Z}_{T^4}[k,l]$ is the contribution to the partition function from the compact bosons on $T^4$ and ${Z}_F[k,l]$ is the fermionic contribution to the partition function. These pieces of the partition function are exactly the same with those presented in section \ref{sec:Partition function}. In addition, ${Z}_{\mathbb{R}^{1,3}}$ is the contribution to the partition function from the non-compact bosons, which reads
\begin{equation}
    {Z}_{\mathbb{R}^{1,3}}= \left(\sqrt{\tau_2}\,\eta\,\bar{\eta}\right)^{-2}\,.
\end{equation}
Finally, ${Z}_{T^2}[k,l]$ is the contribution to the partition function from the compact bosons on $T^2$, on which the orbifold acts as a shift. Let us now analyse this piece of the partition function in detail.

We denote the two real $T^2$ coordinates by $Z_1$ and $Z_2$.  Without loss of generality, we define the orbifold action on $T^2$ as
\begin{equation}\label{shiftz1z2}
\begin{aligned}
&Z_1 \;\rightarrow\; Z_1 + 2\pi \mathcal{R}_5 / p \,,\\
&Z_2 \;\rightarrow\; Z_2\,,
\end{aligned}
\end{equation}
where $\mathcal{R}_5$ is the radius of the $S^1$ on which the orbifold acts as a shift. Also, we denote by $\mathcal{R}_4$ the radius of the $S^1$ which is inert under the orbifold action. Furthermore, we impose the following boundary conditions on the $T^2$ coordinates\footnote{The boundary conditions along the non-compact directions and the $T^4$ directions are the same as in \eqref{boundaryconditions}.} 
\begin{equation}
\begin{aligned}
&Z_1(\sigma^1, \sigma^2+2\pi) = Z_1(\sigma^1, \sigma^2) + 2\pi \mathcal{R}_5 \,(w^1 + k/p) \,, \\
&Z_2(\sigma^1, \sigma^2+2\pi) = Z_2(\sigma^1, \sigma^2) + 2\pi \mathcal{R}_4w^2 \,,
\end{aligned}
\end{equation}
where $w^1,w^2\in \mathbb{Z}$ are the winding numbers associated with $T^2$. 

We mention here that the shift along the one circle of $T^2$ \eqref{shiftz1z2} can be represented by a shift vector 
\begin{equation}
    v=  \begin{pmatrix}
  \frac{1}{p}\\
   0 \\
   0\\
   0
    \end{pmatrix}\,.
    \label{shiftvapp}
\end{equation}
This is written in the basis of winding $(w^i)$ and momentum $(n_i)$ numbers. Recall that in this basis, a vector of the lattice $\Gamma^{2,2}$, associated with $T^2$, can be written as
\begin{equation}\label{latticeP}
    P=  \begin{pmatrix}
  w^1\\
   w^2 \\
   n_1\\
   n_2
    \end{pmatrix}\,.
\end{equation}
We denote the (dimensionful) metric and antisymmetric Kalb-Ramond $B$-field on $T^2$ by $g_{ij}$ and $b_{ij}$, respectively. Here $i,j=1,2$, and  $b_{ij}=b\,\varepsilon_{ij}$, where $b$ is a constant, and $\varepsilon_{12}=-\varepsilon_{21}=1$, $\varepsilon_{11}=\varepsilon_{22}=0$. Given these, we can define the complexified K\" ahler modulus $T$ and the complex structure modulus $U$ of $T^2$ as (see e.g. \cite{Blumenhagen:2013fgp})
\begin{equation}
\begin{aligned}
  &T = T_1+i T_2 = \frac{1}{\alpha'}\left(b + i \sqrt{\text{det}g}\right) \,,\\
& U = U_1+iU_2 = \frac{g_{12}}{g_{11}} + i \frac{\sqrt{\text{det}g}}{g_{11}} \,.
 \end{aligned}
 \label{T,Udefap}
\end{equation}
Then, the $T^2$ metric and Kalb-Ramond field can be expressed in terms of these moduli as
\begin{equation}
    g_{ij} = \alpha'\frac{T_2}{U_2}\begin{pmatrix}
        1&U_1\\
        U_1 & |U|^2
    \end{pmatrix}\,,\qquad \text{and} \qquad b_{ij}= \alpha'\begin{pmatrix}
        0& T_1\\
        -T_1&0
    \end{pmatrix}\,.
    \label{metricandb}
\end{equation}
Note that for a rectangular torus, that is $U_1=0$, and a trivial $B$-field, i.e. $T_1=0$, $T_2=\mathcal{R}_5\mathcal{R}_4/\alpha'$ and $U_2=\mathcal{R}_4/\mathcal{R}_5$. 

The $T^2$ partition function, in the untwisted orbifold sector and without the insertion of the orbifold group element, reads
\begin{equation}
{Z}_{T^2}[0,0]=  \frac{1}{(\eta\bar{\eta})^2}\sum_{\left\{n_i,w_i\right\} \in \mathbb{Z}^4}\, \bar{q}^{\frac{1}{2}p_{L}^2}\,q^{\frac{1}{2}p_{R}^2}\,, \qquad i=1,2\,,
\end{equation}
where\footnote{Here and below we label both momentum and winding numbers by a subscript for simplicity of the formulae.} 
\begin{equation}
    \begin{aligned}
        &p^2_{L} = \frac{1}{2T_2U_2}|n_2-Un_1 +\bar{T}w_1 +\bar{T}Uw_2|^2\,,\\
         &p^2_{R} = \frac{1}{2T_2U_2}|n_2-Un_1 +{T}w_1 +{T}Uw_2|^2\,.
    \end{aligned}
    \label{momentauntwisted}
\end{equation}
Now, the shift vector $v$ in \eqref{shiftvapp} induces a shift in the winding number $w_1$, that is $w_1\to w_1+k/p$. Moreover, states carrying momentum number $n_1$ will pick up a phase $e^{{2\pi i ln_1 }/{p}}$ under the orbifold action. Consequently, the $T^2$ partition function in a $k$-sector, with the insertion of the orbifold group element, characterized by $l$, will be given by\footnote{A general discussion on similar lattice sums can be found in e.g. \cite{Kiritsis:1997ca}.} 
\begin{equation}
    {Z}_{T^2}[k,l]=  \frac{1}{(\eta\bar{\eta})^2}\sum_{\left\{n_i,w_i\right\} \in \mathbb{Z}^4}\,e^{\frac{2\pi i ln_1 }{p}} \,\bar{q}^{\frac{1}{2}p_{L}^2(k)}\,q^{\frac{1}{2}p_{R}^2(k)}\,, \qquad i=1,2\,,
    \label{shiftedt2}
\end{equation}
where
\begin{equation}
    \begin{aligned}
        &p^2_{L}(k) = \frac{1}{2T_2U_2}|n_2-Un_1 +\bar{T}\left(w_1+\tfrac{k}{p}\right) +\bar{T}Uw_2|^2\,,\\
         &p^2_{R}(k) = \frac{1}{2T_2U_2}|n_2-Un_1 +{T}\left(w_1+\tfrac{k}{p}\right) +{T}Uw_2|^2\,.
    \end{aligned}
    \label{shifted momenta}
\end{equation}
We can rewrite the left and right-moving momenta in terms of the background fields of $T^2$, i.e. the metric $g_{ij}$ and antisymmetric tensor $b_{ij}$, as
\begin{equation}
\begin{aligned}
   & p_{L}^2(k) = \frac{\alpha'}{2}n_ig^{-1}_{ij}n_j+\frac{1}{2\alpha'}\hat{w}_i(g-bg^{-1}b)_{ij}\hat{w}_j+\hat{w}_i(bg^{-1})_{ij}n_j + n_i\hat{w}_i \,,\\
   & p_{R}^2(k) = \frac{\alpha'}{2}n_ig^{-1}_{ij}n_j+\frac{1}{2\alpha'}\hat{w}_i(g-bg^{-1}b)_{ij}\hat{w}_j+\hat{w}_i(bg^{-1})_{ij}n_j - n_i\hat{w}_i \,,
    \end{aligned}
    \label{momentabgtwistedapp}
\end{equation}
where, $i,j=1,2$ and summation over repeated indices is implied. Also, we have defined $\hat{w}_i \equiv w_i+k_i/p$, where $\vec{k}=(k_1,k_2)=(k,0)$. Also, let us define $\vec{l}=(l_1,l_2)=(l,0)$. Then, by performing a Poisson resummation (cf. \eqref{Poissonresum}) over the momentum vector $\vec{n}=(n_1,n_2)$ we can bring \eqref{shiftedt2} to the equivalent form
\begin{equation}
    {{Z}}_{T^2}[k,l]=\frac{\sqrt{\det g}}{\alpha'\, {\tau_2}\,(\eta\,\bar{\eta})^2}\sum_{\left\{n_i,w_i\right\} \in \mathbb{Z}^4}e^{ \frac{-\pi}{\alpha'\tau_2}\left[n_i-\frac{l_i}{p}+\left(w_i+\frac{k_i}{p}\right)\tau\right]\left(g_{ij}-b_{ij}\right)\left[n_j-\frac{l_j}{p}+\left(w_j+\frac{k_j}{p}\right)\bar{\tau}\right]}\,. 
    \label{shift2gb}
\end{equation}

\section{Duality groups}
\label{duality groups of orbifolded theory}

The compactification of type II string theory on $T^6$ has $\mathcal{N}=8$ supergravity as its low-energy effective field theory. This has  70 massless scalars that parametrise the moduli space
\begin{equation}
    {\cal M}=\frac{\text{E}_{7(7)}}{\text{SU}(8)/\mathbb{Z}_2}\ .
\end{equation}
The 38 scalars from the NS-NS sector parametrise a subspace, which factorises as
\begin{equation}
    {\cal M}_{\text{NS}}=\frac{\text{SL}(2)}{\text{U}(1)}\times \frac{\text{Spin}(6,6)}{\text{Spin}(6)\times \text{Spin}(6)}\ .
\end{equation}
In type IIB, the first factor is parametrised by the axion and a shifted dilaton (the axion is a scalar dual to the $4D$ NS-NS two-form) and the second factor is parameterised by the 36 moduli for the metric and $B$-field  on $T^6$.

The duality group of the effective supergravity theory is $\text{E}_{7(7)}$, which has a maximal subgroup
\begin{equation}
    \text{E}_{7(7)}\supset \text{SL}(2)\times \text{Spin}(6,6)\ .
\end{equation}
In the quantum theory, the  duality group $\text{E}_{7(7)}$ is broken to its discrete U-duality subgroup $\text{E}_{7(7)}(\mathbb{Z})$, so that the 
$\text{SL}(2)\times \text{Spin}(6,6)$  subgroup is broken to a discrete group 
$\text{SL}(2;\mathbb{Z})\times \text{Spin}(6,6;\mathbb{Z})$, forming the S and T-duality groups of type II string theory on $T^6$ \cite{Hull:1994ys}. We mention that this $4D$ S-duality, which will be referred to here as $\text{SL}(2)_S$,  should not be confused with the $10D$ S-duality of type IIB.
Note that the {\bf 56} representation of $\text{E}_{7(7)}$ decomposes into 
$\text{SL}(2)\times \text{Spin}(6,6)$ representations as follows:
\begin{equation}
\textbf{56}\to (\textbf{2},\textbf{12})+(\textbf{1},\textbf{32}).
\label{ 56dec}
\end{equation}
The NS-NS charges are in the $(\textbf{2},\textbf{12})$ representation, consisting of the 6 momenta and 6 winding numbers, which are the perturbative charges, together with 6 NS5-brane charges and 6 KK monopole charges, which are the non-perturbative charges \cite{Hull:1994ys}.

Orbifolding this theory  breaks the U-duality group $\text{E}_{7(7)}(\mathbb{Z})$ to a subgroup. The orbifold is specified by a twist and a shift, and the  U-duality group is broken to the subgroup that preserves both of these, which we will refer to as the orbifold duality group. This is then the subgroup of $\text{E}_{7(7)}(\mathbb{Z})$ that commutes with the twist and preserves the shift vector (up to the addition of a lattice vector and up to a sign; see below).

For the theories we consider in this chapter, with decomposition $T^6=T^4\times T^2$, the orbifold duality group is in fact a subgroup of $\text{SL}(2;\mathbb{Z})\times \text{Spin}(6,6;\mathbb{Z})$.
The subgroup of $\text{Spin}(6,6) $ acting as T-duality on $T^4$ is $\text{Spin}(4,4) $ and, for each orbifold, the twist matrix is in the compact part of this, 
$\text{Spin}(4)\times \text{Spin}(4) $. 
Let $\mathcal{C}$ be the subgroup of
${\rm Spin}(4,4)$ that commutes with the twist. 
For the theories considered here, the twist then breaks
${\rm Spin}(6,6)$ to ${\rm Spin}(2,2)\times \mathcal{C}$ in the supergravity theory, where ${\rm Spin}(2,2)$ is the T-duality group of $T^2$. In the orbifolded string theory, the group $\mathcal{C}$ is broken further to a discrete subgroup $\mathcal{C}(\mathbb{Z})$.

The shift vector \eqref{shiftvapp} leads to a further breaking of the symmetry, as it is preserved only by a subgroup of the T-duality group of $T^2$. Note that, similarly with our arguments in chapter \ref{chap:fao} (see discussion around \eqref{pvector1,1}), adding any integer-valued 4-vector 
$P$ in the lattice $ \mathbb{Z}^4$
to $v$  will give the same orbifold.
Furthermore, $v$ and $-v$ specify physically equivalent orbifolds as the sign of $v$ is changed by the reflection $Z_1\to -Z_1$; such a reflection is an element of the T-duality group $\text{Spin}(6,6,\mathbb{Z})$.

As we have explained in chapter \ref{chap:mod5}, this discussion of duality symmetries applies only to the untwisted orbifold sector, as there can be new duality symmetries relating the untwisted and twisted sectors that do not directly arise from the duality symmetry of the theory before the orbifold. In the next, we will see that at special lines or points in the moduli space our orbifolds exhibit such type of duality enhancement.

Before turning to the subgroup of the T-duality group of $T^2$ that is preserved in the orbifold, 
we discuss an $\text{SL}(2)$ analogue that introduces the relevant groups.
Consider  the simpler problem of finding
the subgroup of $\text{SL}(2;\mathbb{Z})$
that preserves a 2-vector of the form
\begin{equation}
    V=  \begin{pmatrix}
  \frac{1}{p}\\
   0
    \end{pmatrix}
    \label{shiftvvapp}
\end{equation}
up to the addition of a lattice vector $\begin{pmatrix}
  w\\
   n
    \end{pmatrix} \in \mathbb{Z}^2$. It is easy to check that the result is the group
\begin{equation}
  \Gamma_1(p) = \Bigg\{ \begin{pmatrix}
       a &b\\
        c& d
    \end{pmatrix}\in  \text{SL}(2;\mathbb{Z}):\quad a,d=1 \:\text{mod} \:\,p\,,\quad c=0\: \text{mod} \:\, p\Bigg\}\,.
    \label{gamma1}
\end{equation}
The subgroup preserving $V$ up to a sign, i.e.\ taking $V$ to $\pm V$ plus a lattice vector, is 
 \begin{equation}
  \hat \Gamma_1(p) = \Bigg\{ \begin{pmatrix}
       a &b\\
        c& d
    \end{pmatrix}\in  \text{SL}(2;\mathbb{Z}):\quad a,d=\pm 1 \:\text{mod} \:\,p\,,\quad c=0\: \text{mod} \:\, p\Bigg\}\,.
    \label{gamma1h}
\end{equation}
For $p=2$, we have that $-1=1$ mod $2$ and so 
\begin{equation}
    \hat \Gamma_1(2) =  \Gamma_1(2)\, ,
\end{equation}
 while for $p>2$ 
\begin{equation}
    \hat \Gamma_1(p) =  \Gamma_1(p) \times \mathbb{Z}_2\, ,
\end{equation}
 with the $\mathbb{Z}_2$ consisting of the $2\times 2 $ matrices 
 $\{ {1},-{1} \}$. For $\text{SL}(2;\mathbb{R})$, the subgroup preserving $V$ up to a lattice vector is 
 \begin{equation}
  \Delta_1(p) = \Bigg\{ \begin{pmatrix}
       a &b\\
        c& d
    \end{pmatrix}\in  \text{SL}(2;\mathbb{R}):\quad a=1 \:\text{mod} \:\,p\,,\quad c=0\: \text{mod} \:\, p\Bigg\}\, ,
    \label{Del1}
\end{equation}
while the subgroup taking $V$ to $\pm V$ plus a lattice vector is  
 \begin{equation}
 \hat  \Delta_1(p) = \Bigg\{ \begin{pmatrix}
       a &b\\
        c& d
    \end{pmatrix}\in  \text{SL}(2;\mathbb{R}):\quad a=\pm 1 \:\text{mod} \:\,p\,,\quad c=0\: \text{mod} \:\, p\Bigg\}\, .
    \label{Delh1}
\end{equation}
Both of these contain the stability subgroup preserving  $V$, which is the group $\mathbb{R}$ of upper triangular matrices
\begin{equation}
  \begin{pmatrix}
       1 &b\\
        0& 1
    \end{pmatrix}\,.
    \label{uptri}
\end{equation}
We now return to the breaking of the T-duality group of $T^2$, which is 
\begin{equation}
    {\rm Spin}(2,2;\mathbb{Z})\cong\text{SL}(2;\mathbb{Z})_T\times \text{SL}(2;\mathbb{Z})_U\,.
\end{equation}
This group acts on the moduli $T$ and $U$ as follows
\begin{equation}
    \text{SL}(2;\mathbb{Z})_T: \quad\frac{aT+b}{cT+d}  \,,\qquad ad-bc=1\ ,
\end{equation}
\begin{equation}
   \quad \:\, \text{SL}(2;\mathbb{Z})_U: \quad\frac{a'U+b'}{c'U+d'}    \,,\quad \:\:a'd'-b'c'=1\,.
\end{equation}
The T-duality group also acts on the momentum and winding numbers by $\text{Spin}(2,2;\mathbb{Z})$ transformations, leaving $p^2_{{L/R}}$ invariant. The action on the Narain lattice is given by (see e.g. \cite{Bailin:1993wv})
\begin{equation}
     \text{SL}(2;\mathbb{Z})_T: \quad   \begin{pmatrix}
  w_1\\
   w_2 \\
   n_1\\
   n_2
    \end{pmatrix}\to  \begin{pmatrix}
  d&0&0&-c\\
   0&d&c&0 \\
   0&b&a&0\\
   -b&0&0&a
    \end{pmatrix}  \begin{pmatrix}
  w_1\\
   w_2 \\
   n_1\\
   n_2
   \end{pmatrix}\,,
   \label{sl2t}
\end{equation}
\begin{equation}
 \quad   \: \text{SL}(2;\mathbb{Z})_U: \quad   \begin{pmatrix}
  w_1\\
   w_2 \\
   n_1\\
   n_2
    \end{pmatrix}\to  \begin{pmatrix}
  a'&-b'&0&0\\
   -c'&d'&0&0 \\
   0&0&d'&c'\\
   0&0&b'&a'
    \end{pmatrix}  \begin{pmatrix}
  w_1\\
   w_2 \\
   n_1\\
   n_2
   \end{pmatrix}\,,
   \label{sl2u}
\end{equation}
Notice that $\text{SL}(2;\mathbb{Z})_U$ does not mix winding and momenta, but $\text{SL}(2;\mathbb{Z})_T$ does. It is easy to check that the two $\text{SL}(2)$'s commute in the basis of winding and momentum numbers. Also, an element $(g_T,g_U)\in\text{SL}(2;\mathbb{Z})_T \times \text{SL}(2;\mathbb{Z})_U$ can be embedded in $\text{SO}^+(2,2;\mathbb{Z})$ as\footnote{
$\text{SO}^+(2,2;\mathbb{Z})$ is the component of $\text{O}(2,2;\mathbb{Z})$ connected to the identity. Also, $\text{SL}(2;\mathbb{Z}) \times \text{SL}(2;\mathbb{Z}) \cong \text{Spin}(2,2;\mathbb{Z})$ is the double cover of $\text{SO}^+(2,2;\mathbb{Z})$.}
\begin{equation}
(g_T,g_U) \in\text{SL}(2;\mathbb{Z})_T \times \text{SL}(2;\mathbb{Z})_U  \to
     \begin{pmatrix}
        da'&-db'&-cb'&-ca'\\
        -dc'&dd'&cd'&cc'\\
        -bc'&bd'&ad'&ac'\\
        -ba'&bb'&ab'&aa'
    \end{pmatrix}  \in  \text{SO}^+(2,2;\mathbb{Z})\,.
\end{equation}
Consider now the shift vector $v$ given in \eqref{shiftvapp}. Generic $\text{Spin}(2,2;\mathbb{Z})$ T-duality transformations will  transform the shift vector $v$ to another shift vector $v'$, which will not leave the partition function invariant. This means that if we start with a model with shift vector $v$, after a T-duality transformation we will end up with an inequivalent model with shift vector $v'$. Hence, generic T-duality transformations will not be in the orbifold duality group.

However, as in the $\mathrm{SL}(2)$ example discussed before, there exists a subgroup of the T-duality group, $\text{Spin}(2,2;\mathbb{Z})$, that acts on the shift vector in such a way that the partition function remains invariant. This subgroup will be in the orbifold duality group. Firstly, a shift, $v\to v+P$, where $P$ is a lattice vector, corresponds to a shift  ${1}/{p}\to {1}/{p}+\mathbb{Z}$ in the phase factor in \eqref{shiftedt2}, and a shift of the momentum and winding numbers in \eqref{shifted momenta} by integers. Therefore, this is a symmetry of the partition function. In addition, a reflection, $v\to -v$ simply amounts to $l\to -l$ and $k\to -k$ in \eqref{shiftedt2} and \eqref{shifted momenta}, which is also a symmetry of the partition function. So, in order to determine the orbifold duality group associated with $T^2$, we need to find the subgroup of $\text{Spin}(2,2;\mathbb{Z})$ that generates these two symmetries (for a similar discussion see also \cite{Gregori:1997hi}, appendix C).

Let us first focus on the transformations that preserve the shift vector $v$ up to lattice translations. By combining \eqref{shiftvapp} with \eqref{sl2t} and \eqref{sl2u}, we find 
that the duality group
$\text{SL}(2;\mathbb{Z})_T$ is broken to $ \Gamma^1(p)_T$
and
$\text{SL}(2;\mathbb{Z})_U$ is broken to $ \Gamma_1(p)_U $, where
\begin{equation}
  \Gamma^1(p)_T = \Bigg\{ \begin{pmatrix}
       a &b\\
        c& d
    \end{pmatrix}\in  \text{SL}(2;\mathbb{Z})_T: a,d=1 \:\text{mod} \:\,p\,,\quad b=0\: \text{mod} \:\, p\Bigg\}\,,
    \label{gamma1t}
\end{equation}
\begin{equation}
 \qquad\qquad\: \Gamma_1(p)_U = \Bigg\{ \begin{pmatrix}
       a' &b'\\
        c'& d'
    \end{pmatrix}\in  \text{SL}(2;\mathbb{Z})_U: a',d'=1 \:\text{mod} \:\,p\,,\quad c'=0\: \text{mod} \:\, p\Bigg\}\,.
    \label{gamma1u}
\end{equation}
The groups $\Gamma^1(p)$ and $\Gamma_1(p)$ are isomorphic (the isomorphism is simply given by transposition) but are embedded differently in $\text{SL}(2,\mathbb{Z})$.
The appearance of $\Gamma^1(p)_T$ instead of $\Gamma_1(p)_T$ is due to the embedding of $\text{SL}(2,\mathbb{Z})_T$ in $\text{Spin}(2,2,\mathbb{Z})$. It is important to stress that the surviving duality groups depend on the particular shift vector. For instance, a shift along the radius $\mathcal{R}_4$ that shifts $w_2$ instead of $w_1$, would lead to two identical $\Gamma^1(p)_T \times \Gamma^1(p)_U$
subgroups (see e.g. \cite{sen1995dual, ITOYAMA2022115667} or \cite{Gregori:1997hi} for examples).

Regarding the reflection, $v\to -v$, we can see that it can be realized by the values $a=d=-1,b=c=0$ in \eqref{sl2t}, and $a'=d'=-1, b'=c'=0$ in \eqref{sl2u}, which generate another two $\mathbb{Z}_2$ subgroups for $p>2$. Concluding, we have found that the T-duality group of $T^2$ is broken, due to the presence of the shift vector, to
\begin{equation}
\hat \Gamma^1(p)_T\times \hat \Gamma_1(p)_U \subset \text{SL}(2;\mathbb{Z})_T\times \text{SL}(2;\mathbb{Z})_U\,.
\end{equation}
Now, by combining all the above, we conclude that the T-duality group $\text{Spin}(6,6;\mathbb{Z})$ of $T^6$ is broken to the orbifold T-duality group
 \begin{equation}
\hat \Gamma^1(p)_T\times \hat \Gamma_1(p)_U \times
\mathcal{C}(\mathbb{Z})
\subset \text{SL}(2;\mathbb{Z})_T\times \text{SL}(2;\mathbb{Z})_U
\times \text{Spin}(4,4;\mathbb{Z}) \subset
\text{Spin}(6,6;\mathbb{Z}) 
  \label{Tsubg}
\end{equation}
For the supergravity theory, the shift breaks $\text{SL}(2;\mathbb{R})_T\times \text{SL}(2;\mathbb{R})_U$ to the subgroup of matrices
\begin{equation}
\begin{aligned}
\Bigg\{ \begin{pmatrix}
       a &b\\
        c& d
    \end{pmatrix}\in & \text{SL}(2;\mathbb{R})_T, \quad
   \begin{pmatrix}
       a' &b'\\
        c'& d'
    \end{pmatrix}\in  \text{SL}(2;\mathbb{R})_U : \\
    & \qquad
    da'=\pm 1 \:\text{mod} \:\,p\,,\quad 
b=0\: \text{mod} \:\, p, \quad c'=0\: \text{mod} \:\, p,
\Bigg\}\,,
\end{aligned}
    \label{Congr}
\end{equation}
which is the group
\begin{equation}
    (\mathbb{R}\times
\mathbb{Z}_2) \ltimes
\left(\hat  \Delta^1(p)\times  \hat  \Delta_1(p)\right) \,,
\end{equation}
where $\hat{\Delta}^1(p)$ is defined by
\begin{equation}
 \hat  \Delta^1(p) = \Bigg\{ \begin{pmatrix}
       a &b\\
        c& d
    \end{pmatrix}\in  \text{SL}(2;\mathbb{R}):\quad d=\pm 1 \:\text{mod} \:\,p\,,\quad b=0\: \text{mod} \:\, p\Bigg\}\, .
\end{equation}
Then in the supergravity theory, $\text{Spin}(6,6) $ is broken to the product of this group with $\mathcal{C}$.

We now turn to the breaking of the S-duality group $\text{SL}(2)_S$.
Under $\text{SL}(2)_S$,  the perturbative 4-vector $v$ transforms into a non-perturbative 4-vector of NS5-brane charges and KK monopole charges.
 Then, the shift vector transforms as the {\bf 2} representation under $\text{SL}(2)_S$, so that it can be regarded as part of an 8-vector  transforming as the $(\textbf{4},\textbf{2})$ representation of ${\rm Spin}(2,2)\times \text{SL}(2)_S$. However, this ${\rm Spin}(2,2)$ factorises as ${\rm Spin}(2,2)=\text{SL}(2)_T\times \text{SL}(2)_U$, so that  ${\rm Spin}(2,2)\times \text{SL}(2)_S=\text{SL}(2)_T\times \text{SL}(2)_U\times \text{SL}(2)_S$
 and under this the shift vector transforms as a $(\textbf{2},\textbf{2},\textbf{2})$. 
 The non-zero component of (\ref{shiftvapp}) is then part of an $\text{SL}(2)_S$ doublet of the form (\ref{shiftvvapp}).  From the earlier discussion, the subgroup of
 $\text{SL}(2;\mathbb{Z})_S$ preserving this up to a sign and up to a lattice vector is $\hat \Gamma^1(p)_S$, so that the S-duality is broken to this subgroup.
Then
$\mathrm{SL}(2;\mathbb{Z})_T\times \mathrm{SL}(2;\mathbb{Z})_U\times \mathrm{SL}(2;\mathbb{Z})_S$ 
is broken to the subgroup $\hat \Gamma^1(p)_T \times
\hat \Gamma_1(p)_U\times
\hat \Gamma^1(p)_S$ and we have the final result that the U-duality group is broken to 
\begin{equation}\label{Uduality}
    \mathcal{K}(\mathbb{Z}) = \hat \Gamma^1(p)_T \times
\hat \Gamma_1(p)_U\times
\hat \Gamma^1(p)_S\times\mathcal{ C}(\mathbb{Z})\,.
\end{equation}
For the supergravity limit, the subgroup of $\mathrm{SL}(2)_S$ preserving the shift up to a sign and a lattice vector is $ \hat  \Delta^1(p)$. However, the subgroup of 
$\mathrm{SL}(2)_T\times \mathrm{SL}(2)_U\times \mathrm{SL}(2)_S$ preserving the shift is slightly larger than the product of the T-duality group given above.
 It is the group of matrices
 \begin{equation}
 \begin{aligned}
\Bigg\{ \begin{pmatrix}
       a &b\\
        c& d
    \end{pmatrix}\in  &\text{SL}(2;\mathbb{R})_T, ~  
    \begin{pmatrix}
       a' &b'\\
        c'& d'
    \end{pmatrix}\in  \text{SL}(2;\mathbb{R})_U 
    , ~
    \begin{pmatrix}
       a'' &b''\\
        c''& d''
    \end{pmatrix}\in  \text{SL}(2;\mathbb{R})_S
    :  \\
  & \qquad da'd''=\pm 1 \:\text{mod} \:\,p\,,~ 
b=0\: \text{mod} \:\, p, ~ c'=0\: \text{mod} \:\, p,~b''=0\: \text{mod} \:\, p
\Bigg\}\,.
\end{aligned}
    \label{Congrs}
\end{equation}
The supergravity duality group is then the product of this with $\mathcal{C}$.

\section{An $STU$-like  $\mathbb{Z}_{6}$ model}
\label{z12 with nv=3 and nh=0}
In this section we discuss an $\mathcal{N}=2, D=4$ model, with three vector multiplets and two hypermultiplets arising in the massless untwisted orbifold sector. This model can be obtained from a circle reduction of an $\mathcal{N}=2, D=5$ freely acting $\mathbb{Z}_{6}$ orbifold model of IIB string theory on $T^5$ studied in chapter \ref{chap:mod5}; see the first example of section \ref{z6 orbifold with nv=vh=2}.

The mass parameters and twist vectors for this model are given by
\begin{equation}
    m_1={\pi}\ , \qquad m_2= \frac{\pi}{3}\ ,\qquad m_3=\frac{2\pi}{3} \ ,\qquad m_4=0\ .
\end{equation}
\begin{equation}
    \tilde{u}=\left(\frac{5}{6},\frac{1}{6}\right)\ ,\qquad u=\left(\frac{1}{6},\frac{1}{6}\right)\,.
    \label{twist vectors for primary example2}
\end{equation}
Here, we choose the lattice $\Gamma^{4,4}(A_2\oplus A_2)\oplus \Gamma^{2,2}$.  Notice that only $\Gamma^{2,2}$ is left invariant under the twist. Also, the conditions for modular invariance can be verified similarly with the $5D$ example of section \ref{z6 orbifold with nv=vh=2}. 

In the following, we will discuss the massless spectrum of our model in the untwisted sector, and the spectrum of the lightest states in the twisted sectors. Due to the orbifold shift, states in the twisted sectors are generically massive. However, as we will see in section \ref{moduli and swampland}, there exist special lines in the interior of the moduli space in which a finite number of massive twisted states become massless. Also, there exist points at infinite distance, where infinite towers of states become massless.  

\subsection{Untwisted sector}

The massless spectrum in the untwisted sector of our $\mathbb{Z}_6$ model can be obtained by circle reduction of the five-dimensional model, with the same mass parameters and twist vectors, discussed in section \ref{z6 orbifold with nv=vh=2}. The untwisted massless spectrum of the five-dimensional model consists of the $\mathcal{N}=2$ gravity multiplet, two vector multiplets, and two hypermultiplets, and all the fields come from the NS-NS and NS-R sectors. Upon reduction on a circle, we simply get one additional vector multiplet. So, the massless untwisted orbifold spectrum consists of the  $\mathcal{N}=2$, $D=4$ gravity multiplet coupled to three vector multiplets and two hypermultiplets.

Note that the scalars in the three vector multiplets are the complex $S,\,T$ and $U$ moduli, where $T$ and $U$ are the $T^2$ moduli, defined in \eqref{T,Udefap}, and $S$ is defined by
\begin{equation}
    S = a + i e^{-2\phi_4}\,.
\end{equation}
Here, $a$ is a scalar that is dual to the NS-NS $B$-field in four dimensions, and is usually referred to as the axion, and $\phi_4$ is a scalar parametrising the four-dimensional string coupling by $\lambda_4=\langle e^{\phi_4}\rangle$. There are also 4 complex scalars in the two hypermultiplets.

The spectrum of the $4D$ model can be also obtained using the general procedure presented in chapter \ref{chap:spectrum}, but with two modifications. In particular, the $S^1$ momenta $\frac{\alpha'}{2}P^2_{L/R}(k)$ in \eqref{untwisted masses} and \eqref{twisted masses} should be replaced by the $T^2$ momenta given in \eqref{momentauntwisted} and \eqref{shifted momenta}, respectively. Also, in $4D$ string states can be classified based on their helicity. The helicity of a state characterised by a generic $\text{SO}(8)$ weight vector, $(\tilde{r}_1,\tilde{r}_2,\tilde{r}_3,\tilde{r}_4)\times (r_1,r_2,r_3,r_4)$, is equal to $\tilde{r}_1-r_1$ \cite{font2005introduction}. For completeness, we list in \autoref{untwisted stu states} the NS and R-sector $\text{SO}(8)$ weight vectors for the lightest left and right-moving states that survive the GSO projection in the untwisted sector; all of these are massless in the absence of momentum and/or winding modes. 
\renewcommand{\arraystretch}{2}
\begin{table}[h!]
\centering
 \begin{tabular}{|c|c|}
    \hline
    Sector & SO(8) weight  \\
    \hline
    \hline
  NS & $(\pm1,0,0,0)$,\: $(0,\underline{0,0,\pm1})$  \\
  \hline
   \multirow{2}{*}{R}  & $\pm(\frac{1}{2},\frac{1}{2},\frac{1}{2},\frac{1}{2})$,\: $\pm(-\frac{1}{2},-\frac{1}{2},\frac{1}{2},\frac{1}{2})$ \\
   \cline{2-2}
   & $\pm(\frac{1}{2},-\frac{1}{2},\frac{1}{2},-\frac{1}{2})$,\: $\pm(-\frac{1}{2},\frac{1}{2},\frac{1}{2},-\frac{1}{2})$ \\
   \hline
    \end{tabular}
\captionsetup{width=.9\linewidth}
\caption{\textit{Here we list the weight vectors of the lightest left and right-moving states in the untwisted sector. Underlying denotes permutation; e.g. $(0,\underline{0,0,1})$ denotes the states $(0,1,0,0)$, $(0,0,1,0)$ and $(0,0,0,1)$.}}
\label{untwisted stu states}
\end{table}
\renewcommand{\arraystretch}{1}

Finally, we mention here that all the aforementioned states in the NS-NS and NS-R sectors come with infinite towers of momentum and winding modes along $T^2$, and their masses are given by (cf. \eqref{momentauntwisted} and \eqref{untwisted masses})
\begin{equation}
\begin{aligned}
    &\frac{\alpha'm^2_{L}(0)}{2}=\frac{1}{2}p^2_{L}(0)=\frac{1}{4T_2U_2}|n_2-Un_1 +\bar{T}w_1 +\bar{T}Uw_2|^2\,,\\
    &\frac{\alpha'm^2_{R}(0)}{2}=\frac{1}{2}p^2_{R}(0)=\frac{1}{4T_2U_2}|n_2-Un_1 +{T}w_1 +{T}Uw_2|^2\,.
    \label{toweruntwisted}
\end{aligned}
 \end{equation}
Note that in these formulae the momentum number $n_1$ should obey $n_1=0$ mod $6$, such that the states are orbifold invariant\footnote{Recall that massless states in the untwisted sector must carry a trivial orbifold charge.}. Also, physical states should satisfy the level-matching condition, $m^2_{L}=m^2_{R}$, which in this case reads
\begin{equation}
    n_1w_1+n_2w_2=0\,.
\end{equation}

\subsection{Twisted sectors}
Here we focus on the spectrum of the orbifold twisted sectors. For the construction of states we will follow the techniques developed in chapter \ref{chap:spectrum} and, in particular, section \ref{closed string spectrum}. We recall here that the spectrum in an orbifold $k$-twisted sector is identical with the spectrum of the $(p-k)$-twisted sector.

\subsubsection{$k=1$ sector}
Let us now work out the spectrum of lightest states in the $k=1$ sector of our orbifold model, which is the same with the spectrum in the $k=5$ sector. First, we compute the shift in the zero point energies (cf. \eqref{twist vectors for primary example2}, \eqref{shifted energy})
\begin{equation}
    \tilde{E}_1 =E_1=\frac{5}{36}\,,
\end{equation}
and the degeneracy of a state characterised by generic weight vectors $\tilde{r}$ and $r$ and without bosonic oscillator excitations (cf. \eqref{twist vectors for primary example2}, \eqref{fixed points}, \eqref{shift phase factor}, \eqref{degeneracy twisted})
\begin{equation}
    D(1)=\frac{1}{6}\sum_{l=0}^{5} e^{\frac{2\pi il}{6}[5\tilde{r}_3+\tilde{r}_4-(r_3+r_4) +2 +n_1] }\,.
    \label{degeneracy1}
\end{equation}
We list the weight vectors for the lightest left and right-moving states in the $k=1$ sector in \autoref{k=1 stu states}.
\renewcommand{\arraystretch}{2}
\begin{table}[h!]
\centering
 \begin{tabular}{|c|c|c|}
    \hline
    Sector &  $\tilde{r}$ &   ${r}$ \\
    \hline
    \hline
  NS & $(0,0,-1,0)$ &$(0,0,\underline{0,-1})$ \\
  \hline
   \multirow{2}{*}{R}  & $(\frac{1}{2},\frac{1}{2},-\frac{1}{2},-\frac{1}{2})$& $(\frac{1}{2},\frac{1}{2},-\frac{1}{2},-\frac{1}{2})$ \\
   \cline{2-3}
   & $(-\frac{1}{2},-\frac{1}{2},-\frac{1}{2},-\frac{1}{2})$& $(-\frac{1}{2},-\frac{1}{2},-\frac{1}{2},-\frac{1}{2})$ \\
   \hline
    \end{tabular}
\captionsetup{width=.9\linewidth}
\caption{\textit{Here we list the weight vectors of the lightest left and right-moving states in the $k=1$ twisted sector.}}
\label{k=1 stu states}
\end{table}
\renewcommand{\arraystretch}{1}

\noindent Let us start with the construction of states in the R-R sector, in which we have 
\begin{equation}
\begin{aligned}
    & (\tfrac{1}{2},\tfrac{1}{2},-\tfrac{1}{2},-\tfrac{1}{2})\times (\tfrac{1}{2},\tfrac{1}{2},-\tfrac{1}{2},-\tfrac{1}{2}) : (0)\,,\\
    & (\tfrac{1}{2},\tfrac{1}{2},-\tfrac{1}{2},-\tfrac{1}{2})\times (-\tfrac{1}{2},-\tfrac{1}{2},-\tfrac{1}{2},-\tfrac{1}{2}) : (1)\,,\\
    & (-\tfrac{1}{2},-\tfrac{1}{2},-\tfrac{1}{2},-\tfrac{1}{2})\times (\tfrac{1}{2},\tfrac{1}{2},-\tfrac{1}{2},-\tfrac{1}{2}) : (-1)\,,\\
     &(-\tfrac{1}{2},-\tfrac{1}{2},-\tfrac{1}{2},-\tfrac{1}{2})\times (-\tfrac{1}{2},-\tfrac{1}{2},-\tfrac{1}{2},-\tfrac{1}{2}) : (0)\,.
\end{aligned}
\end{equation}
Here we denote a state of helicity $h$ by $(h)$. Using \eqref{degeneracy1}, we can see that the above states are orbifold invariant for $n_1=0$ mod $6$, and have degeneracy $1$. From the spacetime point of view, we have the helicities that correspond to 1 massive vector $(\pm 1,0)$ and 1 scalar $(0)$. Also, from the mass formulae \eqref{twisted masses} we get (using \eqref{shifted momenta} and $\tilde{N}=N=0$)
\begin{equation}
\begin{aligned}
    &\frac{\alpha'm^2_{L}(1)}{2}=\frac{1}{2}p^2_{L}(1)=\frac{1}{4T_2U_2}\left|n_2-Un_1 +\bar{T}\left(w_1+\tfrac{1}{6}\right) +\bar{T}Uw_2\right|^2\,,\\
    &\frac{\alpha'm^2_{R}(1)}{2}=\frac{1}{2}p^2_{R}(1)=\frac{1}{4T_2U_2}\left|n_2-Un_1 +{T}\left(w_1+\tfrac{1}{6}\right) +{T}Uw_2\right|^2\,,
\end{aligned}
    \label{vmk1}
\end{equation}
where now $n_1=0$ mod $6$. Of course, physical states should also satisfy level-matching, that is $m^2_{L}(1)=m^2_{R}(1)$, which yields
\begin{equation}
    n_1\left(w_1+\frac{1}{6}\right) + n_2w_2 = 0\,.
    \label{levelvmk1}
\end{equation}
In the R-NS sector we find the following states
\begin{equation}
    \begin{aligned}
       & (\tfrac{1}{2},\tfrac{1}{2},-\tfrac{1}{2},-\tfrac{1}{2})\times (0,0,\underline{0,-1}) : 2\times (\tfrac{1}{2})\,,\\
    & (-\tfrac{1}{2},-\tfrac{1}{2},-\tfrac{1}{2},-\tfrac{1}{2})\times(0,0,\underline{0,-1}) : 2\times (-\tfrac{1}{2})\,. 
    \end{aligned}
\end{equation}
Similarly with the R-R sector, these states are orbifold invariant for $n_1=0$ mod $6$, and have degeneracy $1$. They  correspond to 2 dilatini with helicity $(\pm \tfrac{1}{2})$ and mass given in \eqref{vmk1}. The fields from the R-R and R-NS sectors form 1 tower of $4D$ massive vector multiplets, with mass \eqref{vmk1}.

Let us now move on to the NS-NS sector, in which we have
\begin{equation}
    (0,0,-1,0)\times (0,0,\underline{0,-1}) : 2\times (0)\,.
    \label{NSNSk=1}
\end{equation}
From \eqref{degeneracy1}, we can see that the above states are orbifold invariant for $n_1=2$ mod $6$, and have degeneracy $1$. So, in the NS-NS sector we find 2 scalars, denoted by $2\times (0)$. From the mass formulae \eqref{twisted masses} we get (using \eqref{shifted momenta} and $\tilde{N}=N=0$)
\begin{equation}
\begin{aligned}
    &\frac{\alpha'm^2_{L}(1)}{2}=\frac{1}{2}p^2_{L}(1)-\frac{1}{3}=\frac{1}{4T_2U_2}|n_2-Un_1 +\bar{T}(w_1+\tfrac{1}{6}) +\bar{T}Uw_2|^2-\frac{1}{3}\,,\\
    &\frac{\alpha'm^2_{R}(1)}{2}=\frac{1}{2}p^2_{R}(1)=\frac{1}{4T_2U_2}|n_2-Un_1 +{T}(w_1+\tfrac{1}{6}) +{T}Uw_2|^2\,,
\end{aligned}
\label{hmk1}
\end{equation}
and the level-matching condition becomes
\begin{equation}
    n_1\left(w_1+\frac{1}{6}\right) + n_2w_2 = \frac{1}{3}\,,
    \label{levelhmk1}
\end{equation}
which gives $n_1=2$ mod $6$. 

In the NS-R sector we find the following states
\begin{equation}
    \begin{aligned}
  &   (0,0,-1,0)\times (\tfrac{1}{2},\tfrac{1}{2},-\tfrac{1}{2},-\tfrac{1}{2}) :  (-\tfrac{1}{2})\,,\\
  &  (0,0,-1,0)\times (-\tfrac{1}{2},-\tfrac{1}{2},-\tfrac{1}{2},-\tfrac{1}{2}) :  (\tfrac{1}{2})\,.
    \end{aligned}
    \label{NSRk=1}
\end{equation}
As in the NS-NS sector, these states are orbifold invariant for $n_1=2$ mod $6$, and have degeneracy $1$. They correspond to 1 dilatino $(\pm\tfrac{1}{2})$ with mass \eqref{hmk1}. The fields from the NS-NS and NS-R sectors in the $k=1$ sector, constitute half of the field content of a hypermultiplet\footnote{The other half arises from the $k=5$ twisted sector, which, as we have already mentioned, is equivalent to the $k=1$ sector.}. 

Now, we can also act on the NS-NS states \eqref{NSNSk=1} with the left-moving creation operators $\tilde{a}^2_{-{1}/{6}}$ or $\bar{\tilde{a}}^1_{-{1}/{6}}$ (cf. \eqref{shifted oscillators}, \eqref{twist vectors for primary example2}). These will contribute to $m_{L}^2(1)$ in \eqref{hmk1} by a factor of $\tilde{N}=1/6$, and to the degeneracy of those states \eqref{degeneracy1} by a factor $e^{2\pi i l/6}$. So, these states will be orbifold invariant for $n_1=1$ mod $6$. Putting everything together, we find 4 scalars with mass
\begin{equation}
\begin{aligned}
    &\frac{\alpha'm^2_{L}(1)}{2}=\frac{1}{2}p^2_{L}(1)-\frac{1}{3}+\frac{1}{6}=\frac{1}{4T_2U_2}|n_2-Un_1 +\bar{T}(w_1+\tfrac{1}{6}) +\bar{T}Uw_2|^2-\frac{1}{6}\,,\\
    &\frac{\alpha'm^2_{R}(1)}{2}=\frac{1}{2}p^2_{R}(1)=\frac{1}{4T_2U_2}|n_2-Un_1 +{T}(w_1+\tfrac{1}{6}) +{T}Uw_2|^2\,,
\end{aligned}
\label{hmk1-case2}
\end{equation}
and the level-matching condition becomes
\begin{equation}
    n_1\left(w_1+\frac{1}{6}\right) + n_2w_2 = \frac{1}{6}\,,
    \label{levelhmk1-case2}
\end{equation}
with $n_1=1$ mod $6$. The above discussion applies also to the NS-R states \eqref{NSRk=1}, where we find 2 dilatini with mass \eqref{hmk1-case2}. So, in the $k=1$ sector we also find 1 tower of massive hypermultiplets with mass \eqref{hmk1-case2}.

Finally, we can act on the NS-NS states \eqref{NSNSk=1} with combinations of two left-moving creation operators $\tilde{a}^2_{-{1}/{6}}$ and/or $\bar{\tilde{a}}^1_{-{1}/{6}}$ (cf. \eqref{shifted oscillators}, \eqref{twist vectors for primary example2}). These will contribute to $\alpha'm^2(1)_{L}/2$ in \eqref{hmk1} by a factor of $\tilde{N}=1/3$, and to the degeneracy of those states \eqref{degeneracy1} by a factor $e^{2\pi i l/3}$. So, these states will be orbifold invariant for $n_1=0$ mod $6$. Putting everything together, we find 6 scalars with mass
\begin{equation}
\begin{aligned}
    &\frac{\alpha'm^2_{L}(1)}{2}=\frac{1}{2}p^2_{L}(1)-\frac{1}{3}+\frac{1}{3}=\frac{1}{4T_2U_2}|n_2-Un_1 +\bar{T}(w_1+\tfrac{1}{6}) +\bar{T}Uw_2|^2\,,\\
    &\frac{\alpha'm^2_{R}(1)}{2}=\frac{1}{2}p^2_{R}(1)=\frac{1}{4T_2U_2}|n_2-Un_1 +{T}(w_1+\tfrac{1}{6}) +{T}Uw_2|^2\,,
\end{aligned}
\label{hmk1-case3}
\end{equation}
and the level-matching condition becomes
\begin{equation}
    n_1\left(w_1+\frac{1}{6}\right) + n_2w_2 =0\,,
    \label{levelhmk1-case3}
\end{equation}
with $n_1=0$ mod $6$. Similarly, we can act with the left-moving oscillators $\tilde{a}^2_{-{1}/{6}}$ and $\bar{\tilde{a}}^1_{-{1}/{6}}$ on the NS-R states \eqref{NSRk=1}. Then we find 3 dilatini with mass \eqref{hmk1-case3}. In this case, the fields from the NS-NS and NS-R sectors constitute the field content of 1 and a half hypermultiplet\footnote{The other 1 and a half hypermultiplet arises from the $k=5$ twisted sector.}. 

\subsubsection{$k=2$ sector}
Now we move on to the construction of states in the $k=2$ sector, which is equivalent to the $k=4$ sector. The shift in the zero point energies is (cf. \eqref{shifted energy}) 
\begin{equation}
    \tilde{E}_2=E_2=\frac{2}{9}\,,
\end{equation}
and the degeneracy of a generic state is given by\footnote{Here, we have introduced an additional phase factor $e^{\pi il}$, in order to account for an extra minus sign arising from the fixed points.} (cf. \eqref{fixed points}, \eqref{shift phase factor}, \eqref{degeneracy twisted})
\begin{equation}
    D(2)= \frac{1}{6}\sum_{l=0}^{5}{\chi} [2,l]\,\tilde{\chi}[2,l] e^{\frac{2\pi il}{6}[5\tilde{r}_3+\tilde{r}_4-(r_3+r_4) -2 +n_1] }\,,
    \label{deg2}
\end{equation}
where
\begin{equation}
   {\chi} [2,l]\, \tilde{\chi}[2,l]= 
\begin{dcases}
   \: 9& \text{for}\quad l=0,2,4\,.\\
   \: 1& \text{for} \quad l=1,3,5\,.
\end{dcases}
\label{chis2}
\end{equation}
We list the weight vectors for the lightest left and right-moving states of the $k=2$ sector in \autoref{k=2 stu states}.
\renewcommand{\arraystretch}{2}
\begin{table}[h!]
\centering
 \begin{tabular}{|c|c|c|}
    \hline
    Sector &  $\tilde{r}$ &   ${r}$ \\
    \hline
    \hline
  \multirow{2}{*}{NS} & $(0,0,-1,0)$ &$(0,0,-1,0)$ \\
  \cline{2-3}
  & $(0,0,-2,-1)$ &$(0,0,0,-1)$ \\
  \hline
   \multirow{2}{*}{R}  & $(\frac{1}{2},-\frac{1}{2},-\frac{3}{2},-\frac{1}{2})$& $(\frac{1}{2},\frac{1}{2},-\frac{1}{2},-\frac{1}{2})$ \\
   \cline{2-3}
   & $(-\frac{1}{2},\frac{1}{2},-\frac{3}{2},-\frac{1}{2})$& $(-\frac{1}{2},-\frac{1}{2},-\frac{1}{2},-\frac{1}{2})$ \\
   \hline
    \end{tabular}
\captionsetup{width=.9\linewidth}
\caption{\textit{Here we list the weight vectors of the lightest left and right-moving states in the $k=2$ twisted sector.}}
\label{k=2 stu states}
\end{table}
\renewcommand{\arraystretch}{1}

\noindent Now, we take tensor products between left and right-movers from \autoref{k=2 stu states}. In the NS-NS sector we find 
\begin{equation}
    \begin{aligned}
      &(0,0,-1,0) \times  (0,0,\underline{0,-1}) : 2\times (0)\,,\\
       &(0,0,-2,-1) \times  (0,0,\underline{0,-1}) : 2\times (0)\,.
    \end{aligned}
    \label{k=2 NSNS}
\end{equation}
From \eqref{deg2} and \eqref{chis2}, we see that these states survive the orbifold projection for $n_1=0$ mod $6$, and have degeneracy 5. So, we find 20 scalars (0) with mass (cf. \eqref{twisted masses} with $\tilde{N}=N=0$, and \eqref{shifted momenta} )
\begin{equation}
\begin{aligned}
    &\frac{\alpha'm^2_{L}(2)}{2}=\frac{1}{2}p^2_{L}(2)=\frac{1}{4T_2U_2}|n_2-Un_1 +\bar{T}(w_1+\tfrac{1}{3}) +\bar{T}Uw_2|^2\,,\\
    &\frac{\alpha'm^2_{R}(2)}{2}=\frac{1}{2}p^2_{R}(2)=\frac{1}{4T_2U_2}|n_2-Un_1 +{T}(w_1+\tfrac{1}{3}) +{T}Uw_2|^2\,,
\end{aligned}
    \label{hmk2}
\end{equation}
where $n_1=0$ mod $6$, and the level-matching condition reads
\begin{equation}
     n_1\left(w_1+\frac{1}{3}\right) + n_2w_2 = 0\,.
     \label{levelmatch2}
\end{equation}
Note that the states in \eqref{k=2 NSNS}, can also survive the orbifold projection for $n_1=3$ mod $6$, and, in this case, they come with degeneracy $4$. The construction of the lightest states in the NS-R sector is similar with the NS-NS sector, so we omit the details. We find 10 dilatini $(\pm\tfrac{1}{2})$ with mass \eqref{hmk2}. So, from the NS-NS and NS-R sectors we obtain 5 towers of hypermultiplets with mass \eqref{hmk2}, where $n_1=0$ mod $6$, and 4 towers of hypermultiplets with mass \eqref{hmk2}, where $n_1=3$ mod $6$.

Let us now discuss the spectrum in the R-R sector, in which we have
\begin{equation}
  \begin{aligned}
      &(\tfrac{1}{2},-\tfrac{1}{2},-\tfrac{3}{2},-\tfrac{1}{2})\times (\tfrac{1}{2},\tfrac{1}{2},-\tfrac{1}{2},-\tfrac{1}{2}): (0)\,,\\
      &(\tfrac{1}{2},-\tfrac{1}{2},-\tfrac{3}{2},-\tfrac{1}{2})\times (-\tfrac{1}{2},-\tfrac{1}{2},-\tfrac{1}{2},-\tfrac{1}{2}): (1)\,,\\
      &(-\tfrac{1}{2},\tfrac{1}{2},-\tfrac{3}{2},-\tfrac{1}{2})\times (\tfrac{1}{2},\tfrac{1}{2},-\tfrac{1}{2},-\tfrac{1}{2}): (-1)\,,\\
      &(-\tfrac{1}{2},\tfrac{1}{2},-\tfrac{3}{2},-\tfrac{1}{2})\times (-\tfrac{1}{2},-\tfrac{1}{2},-\tfrac{1}{2},-\tfrac{1}{2}): (0)\,.
  \end{aligned}  
  \label{k=2 RR}
\end{equation}
From \eqref{deg2} and \eqref{chis2}, we see that these states survive the orbifold projection for $n_1=0$ mod $6$, and have degeneracy 4. So, we find 4 vectors $(\pm1,0)$ and 4 scalars $(0)$ with the same mass and level-matching condition as in \eqref{hmk2} and \eqref{levelmatch2}. Similarly with the NS-NS (and NS-R) sector, the states in \eqref{k=2 RR} can also survive the orbifold projection for $n_1=3$ mod 6, and then, they have degeneracy 5. States in the R-NS sector are constructed similarly with the R-R sector. In the R-NS sector we find 8 dilatini $(\pm\tfrac{1}{2})$. In total, the lightest states from the R-R and R-NS sectors form 4 towers of vector multiplets with mass \eqref{hmk2}, where $n_1=0$ mod $6$, and 5 towers of vector multiplets with mass \eqref{hmk2}, where $n_1=3$ mod $6$.

\subsubsection{$k=3$ sector}
Finally, we discuss the spectrum in the $k=3$ sector.  The shift in the zero point energies is (cf. \eqref{shifted energy}) 
\begin{equation}
    \tilde{E}_3=E_3=\frac{1}{4}\,,
\end{equation}
and the degeneracy of a generic state is given by (cf. \eqref{fixed points}, \eqref{shift phase factor}, \eqref{degeneracy twisted})
\begin{equation}
    D(3)= \frac{1}{6}\sum_{l=0}^{5}{\chi} [3,l]\,\tilde{\chi}[3,l] e^{\frac{2\pi il}{6}[5\tilde{r}_3+\tilde{r}_4-(r_3+r_4) +n_1] }\,,
    \label{deg3}
\end{equation}
where
\begin{equation}
   {\chi} [3,l]\, \tilde{\chi}[3,l]= 
\begin{dcases}
   \: 16& \text{for}\quad l=0,3\,.\\
   \: 1& \text{for} \quad l=1,2,4,5\,.
\end{dcases}
\label{chis3}
\end{equation}
We list the weight vectors for the lightest left and right-moving states of the $k=3$ sector in \autoref{k=3 stu states}.
\renewcommand{\arraystretch}{2}
\begin{table}[h!]
\centering
 \begin{tabular}{|c|c|c|}
    \hline
    Sector &  $\tilde{r}$ &   ${r}$ \\
    \hline
    \hline
  \multirow{2}{*}{NS} & $(0,0,-3,0)$ &$(0,0,-1,0)$ \\
  \cline{2-3}
  & $(0,0,-2,-1)$ &$(0,0,0,-1)$ \\
  \hline
   \multirow{2}{*}{R}  & $(\frac{1}{2},\frac{1}{2},-\frac{5}{2},-\frac{1}{2})$& $(\frac{1}{2},\frac{1}{2},-\frac{1}{2},-\frac{1}{2})$ \\
   \cline{2-3}
   & $(-\frac{1}{2},-\frac{1}{2},-\frac{5}{2},-\frac{1}{2})$& $(-\frac{1}{2},-\frac{1}{2},-\frac{1}{2},-\frac{1}{2})$ \\
   \hline
    \end{tabular}
\captionsetup{width=.9\linewidth}
\caption{\textit{Here we list the weight vectors of the lightest left and right-moving states in the $k=3$ twisted sector.}}
\label{k=3 stu states}
\end{table}
\renewcommand{\arraystretch}{1}

\noindent Let us start with the construction of states in the NS-NS sector, in which we find
\begin{equation}
    \begin{aligned}
       & (0,0,-3,0)\times (0,0,\underline{0,-1}): 2\times (0)\,,\\
       & (0,0,-2,-1)\times (0,0,\underline{0,-1}): 2\times (0)\,.
    \end{aligned}
    \label{k=3 NS-NS}
\end{equation}
As we can see from \eqref{deg3} and \eqref{chis3}, these states survive the orbifold projection for $n_1=0$ mod $6$, and have degeneracy 5. So, we find 20 scalars (0) with mass (cf. \eqref{twisted masses} with $\tilde{N}=N=0$, and \eqref{shifted momenta})
\begin{equation}
\begin{aligned}
    &\frac{\alpha'm^2_{L}(3)}{2}=\frac{1}{2}p^2_{L}(3)=\frac{1}{4T_2U_2}|n_2-Un_1 +\bar{T}(w_1+\tfrac{1}{2}) +\bar{T}Uw_2|^2\,,\\
    &\frac{\alpha'm^2_{R}(3)}{2}=\frac{1}{2}p^2_{R}(3)=\frac{1}{4T_2U_2}|n_2-Un_1 +{T}(w_1+\tfrac{1}{2}) +{T}Uw_2|^2\,,
\end{aligned}
    \label{hmk3}
\end{equation}
where $n_1=0$ mod $6$, and the level-matching condition reads
\begin{equation}
     n_1\left(w_1+\frac{1}{2}\right) + n_2w_2 = 0\,.
     \label{levelmatch3}
\end{equation}
Similarly with the $k=2$ twisted sector, the states in \eqref{k=3 NS-NS} can also survive the orbifold projection for $n_1=2$ mod $6$ and $n_1=4$ mod $6$ and, in these cases, they come with degeneracy 5 or 6.

The construction of the lightest states in the NS-R sector is similar with the NS-NS sector, and we find 10 dilatini $(\pm\tfrac{1}{2})$ with mass \eqref{hmk3}. So, from the NS-NS and NS-R sectors we obtain 5 towers of hypermultiplets with mass \eqref{hmk3}, when $n_1=0$ mod $6$, and 11 towers of hypermultiplets with mass \eqref{hmk3}, from both $n_1=2$ mod $6$ and $n_1=4$ mod $6$.

Now, we move on to the R-R sector, in which we find
\begin{equation}
    \begin{aligned}
    &(\tfrac{1}{2},\tfrac{1}{2},-\tfrac{5}{2},-\tfrac{1}{2})\times (\tfrac{1}{2},\tfrac{1}{2},-\tfrac{1}{2},-\tfrac{1}{2}): (0)\,,\\
      &(\tfrac{1}{2},\tfrac{1}{2},-\tfrac{5}{2},-\tfrac{1}{2})\times (-\tfrac{1}{2},-\tfrac{1}{2},-\tfrac{1}{2},-\tfrac{1}{2}): (1)\,,\\
      &(-\tfrac{1}{2},-\tfrac{1}{2},-\tfrac{5}{2},-\tfrac{1}{2})\times (\tfrac{1}{2},\tfrac{1}{2},-\tfrac{1}{2},-\tfrac{1}{2}): (-1)\,,\\
      &(-\tfrac{1}{2},-\tfrac{1}{2},-\tfrac{5}{2},-\tfrac{1}{2})\times (-\tfrac{1}{2},-\tfrac{1}{2},-\tfrac{1}{2},-\tfrac{1}{2}): (0)\,.
    \end{aligned}
    \label{k=3 RR}
\end{equation}
From \eqref{deg3} and \eqref{chis3}, we can see that these states survive the orbifold projection for $n_1=0$ mod $6$, and have degeneracy 6. So, we find 6 vectors $(\pm1,0)$ and 6 scalars $(0)$ with the same mass and level-matching condition as in \eqref{hmk3} and \eqref{levelmatch3}. States in the R-NS sector are constructed similarly with the R-R sector. In the R-NS sector we find 12 dilatini $(\pm\tfrac{1}{2})$. In total, the lightest states from the R-R and R-NS sectors form 6 towers of vector multiplets with mass \eqref{hmk3} and $n_1=0$ mod 6. Finally, we mention here that the states in the R-R and R-NS sectors can also survive the orbifold projection with the addition of momentum number $n_1=2$ or $4$ mod $6$ and in these cases they come with degeneracy 5. So, we also find 10 towers of vector multiplets with mass \eqref{hmk3} and $n_1=2$ or $4$ mod $6$.

\section{Moduli space and the swampland}
\label{moduli and swampland}

In this section, we will determine the vector multiplet and hypermultiplet moduli space of our $\mathbb{Z}_6$ orbifold model. In order to achieve this, we have to compute the commutant of the twist matrix in the $4D$ T-duality group $\text{SO}(6,6)$. 

First, we recall that in section \ref{z6 orbifold with nv=vh=2} we computed the commutant of the same twist matrix in the $5D$ T-duality group $\text{SO}(5,5)$ and we found  (cf. \eqref{comm5dcase})
\begin{equation}
    \mathcal{C}^{(5)}=\text{SO}(1,1)\times \text{SU}(2,2)\times \text{U}(1)\,.
\end{equation}
For the $4D$ model, since the orbifold acts as a symmetry of $T^4$ together with a shift on $S^1_{{\cal R}_5}$, it is straightforward to verify that the commutant is enhanced as follows
\begin{equation}
     \mathcal{C}^{(5)}\to \mathcal{C}^{(4)}=\text{SO}(2,2)\times \text{SU}(2,2)\times \text{U}(1)\ .
\end{equation}
The hypermultiplet scalars form the quaternion-K{\"a}hler manifold of real dimension 8 (with $\text{SU}(2,2)\simeq \text{SO}(4,2)$),
\begin{equation}
{\cal M}_H=\frac{\text{SU}(2,2)}{\text{SU}(2)\times \text{SU}(2)\times \text{U}(1)}\simeq \frac{\text{SO}(4,2)}{\text{SO}(4)\times \text{SO}(2)}\ .
\end{equation}
Regarding the vector multiplet moduli space, note that since the monodromy is in the T-duality group, it also commutes with the classical $4D$ S-duality group $\text{SL}(2)_S$. 
Then, using $\text{SO}(2,2)\sim \text{SL}(2) \times \text{SL}(2)$, we find
\begin{equation}\label{modspace3}
    {\cal M}_V= \frac{\text{SL}(2)}{\text{U}(1)}\times\frac{\text{SL}(2)}{\text{U}(1)}\times\frac{\text{SL}(2)}{\text{U}(1)}\,.
\end{equation}
The vector multiplet moduli space consists of 3 complex NS-NS scalar fields, which are the $S, \,T$ and $U$ moduli. This is all consistent with the supergravity dimensional reduction of the $5D$ theory to four dimensions, and is given by the so-called $r$-map. Furthermore, as we have explained in section \ref{duality groups of orbifolded theory}, the duality group of our $\mathbb{Z}_6$ orbifold is further broken to a subgroup
\begin{equation}
    \hat{\Gamma}^1(6)_S \times \hat{\Gamma}^1(6)_T \times \hat{\Gamma}_1(6)_U\,,
\end{equation}
due to the orbifold shift. Therefore, each of the $\text{SL}(2)$ factors in \eqref{modspace3} should be modded out by the corresponding congruence subgroup of $\mathrm{SL}(2,\mathbb{Z})$.

Now, the action of an $\mathcal{N}=2$ supergravity theory coupled to $n$ vector multiplets in $4D$ can be specified by the prepotential $F(X)$, which is a holomorphic and homogeneous function of second degree in the variables $X^I$, $I=0,\ldots,n$.
For the model at hand, the corresponding prepotential governing the $4D$ vector multiplets is given by 
\begin{equation}
    F(X)=id_{ABC}\frac{X^AX^BX^C}{X^0}\,.
\end{equation}
Here $A,B,C=1,2,3$, and the $d$-symbols are the same as in the $5D$ theory, namely
\begin{equation}
      d_{122}=1\ ,\qquad d_{133}=-1\ ,
\end{equation}
while all other $d$-symbols vanish. So, we obtain
\begin{equation}
    F(X)= 3i\frac{X^1}{X^0}\left[(X^2)^2-(X^3)^2\right]\ .
\end{equation}
The complex coordinates $X^1,X^2,X^3$ are the complexifications of the $5D$ real coordinates denoted by $h^1,h^2,h^3$ in section \ref{8sup}, as any $5D$ vector yields a scalar in $4D$ which pairs up with the $5D$ real scalar (reduced to $4D$) to make up the complex variables as part of the $4D$ vector multiplet \cite{Gunaydin:1983rk}. By applying a symplectic transformation we can bring the prepotential to the equivalent form 
\begin{equation}
     F(X)=i\frac{X^1X^2X^3}{X^0}\,.
\end{equation}
In this basis the only non-vanishing $d$-symbol is $d_{123}=1/6$. Identifying 
\begin{equation}
S=\frac{X^1}{X^0}\ ,\qquad  T=\frac{X^2}{X^0}\ ,\qquad U=\frac{X^3}{X^0}\ ,
\end{equation}
yields the classical prepotential
\begin{equation}
      F(X)= iSTU(X^0)^2\,.
      \label{classicalSTU}
\end{equation}
From the prepotential, one can also compute the K\"ahler potential $K$,  which is given by 
\begin{equation}
    K = -\ln Y = -\ln \left(\tfrac{1}{2}X^I\bar{F}_I + \tfrac{1}{2}\bar{X}^I{F}_I\right)\,,
    \label{defY}
\end{equation}
where $F_I=\partial F/\partial X^I$ and $I=0,\dots,3$.
 In our case, we find  
\begin{equation}
    Y = \frac{1}{2i}(S-\bar{S})(T-\bar{T})(U-\bar{U})\left|X^0\right|^2\,.
    \label{kahlerY}
\end{equation}
The vector multiplet moduli space is a K\"ahler manifold, and its metric is determined by the second derivative of the K\"ahler potential \eqref{defY}, which is
\begin{equation}
\begin{aligned}
    K_{I\bar{J}}\ {\rm d} z^I \text{d}\bar{z}^{\bar{J}}&=-\frac{1}{(S-\bar{S})^2}\text{d} S\text{d} \bar{S} - \frac{1}{(T-\bar{T})^2}\text{d} T\text{d} \bar{T} - \frac{1}{(U-\bar{U})^2}\text{d} U\text{d} \bar{U}\\&=\frac{1}{4S_2^2}\left(\text{d} S_1^2+\text{d} S_2^2\right)+\frac{1}{4T_2^2}\left(\text{d} T_1^2+\text{d} T_2^2\right)+\frac{1}{4U_2^2}\left(\text{d} U_1^2+\text{d} U_2^2\right)\\
    &=\frac{1}{4}e^{-2\sqrt{2}\phi_S}\text{d} S_1^2 +\frac{1}{2}\text{d}\phi_S^2 + \frac{1}{4}e^{-2\sqrt{2}\phi_T}\text{d} T_1^2 +\frac{1}{2}\text{d}\phi_T^2+\\&\:\:\:\:\,\,\frac{1}{4}e^{-2\sqrt{2}\phi_U}\text{d} U_1^2 +\frac{1}{2}\text{d}\phi_U^2.
\end{aligned}
\end{equation}
Here $\phi_I,\,I=S,T,U$, are the corresponding normalized moduli, defined by
\begin{equation}
    S_2 = e^{\sqrt{2}\phi_S},\quad T_2 = e^{\sqrt{2}\phi_T},\quad U_2 = e^{\sqrt{2}\phi_U}.
    \label{normalizedmoduli}
\end{equation}
These give the normalized classical moduli space metric.

\subsection{Massless states at finite distance}
\label{Massless states at finite distance}
In this section, we will analyse the special lines and points at finite distance in the interior of the moduli space where generically massive states become massless. It is important to mention here that in the interior of the moduli space, only a finite number of states become massless and no infinite towers become massless. 

Recall that in  the $k=1$ sector we found half a hypermultiplet\footnote{The other half arises in the $k=5$ sector.} with mass
\begin{equation}
\begin{aligned}
    &\frac{\alpha'm^2_{L}(1)}{2}=\frac{1}{2}p^2_{L}(1)-\frac{1}{3}=\frac{1}{4T_2U_2}|n_2-Un_1 +\bar{T}(w_1+\tfrac{1}{6}) +\bar{T}Uw_2|^2-\frac{1}{3}\,,\\
    &\frac{\alpha'm^2_{R}(1)}{2}=\frac{1}{2}p^2_{R}(1)=\frac{1}{4T_2U_2}|n_2-Un_1 +{T}(w_1+\tfrac{1}{6}) +{T}Uw_2|^2\,,
\end{aligned}
\label{hmk1new}
\end{equation}
with the constraint $n_1=2$ mod $6$, and the level-matching condition
\begin{equation}
    n_1\left(w_1+\frac{1}{6}\right) + n_2w_2 = \frac{1}{3}\,.
    \label{levelhmk1new}
\end{equation}
From the mass formulae \eqref{hmk1new}, we see that massless states can appear if
\begin{equation}
    T =\frac{n_1U-n_2}{w_2U+(w_1+\tfrac{1}{6})}\,,
\end{equation}
where $n_1=2$ mod $6$ and the level-matching condition \eqref{levelhmk1new} should be satisfied. We can rewrite the above expression together with the level-matching condition in a more convenient form as follows
\begin{equation}
    \frac{T}{6} =\frac{n_1U-n_2}{6w_2U+(6w_1+1)}\,,\qquad \frac{n_1}{2}(6w_1+1) + 3n_2w_2 = 1\,.
\end{equation}
By setting $\alpha=\frac{n_1}{2}\in 1+3\mathbb{Z}$,  $\beta=-n_2\in \mathbb{Z}$, $\gamma=3w_2\in 3\mathbb{Z}$ and $\delta=1+6w_1\in 1+6\mathbb{Z}$, we obtain
\begin{equation}
    \frac{T}{6}=\frac{\alpha2U+\beta}{\gamma 2U+\delta}\,,\qquad \alpha\delta-\beta\gamma=1\,.
\end{equation}
So, massless states will appear when
\begin{equation}
    \frac{T}{6} = g(2U)\,,\qquad  g\in  \mathcal{S}\equiv \left\{\alpha\delta-\beta\gamma=1:\,\alpha\in 1+3\mathbb{Z},\,\beta\in\mathbb{Z},\,\gamma\in3\mathbb{Z},\,\delta\in1+6\mathbb{Z}\,\right\}.
     \label{t12branch}
\end{equation}
It can be easily proven that all solutions of \eqref{t12branch} can be obtained from the solution  $\frac{T}{6}=2U$ by applying $\Gamma^1(6)_T\times \Gamma_1(6)_U$ transformations. Starting from the solution $\frac{T}{6}=2U$ and applying a generic $\Gamma^1(6)_T\times \Gamma_1(6)_U$ transformation we obtain (cf. \eqref{gamma1t} and \eqref{gamma1u})
\begin{equation}
    \frac{a\frac{T}{6}+\frac{b}{6}}{cT+d}=\frac{a'2U+2b'}{c'U+d'}\implies \frac{a\frac{T}{6}+\frac{b}{6}}{6c\frac{T}{6}+d}=\frac{a'2U+2b'}{\frac{c'}{2}2U+d'}\,.
\end{equation}
Here $a,d,a',d'\in 1+6\mathbb{Z}$, $b,c'\in6\mathbb{Z}$ and $c,b'\in\mathbb{Z}$. Solving for $\frac{T}{6}$ yields
\begin{equation}
    \frac{T}{6} = \frac{(da'-\frac{bc'}{12})2U+(2db'-\frac{bd'}{6})}{(\frac{ac'}{2}-6a'c)2U+(ad'-12b'c)}\,.
    \label{t12ug1tg1u}
\end{equation}
Now, we note that
\begin{equation}
\begin{aligned}
    &\alpha\equiv da'-\frac{bc'}{12}\in1+3\mathbb{Z}\,,\qquad \beta \equiv 2db'-\frac{bd'}{6} \in \mathbb{Z}\,,\\
    &\gamma \equiv \frac{ac'}{2}-6a'c \,\in \,3\mathbb{Z}\,,\qquad \delta \equiv ad'-12b'c \,\in 1+6\mathbb{Z}\,,\qquad\alpha \delta - \beta\gamma = 1\,.
    \end{aligned}
\end{equation}
So, we can rewrite \eqref{t12ug1tg1u} as
\begin{equation}
     \frac{T}{6}=\frac{\alpha2U+\beta}{\gamma 2U+\delta}\,,\qquad \alpha\delta-\beta\gamma=1\,,
\end{equation}
which exactly  reproduces \eqref{t12branch}. 

As we have already discussed, the sectors $k=1$ and $k=5$ are equivalent. So, we conclude that at the critical line $\frac{T}{6}=2U$, modulo $\Gamma^1(6)_T\times \Gamma_1(6)_U$ transformations, 1 hypermultiplet becomes massless.

\begin{figure}[h]
\centering
\includegraphics[width=0.5\textwidth]{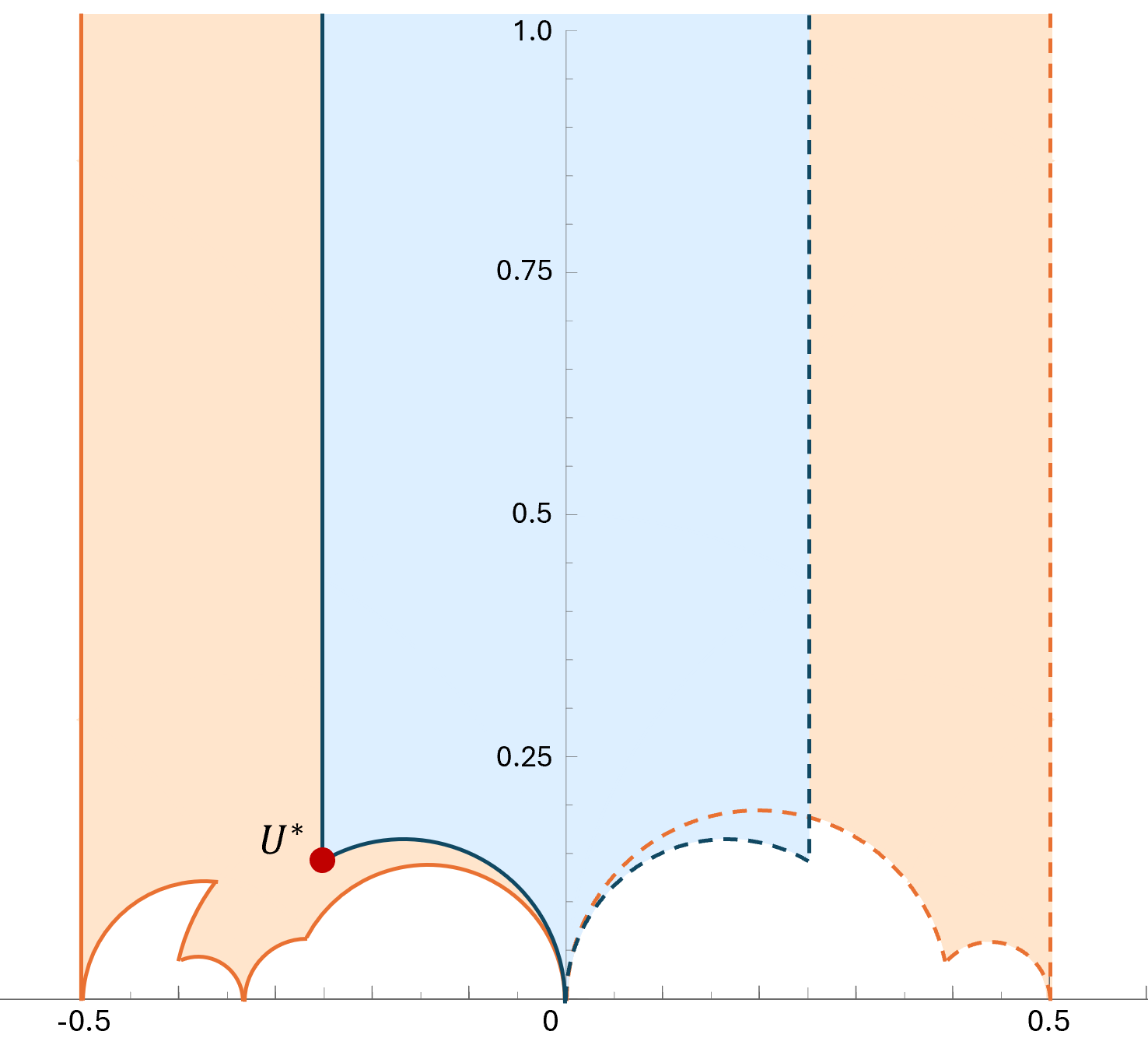}
\caption{\textit{The location of $U^*$, represented by the red dot. The orange region is the fundamental domain of $\Gamma_1(6)_U$, whereas the blue region is the fundamental domain of $\Gamma_1(3)_{2U}$. The shapes of the fundamental domains are obtained by the Mathematica package} \texttt{DrawFunDoms.m}.}
\label{funddom63}
\end{figure}

Now, as in heterotic string constructions (see e.g. \cite{LopesCardoso:1994ik}), whenever two (or more) critical lines intersect, two (or more) hypermultiplets will become massless. Starting from \eqref{t12branch}, it is easy to show that the only lines intersecting inside the fundamental domain of $\Gamma^1(6)_T\times \Gamma_1(6)_U$ are the lines $ \frac{T}{6}=2U$ and $\frac{T}{6}=-(4U+1)/(6U+1)$. These lines intersect at the point 
$(T,U)=(12U^*,U^*)$, where $U^*= -\frac{1}{4}+i\frac{\sqrt{3}}{12}$, and it is depicted in \autoref{funddom63}. Hence, at this point two hypermultiplets become massless.

\noindent Finally, in the $k=1$ sector we also found one tower of hypermultiplets with mass
\begin{equation}
\begin{aligned}
    &\frac{\alpha'm^2_{L}(1)}{2}=\frac{1}{2}p^2_{L}(1)-\frac{1}{3}+\frac{1}{6}=\frac{1}{4T_2U_2}|n_2-Un_1 +\bar{T}(w_1+\tfrac{1}{6}) +\bar{T}Uw_2|^2-\frac{1}{6}\,,\\
    &\frac{\alpha'm^2_{R}(1)}{2}=\frac{1}{2}p^2_{R}(1)=\frac{1}{4T_2U_2}|n_2-Un_1 +{T}(w_1+\tfrac{1}{6}) +{T}Uw_2|^2\,,
\end{aligned}
\label{hmk1-case2new}
\end{equation}
with the constraint $n_1=1$ mod $6$ and the level-matching condition 
\begin{equation}
    n_1\left(w_1+\frac{1}{6}\right) + n_2w_2 = \frac{1}{6}\,.
    \label{levelhmk1-case2new}
\end{equation}
From the mass formulae \eqref{hmk1-case2new}, we see that massless states can appear if
\begin{equation}
    T =\frac{n_1U-n_2}{w_2U+(w_1+\tfrac{1}{6})}\,,
\end{equation}
where $n_1=1$ mod $6$ and the level-matching condition \eqref{levelhmk1-case2new} should be satisfied. Again, we can rewrite the above expression together with the level-matching condition in a more convenient form as follows
\begin{equation}
    \frac{T}{6} =\frac{n_1U-n_2}{6w_2U+(6w_1+1)}\,,\qquad n_1(6w_1+1) + 6n_2w_2 = 1\,.
\end{equation}
By setting $a=n_1\in 1+6\mathbb{Z}$, $b=-n_2\in \mathbb{Z}$, $c=6w_2\in 6\mathbb{Z}$, and $d=1+6w_1\in 1+6\mathbb{Z}$ we obtain
\begin{equation}
   \frac{T}{6} =\frac{a{U}+b}{c{U}+d}\,,\qquad ad-bc=1\,.
   \label{Tmod6U}
\end{equation}
So, massless states will appear when
\begin{equation}
   \frac{T}{6}= gU\,,\qquad g\in \Gamma_1(6)_U\,.
\end{equation}
 or, equivalently,
 \begin{equation}
      {hT}= 6U\,,\qquad h\in \Gamma^1(6)_T\,.
 \end{equation}
We mention here that for the branch of solutions \eqref{Tmod6U}, there are no intersecting lines. As we have already mentioned the sectors $k=1$ and $k=5$ are equivalent. Thus, at $\frac{T}{6}=U$ (mod $\Gamma^1(6)_T\times \Gamma_1(6)_U$) we get another massless hypermultiplet from the $k=5$ sector. 

Concluding, we have found that there exist 2 special lines and 1 special point in the interior of the moduli space, where massive hypermultiplets become massless. In particular, at $\frac{T}{6}=U$, modulo $\Gamma^1(6)_T\times \Gamma_1(6)_U$ transformations, 2 hypermultiplets become massless, and at $\frac{T}{6}=2U$, modulo $\Gamma^1(6)_T\times \Gamma_1(6)_U$ transformations, 1 hypermultiplet becomes massless. At the special point $(T,U)=(12U^*,U^*)$, where $U^*= -\frac{1}{4}+i\frac{\sqrt{3}}{12}$, two critical lines intersect, and 2 hypermultiplets become massless. Note that  the hypermultiplets that become massless carry momentum and winding numbers along the $T^2$ directions, so they are charged under the corresponding vector fields.

\subsection{Massless states at infinite distance}
\label{Massless states at infinite distance}
In this section, we will study the spectrum of states at all different asymptotic limits in the $T-U$ plane of the moduli space. First, we will discuss these limits separately for $T$ and $U$, i.e. we will fix $U$ in the bulk of the moduli space and we will consider the infinite distance points in the $T$-plane, and vice-versa. These will be referred to as single cusps. Then we will consider the asymptotic limits simultaneously for both $T$ and $U$, which will be referred to as double cusps.

\subsubsection{Single cusps}

\begin{figure}[h]
\centering
\begin{subfigure}{.7\textwidth}
  \centering
  \includegraphics[width=.86\linewidth]{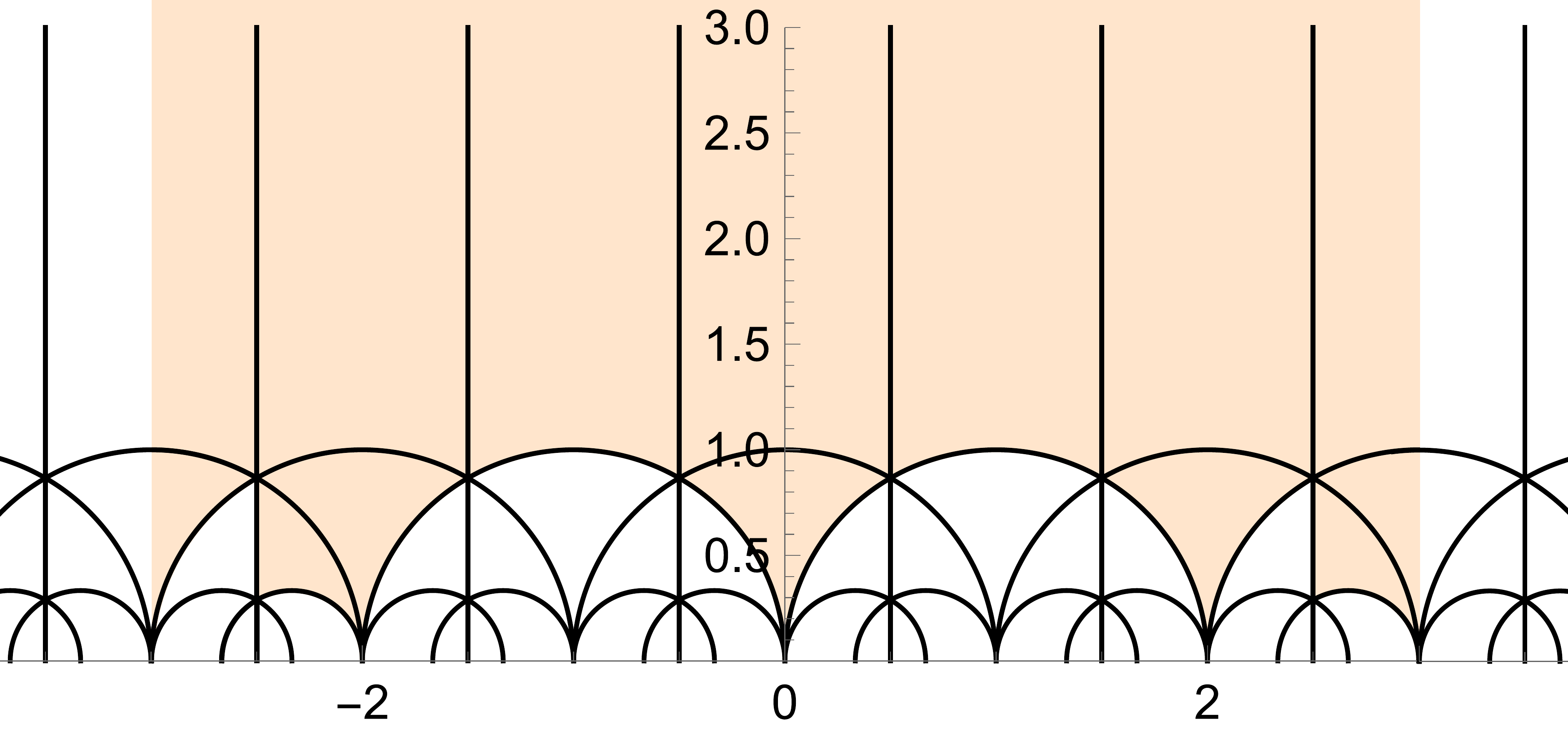}
\end{subfigure}%
\begin{subfigure}{.3\textwidth}
  \centering
  \includegraphics[width=0.75\linewidth]{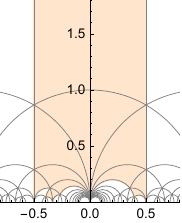}
\end{subfigure}
\caption{\textit{The fundamental domains of the congruence subgroups $\Gamma^1(6)_T$ and $\Gamma_1(6)_U$. The shaded region of the left and right figure corresponds to the fundamental domain of $\Gamma^1(6)_T$ and $\Gamma_1(6)_U$, respectively. The figures are obtained by the Mathematica package} \texttt{DrawFunDoms.m}.}
\label{funddom}
\end{figure}

The fundamental domains of $\Gamma^1(6)_T$ and $\Gamma_1(6)_U$ are shown in \autoref{funddom}. There are 4 inequivalent cusps in the fundamental domain of $\Gamma^1(6)_T$, at the points $T=-3,-2,0$ and $i\infty$. Similarly, there exist 4 inequivalent cusps in the fundamental domain of $\Gamma_1(6)_U$, at the points $U=-\frac{1}{2},-\frac{1}{3},0$ and $i\infty$.

\renewcommand{\arraystretch}{2}
\renewcommand{\arraystretch}{1}

\noindent The cusp points correspond to singular geometric structures of the torus. In the limit $T_2\to 0 $, with $U$ constant and finite, the volume of $T^2$ goes to zero, while the complex structure remains regular. This implies that the torus shrinks to a point. For the limit $U_2\to 0 $, with $T$ constant and finite, the torus, as an elliptic curve, has zero discriminant and divergent modular invariants, so that it becomes a nodal curve. In the special case in which $U_1=0$, the nodal curve is described by  $\mathcal{R}_4\to0$ and  $\mathcal{R}_5\to\infty$. 

We should highlight here that all cusp points are at the asymptotic boundary of the fundamental domain. Hence, it is worth studying the spectrum and physical interpretation around these infinite distance points, and investigating whether the Swampland Distance Conjecture \cite{Ooguri:2006in} is valid. The Swampland Distance Conjecture (SDC) proposes that as we approach an asymptotic limit on the moduli space, there must be at least one tower of states, whose mass scale decreases exponentially as
\begin{equation}\label{SDC}
    m= m_0 e^{-\lambda |\Delta\phi|}\,,
\end{equation}
where $\phi$ is a normalized modulus and $\lambda\sim\mathcal{O}(1)$. To verify this conjecture, let us analyse the spectrum of lightest states of all sectors with $\tilde{N}=N=0$.

The mass formulae for the lightest states that we found in the previous section  (cf. \eqref{toweruntwisted}, \eqref{vmk1}, \eqref{hmk2}, and \eqref{hmk3}) can be summarized as follows:
\begin{equation}
\begin{aligned}
    &\frac{\alpha'm^2_{L}(k)}{2}=\frac{1}{4T_2U_2}\left|n_2-Un_1 +\bar{T}\left(\hat{w}_1 +Uw_2\right)\right|^2=\frac{1}{4T_2U_2}\left|n_2+T\hat{w}_1+\bar{U}\left(-n_1+Tw_2\right)\right|^2\,,\\
    &\frac{\alpha'm^2_{R}(k)}{2}=\frac{1}{4T_2U_2}\left|n_2-Un_1 +{T}\left(\hat{w}_1 +Uw_2\right)\right|^2=\frac{1}{4T_2U_2}\left|n_2+T\hat{w}_1+U\left(-n_1+Tw_2\right)\right|^2\,,
\end{aligned}
    \label{mk}
\end{equation}
and the level matching condition reads
\begin{equation}
    n_1 \hat{w}_1+n_2 w_2=0\,,\quad \text{where }\quad \hat{w}_1\equiv w_1 +\frac{k}{6}\,, \qquad k=0,1,\ldots,5\,.
\end{equation}
Utilising this level matching condition, it can be derived from \eqref{mk} that
\begin{equation}\label{mkfinal}
\alpha'm^2(k)=\frac{1}{T_2U_2w_2^2}\left|-n_1+Tw_2\right|^2\left|\hat{w}_1+Uw_2\right|^2=\frac{1}{T_2U_2n_2^2}\left|n_2+T\hat{w}_1\right|^2\left|-n_2+Un_1\right|^2\,.
\end{equation}
Given these, we can solve the equation $m^2=m^2_{L}(k)+m^2_{R}(k)=0$, as we approach the infinite distance points in the $T-U$ moduli space. There are 4 inequivalent sets of solutions: 
\begin{itemize}
    \item $T_2\to0$: $T=T_1=\frac{n_1}{w_2}=-\frac{n_2}{\hat{w}_1}$, while $U$ is in the bulk.
    \item $U_2\to0$: $U=U_1=\frac{n_2}{n_1}=-\frac{\hat{w}_1}{w_2}$, while $T$ is in the bulk.
    \item $T_2\to\infty$: $\hat{w}_1=w_2=0$, while $U$ is in the bulk.
    \item $U_2\to\infty$: $n_1=w_2=0$, while $T$ is in the bulk.
\end{itemize}

\renewcommand{\arraystretch}{1.7}
\begin{table}[t!]
\centering
 \begin{tabular}[c]{|c|l|l|c|c|}
    \hline
    Sector & \multicolumn{2}{|c|}{Cusp} & Lattice constraints & Mass of tower \\
    \hline
    \hline
  \multirow{6}{*}{$k=0$}& \multirow{3}{*}{$T_2\to0$} & $T_1=0$ &$n_1=n_2=0$ & \multirow{3}{*}{$\frac{1}{\sqrt{\alpha'U_2}}\left|w_1+Uw_2\right|\sqrt{T_2}$} \\
  \cline{3-4}
  &  & $T_1=-2$ & $n_1=-2w_2,\,n_2=2w_1$ &  \\
  \cline{3-4}
  & & $T_1=-3$ & $n_1=-3w_2,\,n_2=3w_1$ & \\
  \cline{2-5}
  & \multirow{3}{*}{$U_2\to0$} & $U_1=0$ & $n_2=w_1=0$ & \multirow{3}{*}{$\frac{1}{\sqrt{\alpha'T_2}}\left|n_1-Tw_2\right|\sqrt{U_2}$}\\
  \cline{3-4}
  & & $U_1=-\frac{1}{3}$ & $n_1=-3n_2,\,w_2=3w_1$ & \\
  \cline{3-4}
  & & $U_1=-\frac{1}{2}$ & $n_1=-2n_2,\,w_2=2w_1$ & \\
  \hline
   \multirow{1}{*}{$k=1$} & $T_2\to0$ & $T_1=0$& $n_1=n_2=0$ & $\frac{1}{\sqrt{\alpha'U_2}}\left|w_1+\frac{1}{6}+Uw_2\right|\sqrt{T_2}$ \\
   \hline
   \multirow{3}{*}{$k=2$} & \multirow{2}{*}{$T_2\to0$} & $T_1=0$& $n_1=n_2=0$ & \multirow{2}{*}{$\frac{1}{\sqrt{\alpha'U_2}}\left|w_1+\frac{1}{3}+Uw_2\right|\sqrt{T_2}$} \\
   \cline{3-4}
   & & $T_1=-3$& $n_1=-3w_2,\,n_2=3w_1+1$ &  \\
   \cline{2-5}
   &$U_2\to0$ & $U_1=-\frac{1}{3}$& $n_1=-3n_2,\,w_2=3w_1+1$ & $\frac{1}{\sqrt{\alpha'T_2}}\left|n_1-Tw_2\right|\sqrt{U_2}$ \\
   \hline
   \multirow{3}{*}{$k=3$} & \multirow{2}{*}{$T_2\to0$} & $T_1=0$& $n_1=n_2=0$ & \multirow{2}{*}{$\frac{1}{\sqrt{\alpha'U_2}}\left|w_1+\frac{1}{2}+Uw_2\right|\sqrt{T_2}$} \\
   \cline{3-4}
   & & $T_1=-2$& $n_1=-2w_2,\,n_2=2w_1+1$ & \\
   \cline{2-5}
   &$U_2\to0$ & $U_1=-\frac{1}{2}$& $n_1=-2n_2,\,w_2=2w_1+1$ & $\frac{1}{\sqrt{\alpha'T_2}}\left|n_1-Tw_2\right|\sqrt{U_2}$ \\
   \hline
   \multirow{3}{*}{$k=4$} & \multirow{2}{*}{$T_2\to0$} & $T_1=0$& $n_1=n_2=0$ & \multirow{2}{*}{$\frac{1}{\sqrt{\alpha'U_2}}\left|w_1+\frac{2}{3}+Uw_2\right|\sqrt{T_2}$} \\
   \cline{3-4}
   & & $T_1=-3$& $n_1=-3w_2,\,n_2=3w_1+2$ &  \\
   \cline{2-5}
   &$U_2\to0$ & $U_1=-\frac{1}{3}$& $n_1=-3n_2,\,w_2=3w_1+2$ & $\frac{1}{\sqrt{\alpha'T_2}}\left|n_1-Tw_2\right|\sqrt{U_2}$ \\
   \hline
   \multirow{1}{*}{$k=5$} & $T_2\to0$ & $T_1=0$& $n_1=n_2=0$ & $\frac{1}{\sqrt{\alpha'U_2}}\left|w_1+\frac{5}{6}+Uw_2\right|\sqrt{T_2}$ \\
   \hline
    \end{tabular}
\captionsetup{width=.9\linewidth}
\caption{\textit{All towers of states that become massless for $T_2\to0$ and $U$ fixed in the bulk of the moduli space, or $U_2\to 0$ and $T$ fixed in the bulk of the moduli space.}}
\label{cusptowers}
\end{table}
\renewcommand{\arraystretch}{1}

\noindent In \autoref{cusptowers}, we list all towers of states that become massless as $T_2\to0$ or $U_2\to 0$. Note that there are multiple towers of states becoming massless as $T_2\to 0$ or $U_2\to0$. By using \eqref{normalizedmoduli}, comparing \eqref{SDC} with the last column of \autoref{cusptowers} and identifying $|\Delta\phi|=|\phi_T|$ or $|\phi_U|$, we can see that for all towers $\lambda = \frac{1}{\sqrt{2}}$. Then, as $T_2\to0$ or $U_2\to 0$, that is $\phi_T\to -\infty$ or $\phi_U\to -\infty$, the masses of all towers decrease as
\begin{equation}
     m= m_0 e^{-\frac{1}{\sqrt{2}} |\phi_I|}\,,\qquad I=T,U\,.
\end{equation}
In the limit $T_2\to \infty$, that is $\phi_T\to \infty$, it is easy to see that there exist towers of KK states only in the untwisted sector whose masses decay as
\begin{equation}
  m=  \frac{1}{\sqrt{\alpha'U_2}}\left|n_2-Un_1\right| e^{-\frac{1}{\sqrt{2}} |\phi_T|}\,.
\end{equation}
Finally, in the limit $U_2\to \infty$, namely $\phi_U\to\infty$, there exist towers of states in all sectors whose masses decay exponentially fast. For example, in the untwisted sector we find towers with mass
\begin{equation}
     m=  \frac{1}{\sqrt{\alpha'T_2}}\left|n_2+Tw_1\right| e^{-\frac{1}{\sqrt{2}} |\phi_U|}\,.
\end{equation}
Hence, regarding the single cusps on the $T-U$ moduli space, the SDC is satisfied  due to the existence of all the aforementioned towers of states.

Finally, we mention that the Emergent String Conjecture \cite{Lee:2019wij} proposes that each infinite distance limit on the moduli space 
corresponds to either a decompactification limit in which an infinite tower of Kaluza-Klein modes become massless, or a limit in which a string becomes tensionless and an infinite tower of string modes become massless. For our compactification, we will demonstrate that the infinite distance limits on the $T-U$ moduli space are all decompactification limits.

\subsubsection*{Asymptotic supersymmetry enhancement}
In our 4-dimensional $\mathcal{N}=2$ theory, there are 6 massive gravitini and 2 massless gravitini. These gravitini are in the multiplets of the untwisted sector. The massless gravitini arise in the NS-R sector and survive the orbifold projection if $n_1=0$ mod $6$. The massive gravitini come from the NS-R and R-NS sector and carry non-trivial orbifold charge, as can be easily checked by using \autoref{untwisted stu states} and the formula \eqref{degeneracy untwisted}. This charge can be cancelled by the addition of an appropriate momentum number $n_1$, which makes the gravitini massive. This indicates that supersymmetry is spontaneously broken from $\mathcal{N}=8$ to $\mathcal{N}=2$. The weight vectors of the 6 massive gravitini and their corresponding momentum numbers are:
\begin{equation}\label{massivegravitini}
    \begin{aligned}
    & (\pm 1,0,0,0)\times (\pm\tfrac{1}{2},\pm\tfrac{1}{2},\tfrac{1}{2},\tfrac{1}{2}) :  \left(\pm\tfrac{3}{2},\pm\tfrac{1}{2}\right)\,,\quad & n_1=1\bmod 6\,,\\
    & (\pm 1,0,0,0)\times (\pm\tfrac{1}{2},\pm\tfrac{1}{2},-\tfrac{1}{2},-\tfrac{1}{2}) :  \left(\pm\tfrac{3}{2},\pm\tfrac{1}{2}\right)\,,\quad & n_1=5\bmod 6\,,\\
    & (\pm\tfrac{1}{2},\pm\tfrac{1}{2},\tfrac{1}{2},\tfrac{1}{2})\times (\pm 1,0,0,0) : \left(\pm\tfrac{3}{2},\pm\tfrac{1}{2}\right)\,,\quad & n_1=3\bmod 6\,,\\
    & (\pm\tfrac{1}{2},\pm\tfrac{1}{2},-\tfrac{1}{2},-\tfrac{1}{2})\times (\pm 1,0,0,0) :  \left(\pm\tfrac{3}{2},\pm\tfrac{1}{2}\right)\,,\quad & n_1=3\bmod 6\,,\\
       & (\pm\tfrac{1}{2},\mp\tfrac{1}{2},\tfrac{1}{2},-\tfrac{1}{2})\times (\pm 1,0,0,0) : \left(\pm\tfrac{3}{2},\pm\tfrac{1}{2}\right)\,,\quad & n_1=4\bmod 6\,,\\
    & (\pm\tfrac{1}{2},\mp\tfrac{1}{2},-\tfrac{1}{2},\tfrac{1}{2})\times (\pm 1,0,0,0) : \left(\pm\tfrac{3}{2},\pm\tfrac{1}{2}\right)\,,\quad & n_1=2\bmod 6\,. 
    \end{aligned}
\end{equation}
In summary, 1 gravitino survives the orbifold projection if $n_1=1$ mod $6$, 1 gravitino survives if $n_1=2$ mod $6$, 2 gravitini survive if $n_1=3$ mod $6$,  1 gravitino survives if $n_1=4$ mod $6$ and  1 gravitino survives if $n_1=5$ mod $6$.

The masses of the gravitini can be easily computed using \eqref{untwisted masses} and \eqref{momentauntwisted}. We find 
\begin{equation}
\begin{aligned}
    &\frac{\alpha'm^2_{L}(0)}{2}=\frac{1}{2}p^2_{L}(0)=\frac{1}{4T_2U_2}|n_2-Un_1 +\bar{T}w_1 +\bar{T}Uw_2|^2\,,\\
    &\frac{\alpha'm^2_{R}(0)}{2}=\frac{1}{2}p^2_{R}(0)=\frac{1}{4T_2U_2}|n_2-Un_1 +{T}w_1 +{T}Uw_2|^2\,.
    \label{gravitinimasses}
\end{aligned}
 \end{equation}
Also, the level-matching condition reads
\begin{equation}
    n_1w_1+n_2w_2=0\,.
    \label{levelgravitini}
\end{equation}
It is important to stress here that the quantum number $n_1$ in \eqref{gravitinimasses} and \eqref{levelgravitini} is constrained, according to \eqref{massivegravitini}.  

Now, as we approach the various cusps in the $T-U$ moduli space, all gravitini may remain massive, or some (or all) of them may become massless, as can be easily verified by using \eqref{massivegravitini}, \eqref{gravitinimasses} and \autoref{cusptowers}. In addition, all these infinite distance points can be interpreted as decompactification limits in type IIB theory or in a dual type IIA picture; this will be explained in detail below. This is consistent with the Gravitini Mass (Swampland) Conjecture \cite{Cribiori:2021gbf,Castellano:2021yye}. This conjecture proposes that a massive gravitino can only become massless with an infinite tower of states at an infinite distance on the moduli space, and the gravitino mass is proportional to some power of the KK (or string excitation) mass scale. The power is between 1 and 3; in our case, it is 1.

Moreover, at the single cusps the effective supergravity theory becomes six-dimensional. The massless spectrum follows immediately from the analysis of section \ref{z12 with nv=3 and nh=0}, with the only difference that all fields fall in representations of the massless little group $\text{SU}(2)\times \text{SU}(2)$ in $6D$.  For the supergravity multiplets in $6D$ and our conventions we refer to appendix \ref{6dsugra}.

The various infinite distance limits can be understood by studying the behaviour of the $T^2$ partition function (see section \ref{torpartition})  at all cusps. We first discuss the cusps of the modulus $T$ and we keep $U$ fixed in the bulk. For convenience, we set $U_1=0$. Then,  $T_2=\mathcal{R}_5\mathcal{R}_4/\alpha'$ and $U_2=\mathcal{R}_4/\mathcal{R}_5$. First we study the limit $T_2\to \infty$, that is $\mathcal{R}_5\to \infty$ and $\mathcal{R}_4\to \infty$. In this limit, all gravitini become massless and supersymmetry in enhanced from $\mathcal{N}=2$ to $\mathcal{N}=8$. Also, as we can see from \eqref{metricandb}, $g_{ij}\to \infty$ and the only term that contributes in the sum in \eqref{shift2gb} is the term with $n_1=n_2=w_1=w_2=k=l=0$. So, we find
\begin{equation}
  \lim_{T_2\to\infty}  {{Z}}_{T^2}[0,0]=  
   \frac{\mathcal{R}_4\mathcal{R}_5}{\alpha'\, {\tau_2}\,(\eta\,\bar{\eta})^2}\,,
\end{equation}
while for $k$ and/or $l\neq 0$ the limit is exponentially suppressed. Using that the string length $\ell_s$ is given by $\ell_s=2\pi\sqrt{\alpha'}$, we can rewrite the above expression as
\begin{equation}
   \lim_{T_2\to\infty}  {{Z}}_{T^2}[0,0]=  \frac{4\pi^2\mathcal{R}_4\mathcal{R}_5}{\ell_s^2\, {\tau_2}\,(\eta\,\bar{\eta})^2} = \frac{V}{\ell_s^2} \frac{1}{ {\tau_2}\,(\eta\,\bar{\eta})^2}\,,
   \label{freet2}
\end{equation}
where $V$ is the volume of a very large two-torus. Now, we recognize that the expression \eqref{freet2} is the properly normalized partition function of two non-compact bosons (see e.g. \cite{Blumenhagen:2013fgp}). Combining this result with the orbifold partition function \eqref{generic orbi par 4d}-\eqref{partition function 4d models}, we conclude that in the limit $T\to i\infty$, the resulting theory is type IIB on $\mathbb{R}^{1,5}\times T^4$. The massless fields make up the $\mathcal{N}=8$ gravity multiplet is $6D$.

We mention here that there is a subtlety regarding the volume of the torus that becomes very large. Since the orbifold partition function is divided by the orbifold rank $p$, the volume of the torus that decompactifies is actually $4\pi^2\mathcal{R}_4\mathcal{R}_5/p$. Moreover, it is interesting to note that the radius $\mathcal{R}_5$ on which the orbifold acts by a shift is related to the radius $R_5$ of the corresponding Scherk-Schwarz effective supergravity theory by $\mathcal{R}_5=pR_5$.

Now we focus on the limit $T_2\to 0$, i.e $\mathcal{R}_4\to 0$ and $\mathcal{R}_5\to 0$. Here, we have three different cusps, namely the cusps at $T=0,-2$ and $-3$. We start from the cusp $T_1=0$. In this case, all gravitini remain massive. Also, the partition function \eqref{shift2gb} reads
\begin{equation}
    {{Z}}_{T^2}[k,l]=\frac{\mathcal{R}_4\mathcal{R}_5}{\alpha'\, {\tau_2}\,(\eta\,\bar{\eta})^2}\sum_{\left\{n_i,w_i\right\} \in \mathbb{Z}^4}e^{ \frac{-\pi\mathcal{R}_5^2}{\alpha'\tau_2}\left|n_1-\frac{l}{p}+\left(w_1+\frac{k}{p}\right)\tau\right|^2}e^{ \frac{-\pi\mathcal{R}_4^2}{\alpha'\tau_2}|n_2+w_2\tau|^2}\,. 
    \label{factt2gb}
\end{equation}
By performing a multiple Poisson resummation over all momentum and winding numbers we can bring \eqref{factt2gb} to the equivalent form
\begin{equation}
     {{Z}}_{T^2}[k,l]=\frac{\alpha'}{\mathcal{R}_5\mathcal{R}_4}\frac{1}{\tau_2\,(\eta\,\bar{\eta})^2}\sum_{\left\{n_i,w_i\right\} \in \mathbb{Z}^4}e^{\frac{2\pi i }{p}(n_1l+w_1k)}e^{ \frac{-\pi\alpha'}{\mathcal{R}_5^2\tau_2}|w_1+n_1\tau|^2}e^{ \frac{-\pi\alpha'}{\mathcal{R}_4^2\tau_2}|w_2+n_2\tau|^2}\,. 
     \label{2tdualities}
\end{equation}
It is easy to verify that the above result could also be obtained by performing the T-duality transformation $\mathcal{R}_5\to \alpha'/\mathcal{R}_5$, $\mathcal{R}_4\to \alpha'/\mathcal{R}_4$, $n_1\leftrightarrow w_1$, $n_2\leftrightarrow w_2$. Note that this transformation changes the shift vector $v=(1/p,0,0,0)$ to $\tilde{v}=(0,0,1/p,0)$. As we can see from \eqref{2tdualities}, in the limit $\mathcal{R}_4\to 0$ and $\mathcal{R}_5\to 0$ all terms are exponentially suppressed except for the terms with $n_1=n_2=w_1=w_2=0$ and $k,l=0,\ldots,p$. In particular, we find that for all $k,l$
\begin{equation}
     \lim_{T_2\to 0}{{Z}}_{T^2}[k,l]=\frac{\alpha'}{\mathcal{R}_5\mathcal{R}_4}\frac{1}{\tau_2\,(\eta\,\bar{\eta})^2}\,,
\end{equation}
or, by defining the dual radii $\widetilde{\mathcal{R}}_5=\alpha'/\mathcal{R}_5$, $\widetilde{\mathcal{R}}_4=\alpha'/\mathcal{R}_4$
\begin{equation}
     \lim_{T_2\to 0}{{Z}}_{T^2}[k,l]=\frac{\widetilde{\mathcal{R}}_4\widetilde{\mathcal{R}}_5}{\alpha'\tau_2\,(\eta\,\bar{\eta})^2}\,.
\end{equation}
So, in the limit $T\to 0$ we obtain the partition function of two non-compact bosons, for all values of $k$ and $l$. Then, from \eqref{generic orbi par 4d}-\eqref{partition function 4d models} we can see that in the limit $T\to 0$ the theory becomes type IIB on a non-freely acting asymmetric orbifold $\mathbb{R}^{1,5}\times T^4/\mathbb{Z}_6$, characterized by the twist vectors  $\tilde{u}=\left(\frac{5}{6},\frac{1}{6}\right)$ and $u=\left(\frac{1}{6},\frac{1}{6}\right)$. The massless fields make up the $\mathcal{N}=2$ gravity multiplet in $6D$ coupled to $9$ tensor multiplets, $8$ vector multiplets and $20$ hypermultiplets, satisfying the gravitational anomaly cancellation condition $n_H-n_V=273-29n_T$.

We continue with the cusp $T_1=-2$, at which the two R-NS gravitini carrying momentum number $n_1 = 2\bmod 6$ and $n_1 = 4\bmod 6$ become massless and supersymmetry is enhanced from $\mathcal{N}=2$ to $\mathcal{N}=4$. For the analysis of this cusp we use the partition function given in \eqref{shiftedt2}-\eqref{momentabgtwistedapp}. After a bit of algebra, it is easy to verify that, in the limit $T_2\to 0$, all terms in the partition function are exponentially suppressed unless\footnote{Recall that we study a $\mathbb{Z}_6$ orbifold, hence $p=6$.} 
\begin{equation}
    n_1 = -2w_2\qquad \text{and}\qquad n_2 = 2w_1 +\frac{k}{3}\,.
    \label{t-2cusp}
\end{equation}
Note that \eqref{t-2cusp} has solutions only for $k=0$ and $3$. Now, by plugging \eqref{t-2cusp} back in \eqref{shiftedt2}-\eqref{momentabgtwistedapp} we obtain
\begin{equation}
   \lim_{\substack{T_2\to 0\\ T_1=-2}}{{Z}}_{T^2}[k,l] = \lim_{T_2\to 0} \frac{1}{(\eta\bar{\eta})^2}\sum_{\left\{w_1,w_2\right\} \in \mathbb{Z}^2}\,e^{\frac{-2\pi i lw_2 }{3}}e^{ {-\pi\alpha'\tau_2}(\hat{w}_1^2\mathcal{R}_5^2 + w_2^2\mathcal{R}_4^2)}\,.
\end{equation}
By performing a double Poisson resummation over the winding numbers $w_1$ and $w_2$ we find
\begin{equation}
    \lim_{\substack{T_2\to 0\\ T_1=-2}}{{Z}}_{T^2}[k,l] = \lim_{\substack{\mathcal{R}_5\to 0\\ \mathcal{R}_4\to 0}}\frac{\alpha'}{\mathcal{R}_5\mathcal{R}_4}\frac{1}{\tau_2\,(\eta\,\bar{\eta})^2}\sum_{\left\{w_1,w_2\right\} \in \mathbb{Z}^2}e^{\frac{\pi i k w_1}{3}}e^{ \frac{-\pi\alpha'}{\mathcal{R}_5^2\tau_2}w_1^2} e^{ \frac{-\pi\alpha'}{\mathcal{R}_4^2\tau_2}(w_2+\frac{l}{3})^2}\,.
\end{equation}
From this expression we see that all terms are exponentially suppressed unless 
\begin{equation}
    w_1=0\qquad \text{and} \qquad w_2+\frac{l}{3}=0\,,
\end{equation}
which is satisfied only for $l=0$ and $3$. Putting everything together, we conclude that if $[k,l]=[0,0],[0,3],[3,0]$ or $[3,3]$
\begin{equation}
     \lim_{\substack{T_2\to 0\\ T_1=-2}}{{Z}}_{T^2}[k,l]=\frac{\alpha'}{\mathcal{R}_5\mathcal{R}_4}\frac{1}{\tau_2\,(\eta\,\bar{\eta})^2}\,,
\end{equation}
while for all other values of $[k,l]$ the limit is exponentially suppressed. Now, from \eqref{generic orbi par 4d}-\eqref{partition function 4d models} we can see that we obtain type IIB on a non-freely acting symmetric orbifold $\mathbb{R}^{1,5}\times T^4/\mathbb{Z}_2$, characterized by the twist vectors  $\tilde{u}=u=\left(\frac{1}{2},\frac{1}{2}\right)$. The massless fields make up the $\mathcal{N}=4\,(0,2)$ gravity multiplet and 21 tensor multiplets in $6D$. The same spectrum could also be obtained from type IIB on $\mathbb{R}^{1,5}\times K3$. Also, note that the resulting number of tensor multiplets is exactly the number that is required for the gravitational anomalies to cancel in a chiral $\mathcal{N}=4\,(0,2)$ theory in 6$D$ \cite{townsend1984new}. 

Finally, the analysis of the cusp $T_1=-3$ is completely analogous to the case $T_1=-2$, so we omit the details. The two R-NS gravitini carrying momentum number $n_1 = 3 \bmod 6$ become massless and supersymmetry is enhanced from $\mathcal{N}=2$ to $\mathcal{N}=4$. At this cusp, we find that if $[k,l]=[0,0],[0,2],[0,4],[2,0],[2,2],[2,4],[4,0],[4,2]$ or $[4,4]$
\begin{equation}
     \lim_{\substack{T_2\to 0\\ T_1=-3}}{{Z}}_{T^2}[k,l]=\frac{\alpha'}{\mathcal{R}_5\mathcal{R}_4}\frac{1}{\tau_2\,(\eta\,\bar{\eta})^2}\,,
\end{equation}
while for all other values of $[k,l]$ the limit is exponentially suppressed. In this case, we obtain type IIB on a non-freely acting asymmetric orbifold
$\mathbb{R}^{1,5}\times T^4/\mathbb{Z}_3$, characterized by the twist vectors  $\tilde{u}=\left(-\frac{1}{3},\frac{1}{3}\right)$ and $u=\left(\frac{1}{3},\frac{1}{3}\right)$. The massless fields make up the $\mathcal{N}=4\,(1,1)$ gravity multiplet and 20 vector multiplets in $6D$.

The analysis of the cusps of the modulus $U$, with $T$ fixed in the bulk, proceeds in a similar way; a detailed discussion can be found in \cite{Gkountoumis:2025btc}. Alternatively, the results for the cusps of the modulus $U$ can be simply obtained from those of $T$ by performing the T-duality transformation $T\to 1/\bar{U}$. To be precise, let us denote the element of $\mathrm{O}(2,2;\mathbb{Z})$ that exchanges the moduli $T$ and $ U$ by $\gamma_e$; this element also exchanges type IIB with type IIA theory. Furthermore, we denote the coordinate reflection $Z_1\to -Z_1$, which acts on the moduli as $(T,U)\to (-\bar{T},-\bar{U})$, by $\gamma_r$. These two $\mathbb{Z}_2$'s act on a vector of the lattice $\Gamma^{2,2}$ as
\begin{equation}
   \gamma_e:\quad \begin{pmatrix}
        0&0&1&0\\
        0&-1&0&0\\
        1&0&0&0\\
        0&0&0&-1
    \end{pmatrix},\qquad \gamma_r:\quad \begin{pmatrix}
        -1&0&0&0\\
        0&1&0&0\\
        0&0&-1&0\\
        0&0&0&1
    \end{pmatrix}.
\end{equation}
Finally, we denote the transformation $(T,U)\to (-1/T,U)$ of $\mathrm{SL}(2,\mathbb{Z})_T$ as $\gamma_i$. Then, the T-duality element $\gamma=\gamma_i\gamma_e\gamma_r\gamma_i$ act as
\begin{equation}
     T\leftrightarrow \frac{1}{\bar{U}}, \quad U\leftrightarrow \frac{1}{\bar{T}},\qquad \begin{pmatrix}
        w_1\\ w_2\\n_1\\n_2
    \end{pmatrix}\leftrightarrow\begin{pmatrix}
        w_1\\ n_2\\n_1\\w_2
    \end{pmatrix}\,.
\end{equation}
Note that this duality element leaves the shift vector invariant, and it exchanges the cusps of $T$ and $U$ as follows:
\begin{equation}
    \begin{aligned}
        U\to 0 \qquad&\leftrightarrow&\quad T\to i\infty\,,\\
        T\to 0 \qquad&\leftrightarrow& U\to i\infty\,,\\
        U\to -\frac{1}{2} \qquad&\leftrightarrow& T\to-2\,,\\
        U\to -\frac{1}{3} \qquad&\leftrightarrow& T\to-3\,.
    \end{aligned}
\end{equation}
Finally, it is easy to see that $\text{det}(\gamma)=-1$. Hence, the T-duality element $\gamma$ exchanges type IIB with type IIA theory and it also changes the chirality of the right-moving Ramond vacuum; this is important for understanding the representations of the various supergravity fields in $6D$. So, we can see that the cusps of the modulus $U$ can be interpreted as various decompactification limits of the type IIA theory. In particular, we obtain the following results:
\begin{itemize}
      \item $U\to i\infty$: All gravitini remain massive. The theory becomes type IIA on a non-freely acting asymmetric orbifold $\mathbb{R}^{1,5}\times T^4/\mathbb{Z}_6$, characterized by the twist vectors  $\tilde{u}=\left(\frac{5}{6},\frac{1}{6}\right)$ and $u=\left(\frac{1}{6},\frac{1}{6}\right)$. The massless fields make up the $\mathcal{N}=2$ gravity multiplet in $6D$ coupled to $9$ tensor multiplets, $8$ vector multiplets and $20$ hypermultiplets, satisfying the gravitational anomaly cancellation condition $n_H-n_V=273-29n_T$.
    \item $U\to 0$: All gravitini become massless and supersymmetry is enhanced from $\mathcal{N}=2$ to $\mathcal{N}=8$. The resulting theory is  type IIA theory on $\mathbb{R}^{1,5}\times T^4$. The massless fields make up the $\mathcal{N}=8$ gravity multiplet is $6D$.
    \item $U\to -1/2$: The two R-NS gravitini carrying momentum number $n_1 = 2\bmod 6$ and $n_1 = 4\bmod 6$ become massless and supersymmetry is enhanced from $\mathcal{N}=2$ to $\mathcal{N}=4$. At this cusp we obtain type IIA on a non-freely acting symmetric orbifold $\mathbb{R}^{1,5}\times T^4/\mathbb{Z}_2$, characterized by the twist vectors $\tilde{u}=u=\left(\frac{1}{2},\frac{1}{2}\right)$. The massless fields make up the $\mathcal{N}=4\,(1,1)$ gravity multiplet and 20 vector multiplets in $6D$. The same spectrum could also be obtained from type IIA on $\mathbb{R}^{1,5}\times K3$. 
    \item $U\to -1/3$: The two R-NS gravitini carrying momentum number $n_1 = 3 \bmod 6$ become massless and supersymmetry is enhanced from $\mathcal{N}=2$ to $\mathcal{N}=4$. In this case, we get type IIA theory on a non-freely acting asymmetric orbifold
    $\mathbb{R}^{1,5}\times T^4/\mathbb{Z}_3$, characterized by the twist vectors  $\tilde{u}=\left(-\frac{1}{3},\frac{1}{3}\right)$ and $u=\left(\frac{1}{3},\frac{1}{3}\right)$. The massless fields make up the $\mathcal{N}=4\,(2,0)$ gravity multiplet and 21 tensor multiplets in $6D$, ensuring that gravitational anomalies cancel.
\end{itemize}

\subsubsection{Double cusps }
So far, we have discussed the single cusps on the $T-U$ plane of the moduli space. However, there are limits on the boundary of both spaces: the double cusps. The analysis of these double cusps is similar to the analysis of the single cusps. Hence, we will omit most of the details and we will simply present our results, which we collect in \autoref{doublecusps2}. Note that at all double cusps the effective supergravity theory becomes five-dimensional. Regarding our conventions for the massless spectra at each of the double cusps we refer to chapter \ref{chap:spectrum}.

\renewcommand{\arraystretch}{1.7}
\begin{table}[t!]
\centering
{\small
 \begin{tabular}[c]{|c|c|c|c|c|c|}
    \hline
    $T$ & $U$ & $\alpha'm^2$ & Compactification & Supersymmetry & Massless spectrum \\
    \hline
    \hline
  $ 0 $& $0$ &\multirow{9}{*}{$w_2^2T_2U_2$} &  $\left(T^4\times S^1\right)/\mathbb{Z}_6$ & $\mathcal{N}=2$ IIA & 
 $1\text{GM}+2\text{VM}+x\text{HM}$  \\
  \cline{1-2}\cline{4-6}
  $0$ & $-\frac{1}{2}$ & & $\left(T^4\times S^1\right)/\mathbb{Z}_6$, $v=\left(0,\frac{1}{3}\right)$  & $\mathcal{N}=2$ IIA & $1\text{GM}+8\text{VM}+7\text{HM}$ \\
  \cline{1-2}\cline{4-6}
  $0$ & $-\frac{1}{3}$ & & $\left(T^4\times S^1\right)/\mathbb{Z}_6$, $v=\left(0,\frac{1}{2}\right)$ & $\mathcal{N}=2$ IIA & $1\text{GM}+8\text{VM}+10\text{HM}$   \\
  \cline{1-2}\cline{4-6}
  $-2$ & $0$ & & $ \left(T^4\times S^1\right)/\mathbb{Z}_2$ & $\mathcal{N}=4\,(1,1)$ IIA & $1\text{GM}+5\text{VM}$  \\
  \cline{1-2}\cline{4-6}
  $-3$ & $0$ & & $ \left(T^4\times S^1\right)/\mathbb{Z}_3$ & $\mathcal{N}=4\,(2,0)$ IIA & $1\text{GM}+3\text{VM}$   \\
  \cline{1-2}\cline{4-6}
  $-2$ & $-\frac{1}{2}$ & & $T^4/\mathbb{Z}_2\times S^1$ & $\mathcal{N}=4\,(1,1)$ IIA & $1\text{GM}+21\text{VM}$   \\
  \cline{1-2}\cline{4-6}
  $-3$ & $-\frac{1}{3}$ & & $T^4/\mathbb{Z}_3\times S^1$ & $\mathcal{N}=4\,(2,0)$ IIA & $1\text{GM}+21\text{VM}$   \\
  \cline{1-2}\cline{4-6}
  $-2$ & $-\frac{1}{3}$ & & \multirow{2}{*}{$\left(T^4\times S^1_{\mathcal{R}_5=\sqrt{6}}\right)/\mathbb{Z}_6$} & \multirow{2}{*}{$\mathcal{N}=2$ IIA} & \multirow{2}{*}{$1\text{GM}+2\text{VM}+4\text{HM}$}  \\
  \cline{1-2}
  $-3$ & $-\frac{1}{2}$ & &  &  &    \\
  \hline
  $i\infty$ & $i\infty$ & {$\frac{n_2^2}{T_2U_2}$} & $\left(T^4\times S^1\right)/\mathbb{Z}_6$ & {$\mathcal{N}=2$ IIB} &
 $1\text{GM}+2\text{VM}+x\text{HM}$   \\
    \hline
  $i\infty$ & $0$ & \multirow{3}{*}{$\frac{n_1^2U_2}{T_2}$} & $T^5$ & $\mathcal{N}=8$ IIB & $1\text{GM}$ \\
  \cline{1-2}\cline{4-6}
  $i\infty$ & $-\frac{1}{2}$ &  &$ \left(T^4\times S^1\right)/\mathbb{Z}_2$  & $\mathcal{N}=4\,(0,2)$ IIB & $1\text{GM}+5\text{VM}$  \\
  \cline{1-2}\cline{4-6}
  $i\infty$ & $-\frac{1}{3}$ &  & $\left(T^4\times S^1\right)/\mathbb{Z}_3$  & $\mathcal{N}=4\,(1,1)$ IIB & $1\text{GM}+3\text{VM}$  \\
  \hline
  $0$ & $i\infty$ & \multirow{3}{*}{$\frac{\hat{w}_1^2T_2}{U_2}$} & $T^4/\mathbb{Z}_6\times S^1$ & $\mathcal{N}=2$ IIA & $1\text{GM}+18\text{VM}+20\text{HM}$  \\
  \cline{1-2}\cline{4-6}
  $-2$ & $i\infty$ &  & $\left(T^4\times S^1\right)/\mathbb{Z}_6$, $v=\left(0,\frac{1}{3}\right)$ & $\mathcal{N}=2$ IIA & $1\text{GM}+8\text{VM}+7\text{HM}$  \\
  \cline{1-2}\cline{4-6}
  $-3$ & $i\infty$ &  & $\left(T^4\times S^1\right)/\mathbb{Z}_6$, $v=\left(0,\frac{1}{2}\right)$  & $\mathcal{N}=2$ IIA & $1\text{GM}+8\text{VM}+10\text{HM}$  \\
   \hline
    \end{tabular}}
\captionsetup{width=.9\linewidth}
\caption{\textit{Here we present our results for all double cusps in the $T$-$U$ moduli space. First, we list the double-cups. Then we specify the masses of the towers that become massless and the resulting decompactified theory at the corresponding cusp. In all cases, there are five non-compact directions, $\mathbb{R}^{1,4}$, and five compact directions. Here, GM stands for gravity multiplet, VM stands for vector multiplet, and HM stands for hypermultiplet. Regarding the massless spectrum at the cusps $(T,U)=(0,0)$ and $(i\infty, i\infty)$, if  $T_2/U_2=6$, $x=4$, if $T_2/U_2=12$, $x=3$ and for all other cases $x=2$. Finally, we specify the shift vectors for some freely acting orbifold limits, of which the shift along the circle is not inversely proportional to the orbifold rank.}}
\label{doublecusps2}
\end{table}
\renewcommand{\arraystretch}{1}

There are four $T$-cusps and four $U$-cusps, and by combination there are 16 double-cusp infinite distance limits on the moduli space. The masses of the towers of states that become massless as we approach each double cusp can be derived from \eqref{mkfinal}. As an example, we consider the limit $T\to i\infty$ and $U\to -1/2$. In this case, we find a KK tower with mass 
\begin{equation}\label{m=nU/T}
    m^2 = \frac{n_1^2 U_2}{\alpha' T_2} =\frac{n_1^2}{\alpha' }e^{\sqrt{2}(\phi_U-\phi_T)}\propto e^{-2|\bm{\phi}|},
\end{equation}
for $\hat{w}_1=w_2=0$, and $n_1=-2n_2$. These constraints on the lattice of momenta and windings imply that this tower appears only in the untwisted sector, that is $k=0$, and if $n_1\in 2\mathbb{Z}$. Furthermore, there are two gravitini carrying momentum number $n_1\in 2\mathbb{Z}$, which become massless at this double cusp limit. Hence, supersymmetry is enhanced from $\mathcal{N}=2$ to $\mathcal{N}=4$ in $5D$. We collect all information about the masses of the towers that become zero, the constraints on the momentum and winding numbers and the sectors in which massless towers appear at each of the 16 double cusps in \autoref{doublecusps}.

Note that at the cusps $(T,U)=(-2,-1/3),\,(-3,-1/2)$, there are two additional towers of hypermultiplets that become massless, since we are exactly at the critical line $\frac{T}{6}=U$ (cf. \eqref{Tmod6U}).  Regarding the cusps $(T,U)=(0,0),\,(i\infty,i\infty)$, if the  ratio $T_2/U_2$  as we approach the cusp is 12, we obtain one extra tower of massless hypermultiplets (cf.  \eqref{t12branch}), and if the ratio is 6 we get two towers of massless hypermultiplets (cf. \eqref{Tmod6U}). 

\renewcommand{\arraystretch}{1.7}
\begin{table}[t!]
\centering
 \begin{tabular}[c]{|c|c|c|c|c|c|}
    \hline
    $T$ & $U$ & $\alpha'm^2$ & Lattice constraints & $n_1(\bmod 6)$ & Massless sectors \\
    \hline
    \hline
  $ 0 $& $0$ &\multirow{9}{*}{$w_2^2T_2U_2$} & $n_1=n_2=\hat{w}_1=0$ & $0$ &$0\, (1,5)$  \\
  \cline{1-2}\cline{4-6}
  $0$ & $-\frac{1}{2}$ & & $n_1=-2n_2=0,2\hat{w}_1=w_2$ & $0$ & $0,3$ \\
  \cline{1-2}\cline{4-6}
  $0$ & $-\frac{1}{3}$ & & $n_1=-3n_2=0,3\hat{w}_1=w_2$ & $0$ & $0,2,4$   \\
  \cline{1-2}\cline{4-6}
  $-2$ & $0$ & & $2\hat{w}_1=n_2=0,n_1=-2w_2$ & $0,2,4$ & $0$  \\
  \cline{1-2}\cline{4-6}
  $-3$ & $0$ & & $3\hat{w}_1=n_2=0,n_1=-3w_2$ & $0,3$ & $0$   \\
  \cline{1-2}\cline{4-6}
  $-2$ & $-\frac{1}{2}$ & & $-n_1=2n_2=4\hat{w}_1=2w_2$ & $0,2,4$ & $0,3$   \\
  \cline{1-2}\cline{4-6}
  $-3$ & $-\frac{1}{3}$ & & $-n_1=3n_2=9\hat{w}_1=3w_2$ & $0,3$ & $0,2,4$   \\
  \cline{1-2}\cline{4-6}
  $-2$ & $-\frac{1}{3}$ & & $-n_1=3n_2=6\hat{w}_1=2w_2$ & \multirow{2}{*}{$0$} & \multirow{2}{*}{$0,1,5$}  \\
  \cline{1-2}\cline{4-4}
  $-3$ & $-\frac{1}{2}$ & & $-n_1=2n_2=6\hat{w}_1=3w_2$ &  &    \\
  \hline
  $i\infty$ & $i\infty$ & $\frac{n_2^2}{T_2U_2}$ & $n_1=\hat{w}_1=w_2=0$ & 0 & $0\,(1,5)$   \\
  \hline
  $i\infty$ & $0$ & \multirow{3}{*}{$\frac{n_1^2U_2}{T_2}$} & $\hat{w}_1=w_2=n_2=0$ & All & 0  \\
  \cline{1-2}\cline{4-6}
  $i\infty$ & $-\frac{1}{2}$ &  & $2\hat{w}_1=w_2=0,n_1=-2n_2$ & $0,2,4$ & 0  \\
  \cline{1-2}\cline{4-6}
  $i\infty$ & $-\frac{1}{3}$ &  & $3\hat{w}_1=w_2=0,n_1=-3n_2$ & $0,3$ & 0  \\
  \hline
  $0$ & $i\infty$ & \multirow{3}{*}{$\frac{\hat{w}_1^2T_2}{U_2}$} & $n_1=w_2=n_2=0$ & 0 & All  \\
  \cline{1-2}\cline{4-6}
  $-2$ & $i\infty$ &  & $n_1=-2w_2=0,n_2=2\hat{w}_1$& 0 & $0,3$  \\
  \cline{1-2}\cline{4-6}
  $-3$ & $i\infty$ &  & $n_1=-3w_2=0,n_2=3\hat{w}_1$ & 0 & $0,2,4$  \\
   \hline
    \end{tabular}
\captionsetup{width=.9\linewidth}
\caption{\textit{Here we list the masses of towers that become massless, the constraints on the lattice of momenta and windings and the sectors in which massless states appear at each double cusp. For the cusps $(T,U)=(0,0)$ and $(i\infty,i\infty)$, there are extra towers of massless hypermultiplets arising from the $k=1$ and $5$ sectors if $T=6U$ or $T=12U$ asymptotically.}}
\label{doublecusps}
\end{table}
\renewcommand{\arraystretch}{1}

Moreover, at the cusps $(T,U)=(0,-1/2),\,(0,-1/3),\,(-2,i\infty),\,(-3,i\infty)$ the theory decompactifies to an orbifold of $\mathbb{R}^{1,4}\times S^1 \times T^4$, which acts as a rotation of order $6$ on the torus and as a shift of order $2$ or $3$ on the circle. The fact that the shift is not of order $6$ has important implications for the spectrum of the orbifold. Consider for example the case in which the rotation on the torus is of order $6$ and the shift on the circle is of order $3$. In this orbifold, the winding number along the circle direction will be shifted as $w\to w+k/3$, which implies that states in the $k=3$ twisted sector will not feel the shift. Consequently, there will be massless states coming from this twisted sector. Furthermore, states with momentum number $n$ along the circle direction will pick up a phase $e^{2\pi i n /3}$. So, states with orbifold charge $e^{\pi i  /3}, e^{\pi i }$ or $e^{5\pi i /3}$ will be projected out of the spectrum, since such orbifold charge cannot be cancelled by adding momentum along the circle direction. The situation is similar if the shift along the circle is of order $2$. In this case, states in the $k=2$ and $4$ sectors will not feel the shift, and states with orbifold charge $e^{\pi i  /3}, e^{2\pi i  /3}, e^{4\pi i  /3}$ or $e^{5\pi i /3}$ will be projected out of the spectrum.

As in the example \eqref{m=nU/T}, the masses of all towers that become massless at the double cusps are proportional to both $\exp{\left(\pm\frac{1}{\sqrt{2}}\phi_T\right)}$ and $\exp{\left(\pm\frac{1}{\sqrt{2}}\phi_U\right)}$. Hence, the masses of the towers decrease exponentially with $\lambda=1$ (see \eqref{SDC}), and the SDC is verified also in the case of double cusps in the $T-U$ moduli space.

Finally, we mention here that, in addition to the distance conjecture, there 
are conjectures that the volume of moduli space should be finite or that its asymptotic growth be restricted
 \cite{Ooguri:2006in}. For a recent discussion on this and the relation to dualities see
\cite{Delgado:2024skw,Grimm:2025lip}. 
The volume of the classical moduli space 
for our model is indeed finite, 
 because the hypermultiplet moduli space is a Narain moduli space with a finite volume and the vector multiplet moduli space is the triple product of the fundamental domains of $\hat{\Gamma}^1(6)$ (or $\hat{\Gamma}_1(6)$). The index of $\hat{\Gamma}^1(6)$ is $[\mathrm{SL}(2,\mathbb{Z}):\hat{\Gamma}^1(6)]=12$ (same as $\hat{\Gamma}_1(6)$), so that the volume of the classical vector multiplet moduli space is $(12\text{Vol}(\mathbb{H}/\mathrm{SL}(2,\mathbb{Z})))^3=64\pi^3$.

\chapter{D-branes}
\label{chap:branes}

In this chapter, we determine the conditions for the existence of various D-brane configurations in the freely acting orbifolds, which we discussed in chapters \ref{chap:fao}-\ref{chap:mod5}. First, in section \ref{stapproach} we study this issue using a string-theoretic approach.  Then, in section \ref{branes from supergravity} we perform a supergravity analysis. Also, we show that the two approaches are consistent with each other, as expected. Of course, for freely acting orbifolds, the supergravity analysis only applies to the untwisted orbifold sector, which is the sector in which we will mostly focus on. Finally, in section \ref{sec:branesupersymmetry} we examine if the supersymmetry preserved by the D-branes, is compatible with the supersymmetry preserved by the orbifolds. We mention that the analysis of the upcoming sections can be easily adapted and applied to the orbifolds of chapter \ref{chap:swampland}.

\section{String-theoretic approach}
\label{stapproach}

Before discussing the orbifolds, we recall here the D-brane spectrum of type IIB string theory compactified on $T^5$. The low-energy theory is $D=5$, $\mathcal{N}=8$ supergravity and contains (among other fields) 16 R-R gauge vector fields. These vectors charge the following D-branes: five D1-branes wrapping a one-cycle of $T^5$, ten D3-branes wrapping a three-cycle of $T^5$, and one D5-brane wrapping the $T^5$. Note that all the above branes are pointlike in five dimensions and hence, they are properly charged under vector fields. 

In the orbifolded theory, some (or all) of the 16 R-R vectors will remain massless and the corresponding D-branes, charged under those vectors, will be preserved in the orbifold. In the next, we will determine the various D-brane configurations that survive in our orbifolds, using the boundary state formalism.

\subsection{Boundary states}
\label{branes from string theory}

Here, we will argue for the existence of Dp-branes using the boundary state formalism, originally developed in \cite{Callan:1986bc,Polchinski:1987tu,Ishibashi:1988kg}. In this framework, a Dp-brane is represented by a linear combination of boundary states. We will focus mostly on boundary states in the untwisted orbifold sector, which lead to Dp-branes charged only under untwisted R-R fields \cite{Brunner:1999ce}. The boundary state formalism can be applied in the twisted sectors as well, but in our constructions there are no massless R-R fields in the twisted sectors because the orbifolds are freely acting, and the states coming from the twisted sectors are generically massive. In the next, we will discuss the boundary states briefly and only mention those aspects that are essential for the purposes of this thesis; for more details see e.g. \cite{Bergman:1997rf,DiVecchia:1997vef,Diaconescu:1999dt,DiVecchia:1999mal,DiVecchia:1999fje,Gaberdiel:2000jr}.

Consider first a $10D$ flat non-compact spacetime, with coordinates $X^0,X^1,\ldots X^9$. In such spacetime, an open string ending on a Dp-brane has Neumann boundary conditions ($\partial_{\sigma^2} X=0$) in the extended worldvolume directions of the brane, which we label by $a$, and Dirichlet boundary conditions ($\partial_{\sigma^1} X=0$) in the transverse directions of the brane, which we label by $I$. We  work in lightcone gauge and, following \cite{Green:1996um}, we first perform a Wick rotation $X^{0}\to iX^{0}$, so that we can treat all parallel directions to the brane as spacelike\footnote{All D-branes that we consider in this thesis are extended along the $X^0$ direction.}. Then, we perform another Wick rotation $X^{1}\to iX^{1}$ and we take the lightcone directions to be $X^{\pm}=\tfrac{1}{\sqrt{2}}\left(X^1\pm X^2\right)$, on which we always impose Dirichlet boundary conditions.   

Now, a boundary state, denoted by $\ket{\text{Bp},\eta}$, is constructed such that it satisfies the following conditions 
\begin{equation}
    \begin{aligned}
    p^a\ket{\text{Bp},\eta}_s&=0\,,\qquad a=0,3,\ldots p+2\,, \\
    (x^I-x_0^I)\ket{\text{Bp},\eta}_s&=0\,,\qquad I=+,-, p+3,\ldots ,9\,, \\
    (\alpha_n^{a} + \tilde{\alpha}_{-n}^{a})\ket{\text{Bp},\eta}_s&=0\,,\qquad a=0,3,\ldots p+2\,,\\
    (\alpha_n^{I} - \tilde{\alpha}_{-n}^{I})\ket{\text{Bp},\eta}_s&=0\,,\qquad I= p+3,\ldots ,9\,,\\
    (b_r^{a} + i\eta\tilde{b}_{-r}^{a})\ket{\text{Bp},\eta}_s&=0\,,\qquad a=0,3,\ldots p+2\,,\\
    (b_r^{I} - i\eta\tilde{b}_{-r}^{I})\ket{\text{Bp},\eta}_s&=0\,,\qquad I=p+3,\ldots ,9\,.
    \end{aligned}
    \label{gluing conditions}
\end{equation}
Here, $s\in\{\text{NS},\text{R}\}$  runs over the NS-NS and R-R sectors, where $r\in \mathbb{Z}+\tfrac{1}{2}$ and $r\in \mathbb{Z}$, respectively. Also, $n\in \mathbb{Z}$, and $\eta=\pm 1$ labels the different spin structures of the world-sheet, needed to implement the GSO projection. Moreover, $p^a$ is the momentum operator in the Neumann directions, $x^I$ is the position operation in the Dirichlet directions and $x^I_0$ denotes the position of the brane in the Dirichlet directions. We recall here that we denote the bosonic and fermionic oscillators by $\alpha_n$ and $b_r$, respectively. In addition, a tilde indicates a left-mover.

The set of equations \eqref{gluing conditions} can be solved by means of boundary states\footnote{Here and below, we leave out normalization factors for the boundary states to keep the presentation simple. These factors are important to show open-closed string world-sheet duality and can be easily included. A pedagogical reference is e.g. \cite{Gaberdiel:2000jr}.}
\begin{equation}
    \begin{aligned}
        \ket{\text{Bp},\eta}_{s} \sim\int  \text{d}k_I \, e^{ik_I x_0^I}\,\text{exp}\Bigg\{&\sum_{n>0}\frac{1}{n}\left(-\alpha^{a}_{-n}\tilde{\alpha}^{a}_{-n}+\alpha^I_{-n}\tilde{\alpha}^I_{-n}\right) +\\ i\eta &\sum_{r>0}\left(-b^{a}_{-r}\tilde{b}^{a}_{-r}+b^I_{-r}\tilde{b}^I_{-r} \right)\Bigg\} \ket{\eta,k_I}_{s}\,,
    \end{aligned}
    \label{boundary state definition}
\end{equation}
where a multiple integration over $\text{d}k_I$, with $I=+,-, p+3,\ldots ,9$, is performed. Also, summation over repeated indices is implied, and $\ket{\eta,k_I}_{s}$ denotes the vacuum carrying transverse momenta $k_I$. 

We should highlight here that in the NS-NS sector the vacuum is a scalar and it is unique. However, in the R-R sector the vacuum is a spinor, it is degenerate and it should also solve \eqref{gluing conditions} for the fermionic zero modes,  namely
\begin{equation}
    \begin{aligned}
       (b_0^{a} + i\eta\tilde{b}_{0}^{a})\ket{\eta,k_I}_{\text{R}}&=0\,,\\
    (b_0^{I} - i\eta\tilde{b}_{0}^{I})\ket{\eta,k_I}_{\text{R}}&=0\,. 
    \end{aligned}
    \label{gluing zero modes}
\end{equation}
Later, we will construct explicitly $\ket{\eta,k_I}_{\text{R}}$ for specific examples. Now, a boundary state $\ket{\text{Dp}}$ corresponding to a stable Dp-brane is a GSO invariant\footnote{The GSO projection operator $(-1)^F$ acts in the NS-NS sector as $(-1)^F|\text{Bp},\eta\rangle_{\text{NS}} = -|\text{Bp},-\eta\rangle_{\text{NS}}$, as the NS-vacuum carries odd fermion number. $(-1)^{\tilde{F}}$ acts in the same way. In the R-R sector we have $(-1)^F|\text{Bp},\eta\rangle_{\text{R}} = |\text{Bp},-\eta\rangle_{\text{R}}$ and $(-1)^{\tilde{F}}|\text{Bp},\eta\rangle_{\text{R}} = |\text{Bp},-\eta\rangle_{\text{R}}$, provided that we take p to be even in type IIB.} linear combination of the boundary states $\ket{\text{Bp},\eta}_s$  of the form 
\begin{equation}
    \ket{\text{Dp}} = \left(\ket{\text{Bp},+} - \ket{\text{Bp},-}\right)_{\text{NS}}\: \pm \:\left(\ket{\text{Bp},+} + \ket{\text{Bp}, -}\right)_{\text{R}}\,.
    \label{complete boundary state}
\end{equation}
This particular linear combination of both NS-NS and R-R GSO invariant boundary states in $\ket{\text{Dp}}$ ensures open-closed string world-sheet duality and that the spectrum of an open string starting and ending on a Dp-brane is supersymmetric and free of tachyons. This makes the brane stable and BPS. Also, for type IIB string theory, GSO invariance in the R-R sector leads to odd values of p. Finally, we choose conventions such that the plus sign in \eqref{complete boundary state} corresponds to a brane and the minus sign to an anti-brane.

So far, we only discussed type IIB in flat $10D$ spacetime, where Dp-branes with p odd are stable and BPS. However, in orbifold compactifications it is possible to also have stable non-BPS branes, such as in type 0 string theories \cite{Klebanov:1998yya}, or the non-BPS D-particle \cite{Bergman:1998xv}. This is because the orbifold might break the supersymmetries that the D-brane preserves. 

In order to discuss boundary states in compact spacetimes, let us first consider one compact dimension and see how the boundary state \eqref{boundary state definition} is modified.  Suppose that we compactify $X^5=Z$ on a circle of radius $\mathcal{R}$. For Dirichlet boundary conditions in the circle direction, the transverse momentum is quantized, $k_5=n_5/\mathcal{R}$, with $n_5\in\mathbb{Z}$, and in \eqref{boundary state definition} we have to replace\footnote{Again, we omit normalization factors.}
\begin{equation}
    \int \text{d}k_5 \, e^{ik_5 x_0^5} \longrightarrow \sum_{n_5 \in \mathbb{Z}} e^{i n_5x_0^5/\mathcal{R}}\, \qquad \text{and} \qquad \ket{\eta,k_I}_{s} \longrightarrow \ket{\eta,k_{I'},n_5}_{s}\,,\quad I'=I\setminus\{5\}\,.
\end{equation}
For Neumann boundary conditions in the circle direction, the boundary state \eqref{boundary state definition} should be supplemented by a sum over windings 
\begin{equation}
    \sum_{w^5 \in \mathbb{Z}} e^{i w^5\mathcal{R}\tilde{x}_0^5/\alpha'}\,,
\end{equation}
and the vacuum should also be characterized by the winding number $w^5$. Here $\tilde{x}_0^5$ has the interpretation of a Wilson line on the brane wrapped around the circle; this is the T-dual description of the position of a D-brane along the compact direction (see e.g. \cite{blumenhagen2012basic} for more details). 

It will be also useful to examine how the orbifold $S^1/\mathbb{Z}_2$ acts on $x_0^5$ and $\tilde{x}_0^5$. Consider first the freely acting orbifold, acting as $Z\to Z + \pi \mathcal{R}$. For Neumann boundary condition in the circle direction, this orbifold has a trivial action on $\tilde{x}_0^5$. On the other hand, for Dirichlet boundary conditions, it sends $x_0^5\to x_0^5 + \pi \mathcal{R}$. For the non-freely acting orbifold, acting as $Z\to -Z$, we have $\tilde{x}_0^5\to -\tilde{x}_0^5$  and $x_0^5\to -x_0^5$ for Neumann and Dirichlet boundary condition, respectively (see also \cite{Kawai:2007qd,Becker:2017zai}).

In the rest of this chapter, we will consider Dp-branes in backgrounds of the form $\mathbb{R}^{1,4}\times  S^1\times T^4$ identified under the action of a $\mathbb{Z}_p$ symmetry. The generalization of the above discussion in the $S^1\times T^4$ background is straightforward if the $T^4$ metric is diagonal and the two-form $B$-field is zero (see e.g. \cite{Brunner:1999fj,Gaberdiel:1999ch}). For more general background fields on $T^4$ the construction of boundary states needs slight modification. In particular, in the presence of a non-trivial $B$-field one encounters the so called \say{magnetized} branes (for more details see e.g. \cite{Li:1995pq,Elitzur:1998va, DiVecchia:1999uf, Blumenhagen:2000fp, Pesando:2005df, DiVecchia:2006qeb,DiVecchia:2007dh}).

\subsection{Examples}
\label{examples of boundary states}
In this section, we will explicitly construct boundary states on orbifolds of $\mathbb{R}^{1,4}\times  S^1\times T^4$, using the techniques developed in section \ref{branes from string theory}. We will focus on Dp-branes with p odd and with p Neumann boundary conditions along the compactified directions. The corresponding boundary states are by construction GSO invariant. In the following we will examine under which conditions these boundary states are orbifold invariant. For the examples that we will discuss we choose a diagonal $T^4$ metric and we set the $B$-field to zero. Also, for our notation and conventions regarding the world-sheet fields and the orbifold action we refer to chapter \ref{chap:fao} (see section \ref{sec:the world-sheet fields}).

\subsection*{D5-brane}

Let us start our discussion with a boundary state that describes a D5-brane wrapping the $T^5$. From now on, for convenience, we will not write down the integrals over the transverse momenta, we will omit indicating the dependence of the vacuum on the transverse momenta in the non-compact directions and we will also omit writing down the oscillator part of the boundary state that is not affected by the orbifold action\footnote{Note that the shift along the circle coordinate leaves the corresponding oscillators intact.}.  That being said, for the D5-brane we have
\begin{equation}
\begin{aligned}
    \ket{\text{B5},\eta}_{s}\sim f(w,\tilde{x})\,\text{exp}\Bigg\{&\sum_{n>0}\frac{1}{n}\left(-\sum_{a=6}^9\alpha^{a}_{-n}\tilde{\alpha}^{a}_{-n}\right) +
    i\eta \sum_{r>0}\left(-\sum_{a=6}^9b^{a}_{-r}\tilde{b}^{a}_{-r} \right)\Bigg\} \ket{\eta,w^5,w^l}_{s}\,.
    \end{aligned}
\end{equation}
where $f(w,\tilde{x})$ encodes the contribution of the bosonic zero modes to the boundary state along the compact directions and is given by
\begin{equation}
   f(w,\tilde{x}) =\sum_{w^5 \in \mathbb{Z}}e^{i w^5\mathcal{R}\tilde{x}_0^5/\alpha'}\prod_{l=6}^9\sum_{w^l \in \mathbb{Z}^4} e^{i w^l{R}_l\tilde{x}_0^l/\alpha'} \,.
\end{equation}
In order to make the orbifold action manifest, we rewrite $\ket{\text{B5},\eta}_{s}$ in terms of complex oscillators as
\begin{equation}
    \begin{aligned}
      \ket{\text{B5},\eta}_{s} \sim f(w,\tilde{x})\, \text{exp}\Bigg\{&\sum_{n>0}\frac{1}{n}\left(-\alpha^{i}_{-n}\bar{\tilde{\alpha}}^{i}_{-n}-\bar{\alpha}^i_{-n}\tilde{\alpha}^i_{-n}\right) 
      +\\
      i\eta &\sum_{r>0}\left(-b^{i}_{-r}\bar{\tilde{b}}^{i}_{-r}-\bar{b}^i_{-r}\tilde{b}^i_{-r} \right)\Bigg\} \ket{\eta,w^5,w^l}_{s}\,,  
    \end{aligned}
    \label{D5 boundary state}
\end{equation}
where the index $i=1,2$ labels the complex torus coordinates and the corresponding complex oscillators. Now, let us first focus on the oscillator part of the boundary state and postpone the discussion of the zero modes. From \eqref{orbiaction2} we can immediately see that the exponential in \eqref{D5 boundary state} is orbifold invariant if
\begin{equation}
    \begin{aligned}
          m_1 +m_3-m_2-m_4=0  \quad &\text{mod} \quad 2\pi\,,\\
          m_1 -m_3-m_2+m_4=0  \quad &\text{mod} \quad 2\pi\,.
    \end{aligned}
    \label{D5 boundary state mass conditions}
\end{equation}
Let us now discuss the orbifold action on the vacua $\ket{\eta}_s$; for now, we have suppressed the dependence of the vacua on $w^5,w^l$. In the NS-NS sector the vacuum is a scalar and is invariant under the orbifold action. However, in the R-R sector the vacuum is a spinor, it transforms under the orbifold action and it is should also solve \eqref{gluing zero modes}. For the particular boundary state that we are discussing, we have
\begin{equation}
    \begin{aligned}
       (b_0^{a} + i\eta\tilde{b}_{0}^{a})\ket{\eta}_{\text{R}}&=0\,,\quad a=0,5,6,7,8,9\,,\\
    (b_0^{I} - i\eta\tilde{b}_{0}^{I})\ket{\eta}_{\text{R}}&=0\,,\quad I=3,4. 
    \end{aligned}
    \label{zero modes gluing for D5}
\end{equation}
Following the analysis of \cite{Diaconescu:1999dt}, we will rewrite the above equations in terms of creation and annihilation operators $b_i^{\pm}, i=0,\ldots,3$, where
\begin{equation}
   b_0^{\pm}=\frac{1}{\sqrt{2}}(b_0^0\pm ib_0^5)\,,\quad b_1^{\pm}=\frac{1}{\sqrt{2}}(b_0^3\pm ib_0^4) \,,\quad b_2^{\pm}=\frac{1}{\sqrt{2}}(b_0^6\pm ib_0^7)\,,\quad b_3^{\pm}=\frac{1}{\sqrt{2}}(b_0^8\pm ib_0^9)\,.
   \label{creation-annihilation operators}
\end{equation}
These oscillators satisfy the anti-commutation relations
\begin{equation}
    \{b_i^+, b^-_j\}=\delta_{ij}\,,\qquad  \{b_i^{\pm},b_j^{\pm}\}=0\,.
\end{equation}
Similar expressions hold for the left-moving oscillators. Using the above, \eqref{zero modes gluing for D5} can be written as
\begin{equation}
   \begin{aligned}
       (b_k^{\pm} + i\eta\tilde{b}_{k}^{\pm})\ket{\eta}_{\text{R}}&=0\,,\quad k=0,2,3\,,\\
    (b_1^{\pm} - i\eta\tilde{b}_{1}^{\pm})\ket{\eta}_{\text{R}}&=0\,. 
    \end{aligned}
\end{equation}
These equations are solved by
\begin{equation}
    \ket{\eta}_{\text{R}}= \exp\big\{i\eta \left( -b_k^+\tilde{b}_k^- + b_1^+\tilde{b}_1^- \right) \big\}\ket{\tilde{a}} \otimes \ket{a}\,,
    \label{ramond vacuum for D5}
\end{equation}
where $\ket{\tilde{a}}$ and $\ket{a}$ are defined such that
\begin{equation}
    \tilde{b}^+_i \ket{\tilde{a}} =0\,,\qquad b^-_i\ket{a}=0\,,\qquad i=0,\ldots,3\,,
    \label{ramond vacua a,tilde a for B5}
\end{equation}
and can be expressed by the SO(8) weight vectors
\begin{equation}
    \ket{\tilde{a}} = \ket{\tfrac{1}{2},\tfrac{1}{2},\tfrac{1}{2},\tfrac{1}{2}} \,,\qquad \ket{a} = \ket{-\tfrac{1}{2},-\tfrac{1}{2},-\tfrac{1}{2},-\tfrac{1}{2}}\,.
    \label{SO(8) weights for B5}
\end{equation}
Now, we see that the exponential in \eqref{ramond vacuum for D5} is invariant under the orbifold action if \eqref{D5 boundary state mass conditions} is satisfied. Moreover, we obtain a constraint by demanding invariance of $\ket{\tilde{a}}\otimes \ket{a}$, that is (cf. \eqref{defRvac}-\eqref{transformation of ramond vacua})
\begin{equation}
    m_1-m_2 = 0  \quad \text{mod} \quad 2\pi\,.
\end{equation}
This leads to the additional condition
\begin{equation}
    m_3-m_4 = 0  \quad \text{mod} \quad 2\pi\,.
\end{equation}
So, we conclude that the D5-brane survives in symmetric orbifolds. Moreover, given a specific choice of mass parameters, we can determine the orbifold action on the zero modes.

For example, consider the symmetric $\mathbb{Z}_2$, $\mathcal{N}=4\, (0,2)$ orbifold with $m_1=m_2=\pi$ and $m_3=m_4=0$. From the above discussion it is clear that the oscillator part of the boundary state is orbifold invariant. Regarding the zero mode contribution, we note that the orbifold acts as $w^l\to -w^l, \tilde{x}^l_0\to -\tilde{x}^l_0$ and leaves $w^5, \tilde{x}^5_0$ invariant. So, we get
\begin{equation}
    f(w,\tilde{x})\ket{\eta,w^5,w^l}_{s} \to f(w,\tilde{x})\ket{\eta,w^5,-w^l}_{s}\,.
\end{equation}
Consequently, an orbifold invariant boundary state should be a linear combination of states containing both of the above terms. We write this schematically as
\begin{equation}
    \ket{\text{B5}} =\frac{1}{\sqrt{2}}\left( \ket{\text{B5},w^5,w^l} +  \ket{\text{B5},w^5,-w^l}\right)\,.
\end{equation}
In the next, we will apply the same analysis to different boundary states, corresponding to different Dp-branes, in order to examine which brane configurations survive in our orbifolds. In order to keep the presentation simple, we will focus only on the oscillator part of the boundary states. The contribution from the zero modes can be easily determined based on the discussion of section \ref{branes from string theory} and the D5-brane example above.

\subsection*{D1-branes}

Let us first discuss the various D1-branes, wrapping a one-cycle of the $T^5$. In order to do so, we decompose $T^5=S^1\times T^4$. We start from a D1-brane that wraps the $S^1$. This case is similar to the D5-brane discussed above. In particular, for the D1-brane we have (in terms of complex oscillators)
\begin{equation}
    \begin{aligned}
      \ket{\text{B1},\eta}_{s} \sim \text{exp}\left\{\sum_{n>0}\frac{1}{n}\left(\alpha^{i}_{-n}\bar{\tilde{\alpha}}^{i}_{-n}+\bar{\alpha}^i_{-n}\tilde{\alpha}^i_{-n}\right) +i\eta \sum_{r>0}\left(b^{i}_{-r}\bar{\tilde{b}}^{i}_{-r}+\bar{b}^i_{-r}\tilde{b}^i_{-r} \right)\right\} \ket{\eta}_{s}\,,
    \end{aligned}
    \label{D1 boundary state}
\end{equation}
where $i=1,2$, and $\ket{\eta}_{\text{R}}$ is given by
\begin{equation}
    \ket{\eta}_{\text{R}}= \exp\big\{i\eta ( -b_0^+\tilde{b}_0^- + \sum_{k=1}^3b_k^+\tilde{b}_k^- ) \big\}\ket{\tilde{a}} \otimes \ket{a}\,,
    \label{ramond vacuum for D1}
\end{equation}
with ${\ket{\tilde{a}}}$ and $\ket{a}$ defined as in \eqref{ramond vacua a,tilde a for B5}. Once again, we find that this boundary state survives in symmetric orbifolds.

We now consider D1-branes that wrap a one-cycle of the $T^4$, which we denote by (D1)$_{T^4}$. If the (D1)$_{T^4}$ is extended along $X^6$, then
\begin{equation}
    \begin{aligned}
      \ket{(\text{B1})_{T^4},\eta}_{s} \sim \text{exp}\Bigg\{&\sum_{n>0}\frac{1}{n}\left(-\alpha^{6}_{-n}\tilde{\alpha}^{6}_{-n}+\sum_{I=7}^{9}\alpha^I_{-n}\tilde{\alpha}^I_{-n}\right)+\\
      i\eta &\sum_{r>0}\left(-b^{6}_{-r}\tilde{b}^{6}_{-r}+\sum_{I=7}^{9}b^I_{-r}\tilde{b}^I_{-r} \right)\Bigg\} \ket{\eta}_{s}\,.
    \end{aligned}
    \label{D1T4Y1 boundary state real osc}
\end{equation}
Note that here the boundary state is written in terms of real oscillators. We can rewrite this boundary state using complex oscillators as
\begin{equation}
    \begin{aligned}
    \ket{(\text{B1})_{T^4},\eta}_{s} \sim \text{exp}\Bigg\{&\sum_{n>0}\frac{1}{n}\left(- \alpha^{1}_{-n}  \tilde{\alpha}^{1}_{-n} -  \bar{\alpha}^{1}_{-n}\bar{\tilde{\alpha}}^{1}_{-n} +\alpha^{2}_{-n}\bar{\tilde{\alpha}}^{2}_{-n}+ \bar{\alpha}^{2}_{-n} \tilde{\alpha}^{2}_{-n}\right)+\\
    i\eta &\sum_{r>0}\left(-b^{1}_{-r}  \tilde{b}^{1}_{-r} -     \bar{b}^{1}_{-r}\bar{\tilde{b}}^{1}_{-r} +b^{2}_{-r}\bar{\tilde{b}}^{2}_{-r}+ \bar{b}^{2}_{-r} \tilde{b}^{2}_{-r} \right)\Bigg\} \ket{\eta}_{s}\,.  
    \end{aligned}
    \label{D1T4 BS complex osc}
\end{equation}
The exponential of this boundary state is invariant under a different orbifold action than the one we saw before. From \eqref{orbiaction2} it follows that the boundary state \eqref{D1T4 BS complex osc} is orbifold invariant if
\begin{equation}
    \begin{aligned}
        m_1 +m_3+m_2+m_4 = 0 \quad &\text{mod} \quad 2\pi\,,\\
         m_1 -m_3-m_2+m_4 =0 \quad & \text{mod} \quad 2\pi\,.
    \end{aligned}
    \label{mod 2pi conditions for D1t4}
\end{equation}
Regarding the R-vacuum, it should satisfy the conditions (cf. \eqref{gluing zero modes}, \eqref{creation-annihilation operators})
\begin{equation}
    \begin{aligned}
       (b_k^{\pm} - i\eta\tilde{b}_{k}^{\pm})\ket{\eta}_{\text{R}}&=0\,,\quad k=1,3\,,\\
    (b_l^{\pm} + i\eta\tilde{b}_{l}^{\mp})\ket{\eta}_{\text{R}}&=0\,,\quad l=0,2\,.
    \end{aligned}
\end{equation}
The solution to the above equations is
\begin{equation}
    \ket{\eta}_{\text{R}}= \exp\big\{i\eta \left( b_k^+\tilde{b}_k^- - b_l^+\tilde{b}_l^+  \right) \big\}\ket{\tilde{a}} \otimes \ket{a}\,,
    \label{ramond vacuum for D1T4}
\end{equation}
where $\ket{\tilde{a}}$ should now satisfy
\begin{equation}
    \tilde{b}^+_k \ket{\tilde{a}}=0\,,\quad k=1,3\qquad \text{and} \qquad  \tilde{b}^-_l \ket{\tilde{a}}=0\,,\quad l=0,2\,,
\end{equation}
and $\ket{{a}}$ is again defined as
\begin{equation}
    {b}^-_i \ket{{a}} =0\,,\qquad i=0,\ldots,3\,.
\end{equation}
These vacua can be expressed as
\begin{equation}
     \ket{\tilde{a}} = \ket{-\tfrac{1}{2},\tfrac{1}{2},-\tfrac{1}{2},\tfrac{1}{2}}\,,\qquad \ket{a} = \ket{-\tfrac{1}{2},-\tfrac{1}{2},-\tfrac{1}{2},-\tfrac{1}{2}}\,.
\end{equation}
So, for this boundary state, we find that the exponential in \eqref{ramond vacuum for D1T4} is orbifold invariant if \eqref{mod 2pi conditions for D1t4} holds and we also find that $\ket{\tilde{a}}\otimes \ket{a}$ is invariant if (cf. \eqref{defRvac}-\eqref{transformation of ramond vacua})
\begin{equation}
    m_2+m_3 = 0 \quad \text{mod} \quad 2\pi\,.
    \label{additional conditions for D1t4 1}
\end{equation}
This, combined with \eqref{mod 2pi conditions for D1t4} gives one further condition, namely
\begin{equation}
    m_1+ m_4=  0 \quad \text{mod} \quad 2\pi\,.
    \label{additional conditions for D1T4 2}
\end{equation}
The same holds for a (D1)$_{T^4}$ wrapping $X^7$. For a (D1)$_{T^4}$ wrapping $X^8$ or $X^9$, we find 
\begin{equation}
    \begin{aligned}
        m_1 -m_4=0  \quad &\text{mod} \quad 2\pi\,,\\
         m_2 -m_3=0  \quad &\text{mod} \quad 2\pi\,.
    \end{aligned}
    \label{mod 2pi conditions for D1t4y3}
\end{equation}
Recall that each (D1)$_{T^4}$ brane is located at a point along the circle coordinate $Z$. Under the orbifold action $Z\to Z+2\pi \mathcal{R}/p$, so the position of each brane is shifted. Hence, an orbifold invariant state should be composed of $p$ (D1)$_{T^4}$ branes located at points on the circle separated by a distance of $2\pi \mathcal{R}/p$. We will discuss this issue in more detail in the next section.

\subsection*{D3-branes}

We now focus on the various D3-brane configurations. In order to study D3-branes wrapping the $T^5$, we decompose $T^5=S^1\times T^2 \times T^2$, where the first $T^2$ is parametrised by $W^1$ and the second $T^2$ by $W^2$. We now distinguish three types of D3-branes: those wrapping the $S^1$ and one of the two $T^2$'s, those wrapping the $S^1$ and a one-cycle in both of the $T^2$'s, and those wrapping fully one $T^2$ and only a one-cycle of the other $T^2$; the latter will be denoted by (D3)$_{T^4}$. We start by studying a D3-brane of the first type, namely 
\renewcommand{\arraystretch}{1.3}
\begin{table}[ht!]
\centering
	\begin{tabular}{c|c|cccc|c|cccc}
		\multicolumn{1}{c}{} & \multicolumn{5}{c}{} & \multicolumn{1}{c}{} & \multicolumn{2}{c}{$W^1$} & \multicolumn{2}{c}{$W^2$} \\[-8pt]
		\multicolumn{1}{c}{} & \multicolumn{5}{c}{} & \multicolumn{1}{c}{} & \multicolumn{2}{c}{\downbracefill} & \multicolumn{2}{c}{\downbracefill} \\
		& $X^0$ & $X^1$ & $X^2$ & $X^3$ & $X^4$ & $Z$ & $X^6$ & $X^7$ & $X^8$ & $X^9$ \\ \hline
		\multicolumn{1}{c|}{\,D3\,} & $\checkmark$ & $-$ & $-$ & $-$ & $-$ & $\checkmark$ & $\checkmark$ & $\checkmark$ & $-$ & $-$ \\
	\end{tabular} 
\end{table}
\renewcommand{\arraystretch}{1}

\noindent Here and below, we denote by \say{$\checkmark$} the  directions in which the brane is extended and by \say{$-$} the directions transverse to the brane. For the brane specified above, the construction of the boundary state is similar to the D5-brane. We find (in terms of complex oscillators) 
\begin{equation}
    \begin{aligned}
    \ket{\text{B3},\eta}_{s} \sim \text{exp}\Bigg\{&\sum_{n>0}\frac{1}{n}\left(- \alpha^{1}_{-n}\bar{\tilde{\alpha}}^{1}_{-n}- \bar{\alpha}^{1}_{-n} \tilde{\alpha}^{1}_{-n} +\alpha^{2}_{-n}\bar{\tilde{\alpha}}^{2}_{-n}+ \bar{\alpha}^{2}_{-n} \tilde{\alpha}^{2}_{-n}\right)\\
    +i\eta &\sum_{r>0}\left(- b^{1}_{-r}\bar{\tilde{b}}^{1}_{-r}- \bar{b}^{1}_{-r} \tilde{b}^{1}_{-r} +b^{2}_{-r}\bar{\tilde{b}}^{2}_{-r}+ \bar{b}^{2}_{-r} \tilde{b}^{2}_{-n} \right)\Bigg\} \ket{\eta}_{s}\,,
    \end{aligned}
    \label{D3 BS complex osc}
\end{equation}
and $\ket{\eta}_{\text{R}}$  reads
\begin{equation}
    \ket{\eta}_{\text{R}}= \exp\big\{i\eta ( -b_0^+\tilde{b}_0^- +b_1^+\tilde{b}_1^- -b_2^+\tilde{b}_2^-+ b_3^+\tilde{b}_3^- ) \big\}\ket{\tilde{a}} \otimes \ket{a}\,,
    \label{ramond vacuum for D3}
\end{equation}
with ${\ket{\tilde{a}}}$ and $\ket{a}$ defined as in \eqref{ramond vacua a,tilde a for B5}. It can be easily verified that this boundary state is invariant under the action of a symmetric orbifold. The same result also holds for the other D3-brane of the first type, i.e. the one wrapping $Z$ and $X^8,X^9$.

Now we consider the second type of D3-branes. As an example, we take the following setup
\renewcommand{\arraystretch}{1.3}
\begin{table}[ht!]
\centering
	\begin{tabular}{c|c|cccc|c|cccc}
		\multicolumn{1}{c}{} & \multicolumn{5}{c}{} & \multicolumn{1}{c}{} & \multicolumn{2}{c}{$W^1$} & \multicolumn{2}{c}{$W^2$} \\[-8pt]
		\multicolumn{1}{c}{} & \multicolumn{5}{c}{} & \multicolumn{1}{c}{} & \multicolumn{2}{c}{\downbracefill} & \multicolumn{2}{c}{\downbracefill} \\
		& $X^0$ & $X^1$ & $X^2$ & $X^3$ & $X^4$ & $Z$ & $X^6$ & $X^7$ & $X^8$ & $X^9$ \\ \hline
		\multicolumn{1}{c|}{\,D3\,} & $\checkmark$ & $-$ & $-$ & $-$ & $-$ & $\checkmark$ & $\checkmark$ & $-$ & $\checkmark$ & $-$ \\
	\end{tabular} 
\end{table}
\renewcommand{\arraystretch}{1}

\noindent For this D3-brane we have
\begin{equation}
    \begin{aligned}
     \ket{\text{B3},\eta}_{s} \sim \text{exp}\Bigg\{&\sum_{n>0}\frac{1}{n}\left(-\sum_{a=6,8}\alpha^{a}_{-n}\tilde{\alpha}^{a}_{-n}+\sum_{I=7,9}\alpha^I_{-n}\tilde{\alpha}^I_{-n}\right)\\
     +i\eta &\sum_{r>0}\left(-\sum_{a=6,8}b^{a}_{-r}\tilde{b}^{a}_{-r}+\sum_{I=7,9}b^I_{-r}\tilde{b}^I_{-r} \right)\Bigg\} \ket{\eta}_{s}\,.   
    \end{aligned}
    \label{D3 2nd type boundary state real osc}
\end{equation}
In terms of complex oscillators we find
\begin{equation}
    \begin{aligned}
        \ket{\text{B3},\eta}_{s} \sim \text{exp}\Bigg\{&\sum_{n>0}\left[\frac{1}{n}\left(- \alpha^{i}_{-n}  \tilde{\alpha}^{i}_{-n} -  \bar{\alpha}^{i}_{-n}\bar{\tilde{\alpha}}^{i}_{-n}\right)\right] +i\eta \sum_{r>0}\left[-b^{i}_{-r}  \tilde{b}^{i}_{-r} -     \bar{b}^{i}_{-r}\bar{\tilde{b}}^{i}_{-r}  \right]\Bigg\} \ket{\eta}_{s}\,. 
    \end{aligned}
    \label{D3 2nd type boundary state complex osc}
\end{equation}
The exponential of this boundary state is invariant under the orbifold action if
\begin{equation}
    \begin{aligned}
        m_1 +m_3+m_2+m_4 =0  \quad &\text{mod} \quad 2\pi\,,\\
         m_1 -m_3+m_2-m_4=0  \quad&\text{mod} \quad 2\pi\,.
    \end{aligned}
    \label{mod 2pi conditions for D3 second type}
\end{equation}
For the R-vacuum we have (cf. \eqref{gluing zero modes}, \eqref{creation-annihilation operators})
\begin{equation}
    \begin{aligned}
    (b_0^{\pm} + i\eta\tilde{b}_{0}^{\pm})\ket{\eta}_{\text{R}}&=0\,,\\
    (b_1^{\pm} - i\eta\tilde{b}_{1}^{\pm})\ket{\eta}_{\text{R}}&=0\,,\\
    (b_k^{\pm} + i\eta\tilde{b}_{k}^{\mp})\ket{\eta}_{\text{R}}&=0\,, \quad k=2,3\,.
    \end{aligned}
\end{equation}
The above equations can be solved by
\begin{equation}
    \ket{\eta}_{\text{R}}= \exp\big\{i\eta \left( -b_0^+\tilde{b}_0^- + b_1^+\tilde{b}_1^- - b_k^+\tilde{b}_k^+  \right) \big\}\ket{\tilde{a}} \otimes \ket{a}\,,
    \label{ramond vacuum for D3 second type}
\end{equation}
where  the vacua should satisfy
\begin{equation}
    \tilde{b}^+_0 \ket{\tilde{a}}=\tilde{b}^+_1 \ket{\tilde{a}}=0\,,\qquad \text{and} \qquad  \tilde{b}^-_k \ket{\tilde{a}} =0\,,\quad k=2,3\,.
    \label{vacua D3 second type}
\end{equation}
Also,
\begin{equation}
    {b}^-_i \ket{{a}} =0\,,\qquad i=0,\ldots,3\,.
\end{equation}
These vacua can  be expressed as
\begin{equation}
     \ket{\tilde{a}} = \ket{\tfrac{1}{2},\tfrac{1}{2},-\tfrac{1}{2},-\tfrac{1}{2}}\,,\qquad \ket{a} = \ket{-\tfrac{1}{2},-\tfrac{1}{2},-\tfrac{1}{2},-\tfrac{1}{2}}\,.
\end{equation}
In this case, we find that the exponential in \eqref{ramond vacuum for D3 second type} is orbifold invariant if \eqref{mod 2pi conditions for D3 second type} holds and we also find that $\ket{\tilde{a}}\otimes \ket{a}$ is invariant if (cf. \eqref{defRvac}-\eqref{transformation of ramond vacua})
\begin{equation}
    m_1+m_2 = 0 \quad \text{mod} \quad 2\pi\,,
    \label{D3 second type m1+m2}
\end{equation}
which leads to
\begin{equation}
    m_3+ m_4=  0 \quad \text{mod} \quad 2\pi\,.
    \label{D3 second type m3+m4}
\end{equation}
Note that this boundary state is orbifold invariant if $m_1 = -m_2$ and $m_3 = - m_4$ (mod $2\pi$). One might refer to the orbifold corresponding to this choice of mass parameters as an anti-symmetric orbifold, as it works with an opposite phase on the left and right-moving coordinates, but there is no fundamental difference between this and a symmetric orbifold (they are related by a field redefinition of the coordinates). The same result is obtained for the other three D3-branes of the second type.

Finally, we discuss the third type of D3-branes. We consider the following (D3)$_{T^4}$ configuration
\renewcommand{\arraystretch}{1.3}
\begin{table}[ht!]
\centering
	\begin{tabular}{c|c|cccc|c|cccc}
		\multicolumn{1}{c}{} & \multicolumn{5}{c}{} & \multicolumn{1}{c}{} & \multicolumn{2}{c}{$W^1$} & \multicolumn{2}{c}{$W^2$} \\[-8pt]
		\multicolumn{1}{c}{} & \multicolumn{5}{c}{} & \multicolumn{1}{c}{} & \multicolumn{2}{c}{\downbracefill} & \multicolumn{2}{c}{\downbracefill} \\
		& $X^0$ & $X^1$ & $X^2$ & $X^3$ & $X^4$ & $Z$ & $X^6$ & $X^7$ & $X^8$ & $X^9$ \\ \hline
		\multicolumn{1}{c|}{\,D3\,} & $\checkmark$ & $-$ & $-$ & $-$ & $-$ & $-$ & $\checkmark$ & $\checkmark$ & $-$ & $\checkmark$ \\
	\end{tabular} 
\end{table}
\renewcommand{\arraystretch}{1}

\noindent We find that this brane survives the orbifold if \eqref{mod 2pi conditions for D1t4y3} is satisfied. The same holds for a (D3)$_{T^4}$ wrapping $X^8$ instead of $X^9$. The remaining (D3)$_{T^4}$ configuration (wrapping a one-cycle of the first $T^2$ and fully the second $T^2$) survives the orbifold if \eqref{additional conditions for D1t4 1} and \eqref{additional conditions for D1T4 2} hold. Similarly with the (D1)$_{T^4}$ branes, the position of each (D3)$_{T^4}$ brane is shifted under the orbifold action. So, we need to place $p$ of them at points on the circle separated by a distance of $2\pi \mathcal{R}/p$ in order to build an orbifold invariant boundary state.

\subsection{Bound states}
\label{bound states in boundary state formalism}

So far, we have discussed the conditions under which boundary states, and as a consequence Dp-branes, survive in our orbifolds. However, it is possible that a linear combination of boundary states is orbifold invariant, even if the boundary states separately are not. Denoting the orbifold group element by $g$, with $g^p=1$, an orbifold invariant boundary state can be in general constructed as 
\begin{equation}
    \ket{\text{B}}_{\text{inv}}=\frac{1}{\sqrt{p}}(1+g+\ldots g^{p-1})  \ket{\text{B}}\,.
    \label{construction of bound states}
\end{equation}
We have already encountered examples of such combinations of boundary states when we discussed the D5-brane, and the (D1)$_{T^4}$ and (D3)$_{T^4}$ branes for which we had to place $p$ branes at points on the circle separated by a distance of $2\pi \mathcal{R}/p$ in order to build an orbifold invariant boundary state.

For a more involved example, let us consider a symmetric $\mathbb{Z}_4$, $\mathcal{N}=4\,(2,0)$ orbifold with $m_3=m_4=\pi/2$ and $m_1=m_2=0$. This orbifold acts on the real torus coordinates as
\begin{equation}
    (X^6,X^7,X^8,X^9)\to(-X^7,X^6,X^9,-X^8)\,,
\end{equation}
and on momentum $(n_i)$ and winding ($w^i)$ modes as 
\begin{equation}
    \begin{aligned}
      (n_6,n_7,n_8,n_9)&\to(-n_7,n_6,n_9,-n_8)\,,\\
      (w^6,w^7,w^8,w^9)&\to (-w^7,w^6,w^9,-w^8)\,.
    \end{aligned}
\end{equation}
Now, we consider a D3-brane of the second type wrapping $Z,X^6,X^8$. This brane is allowed to carry quantized winding modes $w^5,w^6,w^8$ along the Neumann directions, $Z,X^6,X^8$, and quantized momentum modes $n_7,n_9$ along the Dirichlet directions, $X^7,X^9$. Let us denote the boundary state corresponding to this D3-brane by $\ket{\text{B3},w^5,w^6,w^8,n_7,n_9}$. As we can see from \eqref{D3 second type m3+m4}, this boundary state is not orbifold invariant. However, note that under the orbifold action this boundary state transforms to a boundary state describing a D3'-brane wrapping $Z,X^7,X^9$ (cf. \eqref{D3 2nd type boundary state real osc}, \eqref{ramond vacuum for D3 second type}-\eqref{vacua D3 second type}). Taking into account the transformation of the quantized momentum and winding modes we can construct an orbifold invariant boundary state as follows
\begin{equation}
\begin{aligned}
    \ket{\text{B3,\,B3'}}=\frac{1}{2}\big(&\ket{\text{B3},w^5,w^6,w^8,n_7,n_9}+\ket{\text{B3'},w^5,-w^7,w^9,n_6,-n_8}+\\ 
    &\ket{\text{B3},w^5,-w^6,-w^8,-n_7,-n_9}+\ket{\text{B3'},w^5,w^7,-w^9,-n_6,n_8}\big)\,.
\end{aligned}
\label{d3-d3' bound state boundary state}
\end{equation}
We can interpret the orbifold invariant boundary state $\ket{\text{B3,B3'}}$ as a state corresponding to a bound state of D3/D3'-branes. 

Now, let us slightly modify the above example by choosing $m_1=m_2=\pi/2$ and $m_3=m_4=0$. This choice of mass parameters leads to a symmetric $\mathbb{Z}_4$,  $\mathcal{N}=4\,(0,2)$ orbifold which acts on the real torus coordinates as $(X^6,X^7,X^8,X^9)\to (-X^7,X^6,-X^9,X^8)$, and with a similar action on momentum and winding modes. If we look again at the D3-brane wrapping $Z,X^6,X^8$, we can see that, in this orbifold, it transforms to an anti-D3'-brane, $\overline{\text{D3}}$', wrapping $Z,X^7,X^9$. The reason for this is that the R-vacuum $\ket{\eta}_{\text{R}}$ picks up a minus sign under the orbifold action (cf. \eqref{ramond vacuum for D3 second type}), which changes the relative sign between the NS-NS and R-R part of the boundary state. Concretely, the orbifold invariant boundary state reads
\begin{equation}
\begin{aligned}
    \ket{\text{B3},\overline{\text{B3}}\text{'}}=\frac{1}{2}\big(&\ket{\text{B3},w^5,w^6,w^8,n_7,n_9}+\ket{\overline{\text{B3}}\text{'},w^5,-w^7,-w^9,n_6,n_8}+\\ 
    &\ket{\text{B3},w^5,-w^6,-w^8,-n_7,-n_9}+\ket{\overline{\text{B3}}\text{'},w^5,w^7,w^9,-n_6,-n_8}\big)\,.
\end{aligned}
\label{d3-anti-d3' bound state boundary state}
\end{equation}
Similarly, we can take other suitable linear combinations of boundary states in order to construct orbifold invariant states. In the next, we will discuss how to determine which branes, or bound states of branes, survive in our orbifolds based on supergravity arguments and we will show that the supergravity results precisely agree with the string-theoretic prediction.

\section{Supergravity perspective}
\label{branes from supergravity}

So far, we have discussed the branes purely from a string-theoretic point of view. However, the orbifolds that we considered in chapters \ref{chap:fao}-\ref{chap:mod5} are the string theory uplifts of the supergravity Scherk-Schwarz reductions studied in \cite{Hull:2020byc}. There, the 5$D$ spectrum was worked out and masses were found for gauge fields coupling to various branes wrapped on the internal manifold. In this section, we will check that the results about which D-branes survive in our orbifold constructions agree with the supergravity calculation. The premise of this check is that a Dp-brane survives in the orbifold when the supergravity gauge field charging the corresponding p-brane remains massless in the Scherk-Schwarz reduction\footnote{The supergravity spectrum arising from a T-duality twist can be found in Table C.2 of \cite{Hull:2020byc}.}. We will see that this supergravity point of view aligns nicely with the results from section \ref{stapproach}. 

Let us first consider the D1-brane wrapping the $S^1$ and the D5-brane wrapping the $T^5$. In \cite{Hull:2020byc} it was found that after diagonalizing the mass matrix, the gauge fields charging the D1 and D5-brane don't obtain masses individually. Instead linear combinations of these fields obtain the masses
\begin{equation}\label{sugratensors}
\begin{aligned}
C_2^{\text{D1}}+C_2^{\text{D5}}\,:\quad |m_1-m_2|\,,\\
C_2^{\text{D1}}-C_2^{\text{D5}}\,:\quad |m_3-m_4|\,.\\
\end{aligned}
\end{equation}
Note that these two-form gauge fields can be dualized to vectors if they are massless, so that they can properly charge pointlike objects (the branes only wrap the internal manifold and are therefore pointlike in 5$D$).
We can immediately see that if we want both the D1 and the D5-brane to survive, i.e. if we want both gauge fields in \eqref{sugratensors} to remain massless, we need $m_1=m_2$ and $m_3=m_4$ (mod $2\pi$). This is in agreement with the results from section \ref{examples of boundary states}. Both lines of reasoning lead to the conclusion that the D1 and the D5-brane survive in symmetric orbifolds.

Similar arguments can be made for the various D3-brane setups. For example, let D3 denote a brane in the $Z, X^6, X^7$ directions, and D3' a brane in the $Z, X^8, X^9$ directions. The supergravity computation yields the following masses
\begin{equation}
\begin{aligned}
C_2^{\text{D3}}+C_2^{\text{D3'}}\,:\quad |m_1-m_2|\,,\\
C_2^{\text{D3}}-C_2^{\text{D3'}}\,:\quad |m_3-m_4|\,.\\
\end{aligned}
\end{equation}
Again, the gauge fields are massless for $m_1=m_2$ and $m_3=m_4$ (mod $2\pi$), corresponding to a symmetric orbifold and therefore in line with the results from section \ref{examples of boundary states}.  In another scenario, where the D3 and D3' lie in the $Z, X^6, X^8$ and $Z, X^7, X^9$ directions, respectively, the masses obtained in the Scherk-Schwarz reduction are
\begin{equation}
\begin{aligned}
C_2^{\text{D3}}+C_2^{\text{D3'}}\,:\quad |m_1+m_2|\,,\\
C_2^{\text{D3}}-C_2^{\text{D3'}}\,:\quad |m_3+m_4|\,.\\
\end{aligned}
\label{sugra tensors D3-D3'}
\end{equation}
In this case, the gauge fields remain massless, if $m_1=-m_2$ and $m_3=-m_4$ (mod $2\pi$). Again, we find agreement between the string theory and supergravity results (recall that the D3-branes in this last example are of the \say{second type} in the terminology of section \ref{examples of boundary states}). 

In all the above examples we saw that linear combinations of gauge fields charge the various branes. So, one could ask what happens to the branes if one linear combination of gauge fields remains massless and another becomes massive. For a concrete example we take the D3-D3' system of branes wrapped on $Z, X^6, X^8$ and $Z, X^7, X^9$, respectively. We consider an $\mathcal{N}=4\,(2,0)$ theory with $m_1=m_2=0$ and $m_3=m_4=\pi/2$ and look at the gauge fields charging this system (cf. \eqref{sugra tensors D3-D3'}). We can see that the linear combination $C_2^{\text{D3}}+C_2^{\text{D3'}}$   remains massless, while $C_2^{\text{D3}}-C_2^{\text{D3'}}$ becomes massive. The massless gauge field  $C_2^{\text{D3}}+C_2^{\text{D3'}}$ is therefore expected to charge some object. We can interpret this object as a bound state of D3/D3'-branes. Such bound states were also discussed in section \ref{bound states in boundary state formalism}. This particular D3/D3' bound state corresponds precisely to the boundary state \eqref{d3-d3' bound state boundary state}. Now, if we choose $m_1=m_2=\pi/2$ and $m_3=m_4=0$, the gauge field $C_2^{\text{D3}}-C_2^{\text{D3'}}$ remains massless and is expected to charge the bound state D3/$\overline{\text{D3}}$', which corresponds to the boundary state \eqref{d3-anti-d3' bound state boundary state}. Once again, we see that the supergravity result is in agreement with the string-theoretic prediction.

\section{Supersymmetry}
\label{sec:branesupersymmetry}

In the previous sections, we determined the conditions under which the various D-branes survive the orbifolding. Here, we wish to examine if the supersymmetry preserved by the (surviving) D-branes, is compatible with the supersymmetry preserved by the orbifolds.

Let us first recall that type IIB supergravity has two $10D$ supersymmetry parameters, which we label by $\epsL$ and $\epsR$. Since type IIB is a chiral theory in ten dimensions, both $\epsL$ and $\epsR$ have the same chirality in 10$D$. We choose conventions such that this can be written as 
\begin{equation}\label{10dchirality}
\Gamma_{11}\,\epsL  = \epsL \,, \qquad\quad \Gamma_{11}\,\epsR  = \epsR \,,
\end{equation}
where $\Gamma_{11} = \Gamma_0 \Gamma_1 \ldots \Gamma_9$ is the chirality operator. For our purposes, chirality will also be important in six dimensions, which is the number of external dimensions upon compactification on $T^4$. Both of the 10$D$ spinors (Majorana--Weyl) decompose into two 6$D$ spinors (symplectic Weyl), one of either chirality. These are the supersymmetry spinors of the 6$D$ $\mathcal{N}= \text{(2,2)}$ theory that is obtained from reducing type IIB on $T^4$. We denote these four spinors by $\epsLR^+$ and $\epsLR^-$, where $i=1,2$ and the $\pm$ sign denotes the 6$D$ chirality, meaning
\begin{equation}\label{6dchirality}
\Gamma_\ast \: \epsLR^+ = \epsLR^+ \,, \qquad\quad \Gamma_\ast \: \epsLR^- = -\, \epsLR^- \,,
\end{equation}
with $\Gamma_\ast = \Gamma_0\Gamma_1\ldots\Gamma_5$. Note that here we have chosen the 10$D$ Dirac matrices such that the matrices $\Gamma_{\hat{\mu}}$, with $\hat{\mu} = 0,1,\dots,5$, form a 6$D$ Clifford algebra. By combining the conditions \eqref{10dchirality} and \eqref{6dchirality}, we find that 
\begin{equation}
\Gamma_{6789} \: \epsLR^+ = \epsLR^+ \,, \qquad\quad \Gamma_{6789} \: \epsLR^- = -\, \epsLR^- \,,
\end{equation}
where we use a $\Gamma$ with multiple indices to denote an anti-symmetrized product of $\Gamma$-matrices. 

Now, both the orbifolds and the presence of D-branes break some (or all) supersymmetries. So, it is natural to ask whether the supersymmetry preserved by the various D-brane configurations is compatible with the supersymmetry preserved by the orbifold. 

In order to address this issue, it will be useful to write the amount of supersymmetry that is preserved by the orbifold in terms of projections on the Killing spinors, $\varepsilon_{1/2}$. Recall that for each mass parameter that we turn on two gravitini become massive and as a result, the corresponding supersymmetries are broken. The gravitini with masses $m_1,m_2$ are in the $(\textbf{3},\textbf{2})$ representation of the massive little group $\text{SU}(2)\times \text{SU}(2)$ in $5D$, while the gravitini with masses $m_3,m_4$ are in the $(\textbf{2},\textbf{3})$ representation. We relate these massive gravitini with the supersymmetry parameters $\varepsilon_i^{\pm}$ as follows
\begin{equation}
    \begin{aligned}
        &\varepsilon_1^- \;\;\leftrightarrow\;\;(\textbf{3},\textbf{2})\:: m_1\,, \qquad  \varepsilon_2^- \;\;\leftrightarrow\;\; (\textbf{3},\textbf{2})\: :m_2\,,\\
        &\varepsilon_1^+ \;\;\leftrightarrow\;\; (\textbf{2},\textbf{3})\:: m_3\,, \qquad  \varepsilon_2^+ \;\;\leftrightarrow\;\;(\textbf{2},\textbf{3})\: : m_4\,.
    \end{aligned}
    \label{gravitini and susy}
\end{equation}
Now we can express the preserved supersymmetries of our various orbifolds in terms of projections on the Killing spinors. 

Take for example a symmetric orbifold with $m_1=m_2\neq0$ and $m_3=m_4=0$. This gives an $\mathcal{N}=4\,(0,2)$ theory in $5D$; the gravitini related to $\epsL^-$ and $\epsR^-$ become massive. In terms of projections on the Killing spinors we can write the preserved supersymmetry as
\begin{equation}\label{susyN=4 (0,2) orbifold}
\mathcal{N}=4: \,(0,2) \quad \Gamma_{6789} \: \epsL = \epsL \quad\text{and}\quad  \Gamma_{6789} \: \epsR = \epsR \,.
\end{equation}
For another example, consider the $\mathcal{N}=4\,(2,0)$ symmetric orbifold with $m_1=m_2=0$ and $m_3=m_4\neq0$. In this case, the gravitini related to $\epsL^+$ and $\epsR^+$ become massive and we have
\begin{equation}\label{susyN=4 (2,0) orbifold}
\mathcal{N}=4: \,(2,0) \quad \Gamma_{6789} \: \epsL = -\epsL \quad\text{and}\quad  \Gamma_{6789} \: \epsR = -\epsR \,.
\end{equation}
Let us now discuss the supersymmetry preserved by D-branes in the absence of a $B$-field. For a Dp-brane extended along the directions $X^0,X^1,\ldots X^p$ the condition for the preserved supersymmetries is 
\begin{equation}
     \Gamma_{01\ldots p} \: \varepsilon_1 = \varepsilon_2\,.
     \label{brane preserved susy}
\end{equation}
In the case of an anti-brane extended along the same directions, the preserved supersymmetry is given by \eqref{brane preserved susy}, but with a minus sign in front of $\varepsilon_2$.

As an example, consider the D1-D5 system of branes with the D1-brane wrapping the $S^1$ and the D5-brane wrapping the $T^5$, which gives the following conditions on the supersymmetries
\begin{equation}\label{D1D5susyconditions}
\text{D1}: \;\; \Gamma_{05}  \: \epsL = \epsR \,, \qquad\quad \text{D5}: \;\; \Gamma_{056789} \: \epsL = \epsR   \,.
\end{equation}
From these conditions we can derive
\begin{equation}
    \Gamma_{6789} \: \epsL = \epsL \quad\text{and}\quad \Gamma_{6789} \: \epsR = \epsR \,.
    \label{D1-D5 and F1-NS5}
\end{equation}
In the unorbifolded theory, this configuration preserves 8 supersymmetries, i.e. it is ${1}/{4}$-BPS: $\epsR$ is determined in terms of $\epsL$ and furthermore, $\epsL$ is halved to $\epsL^+$. 

As we have discussed in the previous sections, both the D1 and the D5-brane survive in symmetric orbifolds. So, we can now examine if the D1-D5 system is BPS in the orbifolded theory. First, we can see that the supersymmetry preserved by the D1-D5 system \eqref{D1-D5 and F1-NS5} is compatible with the supersymmetry preserved by the $\mathcal{N}=4\,(0,2)$ orbifold \eqref{susyN=4 (0,2) orbifold}. So, the D1-D5 system preserves 8 supersymmetries in the $\mathcal{N}=4\,(0,2)$ theory  and therefore, it is ${1}/{2}$-BPS. On the other hand, the supersymmetry preserved by the D1-D5 system \eqref{D1-D5 and F1-NS5} is not compatible with the supersymmetry preserved by the $\mathcal{N}=4\,(2,0)$ orbifold \eqref{susyN=4 (2,0) orbifold}. In this case, the D1-D5 system is not BPS. However, as we will now argue, the D1-D5 system is stable.

First, we recall that an open string ending on a D-brane has Neumann boundary conditions ($\partial_{\sigma^2} X=0$) in the extended worldvolume directions of the brane and Dirichlet boundary conditions ($\partial_{\sigma^1} X=0$) in the transverse directions of the brane. If both ends of the string have the same type of boundary condition (NN or DD), the bosonic modes in that direction are integer-valued, while mixed boundary conditions (ND or DN) give half-integer-valued modes. The fermionic oscillators in the Ramond sector follow the moding of the bosonic modes, and the fermionic modes in the Neveu-Schwarz sector carry the \say{opposite} moding. The open string boundary conditions give a relation between the left and right-moving oscillators. We summarize the consequences of each type of boundary condition in \autoref{open table}.

\begin{table}[h!]
\centering
{\renewcommand*{\arraystretch}{1.4}
\begin{tabular}{|c|c|c|c|c|}
\hline
\;Boundary conditions\; & \;$L$--$R$ relation\; & \;$\alpha_n$\; & \;$b_n$ (NS)\; & \;$b_n$ (R)\; \\ \hline\hline
NN & $\tilde{\alpha}_n = \alpha_n$ & $n\in\mathbb{Z}$ & $n\in\mathbb{Z}+\tfrac{1}{2}$ & $n\in\mathbb{Z}$ \\[2pt]
DD & $\tilde{\alpha}_n = - \alpha_n$ & $n\in\mathbb{Z}$ & $n\in\mathbb{Z}+\tfrac{1}{2}$ & $n\in\mathbb{Z}$ \\[2pt]
ND & $\tilde{\alpha}_n = \alpha_n$ & $n\in\mathbb{Z}+\tfrac{1}{2}$ & $n\in\mathbb{Z}$ & $n\in\mathbb{Z}+\tfrac{1}{2}$ \\[2pt]
DN & $\tilde{\alpha}_n = - \alpha_n$ & $n\in\mathbb{Z}+\tfrac{1}{2}$ & $n\in\mathbb{Z}$ & $n\in\mathbb{Z}+\tfrac{1}{2}$ \\ \hline
\end{tabular}}
\caption{\textit{Here we present the moding of bosonic and fermionic oscillators, as well as the relation between left and right-moving oscillators, depending on the boundary conditions of an open string. The left-right relation for the fermionic modes is the same as for the bosonic ones.} }
\label{open table}
\end{table}

\noindent In addition, the contribution of a real boson to the zero-point energy is  \cite{Johnson:2003glb}
\begin{equation}
   E_B(\theta)= \frac{1}{48}-\frac{\left(2{\theta}-1\right)^2}{16}\,,
   \label{zero point energy real boson}
\end{equation}
where $\theta=0$ for integer modes and $\theta=1/2$ for half-integer modes. Moreover, the contribution of a real fermion to the zero-point energy is $E_F(\theta)=-E_B(\theta)$. Note that in the presence of a non-trivial $B$-field, the moding of the oscillators can be different than $0$ and $1/2$; the same also holds for the values of $\theta$.  With these at hand, it can be shown that for a string with $\nu$ number of mixed boundary conditions, the zero point energy in the NS-sector is given by 
\begin{equation}
    E_{\text{NS}} = -\frac{1}{2} + \frac{\nu}{8}\,.
    \label{NS-sector zero point energy}
\end{equation}
In particular, this implies that for open strings with $\nu\geq 4$ mixed boundary conditions, the ground state energy is non-negative and hence, free of tachyons. Note that in the R-sector the zero point energy is always zero because the bosonic and fermionic oscillators have the same moding irrespectively of the type of boundary condition.

For the D1-D5 system discussed before, we have $\nu=4$. Therefore, the spectrum of an open string stretching between these branes is free of tachyons and, consequently, the brane system is stable. So, even though the D1-D5 system is non-BPS in the $\mathcal{N}=4\,(2,0)$ theory (cf.\eqref{susyN=4 (2,0) orbifold}), it is stable.

Finally, we mention here that in orbifolds which project out both $\varepsilon_1^{+}$ and $\varepsilon_1^{-}$, or similarly, $\varepsilon_2^{+}$ and $\varepsilon_2^{-}$, there can be no BPS D-branes, as the presence of a D-brane relates 
$\varepsilon_1$ with $\varepsilon_2$ (cf. \eqref{brane preserved susy}). Consequently, for theories preserving $\mathcal{N}=4\,(1,1)$, $\mathcal{N}=2$ or $\mathcal{N}=0$ supersymmetry, there exists no BPS D-branes\footnote{As in the D1-D5 example, there might exist stable, non-BPS systems of D-branes.}. On the other hand, for asymmetric orbifolds preserving $\mathcal{N}=6$ supersymmetry there exist bound states of D-branes, as the ones we saw in the previous sections, which can be stable and BPS (for a  detailed analysis we refer to \cite{Bianchi:2008cj} and more recently in  \cite{bianchi2022perturbative,Bianchi:2023hbo}).

\clearpage
\thispagestyle{empty}
\chapter{Conclusion}
\label{chap:conclusion}

\section{Summary}

In this thesis, we have studied various aspects of type IIB string theory compactified on freely acting orbifolds. We focused particularly on asymmetric orbifolds, which are examples of non-geometric string compactifications and constitute an intriguing corner in the string landscape.

In chapter \ref{chap:fao}, we discussed the general construction of freely acting (asymmetric) orbifolds, focusing on the closed string sector. We described the orbifolds in terms of duality twists, we discussed the various possible consistent orbifolds and we analysed in detail the quantization condition on the mass parameters. After that, we discussed the lattice construction; a complementary approach to the construction of orbifolds,  based on even-self dual Lie algebra Narain lattices. Afterwards, we presented the orbifold action on the world-sheet fields and the general procedure for constructing modular invariant orbifold partition functions.

In chapter \ref{chap:spectrum}, we presented explicit examples of orbifolds preserving $\mathcal{N}=6,4,2$ or $0$ supersymmetry in five dimensions and we constructed the lightest closed string states in the untwisted and twisted orbifold sectors. Then, we demonstrated the connection between freely acting orbifolds and Scherk-Schwarz reductions by matching the lightest untwisted orbifold states with those arising form the effective supergravity theory. 

Furthermore, we saw that all string states coming from the twisted sectors are massive at generic points in the moduli space. However, we showed that at special values of the radius of the circle, generically massive twisted states can become massless. In particular, for non-supersymmetric orbifolds there is a critical radius above which the string spectrum is tachyon-free. Moreover, for theories breaking all supersymmetries we computed the one-loop effective potential, which, at the supergravity level, turned out to be negative definite.

In chapter \ref{chap:mod5}, we turned our attention to the orbifold moduli space. In the class of orbifolds studied in this thesis, we saw that there exist only three consistent asymmetric orbifolds, of rank $2,3$ and $4$, preserving $\mathcal{N}=6$ supersymmetry in five dimensions, with moduli space ${\text{SU}^*(6)}/{\text{Sp}(3)}$. For $\mathcal{N}=4$ theories we found that the moduli space is completely determined by the number $n$ of vector multiplets and is given by ${\text{SO}(5,n)}/[{\text{SO}(5)\times \text{SO}(n)}]$. Also, all our orbifold examples yielded an odd number of vector multiplets; this seems to be a common characteristic for a very broad class of $\mathcal{N}=4$ Minkowski vacua in five dimensions. 

Regarding theories preserving $\mathcal{N}=2$ supersymmetry, we determined the classical hypermultiplet and vector multiplet moduli spaces by finding the duality group of the orbifold, that is the subgroup
of the U-duality group of $T^5$ that commutes with the orbifold action. Then, by constructing dual orbifold pairs, we argued that no quantum corrections to the metric on the vector multiplet moduli space arise. Moreover, we presented models with no hypermultiplets and we also reproduced some of the magic square supergravities by using freely acting asymmetric orbifolds.

In chapter \ref{chap:swampland}, we discussed aspects of the swampland program in the context of freely acting asymmetric orbifolds of $\mathbb{R}^{1,3}\times T^2 \times T^4$. We demonstrated that, due to the orbifold shift, the duality group of $T^2$ was reduced to congruence subgroups of the modular group and the fundamental domain contained new interesting points at infinite distance on the real axis bounding the upper half plane. We verified that the swampland distance conjecture is valid in the non-geometric compactifications of string theory studied in this thesis.

Moreover, at points of infinite distance in the moduli space, some (or all) gravitini could become massless, indicating supersymmetry enhancement. Also, at these points, the orbifolded theory decompactified to a different orbifold construction. Remarkably, at points on the real axis of the upper half plane, we saw that an asymmetric freely acting orbifold could decompactify to a non-freely acting symmetric orbifold. Furthermore, at infinite distance points in the moduli space, we found that infinite towers of states became massless. On the other hand, at special points in the interior of the moduli space we found that only a finite number of states became massless.

Finally, in chapter \ref{chap:branes} we focused on the D-brane spectrum of freely acting orbifolds. Using two complementary approaches, i.e. the boundary state formalism and the spectrum of massless Scherk-Schwarz gauge fields, we determined the conditions under which the various D-branes survive the orbifolding. In addition, we showed that even though a single D-brane may not survive in an orbifold, an appropriate combination of branes, that is a bound state of branes, can be orbifold invariant. Then, we examined whether the supersymmetry preserved by various brane configurations is compatible with the supersymmetry preserved by the orbifolds. In particular, we found that there can exist BPS and stable  D-brane systems, as well as non-BPS, stable D-brane systems.

\section{Outlook}

The landscape of freely acting (asymmetric) orbifolds is a very rich corner in string theory and many aspects of these string compactifications require further investigation. We would like to conclude with some open questions and directions for future research.

Firstly, it is worth trying to construct a consistent  $\mathbb{Z}_6$ asymmetric orbifold preserving $\mathcal{N}=6$ supersymmetry in four of five dimensions. If this is impossible, then we should understand why such a model is in the swampland. Secondly, it would be interesting to give a physical interpretation for the appearance of only an odd number of vector multiplets in $\mathcal{N}=4$, $D=5$ theories. Then, this observation could be possibly promoted to a concrete conjecture on the full string landscape with sixteen supersymmetries.

Moreover, we saw that, at the level of supergravity, the effective potential for non-supersymmetric models was negative definite; this result was derived by computing various supertraces. It would be interesting to see how this calculation extends to the full string theory spectrum. For this, a computation of the full partition function, which is generating the cosmological constant, is needed. For specific orbifolds, a vanishing cosmological constant can be still obtained at one-loop, see e.g. \cite{kachru1999vacuum,kachru1998self,harvey1998string,shiu1999bose,angelantonj1999non,Satoh:2015nlc,groot2020tension}, but it is not certain if the cosmological constant vanishes at higher loops. A closely related issue in non-supersymmetric models is to understand if they are perturbatively stable, even if the radius of the circle is large enough, so that tachyons do not appear. 

Regarding $\mathcal{N}=2$, $D=5$ orbifolds, we predicted, using arguments based on duality, that no Chern-Simons terms in the Kaluza-Klein vector are induced by quantum corrections. It would be interesting to verify this by an explicit string theory calculation. Such a calculation could involve integrating out non-perturbative states, as our prediction relies on U-duality. In fact, one can verify that integrating out only low-mass charged states, as in \cite{Hull:2020byc}, does not lead to vanishing Chern-Simons terms in general.

Furthermore, there has been some work with similar mechanisms leading to the absence of quantum corrections \cite{Palti:2020qlc} in (special loci of) the moduli space in four-dimensional theories. While there are some differences with our examples, the main claim of \cite{Palti:2020qlc}, that the absence of quantum corrections beyond those expected from the unbroken supersymmetry must be related to a higher supersymmetric theory, holds for our models too. The precise connection to \cite{Palti:2020qlc} remains to be understood.

Concerning $\mathcal{N}=2$, $D=4$ theories, we found that generically massive, charged hypermultiplets can become massless at special lines and points in the interior of the moduli space. The appearance of massless states in the interior of the moduli space has significant consequences for the structure of the moduli space at the quantum level. As it was shown in \cite{deWit:1995dmj, Antoniadis:1995ct}, the one-loop prepotential exhibits logarithmic singularities exactly at the lines in the moduli space where generically massive, charged states become massless. Hence, the classical moduli space is modified by quantum corrections. The computation of such corrections in our model relies on modularity properties of the prepotential under the congruence subgroups of $\text{SL}(2;\mathbb{Z})$, and it would be very interesting to perform this computation. 

Finally, it would be nice to construct boundary states for D-branes, or bound states of branes, in the presence of a non-trivial background $B$-field and determine if they are BPS. Then, we could also study the fate of e.g. the D1-D5 brane system in the various orbifold theories and its applications to black holes and dual CFT.

Let us now conclude with some general thoughts on the future of the field of orbifolds and, more broadly, string theory. Certainly, the ultimate goal of string theory is to describe and unify all aspects of the universe. A first step in that direction would be the construction of a string-theoretic model that yields a stable non-supersymmetric vacuum with a positive cosmological constant, and no moduli at low energies.

In the class of orbifolds studied in this thesis, we have illustrated that tachyon-free non-supersymmetric stringy vacua can be constructed. However, there are regions in the moduli space in which the tachyon reappears in the spectrum; this presumably indicates instability of the model. Moreover, we have seen that our models fail to generate a positive cosmological constant (at the classical and one-loop level), and there are at least two free moduli associated with the string coupling and the radius of the orbifold circle.

Nevertheless, recent research (see for example \cite{Baykara:2024tjr,Angelantonj:2024jtu}) has shown that stable, tachyon-free non-supersymmetric  vacua can be realised in string theory compactifications on asymmetric orbifolds, at the classical level. In addition, the one-loop contribution to the vacuum energy was found to be positive. These attributes make these models very attractive, and we would like to encourage further research on such orbifold models.

Now, we wish to emphasise that the dilaton, which is the scalar field associated with the string coupling, is always a free modulus in all the aforementioned orbifold models. This is unavoidable in perturbative string theory, and finding a mechanism to stabilise the dilaton is still an open and challenging problem in string theory.

Another crucial question is if and how non-perturbative effects could alter the results obtained in perturbation theory. Some of the non-perturbative effects could be understood by using dualities and studying the D-brane spectrum of string-theoretic models. In this thesis, we have made some progress in understanding dualities and D-branes in the context of (asymmetric) orbifolds. We believe that research on non-perturbative aspects of string theory should be pursued, and a solid step towards this requires further study on string dualities and D-branes.

Finally, significant work on the swampland program is still in progress, and we think that useful insights on the swampland conjectures can be gained by studying (asymmetric) orbifolds. Arguably, it would be very exciting to prove some, or all of the swamplands conjectures. Then, we could confidently say that we fully comprehend certain aspects of quantum gravity.

\section{Samenvatting}

In dit proefschrift, hebben we verschillende aspecten van type IIB-snaartheorie bestudeerd, gecompacteerd op vrij werkende orbifolds. We hebben ons met name gericht op asymmetrische orbifolds, voorbeelden van niet-geometrische snaarcompactificaties en een intrigerende hoek in het snaarlandschap.

In hoofdstuk \ref{chap:fao}, hebben we de algemene constructie van vrij werkende (asymmetrische) orbifolds besproken, met de nadruk op de gesloten snaarsector. We beschreven de orbifolds in termen van dualiteitstwists, we bespraken de verschillende mogelijke consistente orbifolds en we analyseerden gedetailleerd de kwantiseringscondities voor de massaparameters. Daarna bespraken we de roosterconstructie; een complementaire benadering van de constructie van orbifolds, gebaseerd op even-zelf duale Lie-algebra Narain-roosters. Daarna presenteerden we de orbifoldwerking op de wereld-bladvelden en de algemene procedure voor het construeren van modulaire invariante orbifold-partitiefuncties.

In hoofdstuk \ref{chap:spectrum}, presenteerden we expliciete voorbeelden van orbifolds die de supersymmetrie van $\mathcal{N}=6,4,2$ of $0$ in vijf dimensies behouden, en construeerden we de lichtste gesloten snaartoestanden in de ongetwiste en getwiste orbifoldsectoren. Vervolgens demonstreerden we het verband tussen vrij werkende orbifolds en Scherk-Schwarz-reducties door de lichtste ongetwiste orbifoldtoestanden te vergelijken met die welke voortkomen uit de effectieve superzwaartekrachttheorie.

Verder zagen we dat alle snaartoestanden die afkomstig zijn van de getordeerde sectoren massief zijn op generieke punten in de moduliruimte. We toonden echter aan dat bij speciale waarden van de straal van de cirkel, generiek massieve getordeerde toestanden massaloos kunnen worden. In het bijzonder geldt voor niet-supersymmetrische orbifolds een kritische straal waarboven het snaarspectrum tachyonvrij is. Bovendien berekenden we voor theorieën die alle supersymmetrieën verbreken de effectieve potentiaal van één lus, die op superzwaartekrachtniveau negatief definieër bleek te zijn.

In hoofdstuk \ref{chap:mod5}, hebben we onze aandacht gericht op de orbifold-moduliruimte. In de klasse van orbifolds die in dit proefschrift worden bestudeerd, zagen we dat er slechts drie consistente asymmetrische orbifolds bestaan, van rang $2, 3$ en $4$, die de supersymmetrie $\mathcal{N}=6$ in vijf dimensies behouden, met moduliruimte ${\text{SU}^*(6)}/{\text{Sp}(3)}$. Voor $\mathcal{N}=4$-theorieën ontdekten we dat de moduliruimte volledig bepaald wordt door het aantal $n$ vectormultipletten en wordt gegeven door ${\text{SO}(5,n)}/[{\text{SO}(5)\times \text{SO}(n)}]$. Bovendien leverden al onze orbifoldvoorbeelden een oneven aantal vectormultipletten op; dit lijkt een gemeenschappelijk kenmerk te zijn voor een zeer brede klasse van $\mathcal{N}=4$ Minkowski-vacuüms in vijf dimensies.

Wat betreft theorieën die $\mathcal{N}=2$-supersymmetrie behouden, hebben we de klassieke hypermultiplet- en vectormultipletmoduliruimten bepaald door de dualiteitsgroep van de orbifold te vinden, d.w.z. de deelgroep
van de U-dualiteitsgroep van $T^5$ die commuteert met de orbifoldwerking. Vervolgens, door duale orbifoldparen te construeren, betoogden we dat er geen kwantumcorrecties op de metriek op de vectormultipletmoduliruimte optreden. Bovendien presenteerden we modellen zonder hypermultiplets en reproduceerden we enkele van de magisch-vierkante supergravitaties met behulp van vrij werkende asymmetrisch orbifolds.

In hoofdstuk \ref{chap:swampland}, hebben we aspecten van het swampland-programma besproken in de context van vrij werkende asymmetrische orbifolds van $\mathbb{R}^{1,3}\times T^2 \times T^4$. We hebben aangetoond dat, dankzij de orbifoldverschuiving, de dualiteitsgroep van $T^2$ gereduceerd werd tot congruentiedeelgroepen van de modulaire groep en dat het fundamentele domein nieuwe interessante punten bevatte op oneindige afstand op de reële as die het bovenste halfvlak begrenst. We hebben geverifieerd dat de moerasafstandsconjectuur geldig is in de niet-geometrische compactificaties van de snaartheorie die in dit proefschrift worden bestudeerd.

Bovendien konden op punten met oneindige afstand in de moduliruimte sommige (of alle) gravitini massaloos worden, wat duidt op een verbetering van de supersymmetrie. Ook decompacteerde de orbifoldtheorie op deze punten tot een andere orbifoldconstructie. Opmerkelijk genoeg zagen we op punten op de reële as van het bovenste halfvlak dat een asymmetrische, vrij werkende orbifold kon decompacteren tot een niet-vrij werkende symmetrische orbifold. Verder ontdekten we op punten met oneindige afstand in de moduliruimte dat oneindige torens van toestanden massaloos werden. Aan de andere kant ontdekten we op speciale punten in het binnenste van de moduliruimte dat slechts een eindig aantal toestanden massaloos werd.

Ten slotte hebben we ons in hoofdstuk \ref{chap:branes} gericht op het D-braanspectrum van vrij werkende orbifolds. Met behulp van twee complementaire benaderingen, namelijk het randtoestandsformalisme en het spectrum van massaloze Scherk-Schwarz-ijkvelden, bepaalden we onder welke omstandigheden de verschillende D-branen de orbifolding overleven. Bovendien hebben we aangetoond dat, hoewel een enkel D-braan mogelijk niet overleeft in een orbifold, een geschikte combinatie van branen, oftewel een gebonden toestand van branen, orbifold-invariant kan zijn. Vervolgens hebben we onderzocht of de supersymmetrie die behouden blijft door verschillende braanconfiguraties compatibel is met de supersymmetrie die behouden blijft door de orbifolds. We hebben met name vastgesteld dat er BPS- en stabiele D-braansystemen kunnen bestaan, evenals niet-BPS-, stabiele D-braansystemen.

\clearpage
\thispagestyle{empty}



\appendix

\chapter{Group theory}\label{app: group theory}
\lhead[\textit{Group theory}]{}
\rhead[]{\textit{Group theory}}

Here, we collect some (known) group theoretical information and some conventions that are relevant for the main text.

\subsection*{Two frames for SO(\textit{d,d})}\label{app: eta and tau frame}

Canonically, groups of the form SO$(d,d)$ consist of $2d$-dimensional matrices $A$ that satisfy the relation
\begin{equation}\label{etaframe}
    A\,\eta\,A^T = \eta \,,
\end{equation}
where $\eta$ is the indefinite metric
\begin{equation}
    \eta = \begin{pmatrix}
    1 & 0 \\
    0 & -1
    \end{pmatrix} \,.
\end{equation}
Here and everywhere else in this subsection we adopt the notation that $1$ and $0$ denote $d$-dimensional blocks that make up a $2d$-dimensional matrix.

There is a different way of constructing SO$(d,d)$, namely as the group of matrices $\tilde{A}$ satisfying
\begin{equation}\label{tauframe}
    \tilde{A}\,\tau\,\tilde{A}^T = \tau \,,
\end{equation}
where
\begin{equation}
    \tau = \begin{pmatrix}
    0 & 1 \\
    1 & 0
    \end{pmatrix} \,.
\end{equation}
It is straightforward to see that these two ways of constructing SO$(d,d)$ are equivalent, or, one might say, that the groups consisting of the matrices $A$ and $\tilde{A}$ are isomorphic. For example, the map
\begin{equation}\label{conjugation by X}
    A \;\;\rightarrow\;\; \tilde{A} = X\,A\,X^{-1} \,,
\end{equation}
where
\begin{equation}
    X = \frac{1}{\sqrt{2}}\,\begin{pmatrix}
    1 & 1 \\
    1 & -1
    \end{pmatrix} \,,
\end{equation}
takes matrices satisfying \eqref{etaframe} to matrices satisfying \eqref{tauframe}. It can be checked that this map is a proper isomorphism between what we call \say{SO$(d,d)$ in $\eta$-frame} and \say{SO$(d,d)$ in $\tau$-frame}.

We encounter both $\eta$ and $\tau$ frame in this thesis, and we will refer to them by these names.

\subsection*{The isomorphism $\mathfrak{so}(4)\cong\mathfrak{su}(2)\oplus\mathfrak{su}(2)$}

We first discuss the isomorphism between compact Lie algebras before moving on to the non-compact version. The algebra $\mathfrak{so}(4)$ is spanned by six $4d$ anti-symmetric matrices. We denote the generators of this algebra by $M_{ij}$ where the $i,j$ denote which entries in the matrix are non-zero. We have
\begin{equation}
    M_{12} = \begin{pmatrix}
    0 & -1 & 0 & 0 \\
    1 & 0 & 0 & 0 \\
    0 & 0 & 0 & 0 \\
    0 & 0 & 0 & 0
    \end{pmatrix} \;,\qquad
    M_{13} = \begin{pmatrix}
    0 & 0 & -1 & 0 \\
    0 & 0 & 0 & 0 \\
    1 & 0 & 0 & 0 \\
    0 & 0 & 0 & 0
    \end{pmatrix} \;,\qquad \ldots
\end{equation}
Similarly, we construct $M_{14},M_{23},M_{24},M_{34}$. We can compactly write the commutators between these generators as
\begin{equation}\label{commutation relations so4}
    [M_{ij},M_{kl}] = \delta_{ik} M_{jl} - \delta_{il} M_{jk} - \delta_{jk} M_{il} + \delta_{jl} M_{ik} \,,
\end{equation}
where it is understood that $M_{ii}=0$ and $M_{ij}=-M_{ji}$.

The algebra $\mathfrak{su}(2)$ is spanned by three $2d$ anti-Hermitian traceless matrices. We choose the generators
\begin{equation}\label{generators su2}
    N_1 = \frac{1}{2}\begin{pmatrix}
    0 & -1 \\
    1 & 0
    \end{pmatrix} \;, \qquad
    N_2 = \frac{i}{2}\begin{pmatrix}
    0 & 1 \\
    1 & 0
    \end{pmatrix} \;, \qquad
    N_3 = \frac{i}{2}\begin{pmatrix}
    1 & 0 \\
    0 & -1
    \end{pmatrix} \;, \qquad
\end{equation}
for this algebra. This yields the commutator
\begin{equation}\label{commutation relations su2}
    [N_I,N_J] = - \,\varepsilon_{IJK} \,N_K \;,
\end{equation}
where we choose $\varepsilon_{IJK}$ such that $\varepsilon_{123} = 1$.

We have now established sufficient notation to write down the isomorphism $\mathfrak{so}(4)\cong\mathfrak{su}(2)\oplus\mathfrak{su}(2)$. We choose this to be the map
\begin{equation}\label{compact isomorphism}
\begin{aligned}
    M_{12} \;\;&\rightarrow\;\; \big(N_1,\,N_1\big)\,, \\
    M_{13} \;\;&\rightarrow\;\; \big(N_3,\,N_3\big)\,, \\
    M_{14} \;\;&\rightarrow\;\; \big(N_2,\,-N_2\big)\,, \\
    M_{23} \;\;&\rightarrow\;\; \big(N_2,\,N_2\big) \,,\\
    M_{24} \;\;&\rightarrow\;\; \big(\!-\!N_3,\,N_3\big)\,, \\
    M_{34} \;\;&\rightarrow\;\; \big(N_1,\,-N_1\big) \,,
\end{aligned}
\end{equation}
which can readily be checked to preserve the commutator, making it a proper isomorphism.

\subsection*{The isomorphism $\mathfrak{so}(2,2)\cong\mathfrak{sl}(2)\oplus\mathfrak{sl}(2)$}\label{isomorphism so22}

Here, we repeat the previous analysis for the maximally non-compact version of the same isomorphism. On the $\mathfrak{so}(2,2)$ side, this means that we take four of the six generators to be symmetric rather than anti-symmetric. We choose the following generators to span the algebra\footnote{Note that these generators span the algebra $\mathfrak{so}(2,2)$ in $\eta$-frame.}:
\begin{equation}
\begin{alignedat}{6}
    \tilde{M}_{12} &= \begin{pmatrix}
    0 & -1 & 0 & 0 \\
    1 & 0 & 0 & 0 \\
    0 & 0 & 0 & 0 \\
    0 & 0 & 0 & 0
    \end{pmatrix} \;,\qquad
    &\tilde{M}_{13} &= \begin{pmatrix}
    0 & 0 & 1 & 0 \\
    0 & 0 & 0 & 0 \\
    1 & 0 & 0 & 0 \\
    0 & 0 & 0 & 0
    \end{pmatrix} \;,\qquad
    &\tilde{M}_{14} &= \begin{pmatrix}
    0 & 0 & 0 & 1 \\
    0 & 0 & 0 & 0 \\
    0 & 0 & 0 & 0 \\
    1 & 0 & 0 & 0
    \end{pmatrix} \;, \\
    \tilde{M}_{23} &= \begin{pmatrix}
    0 & 0 & 0 & 0 \\
    0 & 0 & 1 & 0 \\
    0 & 1 & 0 & 0 \\
    0 & 0 & 0 & 0
    \end{pmatrix} \;,\qquad
    &\tilde{M}_{24} &= \begin{pmatrix}
    0 & 0 & 0 & 0 \\
    0 & 0 & 0 & 1 \\
    0 & 0 & 0 & 0 \\
    0 & 1 & 0 & 0
    \end{pmatrix} \;,\qquad
    &\tilde{M}_{34} &= \begin{pmatrix}
    0 & 0 & 0 & 0 \\
    0 & 0 & 0 & 0 \\
    0 & 0 & 0 & -1 \\
    0 & 0 & 1 & 0
    \end{pmatrix} \;.
\end{alignedat}
\end{equation}
The algebra $\mathfrak{sl}(2)$ is spanned by traceless real matrices. The generators that we choose are equal to the ones we chose for the $\mathfrak{su}(2)$ algebra \eqref{generators su2}, but without the factors of $i$ in front. We define our three generators as
\begin{equation}
    \tilde{N}_1 = \frac{1}{2}\begin{pmatrix}
    0 & -1 \\
    1 & 0
    \end{pmatrix} \;, \qquad
    \tilde{N}_2 = \frac{1}{2}\begin{pmatrix}
    0 & 1 \\
    1 & 0
    \end{pmatrix} \;, \qquad
    \tilde{N}_3 = \frac{1}{2}\begin{pmatrix}
    1 & 0 \\
    0 & -1
    \end{pmatrix} \;. \qquad
\end{equation}
These algebras satisfy similar commutation relations as their compact counterparts, \eqref{commutation relations so4} and \eqref{commutation relations su2}, with some signs changed. In fact, the signs are changed in such a way that the isomorphism $\mathfrak{so}(2,2)\cong\mathfrak{sl}(2)\oplus\mathfrak{sl}(2)$ can be written down in the exact same way as in \eqref{compact isomorphism}. We simply add the tildes to indicate that we consider the maximally non-compact generators:
\begin{equation}\label{noncompact isomorphism}
\begin{aligned}
    \tilde{M}_{12} \;\;&\rightarrow\;\; \big(\tilde{N}_1,\,\tilde{N}_1\big)\,, \\
    \tilde{M}_{13} \;\;&\rightarrow\;\; \big(\tilde{N}_3,\,\tilde{N}_3\big) \,,\\
    \tilde{M}_{14} \;\;&\rightarrow\;\; \big(\tilde{N}_2,\,-\tilde{N}_2\big) \,,\\
    \tilde{M}_{23} \;\;&\rightarrow\;\; \big(\tilde{N}_2,\,\tilde{N}_2\big)\,, \\
    \tilde{M}_{24} \;\;&\rightarrow\;\; \big(\!-\!\tilde{N}_3,\,\tilde{N}_3\big)\,, \\
    \tilde{M}_{34} \;\;&\rightarrow\;\; \big(\tilde{N}_1,\,-\tilde{N}_1\big) \,.
\end{aligned}
\end{equation}
Again, it is straightforward to check that this map preserves the commutator.

\subsection*{Embedding rotations in Spin(4,4)}
\label{Ap:embedding}

One of the main motivations for adding in the previous discussion was to be able to properly embed rotations in various subgroups of $\spinfourfour$ into an $8d$ matrix. Let us first consider how rotations map through the two isomorphisms that we just discussed. We start with rotations in
\begin{equation}
    \sotwo\times\sotwo\subset\sutwo\times\sutwo\cong\spinfour \,.
\end{equation}
If we take a rotation over an angle $m_1$ in the first $\sotwo$ and a rotation over an angle $m_3$ in the second $\sotwo$, we can use (the inverse of) the isomorphism \eqref{compact isomorphism} to map this to an $\sofour\subset\spinfour$ matrix:
\begin{equation}
\begin{aligned}
    \left(
    \begin{pmatrix}
    \cos m_1 & -\sin m_1 \\
    \sin m_1 & \cos m_1
    \end{pmatrix}\;,\;
    \begin{pmatrix}
    \cos m_3 & -\sin m_3 \\
    \sin m_3 & \cos m_3
    \end{pmatrix}
    \right) \;&=\; \left( e^{2m_1 N_1} , e^{2m_3 N_1} \right) \overset{\text{isomorphism}}{\longrightarrow} \\
     e^{(m_1+m_3)M_{12}+(m_1-m_3)M_{34}} &=
    \begin{pmatrix}
     R(m_1+m_3) & 0\\
     0&R(m_1-m_3)
    \end{pmatrix} \,.
    \end{aligned}
    \label{embedding SO4}
\end{equation}
Here and below we use the shorthand notation $R(x)=\begin{psmallmatrix}\cos x & \,\,\,\,-\sin x \\ \sin x & \,\,\,\,\cos x \end{psmallmatrix}$ for a two by two rotation matrix. 

We can repeat this for the maximally non-compact isomorphism, mapping rotations in the subgroup
\begin{equation}
\sotwo\times\sotwo\subset\sltwo\times
\sltwo\cong\spintwotwo \,,
\end{equation}
to an $\sotwotwo\subset\spintwotwo$ matrix. We find
\begin{equation}
\begin{aligned}
    \left(
    \begin{pmatrix}
    \cos \alpha_1 & -\sin \alpha_1 \\
    \sin \alpha_1 & \cos \alpha_1
    \end{pmatrix} \;,\;
    \begin{pmatrix}
    \cos \alpha_3 & -\sin \alpha_3 \\
    \sin \alpha_3 & \cos \alpha_3
    \end{pmatrix}
    \right) \;&=\; \left( e^{2\alpha_1 \tilde{N}_1} , e^{2\alpha_3 \tilde{N}_1} \right) \overset{\text{isomorphism}}{\longrightarrow} \\
    e^{(\alpha_1+\alpha_3)\tilde{M}_{12}+(\alpha_1-\alpha_3)\tilde{M}_{34}}& = 
   \begin{pmatrix}
     R(\alpha_1+\alpha_3) & 0\\
     0& R(\alpha_1-\alpha_3)
    \end{pmatrix} \,.
    \end{aligned}
    \label{embedding SO22}
\end{equation}
The almost identical structure that we find through the compact and maximally non-compact isomorphisms should not be surprising, as the generators that appear are equal, e.g. $M_{12}=\tilde{M}_{12}$ and $N_1=\tilde{N}_1$. It is the other generators (the non-compact ones) that differ between these algebras.

Now we are ready to embed the four rotation parameters that are relevant in the main text (see section \ref{sec:Orbifold constructions}) in an $\sofourfour\subset\spinfourfour$ matrix.

We consider both the parameters $m_1,\ldots,m_4$ that rotate in the subgroup
\begin{equation}
    \sotwo^4\subset\sutwo^4\cong\spinfour^2\;,
\end{equation}
and the parameters $\alpha_1,\ldots,\alpha_4$ that rotate in the subgroup
\begin{equation}
    \sotwo^4\subset\sltwo^4\cong\spintwotwo^2 \;.
\end{equation}
Both are essentially a repetition of the embeddings shown in \eqref{embedding SO4} and \eqref{embedding SO22}.  The SO$(4)$ matrices can be embedded into SO$(4,4)$ in a block-diagonal way
\begin{equation}
    \begin{pmatrix}
    R(m_1+m_3) & 0 & 0 & 0 \\
    0 & R(m_1-m_3) & 0 & 0 \\
    0 & 0 & R(m_2+m_4) & 0 \\
    0 & 0 & 0 & R(m_2-m_4)
    \end{pmatrix} \in \sofourfour \;,
\end{equation}
while for the SO$(2,2)$ matrices, the embedding into SO$(4,4)$ is given by
\begin{equation}
    \begin{pmatrix}
    R(\alpha_1+\alpha_3) & 0 & 0 & 0 \\
    0 & R(\alpha_2+\alpha_4) & 0 & 0 \\
    0 & 0 & R(\alpha_1-\alpha_3) & 0 \\
    0 & 0 & 0 & R(\alpha_2-\alpha_4)
    \end{pmatrix} \in \sofourfour \;.
\end{equation}
From the way that the two sets of rotation parameters are embedded in $\sofourfour$ we can deduce the relation between them. We find
\begin{equation}
\begin{aligned}
m_1&=\tfrac{1}{2}(\alpha_1+\alpha_2+\alpha_3+\alpha_4) \,,\qquad\quad
&m_2=\tfrac{1}{2}(\alpha_1+\alpha_2-\alpha_3-\alpha_4) \,,\\[4pt]
m_3&=\tfrac{1}{2}(\alpha_1-\alpha_2+\alpha_3-\alpha_4) \,,\qquad\quad
&m_4=\tfrac{1}{2}(\alpha_1-\alpha_2-\alpha_3+\alpha_4) \,.
\end{aligned}
\end{equation}
We use these relations in section \ref{sec:Orbifold constructions} to determine the allowed values for the $m$'s in terms of the allowed values for the $\alpha$'s.

Note that all $\sotwotwo$ and $\sofourfour$ matrices in this subsection are written down in $\eta$-frame. In order to obtain the relevant matrices in $\tau$-frame (which is the frame in which the monodromies are required to be integer-valued), one would have to perform a conjugation \`a la \eqref{conjugation by X}.

\chapter{Modular functions and transformations}
\label{ap B}
\lhead[\textit{Modular functions and transformations}]{}
\rhead[]{\textit{Modular functions and transformations}}

\section*{Conventions and identities}
The Dedekind $\eta$-function is defined as
\begin{equation}
   \eta(\tau)\equiv q^{\frac{1}{24}} \prod_{n=1}^{\infty}(1-q^n)\,,\hspace{0.5cm}q=e^{2\pi i \tau}\,.
\end{equation}
The Jacobi $\vartheta$-function with characteristics $\alpha,\beta$ is given by
\begin{equation}
     \vartheta[\psymbol{ \alpha}{ \beta}](\tau)=\sum_{n\in \mathbb{Z}}q^{\frac{1}{2}(n+\alpha)^2}e^{2\pi i(n+\alpha)\beta}\,.
     \label{14}
\end{equation}
For $-\tfrac{1}{2}\leq \alpha, \beta \leq \tfrac{1}{2}$ there is also a product representation of the $\vartheta$-function, which reads 
\begin{equation}
    \vartheta[\psymbol{ \alpha}{ \beta}](\tau)=\eta(\tau)e^{2\pi i\alpha\beta}q^{\frac{1}{2}\alpha^2-\frac{1}{24}}\prod_{n=1}^{\infty}(1+q^{n+\alpha-\frac{1}{2}}e^{2\pi i\beta})(1+q^{n-\alpha-\frac{1}{2}}e^{-2\pi i\beta})\,.
    \label{product representation}
\end{equation}
Particular $\vartheta$-functions that appear often are 
\begin{equation}
    \vartheta[\psymbol{ 0}{ 0}](\tau)\equiv\vartheta_3(\tau)\,,\hspace{0,2cm} \vartheta[\psymbol{ 0}{ \frac{1}{2}}](\tau)\equiv\vartheta_4(\tau)\,,\hspace{0,2cm} \vartheta[\psymbol{ \frac{1}{2}}{ 0}](\tau)\equiv\vartheta_2(\tau)\,,\hspace{0,2cm} \vartheta[\psymbol{ \frac{1}{2}}{ \frac{1}{2}}](\tau)\equiv\vartheta_1(\tau)\,.
\end{equation}
In addition, two useful $\vartheta$-function identities are\footnote{Various $\vartheta$-function identities can be found in \cite{gannon1992lattices,gannon1992lattices2}.}
\begin{equation}
    \vartheta[\psymbol{ -\alpha}{ -\beta}](\tau)=  \vartheta[\psymbol{ \alpha}{ \beta}](\tau)\,,\hspace{0,3 cm}  \vartheta[\psymbol{ \alpha+m}{ \beta+n}](\tau)= e^{2\pi i n \alpha}\,\vartheta[\psymbol{ \alpha}{ \beta}](\tau)\,,\,\,m,n \in \mathbb{Z}\,.
\end{equation}
Finally, the $d$-dimensional Poisson resummation formula is given by
\begin{equation}
     \sum_{n_i \in \mathbb{Z}^d}e^{-\pi  n_iA_{ij}n_j+\pi B_i n_i}=(\text{det}A)^{-\frac{1}{2}} \sum_{n_i \in \mathbb{Z}^d}e^{-{\pi}\left(n_i+i\frac{B_i}{2}\right)(A^{-1}){ij}\left(n_j+i\frac{B_j}{2}\right)}\,.
     \label{Poissonresum}
\end{equation}
The modular transformations are defined as: $\mathcal{T}\equiv \tau\to \tau+1$ and $\mathcal{S}\equiv \tau \to -{1}/{\tau}$. The Dedekind $\eta$-function transforms as follows 
\begin{equation}
\begin{aligned}
    &\eta(\tau+1)= e^{{\pi i}/{12}}\,\eta(\tau)\,,\\
   & \eta(-{1}/{\tau}) = \sqrt{-i\tau}\,\eta(\tau)\,.
    \end{aligned}
\end{equation}
Note that under both $\mathcal{T}$ and $\mathcal{S}$ transformations the combination $\sqrt{\tau_2}\,\eta(\tau)\,\bar{\eta}(\bar{\tau})$ is invariant. Under modular transformations, the Jacobi $\theta$-function transforms as follows
\begin{equation}
    \begin{aligned}
      & \vartheta[\psymbol{ \alpha}{ \beta}] (\tau+1)=e^{-\pi i(\alpha^2-\alpha)} \, \vartheta[\psymbol{ \alpha}{\alpha+\beta-\frac{1}{2}}](\tau)\,,\\
       & \vartheta[\psymbol{ \alpha}{ \beta}] (-1/\tau) = \sqrt{-i\tau}\,  e^{2\pi i \alpha \beta}\, \vartheta[\psymbol{ -\beta}{ \alpha}](\tau)\,.
    \end{aligned}
\end{equation}
The theta series of the SO(2n) root lattice $D_n$ is
\begin{equation}
    \Theta_{D_n}(\tau)= \frac{1}{2}\left(\vartheta_3(\tau)^n+\vartheta_4(\tau)^n\right)\,.
\end{equation}
For the SU(3) root lattice $A_2$ and its dual $A_2^*$ we have
\begin{equation}
\begin{aligned}
 \Theta_{A_2}(\tau)= \vartheta_3(2\tau)\vartheta_3(6\tau) + \vartheta_2(2\tau)\vartheta_2(6\tau)\,,\\
 \Theta_{A_2^*}(\tau)= \vartheta_3(\tfrac{2\tau}{3})\vartheta_3(2\tau) + \vartheta_2(\tfrac{2\tau}{3})\vartheta_2(2\tau)\,.
 \end{aligned}
\end{equation}
Finally, for a $d$-dimensional lattice $\Lambda$ and its dual $\Lambda^*$ the following expression holds
\begin{equation}
    \Theta_{\Lambda}({-1}/{\tau})= \frac{(-i\tau)^{\frac{d}{2}}}{\text{Vol}(\Lambda)} \Theta_{\Lambda^*}({\tau})\,.
\end{equation}

\chapter{Supergravity multiplets in $6D$}
\label{6dsugra}
\lhead[\textit{Supergravity multiplets in $6D$}]{}
\rhead[]{\textit{Supergravity multiplets in $6D$}}

In this appendix, we discuss the various supergravity fields and supergravity multiplets in $6D$. First of all, we list in \autoref{tablemasslessstatesapp} the weight vectors of the lightest left and right-moving states in the untwisted orbifold sector, and their representations under the massless little group $\text{SU}(2)\times \text{SU}(2)$ in $6D$.  In addition, we table the massless representations that correspond to the various supergravity fields in six dimensions in \autoref{table6Dfieldreps}.
\newline

\renewcommand{\arraystretch}{2}
\begin{table}[h!]
\centering
 \begin{tabular}{|c|c|c|}
    \hline
    Sector &  $\tilde{r}, {r}$   & $\text{SU}(2)\times \text{SU}(2)$ rep\\
    \hline
    \hline
   \multirow{3}{*}{NS}  & $(\underline{\pm 1,0},0,0)$  & $(\textbf{2},\textbf{2})$\\
    \cline{2-3}
    &$(0,0,\pm 1,0)$&  2$\,\times\,(\textbf{1},\textbf{1})$\\
    \cline{2-3}
   & $(0,0,0,\pm 1)$ & 2$\,\times\,(\textbf{1},\textbf{1})$\\
    \hline
    \hline
    \multirow{4}{*}{R}  & $(\pm\frac{1}{2},\pm\frac{1}{2},\frac{1}{2},\frac{1}{2})$ & $(\textbf{2},\textbf{1})$\\
    \cline{2-3}
    & $(\pm\frac{1}{2},\pm\frac{1}{2},-\frac{1}{2},-\frac{1}{2})$ & $(\textbf{2},\textbf{1})$\\
    \cline{2-3}
    & $(\underline{\frac{1}{2},-\frac{1}{2}},\frac{1}{2},-\frac{1}{2})$ & $(\textbf{1},\textbf{2})$\\
    \cline{2-3}
    & $(\underline{\frac{1}{2},-\frac{1}{2}},-\frac{1}{2},\frac{1}{2})$ & $(\textbf{1},\textbf{2})$\\
   \hline
    \end{tabular}
\captionsetup{width=.9\linewidth}
\caption{\textit{The weight vectors of the lightest left and right-moving states in the untwisted sector, and their representations under the massless little group $\text{SU}(2)\times \text{SU}(2)$ in $6D$. Underlying denotes permutation.}}
\label{tablemasslessstatesapp}
\end{table}
\renewcommand{\arraystretch}{1}

\renewcommand{\arraystretch}{1.2}
\begin{table}[h!]
\centering
\begin{tabular}{|c|c|}
\hline
\;\,Massive field\,\; & $\text{SU}(2)\times \text{SU}(2)$ rep \\ \hline\hline
$B_{\mu\nu}^+$ / $B_{\mu\nu}^-$ & $(\textbf{3},\textbf{1})$ / $(\textbf{1},\textbf{3})$ \\
$\psi_\mu^+$ / $\psi_\mu^-$ & \;\:$(\textbf{2},\textbf{3})$ / $(\textbf{3},\textbf{2})$\:\; \\
$A_\mu$ & $(\textbf{2},\textbf{2})$ \\
$\chi^+$ / $\chi^-$ & $(\textbf{1},\textbf{2})$ / $(\textbf{2},\textbf{1})$ \\
$\phi$ & $(\textbf{1},\textbf{1})$ \\ \hline
\end{tabular}
\captionsetup{width=.83\linewidth}
\caption{\textit{Here we show the various massless $6D$ supergravity fields and their representations under the massless little group.}}
\label{table6Dfieldreps}
\end{table}
\renewcommand{\arraystretch}{1}

\noindent Now, we can present the various supergravity multiplets.
\subsection*{$\mathcal{N}=8$ }
All massless fields fit in the gravity multiplet, in the representations
\begin{equation}
\begin{aligned}
&(\textbf{3},\textbf{3})\oplus4\times(\textbf{3},\textbf{2})\oplus4\times(\textbf{2},\textbf{3})\oplus 5\times(\textbf{3},\textbf{1})\oplus 5\times(\textbf{1},\textbf{3})\,\oplus\\
&16\times(\textbf{2},\textbf{2})\oplus20\times(\textbf{2},\textbf{1})\oplus20\times(\textbf{1},\textbf{2})\oplus25\times(\textbf{1},\textbf{1})\,.
\end{aligned}
\end{equation}

\subsection*{$\mathcal{N}=4$ $(0,2)$ }
There are two types of multiplets. The gravity multiplet
\begin{equation}
(\textbf{3},\textbf{3})\oplus4\times(\textbf{2},\textbf{3})\oplus 5\times(\textbf{1},\textbf{3})\,,
\end{equation}
and the tensor multiplet
\begin{equation}
(\textbf{3},\textbf{1})\oplus4\times(\textbf{2},\textbf{1})\oplus 5\times(\textbf{1},\textbf{1})\,.
\end{equation}

\subsection*{$\mathcal{N}=4$ $(1,1)$ }
Again, we have two types of multiplets. The gravity multiplet
\begin{equation}
(\textbf{3},\textbf{3})\oplus2\times(\textbf{3},\textbf{2})\oplus2\times(\textbf{2},\textbf{3})\oplus (\textbf{3},\textbf{1})\oplus (\textbf{1},\textbf{3})\oplus4\times(\textbf{2},\textbf{2})\oplus2\times(\textbf{2},\textbf{1})\oplus2\times(\textbf{1},\textbf{2})\oplus(\textbf{1},\textbf{1})\,,
\end{equation}
and the vector multiplet
\begin{equation} (\textbf{2},\textbf{2})\oplus2\times(\textbf{2},\textbf{1})\oplus2\times(\textbf{1},\textbf{2})\oplus 4\times(\textbf{1},\textbf{1})\,.
\end{equation}

\subsection*{$\mathcal{N}=2$ }
There exist four types of multiplets. The gravity multiplet
\begin{equation}
(\textbf{3},\textbf{3})\oplus2\times(\textbf{2},\textbf{3})\oplus (\textbf{1},\textbf{3})\,,
\end{equation}
the tensor multiplet
\begin{equation}
(\textbf{3},\textbf{1})\oplus2\times(\textbf{2},\textbf{1})\oplus (\textbf{1},\textbf{1})\,,
\end{equation}
the vector multiplet
\begin{equation} (\textbf{2},\textbf{2})\oplus2\times(\textbf{1},\textbf{2})\,.
\end{equation}
and the hypermultiplet
\begin{equation} 2\times(\textbf{2},\textbf{1})\oplus 4\times(\textbf{1},\textbf{1})\,.
\end{equation}

\addcontentsline{toc}{chapter}{Acknowledgments}
\section*{Acknowledgments}

\thispagestyle{plain}
\lhead[]{}
\rhead[]{\textit{}}

First of all, I would like to thank my advisor Stefan Vandoren. You trusted my abilities, offered me a PhD position, and gave me the opportunity to fulfil one of my dreams. Also, you taught me how to do research, read the literature, and write in a proper academic way. I really admire your ability to explain even the most difficult things in a very simple and clear way, your knowledge of many different topics in physics, and your persistence and determination in working and completing our projects. I would also like to thank you for your understanding and flexibility from the COVID times until this day. 

In addition, I would like to thank my copromotor Erik Plauschinn for all the interesting discussions we had. It is also a pleasure to acknowledge Thomas Grimm for all the useful conversations; you were also my string theory professor who motivated me to search for a master's thesis on this subject. Also, I would like to thank Umut G\"ursoy  who was my first professor in high energy physics and really inspired me to pursue a path on this filed of physics.

During my PhD I worked with a lot of bright people. Koen, as a supervisor assistant in my master's thesis you introduced me to the world of Scherk-Schwarz supergravity and freely acting orbifolds. Then, I was fortunate to work with you in the same office for a year, during which I learned a lot and had a great time. Guoen, together we investigated many aspects of the landscape in the context of orbifolds; you were the expert on the swampland conjectures, which you explained to me. Also, we learned some very interesting mathematics, we got some nice results but we didn't have the time to complete our project. Hopefully, one day I will see our results published! Last but not least, I am grateful to Chris Hull who gave me a lot of insight into the world of string dualities, and for all the helpful discussions we had during our meetings.

Except for my collaborators I was really lucky to have some great colleagues. Arno and Nico, we started together our PhDs and we had a fantastic time during the Solvay School. Shradha, Guim and Mick we were teaching assistants in master's thesis courses and we also enjoyed many coffee and lunch breaks. Artim, you were the first student I co-supervised in a master's thesis project and then, I was very happy to see you getting a PhD position. Also, I would like to thank everyone in the institute for creating a positive and friendly workplace, and for the very exciting borrels!

Also, I would like to thank some people from my master's program. Konstantinos Tsagkaris, you were the \say{professor}
of our master's program and you helped me more than anyone in my first steps in theoretical physics. I am sure that you will achieve your goals either in physics or in any other field. Konstantinos Manos, Christoforos Eseroglou 

\clearpage
\thispagestyle{plain}

\noindent and Petros Agridos, I enjoyed studying and working with you and I also enjoyed the time we spent outside of the institute. Finally, I am grateful to all my professors and teaching assistants of my master's program.

Outside of physics, I would like to thank my friends  Anastasia, Paris, Natasa, Nefeli, Nicky and Foteini for the great time we had in Utrecht. I enjoyed our dinners, drinks, board games, quiz nights and, above all, our unforgettable debates around food!

Also, I would like to thank my friends back in Athens, Mitso, Michali, Odyssea and Strato, for always being there for me, no matter how long I was away.

Finally, I would like to thank my brother for all the times I needed help, and he gave it without hesitation (there are also some pictures inside to look at, just like in the old times!).

\clearpage
\newpage
\thispagestyle{plain}

\addcontentsline{toc}{chapter}{About the author}
\section*{About the author}

Georgios Gkountoumis was born on January 22, 1993 in Athens, Greece. He studied physics at the National and Kapodistrian University of Athens and obtained his bachelor's degree in 2017. Then, he continued his studies in 2019 with a master's in theoretical physics at Utrecht University. His master's thesis focused on orbifold compactifications and supersymmetry breaking, and he was supervised by prof. dr. Stefan Vandoren. After obtaining his master's degree in 2021, he started his PhD  at the Institute for Theoretical Physics in Utrecht, under the supervision of prof. dr. Stefan Vandoren. His research focused on freely acting orbifolds of type IIB string theory. The results of his research are presented in this thesis.

\clearpage
\thispagestyle{empty}

\backmatter
\lhead[\textit{\nouppercase{\leftmark}}]{}
\rhead[]{\textit{\nouppercase{\leftmark}}}
\begin{singlespace}
\bibliographystyle{JHEP}
\bibliography{bib}
\end{singlespace}

\end{document}